\documentclass[znote,zbstdefault]{zeus_paper}
\usepackage[english]{babel}
\chardef\usc=95
\chardef\til=126
\catcode`\@=11 
\DeclareRobustCommand\xdotspace{\futurelet\@let@token\@xdotspace}
\def\@xdotspace{%
  \ifx\@let@token.\else
  \ifx\@let@token\bgroup.\else
  \ifx\@let@token\egroup.\else
  \ifx\@let@token\/.\else
  \ifx\@let@token\ .\else
  \ifx\@let@token~.\else
  \ifx\@let@token!.\else
  \ifx\@let@token,.\else
  \ifx\@let@token:.\else
  \ifx\@let@token;.\else
  \ifx\@let@token?.\else
  \ifx\@let@token/.\else
  \ifx\@let@token'.\else
  \ifx\@let@token).\else
  \ifx\@let@token-.\else
  \ifx\@let@token\@xobeysp.\else
  \ifx\@let@token\space.\else
  \ifx\@let@token\@sptoken.\else
   .\space
   \fi\fi\fi\fi\fi\fi\fi\fi\fi\fi\fi\fi\fi\fi\fi\fi\fi\fi}
\catcode`\@=12 

\newcommand{\stru}[2]{%
   \relax\ifmmode\hbox{\vrule height#1 depth#2 width0pt}%
   \else\vrule height#1 depth#2 width0pt\fi}

\newcommand{\Ronum}[1]{\uppercase\expandafter{\romannumeral#1}}
\newcommand{\ronum}[1]{\expandafter{\romannumeral#1}}
\DeclareRobustCommand{\LaTeXZ}{%
  \LaTeX\kern-.05em4\kern-.1em
  {\raisebox{-0.2ex}{$\scriptstyle\text{ZEUS}$}}\xspace}

\DeclareMathAlphabet{\mathbf}{OT1}{cmr}{bx}{sl}
\newcommand{\eVdist}{\kern-0.06667em}

\newcommand{\mb}{\,\text{mb}}

\newcommand{\nb}{\,\text{nb}}
\newcommand{\pb}{\,\text{pb}}

\newcommand{\slashfrac}[2]{%
  \raisebox{0.5ex}{\ensuremath #1}\kern-0.12em/\kern-0.08em
  \raisebox{-.8ex}{\ensuremath #2}}

\newcommand{\sqr}[3]{%
    {\vcenter{\hrule height.#3ex\hbox{\vrule width.#2ex height#1ex
     \kern#1ex\vrule width.#3ex}\hrule height.#2ex}}}

\newcommand{\widebar}[1]{%
   \mkern1.5mu\overline{\mkern-1.5mu#1\mkern-1.mu}\mkern1.mu}
\catcode`\@=11 
\newcommand{\parenbar}{\mathpalette\p@renb@r}
\def\p@renb@r#1#2{\vbox{%
  \ifx#1\scriptscriptstyle \dimen@.7em\dimen@ii.2em\else
  \ifx#1\scriptstyle \dimen@.8em\dimen@ii.25em\else
  \dimen@1em\dimen@ii.4em\fi\fi \offinterlineskip
  \ialign{\hfill##\hfill\cr
    \vbox{\hrule width\dimen@ii}\cr
    \noalign{\vskip-.3ex}%
    \hbox to\dimen@{$\mathchar300\hfil\mathchar301$}\cr
    \noalign{\vskip-.3ex}%
    $#1#2$\cr}}}
\catcode`\@=12 

\newcommand{\pbar}{\widebar{p}}

\newcommand{\qbar}{\widebar{q}}

\newcommand{\cbar}{\widebar{c}}

\newcommand{\IP}{{\rm I$\kern-0.01667em$P}\xspace}

\newcommand{\tot}{{\rm tot}}

\mathchardef\qsm=63
\mathchardef\pls=43
\mathchardef\mns=512
\mathchardef\plm=518
\mathchardef\eql=61
\mathchardef\smallleft=300
\mathchardef\smallright=301
\mathchardef\les=316
\mathchardef\gre=318
\mathchardef\leq=532
\mathchardef\grq=533

\catcode`\@=11 
\newcounter{pict@width}
\newcounter{pict@height}
\newlength{\pict@scale}
\newcommand{\psfigadd}[4]{%
\setcounter{pict@width}{1*\ratio{#2+\pict@scale/2}{\pict@scale}}
\setcounter{pict@height}{1*\ratio{#3+\pict@scale/2}{\pict@scale}}
\setlength{\unitlength}{\pict@scale}
\hbox to #2{\hspace{-\fill}\begin{picture}(\thepict@width,\thepict@height)
\put(0,0){\psfig{figure=#1,width=#2,height=#3,clip=}}
\SetScale{0.283466457}
\SetWidth{1.763889}
{#4}
\end{picture}}
}
\newcounter{pict@widthfst}
\newcounter{pict@widthscd}
\newcounter{pict@widthtot}
\newcommand{\psfigaddtwo}[7]{%
\setcounter{pict@widthfst}{1*\ratio{#2+\pict@scale/2}{\pict@scale}}
\setcounter{pict@widthscd}{1*\ratio{#2+#4+\pict@scale/2}{\pict@scale}}
\setcounter{pict@widthtot}{1*\ratio{#2+#4+#6+\pict@scale/2}{\pict@scale}}
\setcounter{pict@height}{1*\ratio{#3+\pict@scale/2}{\pict@scale}}
\setlength{\unitlength}{\pict@scale}
\hbox{\hspace{-\fill}\begin{picture}(\thepict@widthtot,\thepict@height)
\put(0,0){\psfig{figure=#1,width=#2,height=#3,clip=}}
\put(\thepict@widthscd,0){\psfig{figure=#5,width=#6,height=#3,clip=}}
\SetScale{0.283466457}
\SetWidth{1.763889}
{#7}
\end{picture}}
}
\newcommand{\psfigror}[4]{%
\setcounter{pict@width}{1*\ratio{#2+\pict@scale/2}{\pict@scale}}
\setcounter{pict@height}{1*\ratio{#3+\pict@scale/2}{\pict@scale}}
\setlength{\unitlength}{\pict@scale}
\hbox{\begin{picture}(\thepict@width,\thepict@height)
\put(0,\thepict@height){\psfig{figure=#1,width=#3,height=#2,clip=,angle=270}}
\SetScale{0.283466457}
\SetWidth{1.763889}
{#4}
\end{picture}}
}
\newcommand{\psfigrol}[4]{%
\setcounter{pict@width}{1*\ratio{#2+\pict@scale/2}{\pict@scale}}
\setcounter{pict@height}{1*\ratio{#3+\pict@scale/2}{\pict@scale}}
\setlength{\unitlength}{\pict@scale}
\hbox{\begin{picture}(\thepict@width,\thepict@height)
\put(0,0){\psfig{figure=#1,width=#3,height=#2,clip=,angle=90}}
\SetScale{0.283466457}
\SetWidth{1.763889}
{#4}
\end{picture}}
}
\catcode`\@=12 
\newlength\listtextwidth

\catcode`\@=11 
\newlength{\@tabfninsert}
\newlength{\@tabfnwidth}
\newcommand{\tabfootnote}[2]{%
  \setlength{\@tabfninsert}{0.8em}
  \setlength{\@tabfnwidth}{\textwidth}
  \addtolength{\@tabfnwidth}{-\@tabfninsert}
  \addtolength{\@tabfnwidth}{-0.4em}
  \noindent\makebox[\@tabfninsert][r]{\footnotesize$^{#1}$\hfil}\hfill%
  \parbox[t]{\@tabfnwidth}{\footnotesize #2\hfill}}
\catcode`\@=12 

\usepackage{lineno}
\begin{document}
\newcommand{\pythia}{{\sc Pythia}}
\newcommand{\herwig}{{\sc Herwig}}
\newcommand{\jimmy}{{\sc Jimmy}}
\newcommand{\etcut}{\mbox{$E_T^{\rm CUT}$}}
\newcommand{\etgap}{\mbox{$E_T^{\rm GAP}$}}
\newcommand{\xg}{\mbox{$x_\gamma^{\rm OBS}$}}
\newcommand{\xp}{\mbox{$x_p^{\rm OBS}$}}
\newcommand{\GeV}{\mbox{\rm ~GeV}}
\newcommand{\pom} {I\hspace{-0.2em}P}
\newcommand{\xpom}{\mbox{$x_{_{\pom}}$}}
\newcommand{\reg} {I\hspace{-0.2em}R}

\title{Review of High Energy Diffraction in Real and Virtual Photon Proton scattering at HERA}

\author{G. Wolf\\
Deutsches Elektronen-Synchrotron, DESY, Hamburg, Germany\\
}

\abstract
{The electron-proton collider HERA at DESY opened the door for the study of diffraction in real and virtual photon-proton scattering at center-of-mass energies $W$ up to 250 GeV and for large negative mass squared $-Q^2$ of the virtual photon up to $Q^2 = 1600$ GeV$^2$. At $W = 220$ GeV and $Q^2 = 4$ GeV$^2$, diffraction accounts for about 15\% of the total virtual photon proton cross section decreasing to $\approx 5$\% at $Q^2 = 200$ GeV$^2$. An overview of the results obtained by the experiments H1 and ZEUS on the production of neutral vectormesons and on inclusive diffraction up to the year 2008 is presented. 
\\
\\
\\
\\
\\
\\
\\
\\
This article is dedicated to Volker Soergel, Gustav A. Voss, to the memory of Bj{\"o}rn H.~Wiik, and to Antonino Zichichi, who steered HERA from a plan to reality.
}

\makezeustitle

\include{padifrop-aut}

\nonumber\pagenumbering{arabic}\pagestyle{plain}
\nonumber
\section{Introduction}
\label{sec-int}
Diffraction is a well known phenomenon in optics. Since the sixties, diffraction phenomena are also known from particle physics, for instance from proton-proton scattering. In hadron-hadron scattering these processes are phenomenologically described by the exchange of a virtual, colourless and flavourless neutral object carrying no quantum numbers, called the Pomeron. The Pomeron seemed to couple to quarks~\cite{landspolking,Jaroslans}.
The possibility that the Pomeron may have a partonic structure had been discussed~\cite{pr:d12:169,prl:34:1286,pr:d14:246}.
This point was stressed again~\cite{ingschlein,donnalandsh,streng} 
based on results from UA8~\cite{pl:b211:239,pl:b297:417} which were obtained in $p \pbar$ collisions at a center-of-mass (c.m.) energy of 630 GeV.

The electron (positron) collider HERA at DESY has opened a new avenue towards diffraction. In virtual photon-proton scattering, events can be studied where the spatial resolution provided by the virtual photon is much smaller than the proton radius such that the inner structure of the proton becomes visible. In the analysis of events with energy transfers between photon and proton of more than 10000 GeV - measured in the proton rest frame - the ZEUS collaboration~\cite{pl:b315:481} observed a special class of events, where a hadronic system is emitted close to the direction of the virtual photon, and separated by a large rapidity gap from a proton or low-mass nucleonic system emitted along the direction of the incoming proton. Such events were also detected by the H1 collaboration~\cite{np:b429:477}. For comparison, standard events from deep-inelastic scattering exhibit uniform particle production in the rapidity space between the directions of the incoming virtual photon and the proton. The production characteristics of large rapidity gap events are those of a diffractive process in which no quantum numbers are exchanged in the t-channel, and where the cross section is found to be independent of the c.m. energy, up to logarithmic terms.

On the theoretical side, long before HERA had become a project, the authors of Ref.~\cite{bartelsloewe} had considered the possibility of producing a large mass object such as the Z-boson by diffractive deep-inelastic ep scattering and demonstrated that a viable production rate may be achieved. An extensive study of physics issues, such as diffraction, that could be addressed with a large electron proton collider had been presented by~\cite{grilevrys} before HERA was approved. A group of experimental physicists headed by A. Zichichi had insisted that a detector at HERA must be equipped with a magnetic spectrometer close to the outgoing proton beam for tagging protons which arise from diffractive scattering~\cite{ZEUSLOI}.    

For large negative mass squared $-Q^2$ of the virtual photon ($\gamma^*$), the photon has a small transverse extension of the order $1/\sqrt{Q^2}$; for example, at $Q^2 =10$ GeV$^2$ the transverse dimension of the photon is of the order of a few percent of the proton radius. Since small scales, in general, allow the use of perturbative methods, the data on $\gamma^* p$ scattering from HERA can be used to study diffractive scattering in the framework of QCD perturbation theory.

This article reviews the results obtained on diffraction in high energy photon-proton and deep-inelastic electron-proton scattering by the experiments H1 and ZEUS at HERA, which were in operation for physics between 1992 and 2007.

In order to put the results on diffraction into perspective the report starts with a brief review of the experimental results for the proton structure function $F_2$ and the total $\gamma p$ and $ \gamma^* p$ cross section.

\section{Kinematics of inelastic electron-proton scattering}

\begin{figure}[p]
\begin{center}
\vfill
\vspace*{-1.5cm} \epsfig{file=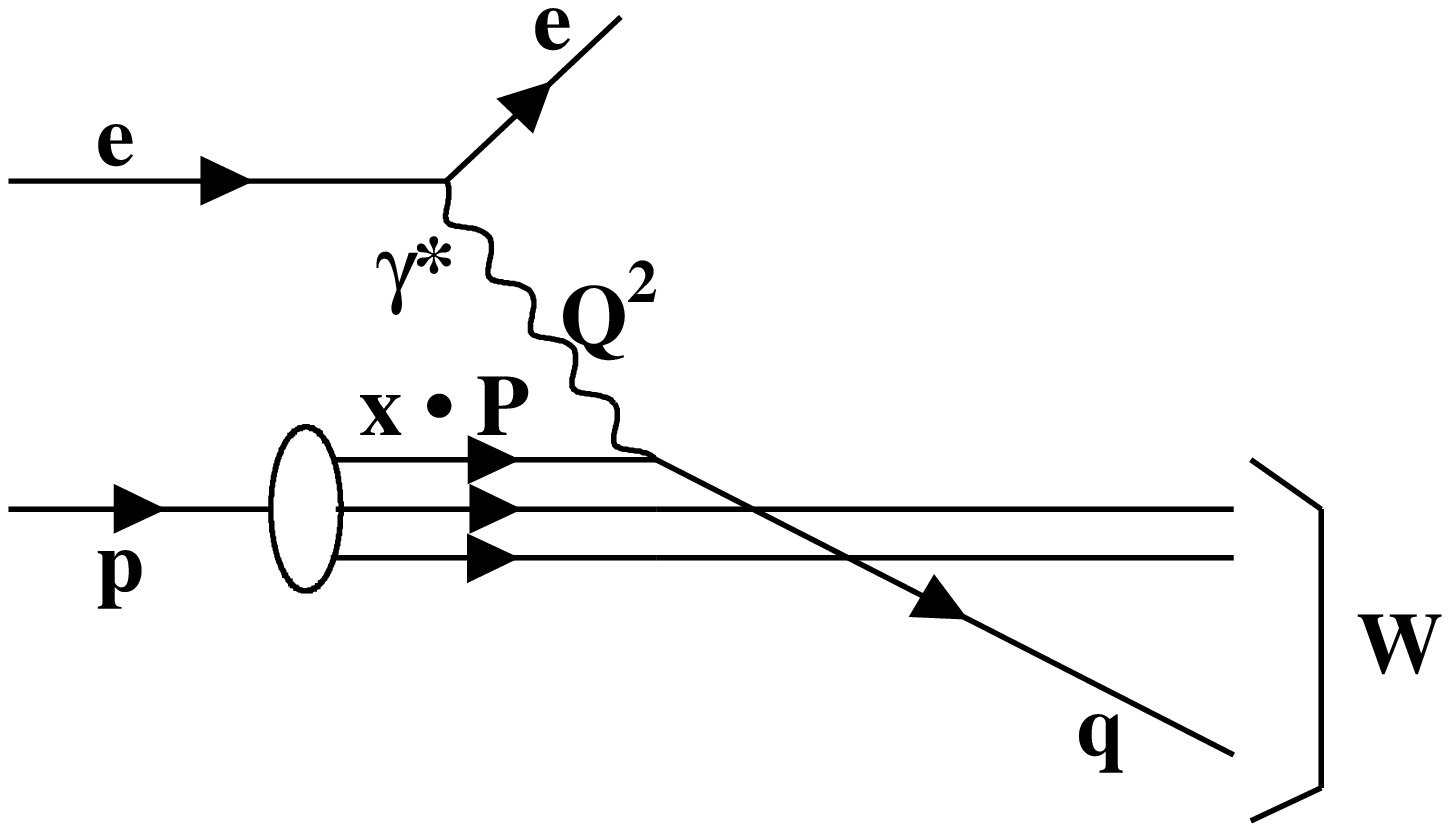,width=11.5cm,clip=}
\end{center}
\vspace*{-1.cm}
\caption{Diagram for non-peripheral deep inelastic scattering.} 
\label{f:nondifdiag}
\vfill
\end{figure}

The reaction
\begin{eqnarray}
e(k) \; p(P) \to e (k^{\prime}) + \rm anything, 
\end{eqnarray}
see Fig.~\ref{f:nondifdiag}, is described in terms of the four momenta $k, (k^{\prime})$ of the incident (scattered) lepton ($e^+$ or $e^-$) and proton ($P$), with beam energies $E_e$ and $E_p$, respectively. At fixed squared centre-of-mass energy, $s = (k+P)^2 \approx 4 E_e E_p$, deep-inelastic scattering is described in terms of $Q^2 \equiv -q^2 = -(k -k^{\prime})^2$ and Bjorken-$x = Q^2/(2 P \cdot q)$. The fractional energy transferred to the proton in its rest system is $y \approx Q^2/(sx)$. The centre-of-mass energy of the hadronic final state, $W$, is given by $W^2 = (P+q)^2 = m^2_p+Q^2(1/x-1)\approx Q^2/x = ys$, where $m_p$ is the mass of the proton.

\begin{figure}[p]
\vfill
\vspace*{-1.cm}
\begin{center}
\epsfig{file=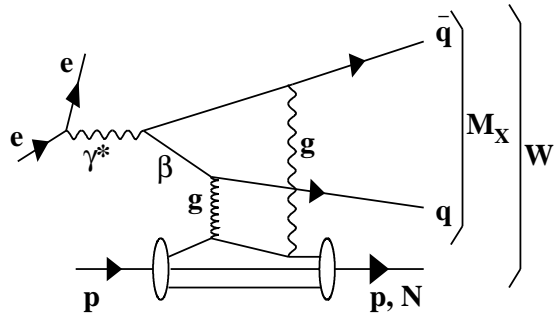,width=9.cm,clip=}
\vspace*{0.5cm}
\epsfig{file=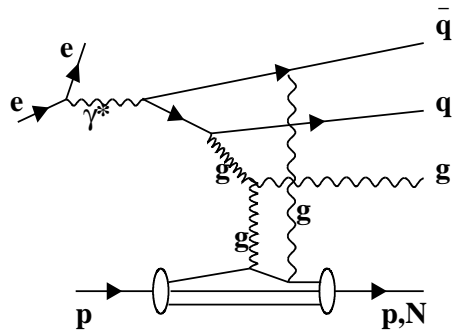,width=9.cm,clip=}
\vfill
\end{center}
\caption{Diagrams of diffractive deep inelastic scattering, $e p \to e X N$, proceeding by the exchange of two gluons.} 
\label{f:difdiag}
\vfill
\end{figure}

In diffraction, proceeding via
\begin{eqnarray} 
\gamma^*  p \to X + N,
\end{eqnarray} 
see Fig.~\ref{f:difdiag}, where $\gamma^* = e - e^{\prime}$, a hadronic system $X$ is produced. The incoming proton undergoes a small perturbation by the emission of a Pomeron, which in lowest order QCD can be represented by a two-gluon system. The proton emerges either intact ($N=p$), or as a low-mass nucleonic state $N$ with mass $M_N$, in both cases carrying a large fraction, $x_L$, of the incoming proton momentum. Diffraction is described in terms of $W$, $Q^2$, $M_X$, $M_N$ and of $t$, the four-momentum transfer squared between the incoming proton and the outgoing system $N,\; t = (P-N)^2$. Alternatively, diffraction is parametrised in terms of $x_{\pom}$, the fraction of the proton momentum carried by the pomeron, and $\beta$, the momentum fraction of the struck quark within the pomeron:
\begin{eqnarray}
x_{\pom} & = & \frac{(P-P')\cdot q}{P \cdot q} = \frac{M^2_X+Q^2-t}{W^2 +Q^2-M^2_p} \approx \frac{M^2_X+Q^2}{W^2 + Q^2}\\
\beta & = & \frac{Q^2}{2(P-P')\cdot q} = \frac{x}{\xpom} = \frac{Q^2}{M^2_X+Q^2-t} \approx \frac{Q^2}{M^2_X+Q^2}\hspace{0.1cm}.
\end{eqnarray}

\section{The H1 and ZEUS detectors} 
\begin{figure}[p]
\begin{center}
\vfill
\epsfig{file=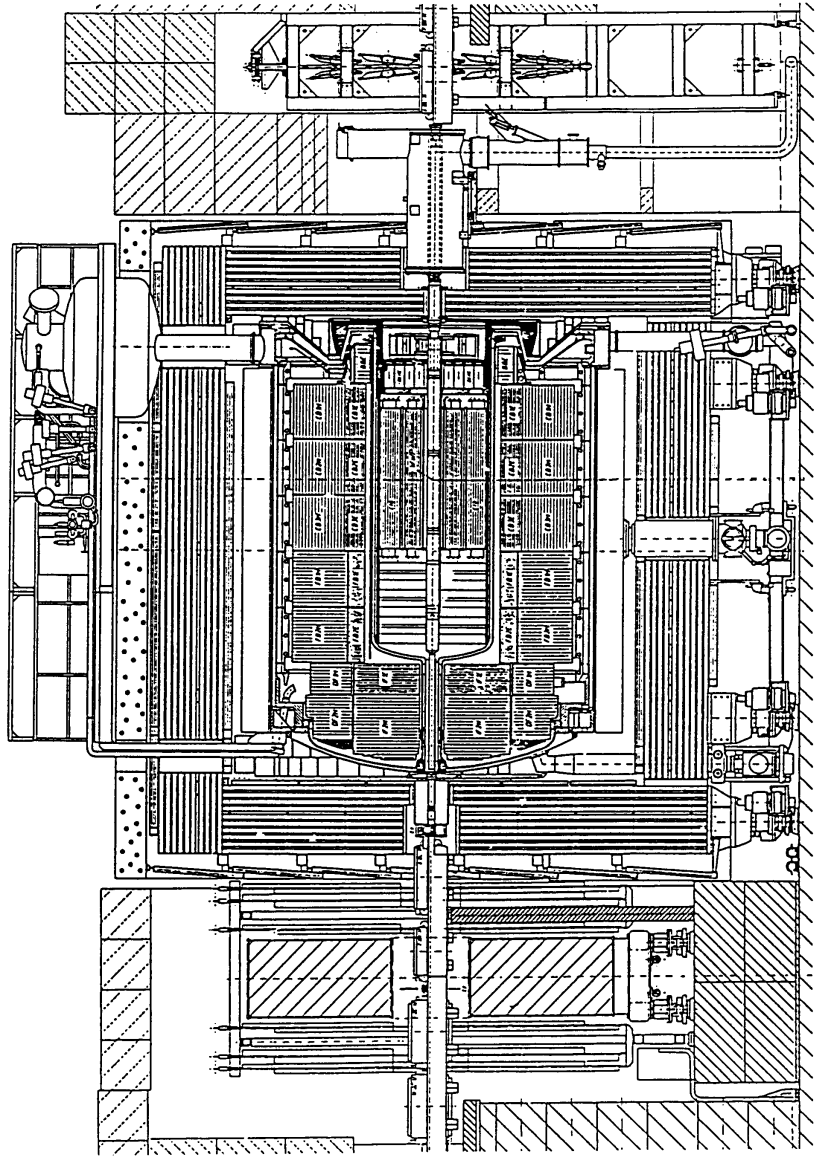,width=12cm}
\end{center}
\caption{Side view of the H1 detector. The proton (electron) beam enters from the right (left).} 
\label{f:h1yzcut}
\vfill
\end{figure}

\begin{figure}[p]
\begin{center}
\vfill
\vspace*{-3.3cm} \epsfig{file=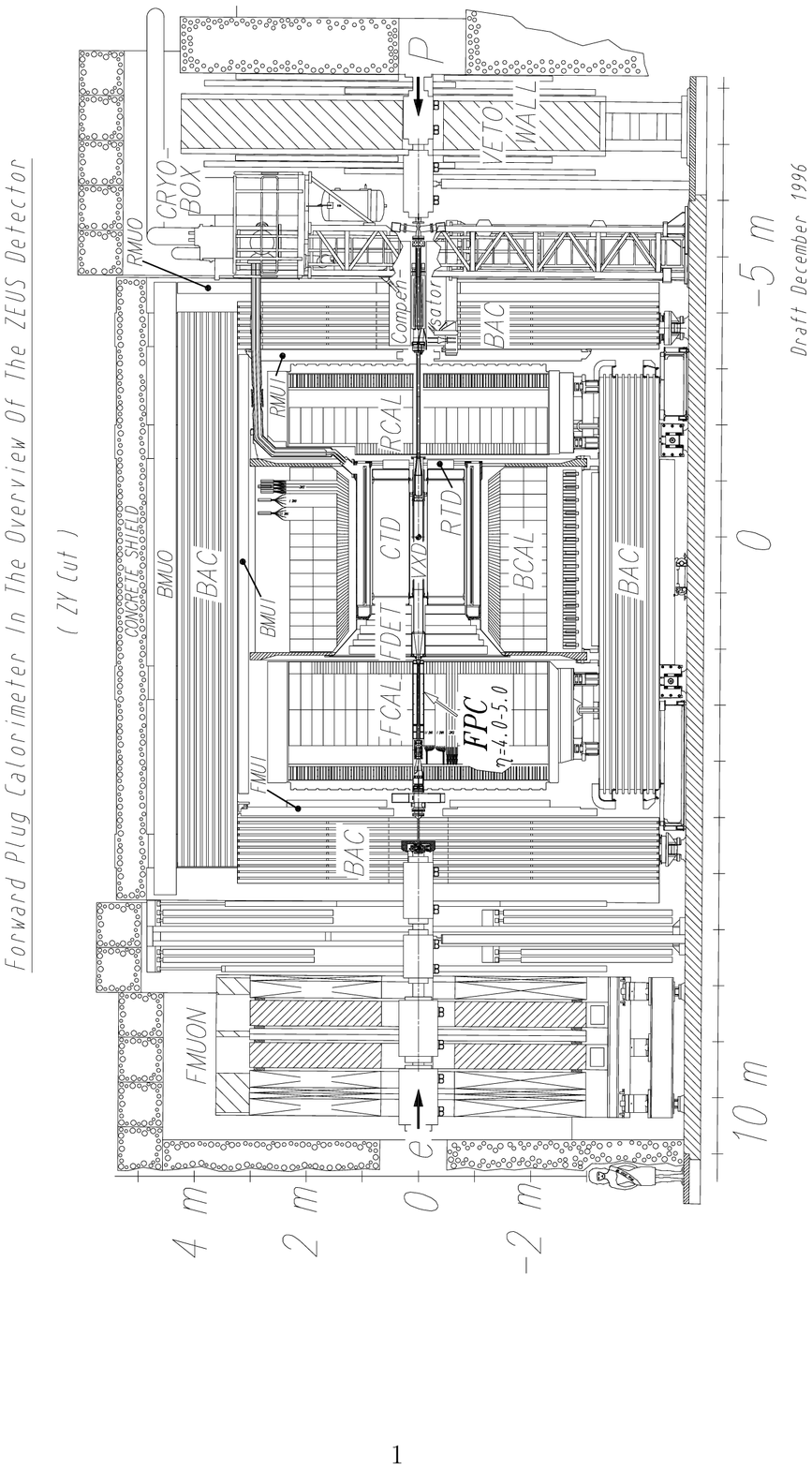,width=16cm,clip=}
\end{center}
\caption{Side view of the ZEUS detector with the FPC integrated in the ZEUS Calorimeter.The proton (electron) beam enters from the right (left).} 
\label{f:zeusyzcut}
\vfill
\end{figure}

For the study of diffraction and of the total photon-proton cross section at HERA it was essential that the detectors of H1~\cite{nim:a386:310,nim:a386:348} and ZEUS~\cite{bluebook,pl:b293:465} covered the full solid angle (up to the beam pipe) with calorimetry. Figures~\ref{f:h1yzcut} and~\ref{f:zeusyzcut} 
show side views of the two detectors. The H1 detector had tracking detectors around the interaction region and a large liquid argon calorimeter equipped with lead plates as absorber for particle detection. The ZEUS detector employed tracking detectors and a large, almost hermetic calorimeter consisting of Uranium plates interspaced with scintillator for signal collection, which provided equal response to electrons, photons and hadrons of the same momentum. 

For a substantial fraction of the data taken by ZEUS, the forward area close to the outgoing proton beam (pseudorapidities $\eta = 4 - 5$, where $\eta = -\ln {\rm tan}(\theta)/2$ and $\theta$ is the polar angle in radians) was instrumented with the Forward Plug Calorimeter (FPC), which had been inserted in the Uranium-scintillator calorimeter leaving a hole of only 3.15 cm radius for the passage of the beams. The FPC was a lead scintillator sandwich calorimeter with 5.4 nuclear absorption lengths that provided equal response to electrons and hadrons of the same energy~\cite{nim:a450:235}. 

Close to the direction of the outgoing proton beam, at a distance of about 100m from the central detector, H1 and ZEUS were equipped with spectrometers and calorimeters (not shown) for the detection of leading protons and neutrons.

\section{The structure function $F_2(x,Q^2)$ of the proton}
The differential cross section for inclusive ep scattering,
\begin{eqnarray}
e(k) p(P) \to e(k^{'})+{\rm anything},
\end{eqnarray}
mediated by virtual photon exchange, 
is given in terms of the structure functions $F_2$ and $F_L$ of the proton by
\begin{eqnarray}
\frac{d^2 \sigma^{e^+ p}}{dx dQ^2} = \frac{2 \pi \alpha^2}{x Q^4} [Y F_2(x,Q^2) - y^2 F_L(x,Q^2)](1 + \delta_r(x,Q^2)),
\end{eqnarray}
where $Y = 1 + (1-y)^2$. In general, the structure function $F_2$ represents the main component of the cross section. In the deep-inelastic scattering (DIS) factorisation scheme, $F_2$ corresponds to the sum of the momentum densities of the quarks and antiquarks weighted by the squares of their electric charges; $F_L$ is the longitudinal structure function and $\delta_r$ is a term accounting for radiative corrections. The contribution of $F_L$ to the cross section relative to that from $F_2$ is given by $(y^2/Y) \cdot (F_L/F_2)$. For the determination of $F_2$ by ZEUS~\cite{np:b713:3,np:b800:1}, the $F_L$ contribution was taken from the ZEUS NLO QCD fit~\cite{pr:d67:12007}. The contribution of $F_L$ to the cross section in the highest $y$ (= lowest $x$) bin of this analysis was 3.2\%, decreasing to 1.3\% for the next highest $y$ bin. For the other bins, the $F_L$ contribution is below 1\%. The resulting uncertainties on $F_2$ are below 1\%~\cite{np:b713:3,np:b800:1}.  

In the $Q^2$ range considered here ($Q^2 \le 500$ GeV$^2$), the contributions from $Z^0$ exchange and $Z^0$ - $\gamma$ interference were estimated to be at most 0.4\% ~\cite{np:b713:3,np:b800:1} and were ignored.

\begin{figure}[p]
\begin{center}
\vfill
\vspace*{-1cm} \epsfig{file=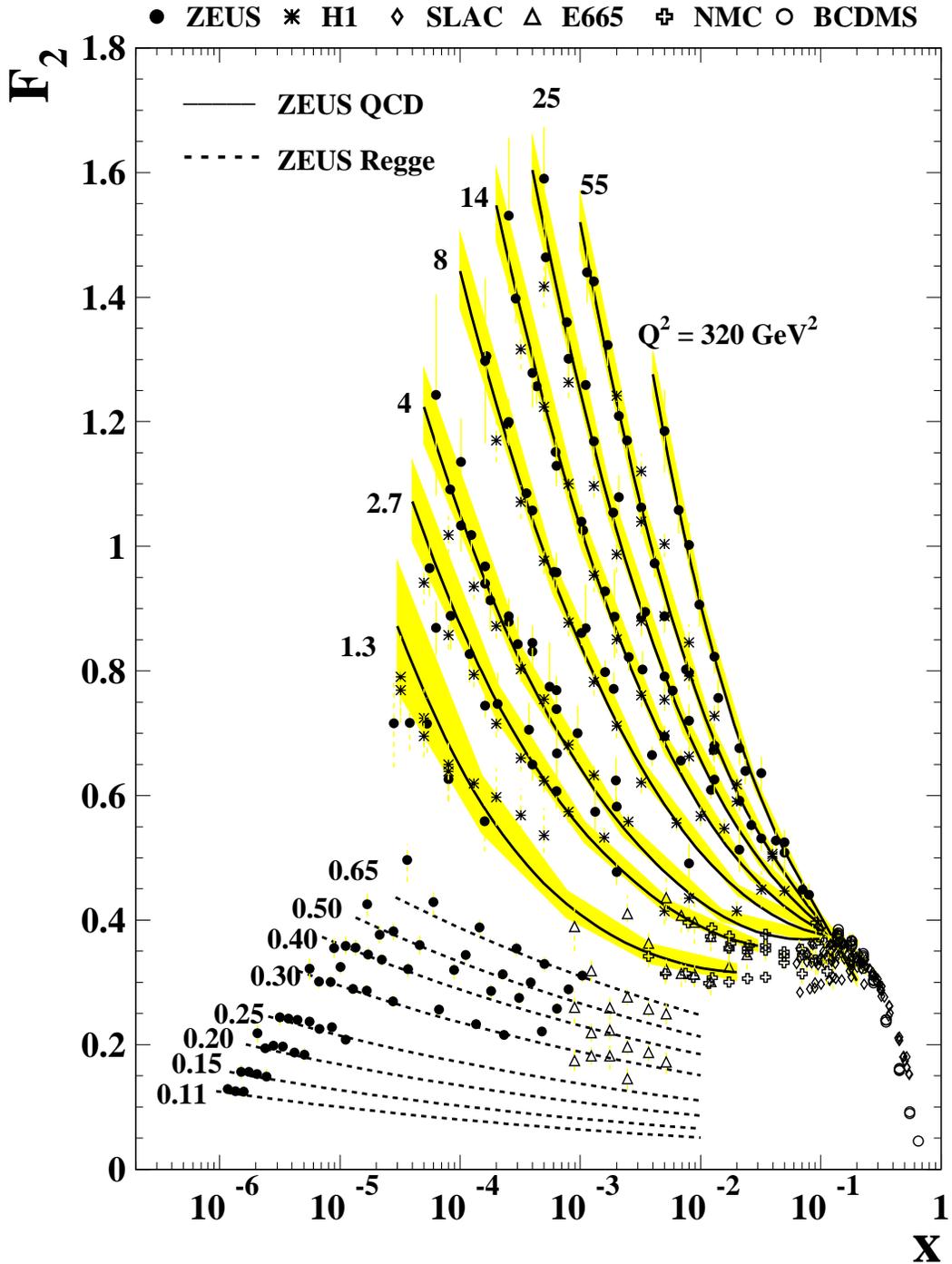,width=15cm,clip=}
\end{center}
\caption{The proton structure function $F_2$ versus $x$ for the $Q^2$ values indicated as measured by H1, ZEUS and the fixed target experiments SLAC, E665, NMC and BCDMS. The dashed curves represent the ZEUS Regge fit to the low $Q^2$ data from ZEUS. The solid curves show the QCD fit to the ZEUS data at higher $Q^2$; the shaded bands represent the uncertainties of the fit.} 
\label{f:f2vsxallxpt}
\vfill
\end{figure}

The structure function $F_2$ is shown in Fig.~\ref{f:f2vsxallxpt} as a function of $x$ for fixed $Q^2$. The results by H1~\cite{epj:c21:33} and ZEUS~\cite{np:b713:3,np:b800:1} are presented together with the results from the fixed target experiments~\cite{pl:b282:475}, E665~\cite{pr:d54:3006}, NMC~\cite{np:b483:3} and BCDMS~~\cite{pl:b223:485}. For $Q^2$ between 0.11 and 0.65 GeV$^2$ and $x<0.01$, the structure function $F_2$ exhibits a modest rise as $x \to 0$. At larger values of $Q^2$, starting from $x = 0.7$, $F_2$ is rising with decreasing $x$, reaching a plateau around $x = 0.1$. Below $x \approx 0.1 -0.01$ and $Q^2 \ge 1.3$ the HERA measurements show $F_2$ growing rapidly as $x \to 0$; the growth accelerates as $Q^2$ increases from 1.3 to 320 GeV$^2$.

\begin{figure}[p]
\begin{center}
\vfill
\vspace*{-1cm} \epsfig{file=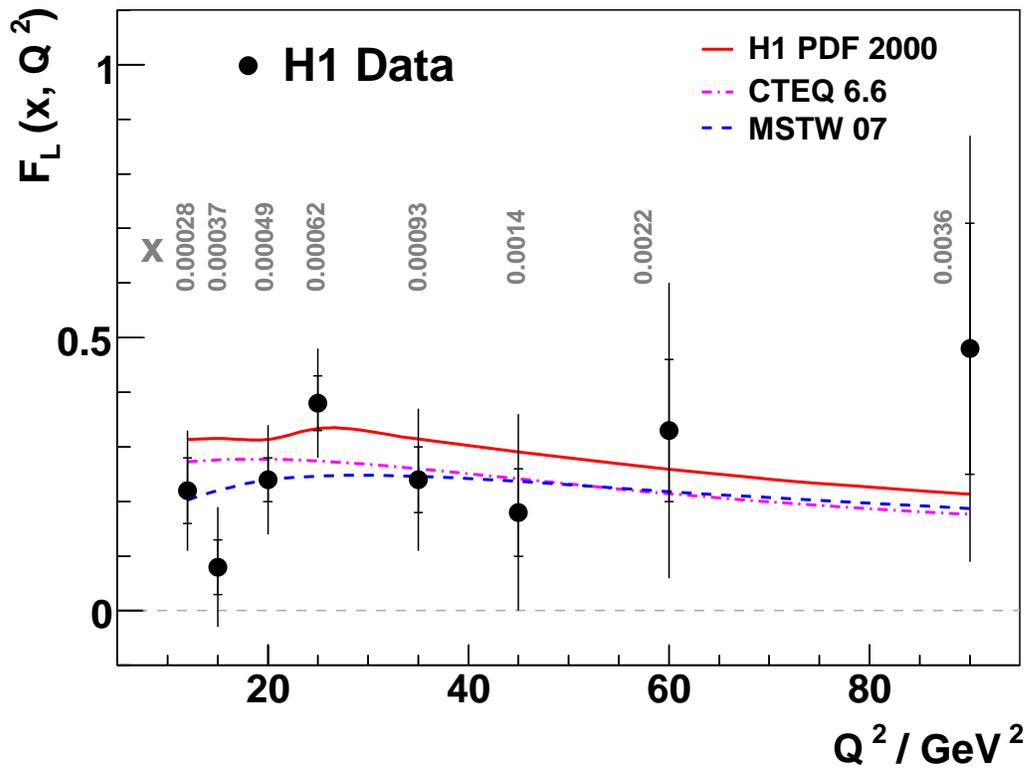,width=15cm,clip=}
\end{center}
\caption{The longitudinal structure function $F_L$ of the proton versus $Q^2$ for the $x$ values indicated as measured by H1. The solid curve decribes the expectation from the H1 PDF 2000 fit using NLO QCD. The dashed (dashed-dotted) curve shows the expectation of the MSTW (CTEQ) group using NNLO (NLO) QCD; from H1} 
\label{f:flvsxh1}
\vfill
\end{figure}

The most recent measurement of the longitudinal structure function $F_L$ at HERA, as provided by H1~\cite{pl:b665:139}, is shown in Fig.~\ref{f:flvsxh1}. Within errors, $F_L = 0.2 - 0.4$ for the $x,Q^2$ values measured.          

\begin{figure}[p]
\begin{center}
\vfill
\vspace*{-6cm} \epsfig{file=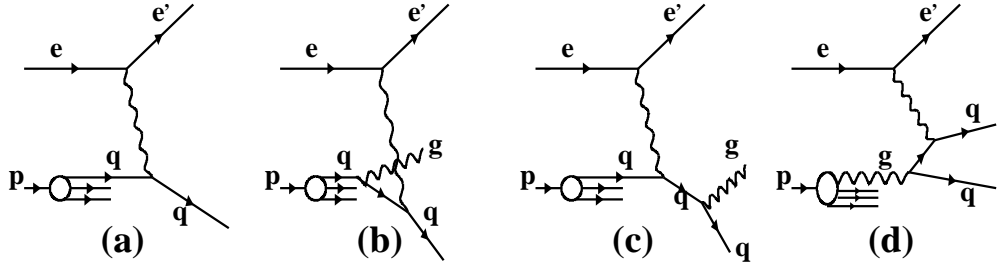,width=17cm,clip=}
\end{center}
\vspace*{-2cm}
\caption{Diagrams for deep-inelastic scattering in leading and next-to-leading order QCD.} 
\label{f:epdisdiags}
\vfill
\end{figure}

\begin{figure}[p]
\begin{center}
\vfill
\epsfig{file=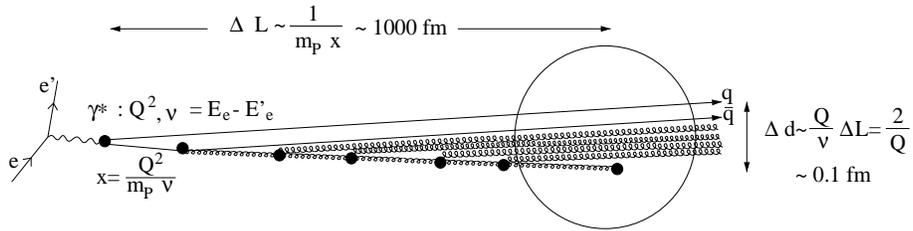,height=3.0cm,width=12cm,clip=}
\end{center}
\caption{Deep inelastic scattering: QCD cascade developing at large $Q^2$ and small $x$, as seen in the proton rest system.}
\label{f:f2qcdcascade}
\vfill
\end{figure}

The leading order (LO) and next-to-leading order (NLO) QCD diagrams for deep-inelastic electron proton scattering are shown in Fig.~\ref{f:epdisdiags}. The QCD-cascade which develops at large energies is illustrated by Fig.~\ref{f:f2qcdcascade}. The solid curves in Fig.~\ref{f:f2vsxallxpt} show the ZEUS next-to-leading order QCD fit~\cite{pr:d67:12007} to data from ZEUS and fixed target experiments. The dashed curves were determined by a Regge fit to the low $Q^2$ data from ZEUS~\cite{pl:b487:53}. 

\begin{figure}[p]
\begin{center}
\vfill
\epsfig{file=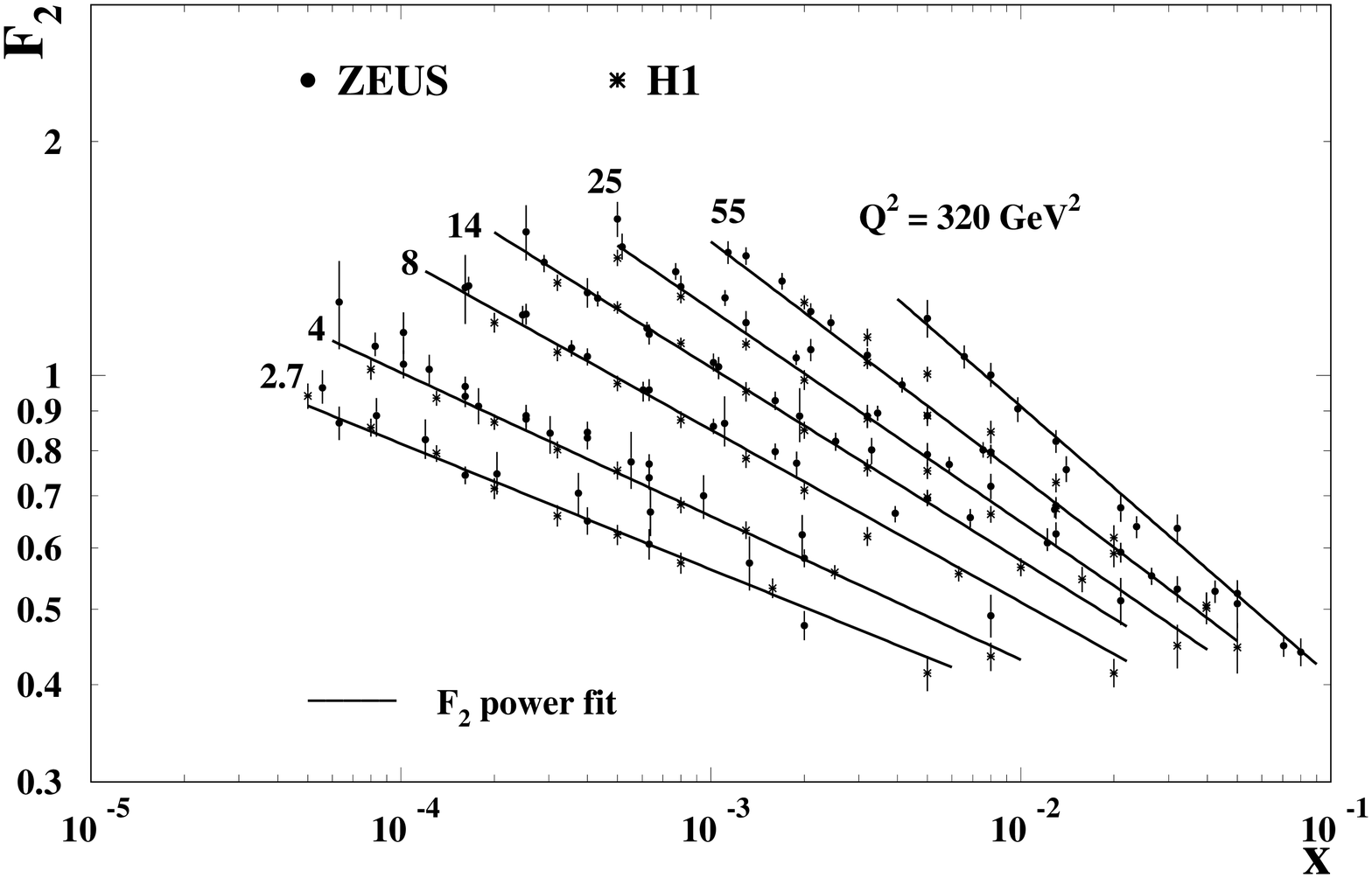,width=15cm,clip=}
\end{center}
\caption{The proton structure function $F_2$ versus $x$ for the $Q^2$ values indicated as measured by H1 and ZEUS. The solid curves show the fit Eq.~\ref{eq:f2fit}, see text.} 
\label{f:logf2vslogxallq}
\vfill
\end{figure}

The HERA data for $F_2$ can be represented in a compact form:
\begin{eqnarray}
\hspace*{-3cm}\log_{10}F_2(x,Q^2) =  c_1+c_2\cdot \log_{10} x +c_3\cdot (\log_{10}x) \cdot \log_{10}(Q^2/Q^2_0) \nonumber 
\\
\hspace*{2cm} + c_4 \cdot(\log_{10} x)\cdot (\log_{10}(Q^2/Q^2_0))^2
\label{eq:f2fit}
\end{eqnarray}
where $Q^2$ in units of GeV$^2$. A fit to the HERA data for $Q^2 \ge 2.7$ GeV$^2$ with 389 data points yielded $c(1)  =   -0.735 \pm 0.004$, $c(2)  =   -0.098 \pm 0.002$, $c(3)  =   -0.158 \pm 0.002$, and $c(4)  =  0.023 \pm 0.001$, with $\chi^2 =368$. The fit gives a good representation of the data, see Fig.~\ref{f:logf2vslogxallq}.

The $F_2$ data will be presented below also in terms of the total cross section for virtual photon proton scattering.

\section{The total cross section for real and virtual photon-proton scattering} 
\subsection{Real photons, $Q^2 = 0$}
\begin{figure}[p]
\begin{center}
\vfill
\epsfig{file=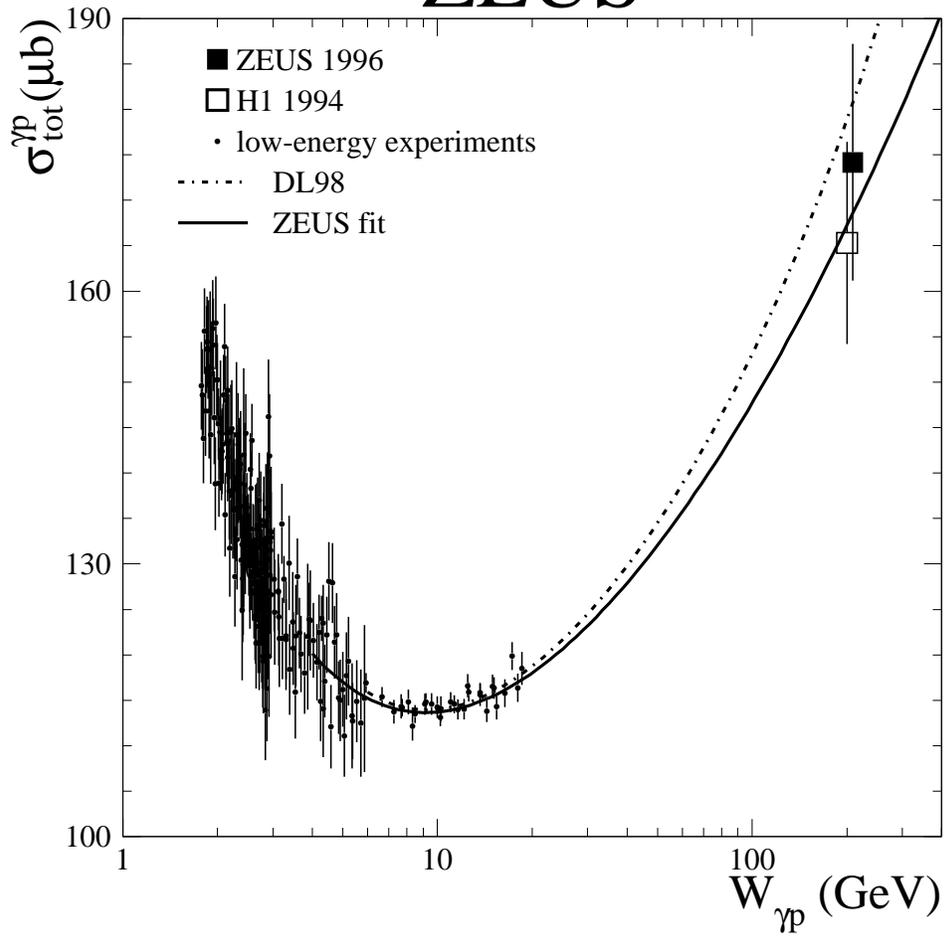,width=14cm}
\end{center}
\caption{The photon proton total cross section for real photons ($Q^2 = 0$) as a function of the photon-proton center-of-mass energy $W_{\gamma p}$, as measured by fixed target experiments ($W_{\gamma p}< 20$\GeV) and by  H1 and ZEUS ($W_{\gamma p} = 200$\GeV). The dashed-dotted and solid curves show fits to the data, see text.} 
\label{f:zhlowesigtot}
\vfill
\end{figure}

The total photon-proton cross section for real photons, $\sigma^{\gamma p}_{tot}(Q^2 = 0)$, was measured by ZEUS~\cite{pl:b293:465,zfp:c63:391,np:b627:3} and H1~\cite{pl:b299:374,zfp:c69:27} at several stages of the two experiments. The final results at the total photon proton center-of-mass energy $W$ are\\
\\
\hspace*{1cm} H1: \hspace*{0.3cm} $W = 200 \GeV$: $\sigma^{\gamma p}_{\tot} = 165\pm 2(stat.) \pm 11 (syst.) {\mu}b$ \\
\hspace*{1cm} ZEUS: $W = 209\GeV$: $\sigma^{\gamma p}_{\tot} = 174\pm 1(stat.) \pm 13 (syst.) {\mu}b$ \\ 
\\
The HERA measurements are presented in Fig.~\ref{f:zhlowesigtot} together with the low-energy results~\cite{caldwell78,alekhin87}. For $W_{\gamma p} < 6 \GeV$, the total-photon proton cross section exhibits a steep descent - on a logarithmic scale - with $W_{\gamma p}$, followed by a rapid increase to the values measured at HERA. The solid curve shows a fit of the data to the form~\cite{np:b627:3}\\
\\
\hspace*{4cm} $\sigma^{\gamma p}_{\tot} = A \cdot W^{\delta}_{\gamma p} + B \cdot W^{-\eta}_{\gamma p}$ \\
\\
with $W$ in units of GeV, $A = 57\pm 5 {\mu}b$, $B = 121\pm 13 {\mu}b$, ${\delta} = 0.200\pm 0.024$; $\eta$ had been fixed to the value $0.716\pm0.030$, found by~\cite{pr:d61:034019,pr:d63:059901}. The dot-dashed curve shows a fit which includes a soft and a hard Pomeron trajectory~\cite{DL98}.

\subsection{Virtual photons, $Q^2 > 0$}

\begin{figure}[p]
\begin{center}
\vfill
\epsfig{file=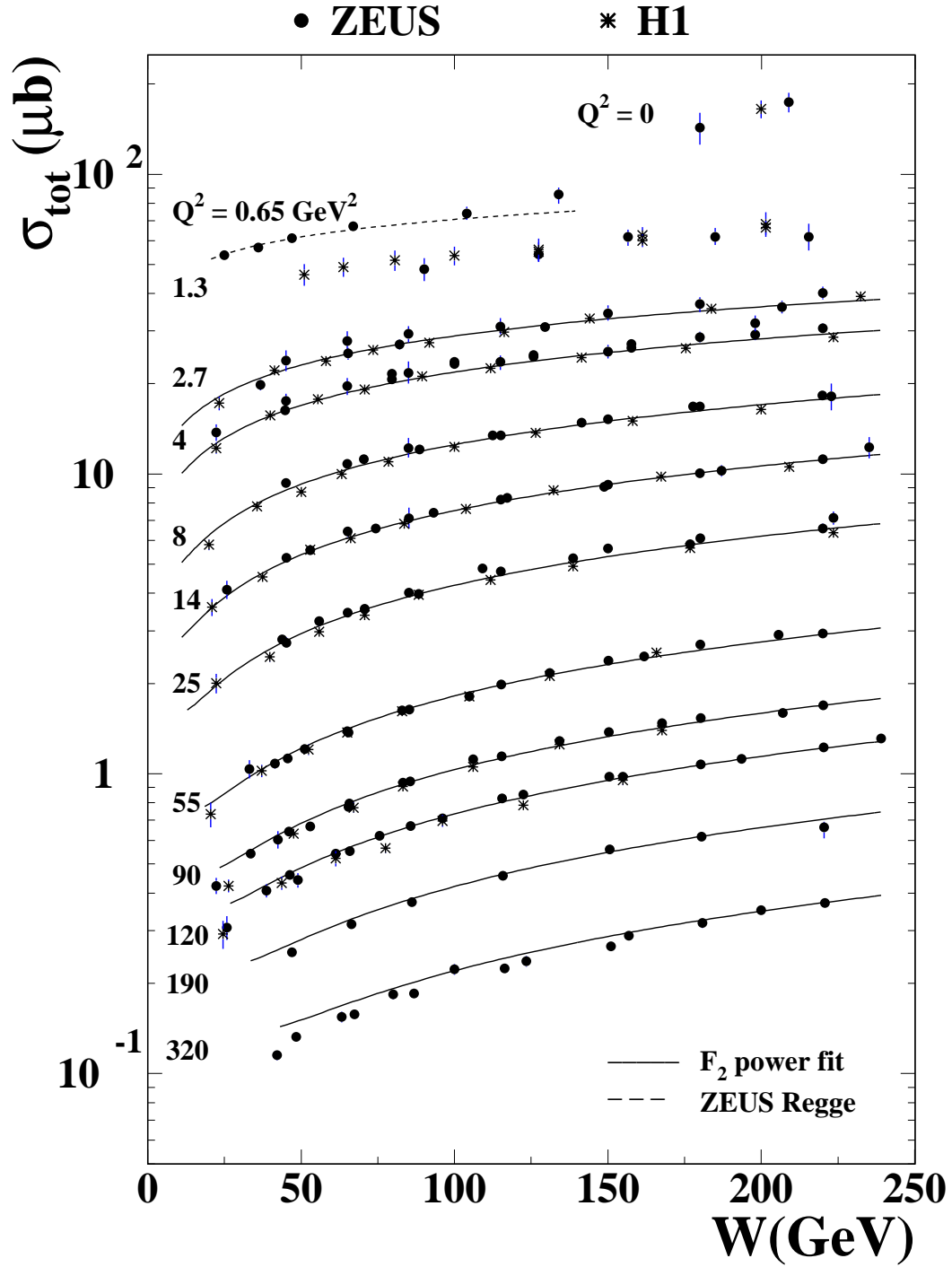,width=15cm}
\end{center}
\caption{The photon proton total cross section for real ($Q^2 =  0$) and virtual photons ($Q^2 > 0$) as a function of the photon-proton center-of-mass energy, $W$, measured by H1 and ZEUS. Also shown is the ZEUS Regge fit for $Q^2 = 0.65$\GeV$^2$ (dashed curve), and the result of the $F_2$ power fit (solid curves) for $Q^2 \ge 2.7$\GeV$^2$.
} 
\label{f:zhsigtotdis}
\vfill
\end{figure}

The total cross section for virtual photon-proton scattering, $\sigma^{\rm tot} _{\gamma^{\ast} p} \equiv \sigma_T(x,Q^2) + \sigma_L(x,Q^2)$, where $T(L)$ stands for transverse (longitudinal) photons,
was extracted from the measurement of $F_2$ using the relation
\begin{eqnarray}
\sigma^{\rm tot} _{\gamma^{\ast} p} = \frac{4 \pi^2 \alpha}{Q^2(1-x)} F_2(x,Q^2),
\end{eqnarray}
which is valid for $4m^2_p x^2 \ll Q^2$~\cite{pr:129:1834}.
The total cross section for virtual photons is shown in Fig.~\ref{f:zhsigtotdis} as a function of $W$ for fixed $Q^2$ between zero and 320 GeV$^2$. The lines show the power fit presented above for the stucture function $F_2$, see Eq.~\ref{eq:f2fit}. The power $\delta$, decribing the rise $\sigma^{\rm tot} _{\gamma^{\ast} p} \propto W^{\delta}$, is increasing with $Q^2$ from $\delta = 0.2$ at $Q^2 =0$ to $\delta \approx 0.8$ at $Q^2 = 320$ GeV$^2$; the latter value of $\delta$ implies an almost linear rise with $W$. Such a rise with $W$ has not been seen in hadron-hadron interactions. The NLO QCD fit presented above for $F_2(x,Q^2)$ gives also a good description of $\sigma^{\rm tot} _{\gamma^{\ast} p}$.


\section{Diffraction}
\label{sec-inclusdiff}
\begin{figure}[p]
\begin{center}
\vfill
\vspace*{-3cm} 
\hspace*{0.5cm}\epsfig{file=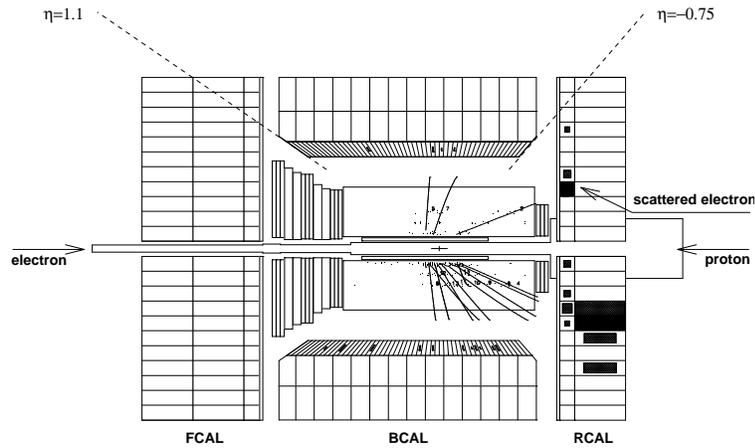,width=10cm,clip=}
\end{center}
\vspace*{-0.2cm}
\caption{One of the first events from deep-inelastic ep scattering with a large rapidity gap, as recorded in the ZEUS detector.} 
\label{f:zeusdifevt}
\vfill
\end{figure}

\begin{figure}[p]
\begin{center}
\vfill
\vspace*{-0.4cm}
\hspace*{+0.5cm}\epsfig{file=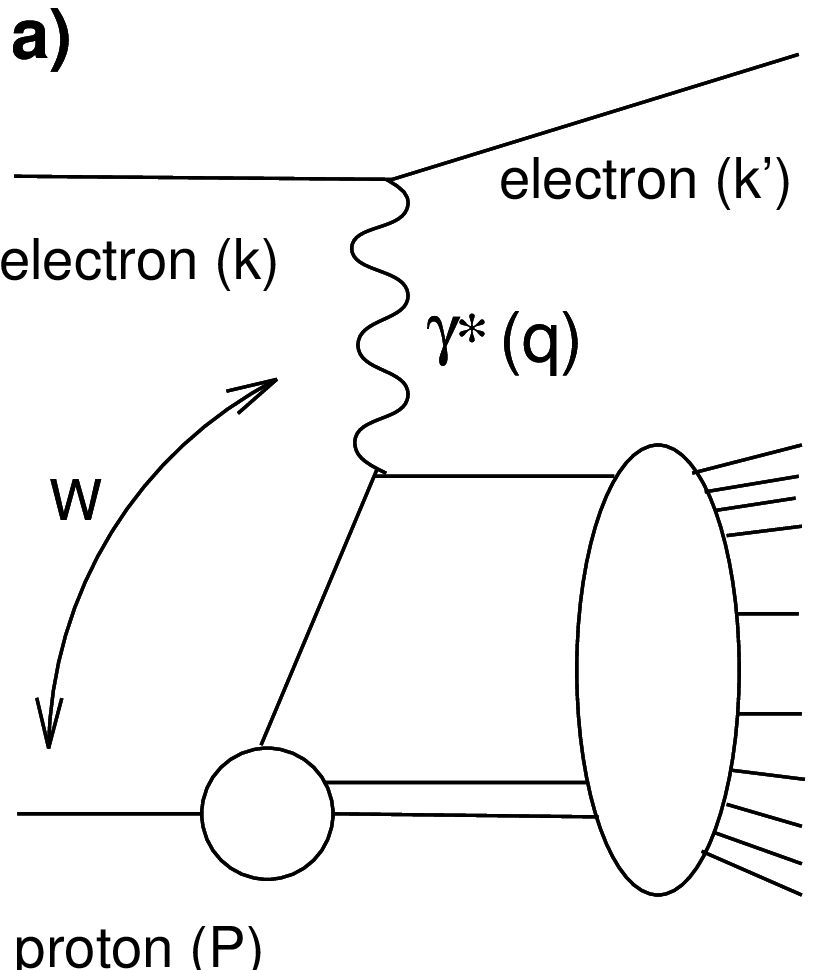,width=3cm}\hspace*{3cm}\epsfig{file=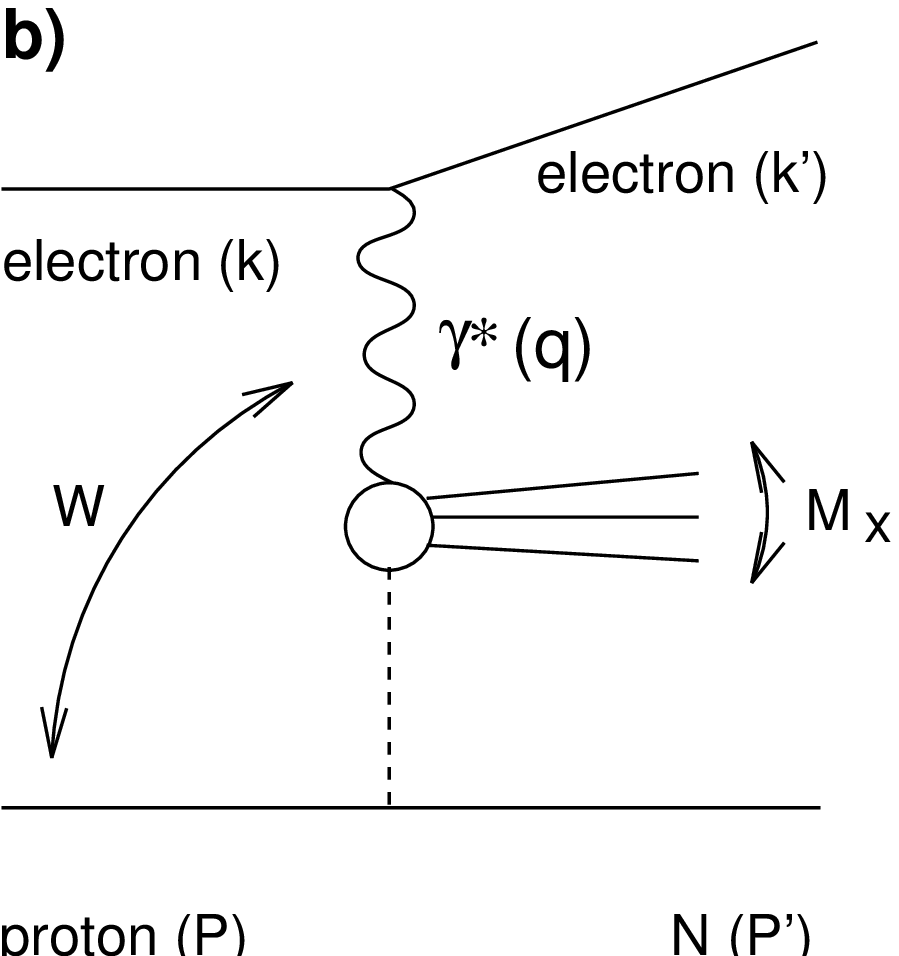,width=3.3cm}
\end{center}
\vspace*{-0.2cm}
\caption{Diagrams illustrating particle production (a) by deep-inelastic inclusive scattering; (b) by deep-inelastic diffractive scattering.} 
\label{f:zdifndif}
\vfill
\end{figure}

First evidence for a substantial contribution from diffraction in deep-inelastic scattering was reported by ZEUS~\cite{pl:b315:481}, observing events in which no energy is deposited close to the direction of the proton beam, i.e. at large angle $\theta$, see Fig.~\ref{f:zeusdifevt}. Described in terms of pseudorapidity $\eta$, where
\begin{eqnarray}
\eta = - {\rm ln}({\rm tan}  \theta/2),
\end{eqnarray}
nondiffractive events show uniform particle production in $\eta$, see Fig.~\ref{f:zdifndif}(a), while diffractive events exhibit a sizeable gap between the pseudorapidity of the smallest calorimeter angle ($\theta = 1.5^0$, $\eta = 4.3$) and the pseudorapidity of the hadron(s) observed closest to the proton direction, see Fig.~\ref{f:zdifndif}(b): in diffractive scattering particle emission is concentrated around the directions of the incoming photon and proton.

\begin{figure}[p]
\begin{center}
\vfill
\vspace*{-0.4cm}
\epsfig{file=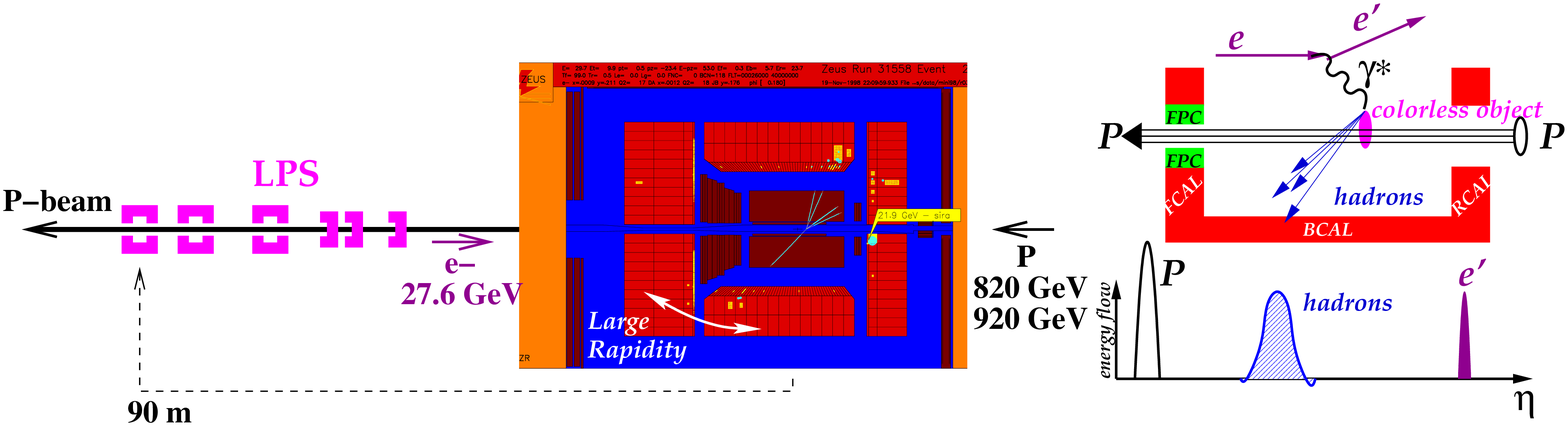,height=40.mm}
\hspace*{+3.7cm}\epsfig{file=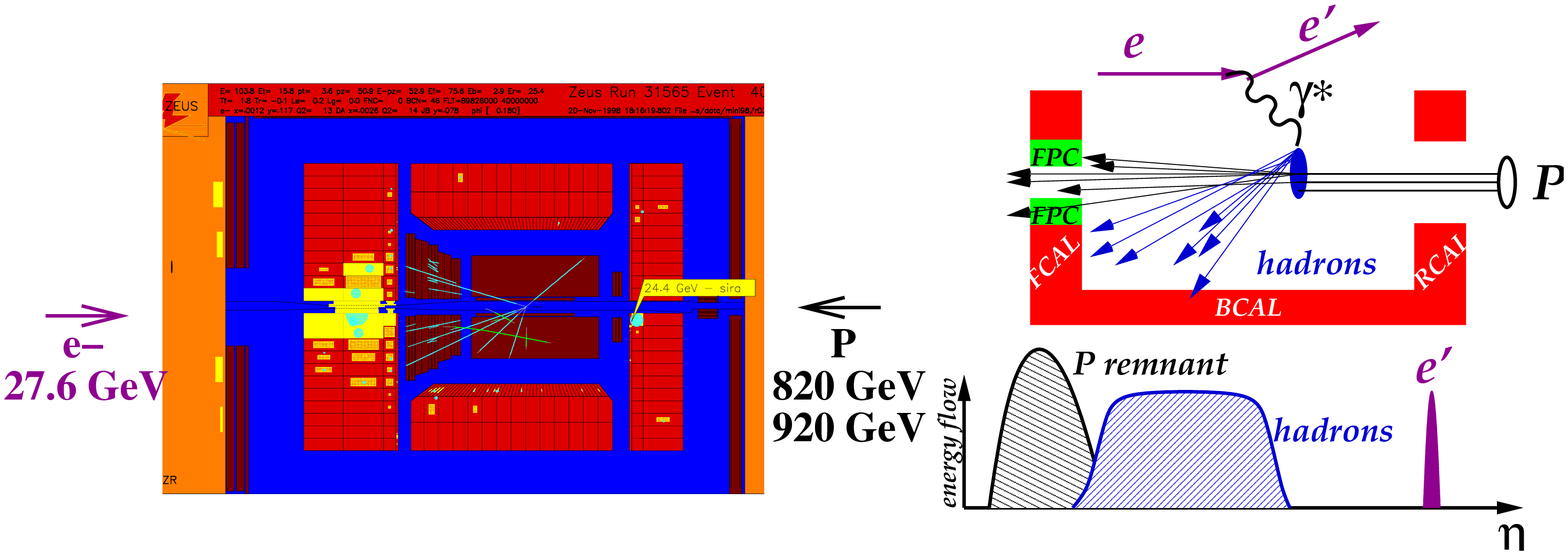,width=11.1cm}
\end{center}
\vspace*{-0.1cm}
\caption{Event topologies of diffractive (top) and nondiffractive deep inelastic scattering (bottom) as seen in the ZEUS detector. The top figure shows also the leading proton spectrometer (LPS) which detected forward scattered protons. Shown are the tracks of charges particles in the central tracking chamber and the energy deposits in the uranium-scintillator calorimeter.}
\label{f:difndifkin}
\vfill
\end{figure}

Figure~\ref{f:difndifkin} illustrates the different topologies of nondiffractive events (bottom) and diffractive events (top), as registered in the ZEUS detector.

\begin{figure}[p]
\begin{center}
\vfill
\epsfig{file=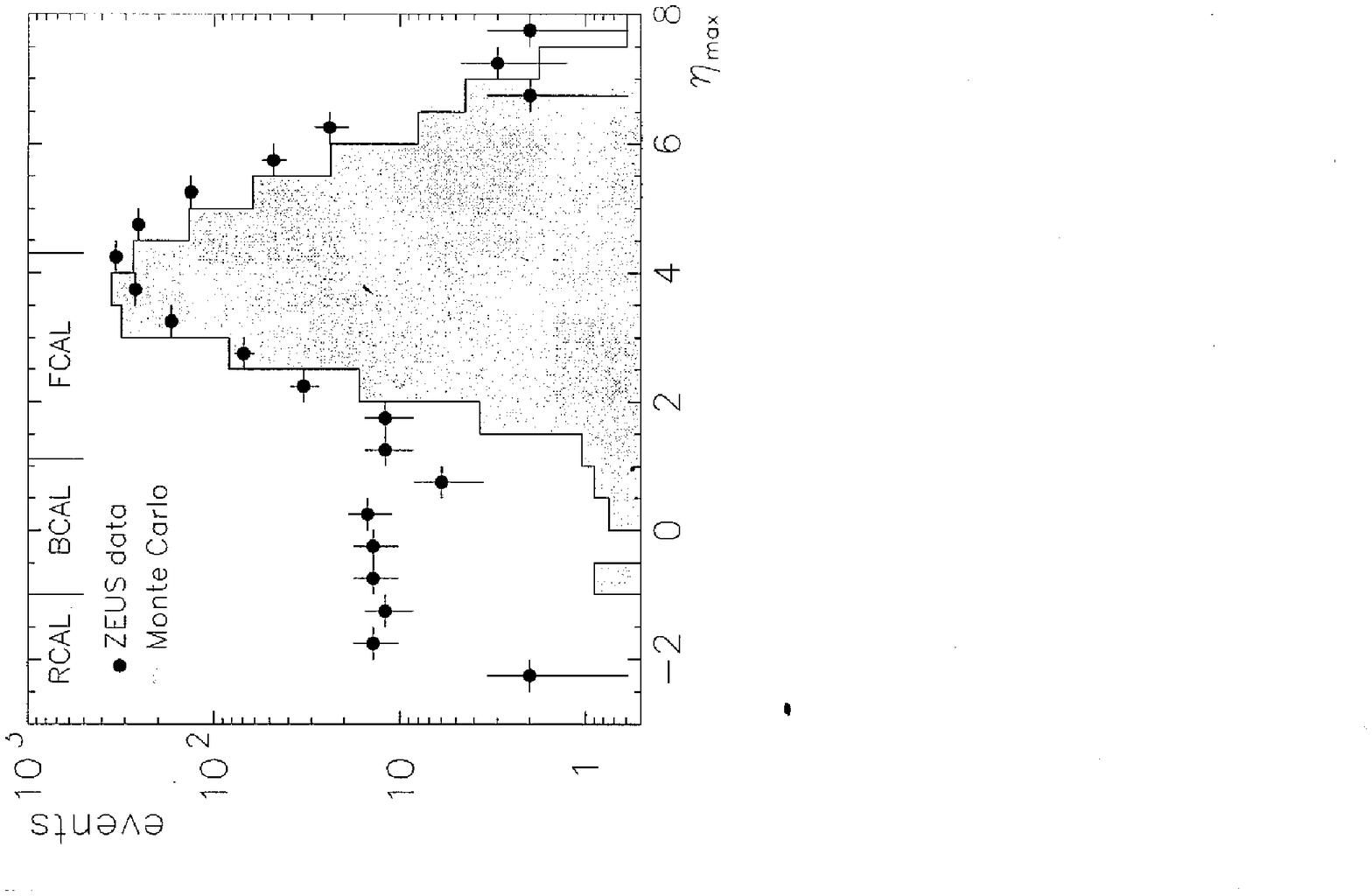,angle=270,width=10cm,clip=}
\end{center}
\vspace*{-7cm}
\caption{Distribution of $\eta_{max}$, the maximum rapidity of a calorimeter cluster in an event for data, and for Monte Carlo events generated for nondiffractive scattering (shaded).} 
\label{f:zetamax}
\vfill
\end{figure}

Figure~\ref{f:zetamax} shows the distribution of the maximum rapidity ($\eta_{max}$) observed in the calorimeter for events from deep-inelastic scattering with $Q^2 > 10$ GeV$^2$. There is a large excess of events at $\eta_{max} <2$ compared to the expectation from nondiffractive deep-inelastic scattering (shaded histogram).

Different methods have been employed by H1 and ZEUS for isolating diffractive contributions. In the case of vectormeson production, $\gamma^* p \to VN$, resonance signals in the decay mass spectrum combined with the absence of other substantial activity in the detector have been used. The contribution from inclusive diffraction has been extracted using the presence of a large rapidity gap, the detection of the leading proton or the hadronic mass spectrum observed in the central detector ($M_X$ method~\cite{kowrocwlf,zfp:c70:391}). Event samples selected on the basis of a large rapidity gap or a leading proton may include additional contributions from Reggeon exchange. Such contributions are exponentially suppressed when using the $M_X$ method.

\section{Deeply virtual Compton scattering}

\begin{figure}[p]
\begin{center}
\vfill
\vspace*{-1cm}
\epsfig{file=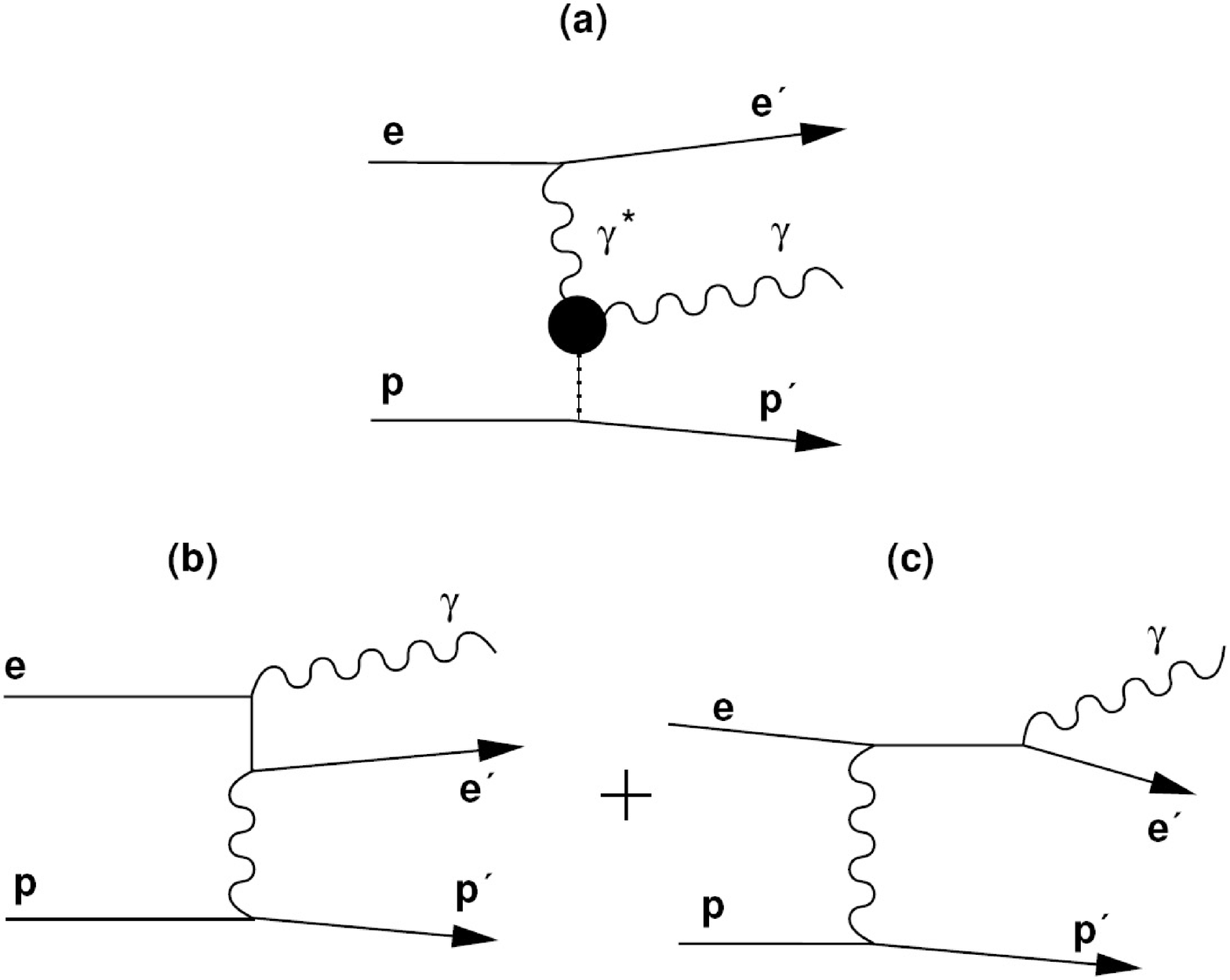,width=10cm}
\end{center}
\vspace*{1cm}
\caption{Diagrams for deeply-virtual Compton scattering (DVCS) (a) and for photon production via the Bethe-Heitler process (b,c).}
\label{f:dvcsbh}
\vfill
\end{figure}

In deeply virtual Compton scattering, DVCS, the incoming electron emits a virtual photon with mass squared $-Q^2$, which in turn scatters on the proton and is emitted as a real photon:
\begin{eqnarray}
ep  \to e^{\prime}+\gamma^* + p \to e^{\prime} + \gamma + p.
\end{eqnarray}
This process can be regarded as the (quasi) elastic scattering of a virtual photon off the proton, viz. $\gamma^* + p \to \gamma + p$. Figure~\ref{f:dvcsbh} shows the diagram for DVCS together with the diagrams for the Bethe-Heitler contribution, which is purely electromagnetic and leads to the same final state.

\begin{figure}[p]
\begin{center}
\vfill
\vspace*{-2cm}
\hspace*{0.5cm}\epsfig{file=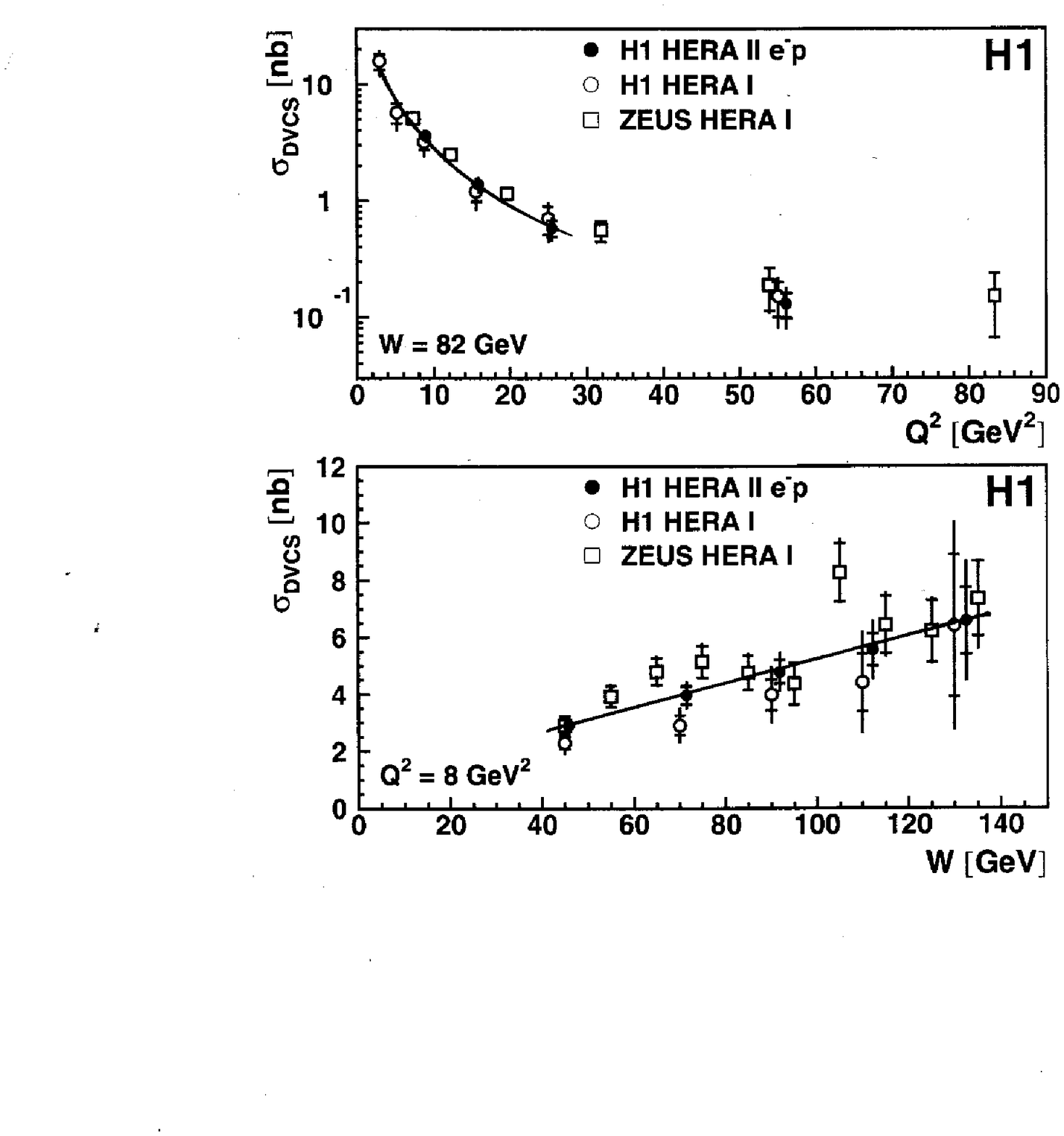,angle=270,width=15cm}
\end{center}
\vspace*{1cm}
\caption{The DVCS cross section as a function of $Q^2$ for $W = 82$\GeV (top), and as a function of $W$ for $Q^2 = 8$\GeV$^2$ (bottom), as measured by H1 and ZEUS. The curves show the calculations by Freund.}
\label{f:dvcsvsqw}
\vfill
\end{figure}

\begin{figure}[p]
\begin{center}
\vfill
\epsfig{file=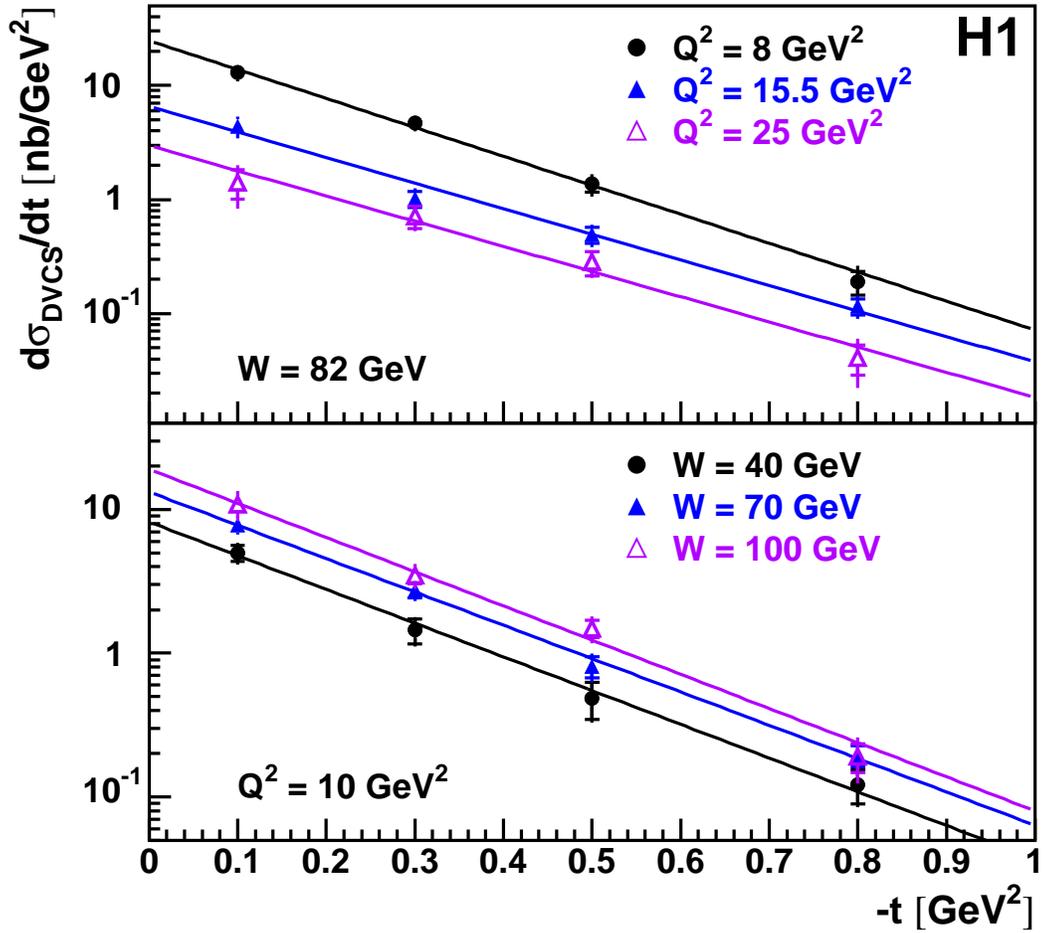,width=14cm}
\end{center}
\vspace*{1cm}
\caption{The DVCS cross section as a function of $t$ for $W = 82$\GeV and $Q^2 = 8,15.5,25$\GeV$^2$ (top), and for $W = 40,80,100$\GeV and $Q^2 = 10$\GeV$^2$ (bottom), as measured by H1.}
\label{f:dvcsvt}
\vfill
\end{figure}

\begin{figure}[p]
\begin{center}
\vfill
\epsfig{file=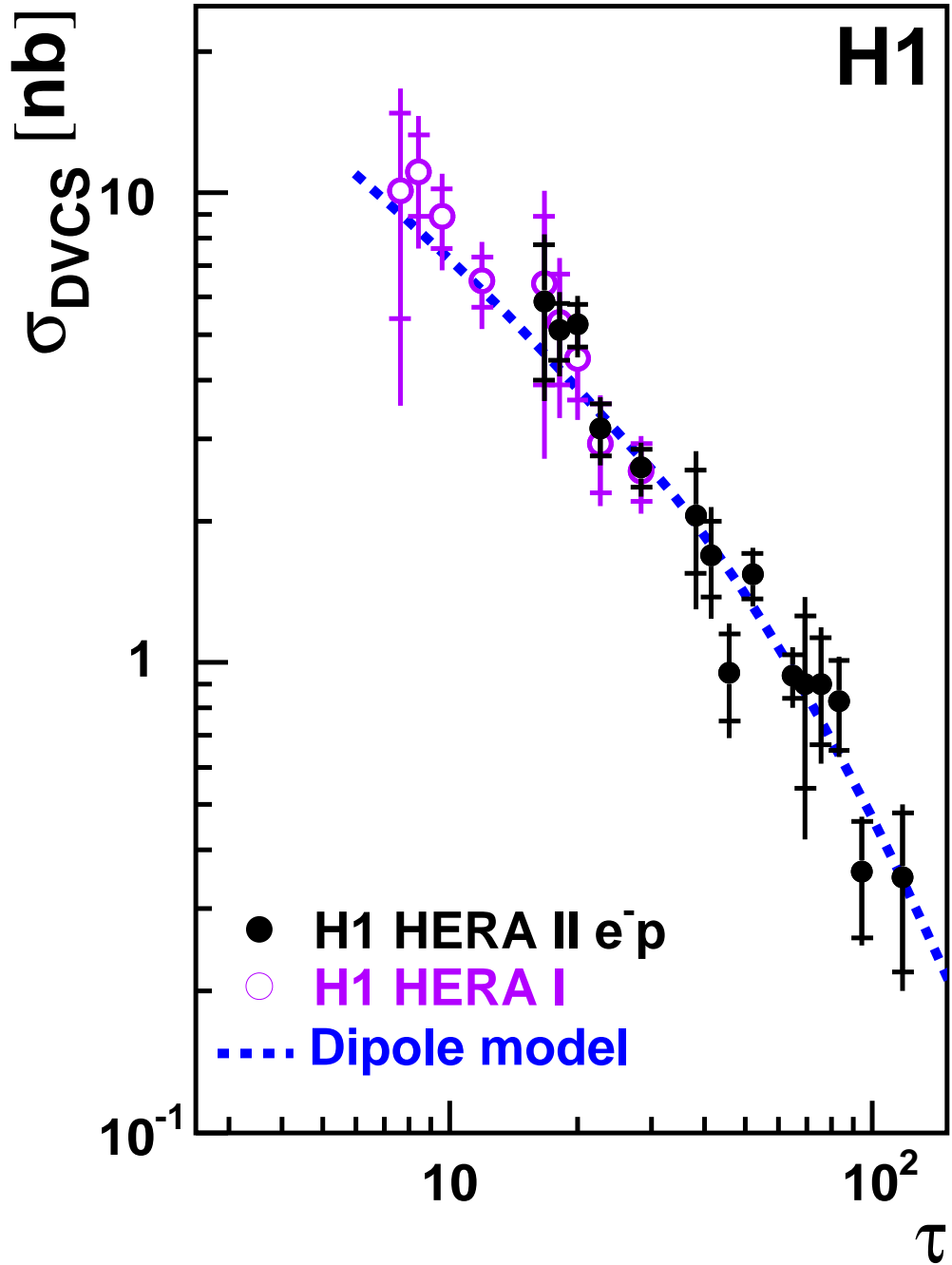,width=14cm}
\end{center}
\vspace*{1cm}
\caption{The DVCS cross section as a function of the scaling variable $\tau = Q^2/Q^2_s(x)$, for $W = 82$\GeV and $Q^2 = 8,15.5,25$\GeV$^2$ (top), and for $W = 40,80,100$\GeV, $Q^2 = 10$\GeV$^2$ (bottom), as measured by H1. The dashed curve shows the prediction of the dipole model, see text.}
\label{f:dvcsvtau}
\vfill
\end{figure}

\begin{figure}[p]
\begin{center}
\vfill
\vspace*{-2cm}
\hspace*{1cm}\epsfig{file=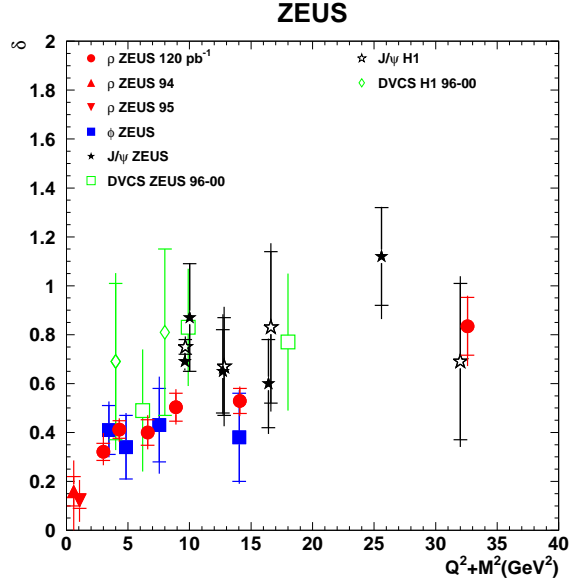,width=8cm}
\end{center} 
\caption{The power $\delta$ determined from a fit of the cross section for $\sigma(\gamma^* p \to Vp) \propto W^{\delta}$ as a function of $(Q^2+M_V^2)$, where $M_V$ is the mass of the vectormeson indicated; for DVCS: $M_V = 0$.  The data stem from H1 and ZEUS.}
\label{f:deltavsqmhz}
\vfill
\end{figure}

Measurements of DVCS at HERA were reported by H1~\cite{pl:b517:47,pl:b659:796} and ZEUS~\cite{pl:b573:46,desy:08:132}. The combined data presented in Fig.~\ref{f:dvcsvsqw} show that for fixed $W = 82$ GeV the DVCS cross section, $\sigma_{DVCS}$, falls rapidly with increasing $Q^2$, while for fixed $Q^2$ the cross section is rising proportional to $W$. For fixed $Q^2$ and $W$, $\sigma_{DVCS}$ is falling exponentially with $|t|$, the four-momentum transfer squared between incoming and outgoing proton: $d\sigma/dt \propto e^{-b|t|}$ (see Fig.~\ref{f:dvcsvt}). A fit to the H1 data at average values of $Q^2 = 8$ GeV$^2$ and $W = 82$ GeV, yielded $b = 6.02 \pm 0.35(stat.) \pm 0.39(syst.)$ GeV$^{-2}$. The slope $b$ is shown below (Fig.~\ref{f:deltavsqmhz}) as a function of ($Q^2+M^2_V$) (where $M^2_V = 0$ for DVCS) together with the $b$-values for the production of vectormesons with mass $M_V$: within errors, the $b$ values from DVCS and from the production of vectormesons lie on a universal curve. The H1 data show that the DVCS cross section scales as a function of $\tau = Q^2/Q^2_s(x)$ where $Q_s(x) = Q_0(x_0/x)^{-\lambda/2}$ with $Q_0 = 1$ GeV, $\lambda = 0.25$ and $x_0 = 2.7 \cdot 10^{-5}$, see Fig.~\ref{f:dvcsvtau}.

The $Q^2$ dependence of the DVCS cross section for fixed $W$, as well as its $W$ dependence for fixed $Q^2$, have been compared with QCD calculations by~\cite{pr:68:09006} which assume that the virtual photon fluctuates into a $q \qbar$ system which in turn interacts with the target proton. The observed increase of the DVCS cross section with $W$ at fixed $Q^2$ is due to the increase of the gluon content of the proton with decreasing Bjorken-x. The QCD calculations are based on twist-2 generalised parton distributions (GPDs) which have been extracted from experimental data. Comparison with the H1 and ZEUS data shows that leading order parton distributions (MRST2001~\cite{epj:c23:73}) lead to a good description of the data, see Fig.~\ref{f:dvcsvsqw}: here the calculated DVCS cross section (curves) is compared with the HERA data for fixed $W = 82$ GeV as a function of $Q^2$, and for fixed $Q^2 = 8$ GeV$^2$ as a function of $W$. 

\section{Production of vectormesons by real and virtual photons}

\subsection{$\rho^0$ production}

\subsubsection{$\gamma p\to \rho^0 p$}
\begin{figure}[p]
\begin{center}
\vfill
\vspace*{-1cm}
\epsfig{file=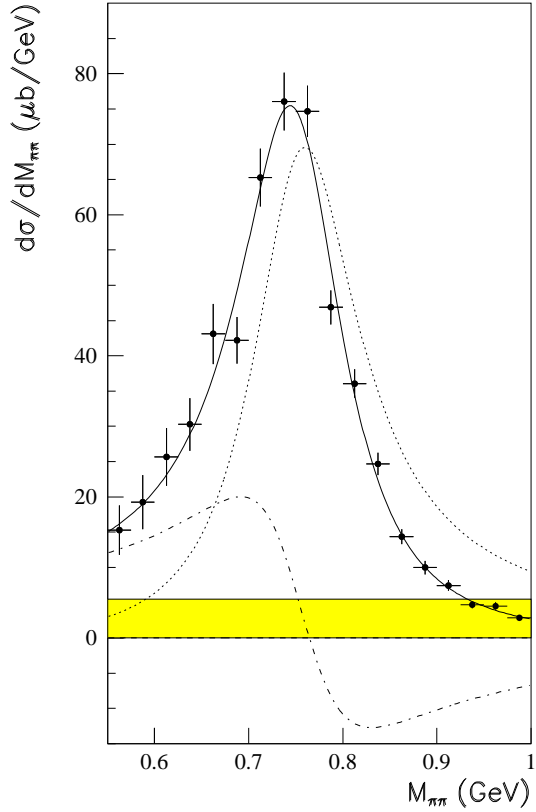,width=7cm}
\epsfig{file=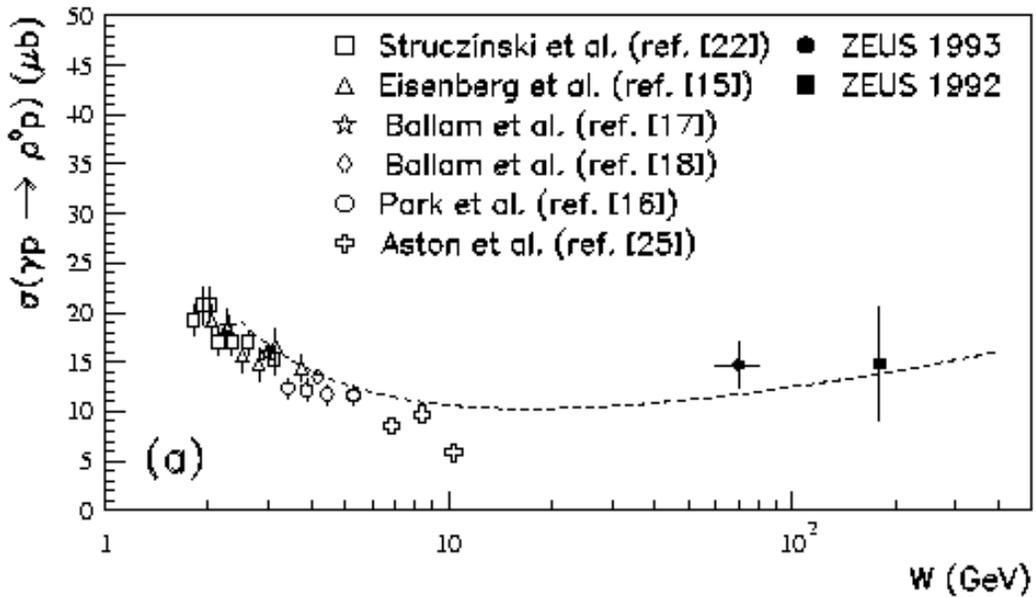,width=14cm}
\end{center}
\caption{First observation of $\gamma p \to \rho^0 p$. (top): $\pi^+ \pi^-$ mass distribution for $60<W<80$\GeV, and $|t| > 0.5$\GeV$^2$. The curves show the results of a fit to the data: the dotted line shows the Breit-Wigner term and the dash-dotted line the contribution from the interference between a Drell-type background and the resonance contribution. (bottom): The integrated cross section $\sigma_{\gamma p \to \rho^0 p}$ as a function of the c.m. energy $W$. Shown are the results from fixed target experiments and from ZEUS.} 
\label{f:frstphotrhoz}
\vfill
\end{figure}

First results from HERA on $\rho^0$ production by real photons, $\gamma p \to \rho^0 p$, were presented in Refs.~\cite{zfp:c63:391,zfp:c69:39}. The second analysis is based on 6000 events of the type $ep \to ep \pi^+ \pi^-$ at an average energy $W=70$ GeV, obtained with an integrated luminosity of $0.55$ pb$^{-1}$. The final state electron and proton were not detected. Figure~\ref{f:frstphotrhoz}(top) shows the $\pi^+ \pi^-$ mass spectrum in terms of the differential cross section $d\sigma/dM_{\pi\pi}$ for events with four-momentum transfer squared between incoming photon an proton, $|t|<0.5$ GeV$^2$. The mass spectrum is skewed compared to a Breit-Wigner distribution which can be understood in terms of the interference between the resonant $\pi^+ \pi^-$ production and a non-resonant ($\gamma p \to \pi^+ \pi^- p$) Drell-type background~\cite{prl:5:278}, as explained by S\"oding~\cite{pl:19:702}. 

\begin{figure}[p]
\begin{center}
\vfill
\vspace*{-1cm}
\hspace*{1cm}\epsfig{file=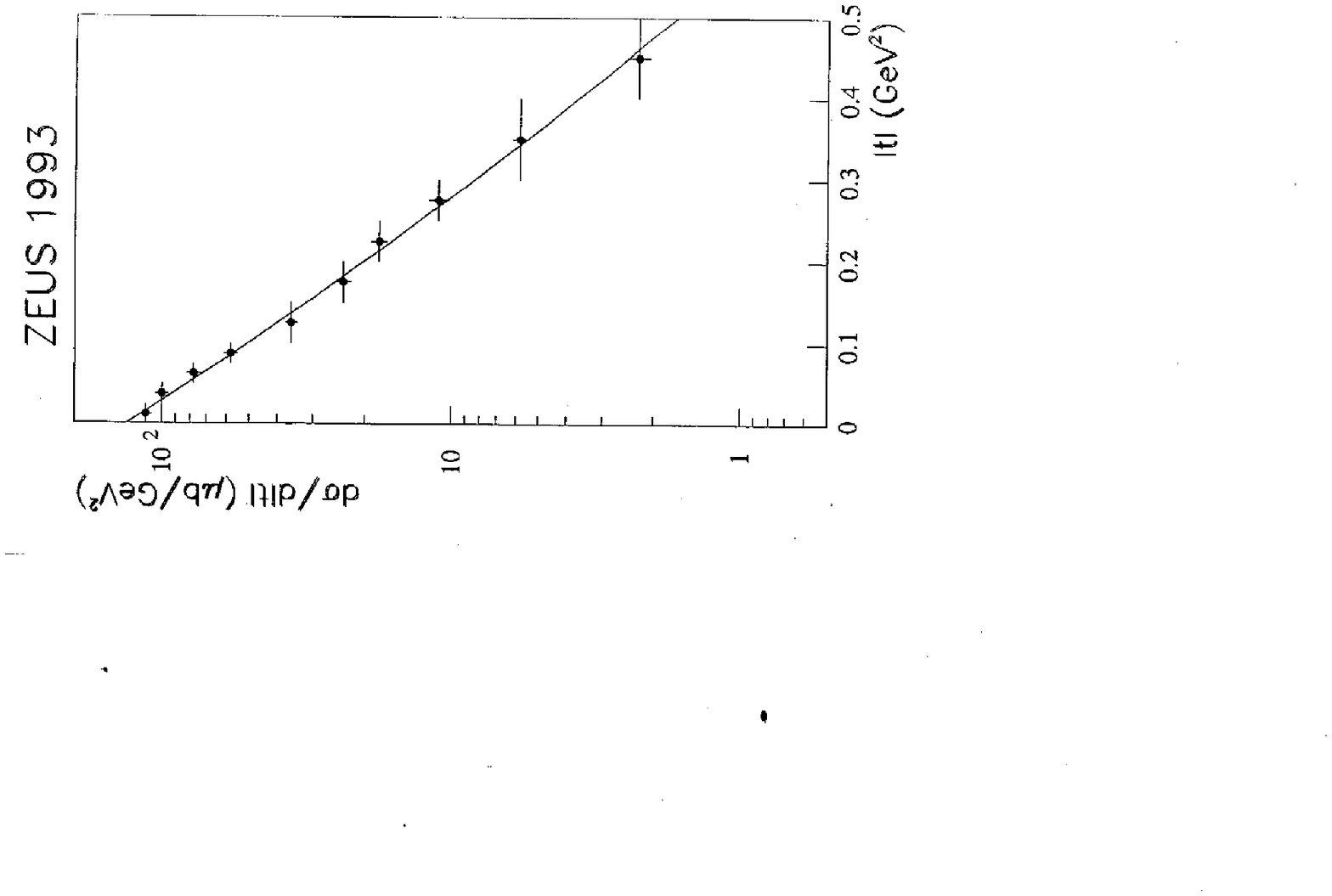,angle=270,width=6cm}
\epsfig{file=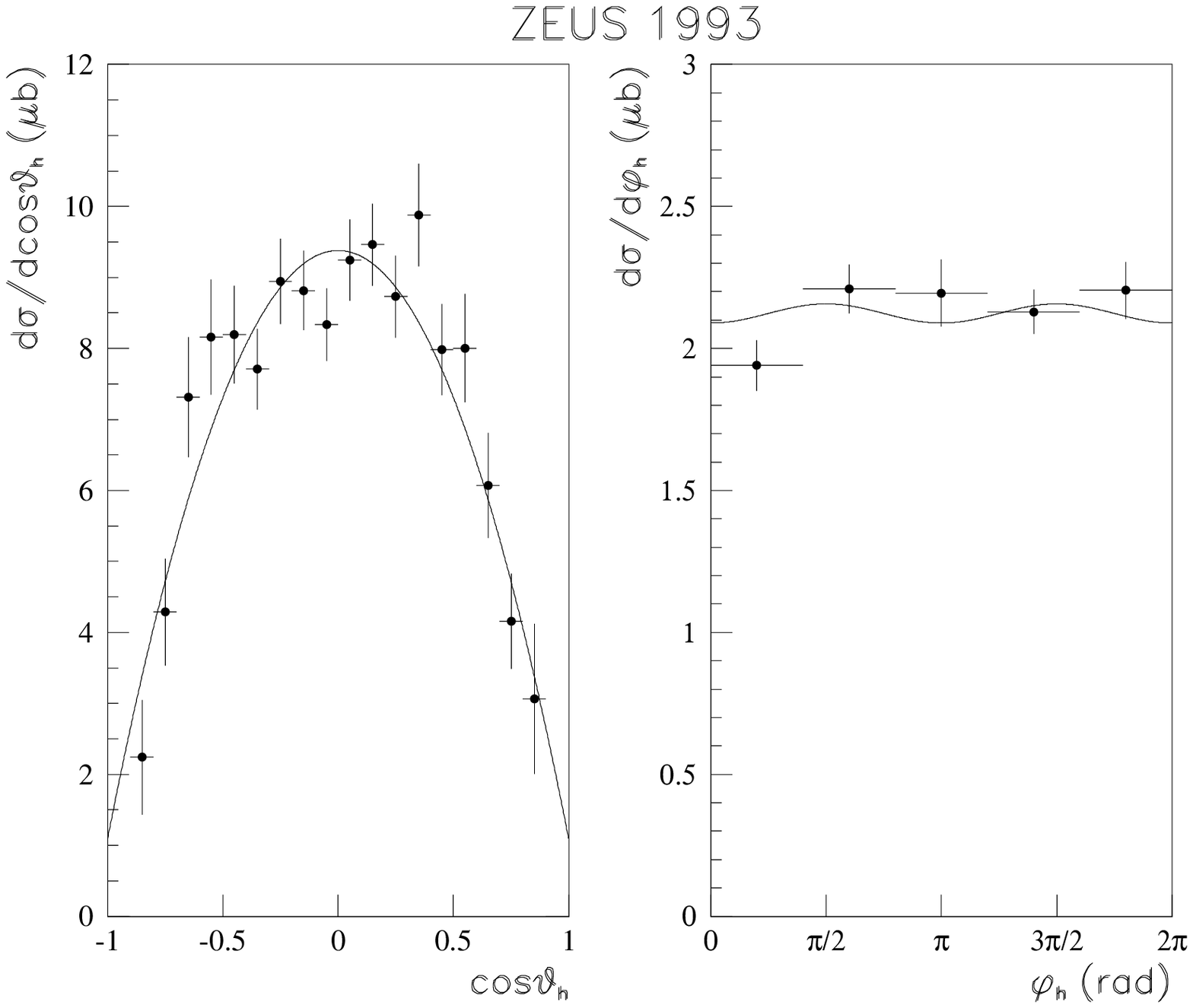,width=12cm}
\end{center} 
\caption{(top): The differential cross section $d\sigma/d|t|$ for $\gamma p \to \rho^0 p$. The line shows the result of the fit $d\sigma/d|t| = d\sigma/d|t|(0) \cdot e^{-b|t|+ct^2}$. (bottom): The cross section for $\gamma p \to \rho^0 p$ in terms of the $\rho^0$ decay angles $\theta$ and $\phi$ measured in the s-channel helicity frame; from ZEUS.}
\label{f:photrhodsdth}
\vfill
\end{figure}

The $W$ dependence of the $\rho^0$ cross section is shown in Fig.~\ref{f:frstphotrhoz}(bottom) which includes measurements from fixed-target experiments~\cite{np:b108:45,pr:d5:15,pr:d5:545,np:b36:404,pr:d7:3150,np:b209:56}. The ZEUS data indicate a substantial rise of $\sigma_{\gamma p \to \rho^0 p}$ for $W > 60$ GeV. The differential cross section, $d\sigma/d|t|$ - see Fig.~\ref{f:photrhodsdth} (top) - is proportional to $e^{-b|t|}$ with $b = 10.4 \pm 0.6(stat.) \pm 1.1(syst.)$ GeV$^{-2}$ for $|t| <0.15$ GeV$^{-2}$. In a measurement of $\rho^0$ production by ZEUS, where the four-momentum of the scattered proton was determined directly, within errors the same value was obtained, viz. $b = 9.8 \pm 0.8(stat.) \pm 1.1(syst.)$ GeV$^{-2}$~\cite{zfp:c73:253}. A similar behaviour of  $d\sigma/d|t|$ was observed by H1~\cite{np:b463:3} where $b = 10.9 \pm 2.4(stat.) \pm 1.1(syst.)$ GeV$^{-2}$ was obtained. The value of the slope $b$ is comparable to what is found for elastic $\pi p$ scattering.  

The decay angular distribution of the $\rho^0$ mesons has been used to determine the orientation of the $\rho^0$ spin~\cite{schillseybwolf,schillwolf}. The exclusive electroproduction of $\rho^0$ mesons and their decay into $\pi^+\pi^-$ is described by three angles: $\Phi_h$ - the angle between the $\rho^0$ production plane and the electron scattering plane in the $\gamma^*p$ c.m. system, and $\theta_h, \phi_h$ - the polar and azimuthal angles of the $\pi^+$ w.r.t. the proton direction, all determined in the $\rho^0$ rest system. The $\rho^0$ decay angular distribution, $W(\rm{cos}\theta_h,\phi_h,\Phi_h)$, can be parametrised by the $\rho^0$ spin-density matrix elements, $\rho^\alpha_{i,k}$, with i,k = -1,0,1, and $\alpha = 0,1,.,6$. The $\rho^0$ s-channel helicity decay angular distribution w.r.t. $\theta_h$ and $\phi_h$ can be written as follows:
\begin{eqnarray}
\frac{1}{N} \frac{dN}{d{\rm cos}\theta_h} & = & \frac{3}{4}[1-r^{04}_{00} + (3r^{04}_{00} -1) \cdot {\rm cos}^2\theta_h]\\
\frac{1}{N} \frac{dN}{d\phi_h} & = & \frac{1}{2 \pi}[1-2r^{04}_{1-1} {\rm cos} 2\phi_h]
\end{eqnarray}
where the density matrix element $r^{04}_{00}$  represents the probability that the $\rho^0$ is produced longitudinally polarised. 

The distributions of ${\rm cos}\theta_h$ and $\phi_h$ for $\rho^0$ production by $ real$ photons~\cite{zfp:c69:39} are shown in Fig.~\ref{f:photrhodsdth}. The dominant contribution is proportional to ${\rm sin}^2 \theta_h$ which shows that the $\rho^0$ mesons are mostly transversely polarised, as verified by the small value of $r^{04}_{00} = 0.055\pm 0.028$. The $\phi_h$ distribution is constant, yielding a value of $r^{04}_{1-1} = 0.008\pm 0.014$. The values of $r^{04}_{00}$ and $r^{04}_{1-1}$ are compatible with zero as expected for s-channel helicity conservation (SCHC): the $\rho^0$ in the final state has the same helicity as the photon in the initial state.

\subsubsection{$\gamma^* p\to \rho^0 p$}

\begin{figure}[p]
\begin{center}
\vfill
\vspace*{-2cm}
\epsfig{file=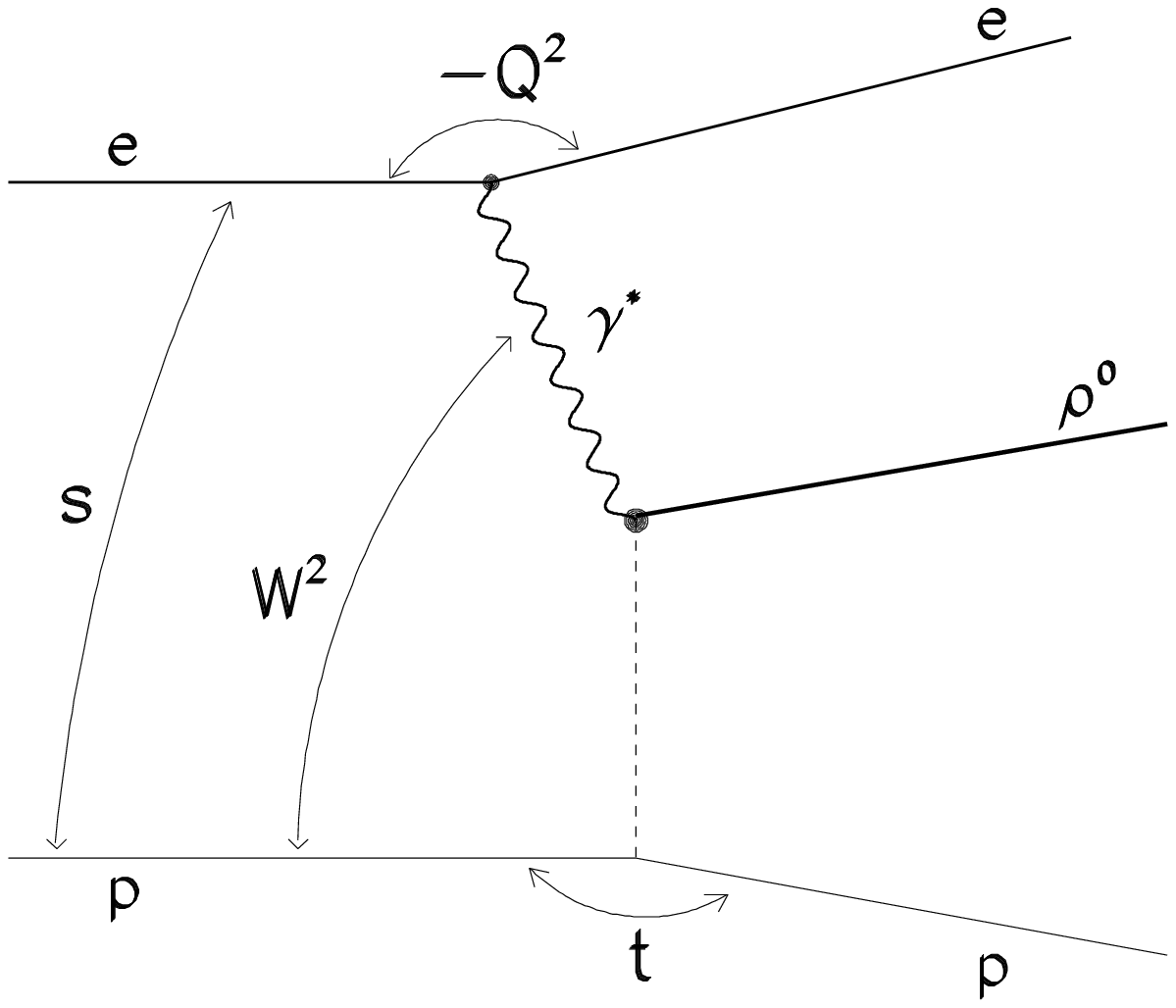,width=8cm}
\end{center}
\caption{Diagram for $ep \to e \rho^0 p$}
\label{f:diaeptrhop}
\end{figure}

\begin{figure}[p]
\begin{center}
\vfill
\vspace*{-2cm}
\epsfig{file=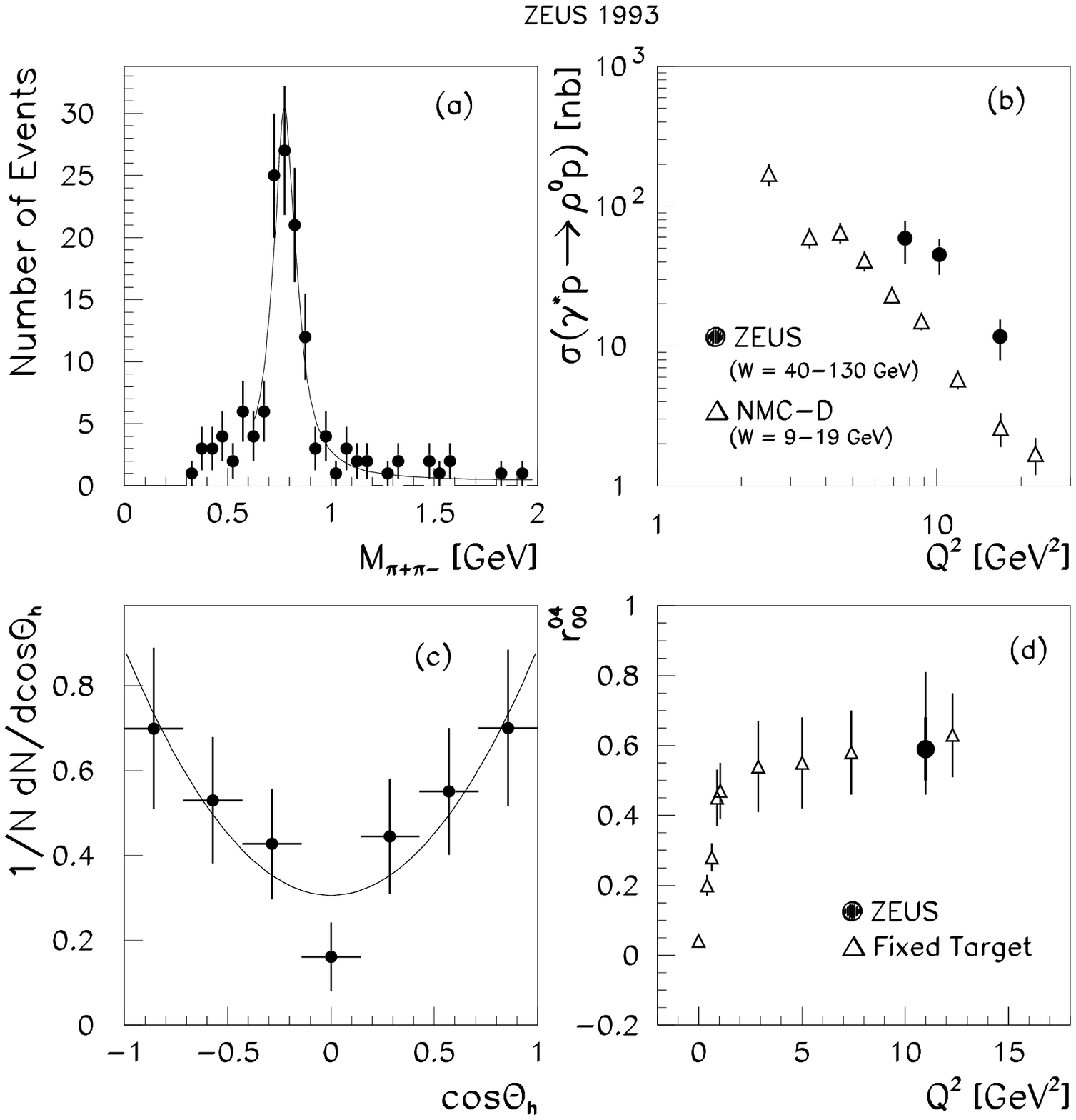,width=10cm}
\epsfig{file=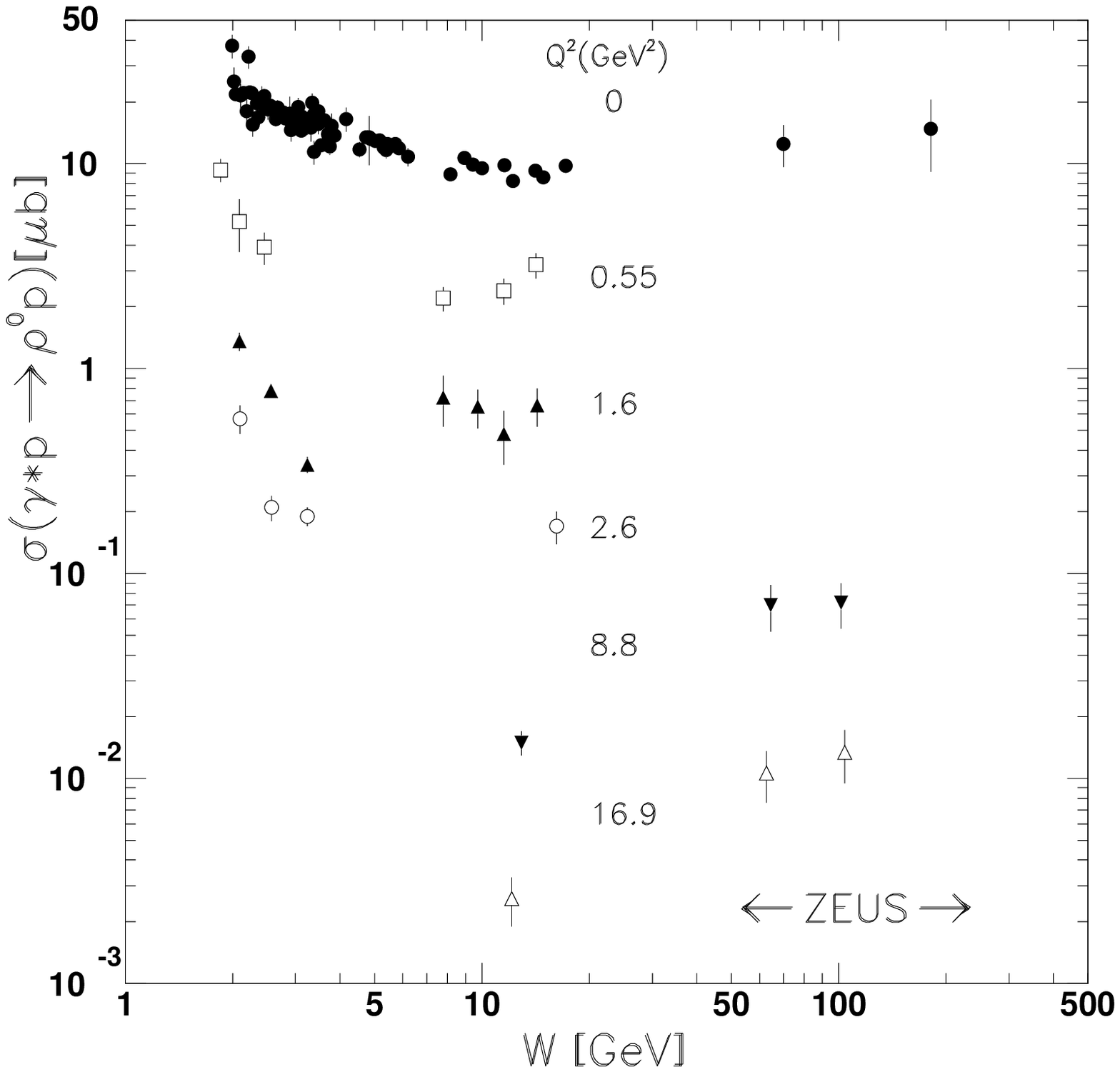,width=10cm}
\end{center} 
\caption{First observation of $\gamma^*p \to \rho^0 p$ for $30<W<130$\GeV, $7<Q^2<25$\GeV$^2$, as measured by ZEUS. (top): (a) $\pi^+ \pi^-$ mass distribution; (b) the cross section for $\gamma^* p \to \rho^0 p$ as a function of $Q^2$: solid points from ZEUS; open triangles: results for $9< W < 19$\GeV from the fixed target experiment NMC-D; (c) the $cos \theta_h$ distribution for the decay $\pi^+$, in the s-channel helicity system; (d) the $\rho^0$ density matrix element $r^{04}_{00}$ compared with results from fixed target experiments. (bottom): The cross section for $\gamma^* p \to \rho^0 p$ as a function of $W$ for different different values of $Q^2$ from fixed target experiments and from ZEUS.}
\label{f:frstdisrhoz}
\vfill
\end{figure}

The first observation at HERA of $\rho^0$ production by $virtual$ photons, $\gamma^*p \to \rho^0 p$ (see Fig.~\ref{f:diaeptrhop}) leading to the final state $\pi^+ \pi^- p$ ~\cite{pl:b356:601}, is shown in Fig.~\ref{f:frstdisrhoz}. The $M_{\pi^+ \pi^-}$ mass spectrum is completely dominated by the production of $\rho$ mesons. The cross section for $\gamma^* p \to \rho^0 p$ shows a fast decrease with $Q^2$, viz. $\sigma(\gamma^*p \to \rho^0 p) \propto Q^{-4.2 \pm 0.8^{+1.4}_{-0.5}}$. For fixed $Q^2 >8$ GeV$^2$, the cross section rises by about a factor of five from $W = 12$ GeV to $ W = 60$ GeV. 

The distribution of the polar angle for the decay $\rho^0 \to \pi^+ \pi^-$ is shown in Fig.~\ref{f:frstdisrhoz}c. The dominant ${\rm cos}^2 \theta$-type distribution shows that the $\rho$-mesons are mostly longitudinally aligned.
Further results on $\gamma (\gamma^*) p \to \rho^0 p$ from HERA were reported in Refs.~\cite{np:b468:3,zfp:c75:607,epj:c6:603}.

A first precise measurement of the helicity amplitudes for $\gamma^* p \to \rho^0 p$ at HERA was provided by H1~\cite{epj:c13:371}. It showed that the helicity is conserved in the s-channel except for a small but nonzero helicity flip amplitude at the level of $8\pm 3\%$ of the nonflip amplitudes. This is evidence for a (small) contribution of longitudinal $\rho^0$ mesons produced by transverse photons. The finding was corroborated by results from ZEUS~\cite{epj:c07:393}. For comparison, in a fixed target experiment at low $W = 1.7 - 2.8$ GeV and low $Q^2 = 0.3 - 1.4$ GeV$^2$, the ratio of single flip to nonflip amplitudes had been found to be of order 15-20$\%$~\cite{joos1976}.

\begin{figure}[p]
\begin{center}
\vfill
\vspace*{-1cm}
\hspace*{1cm}\epsfig{file=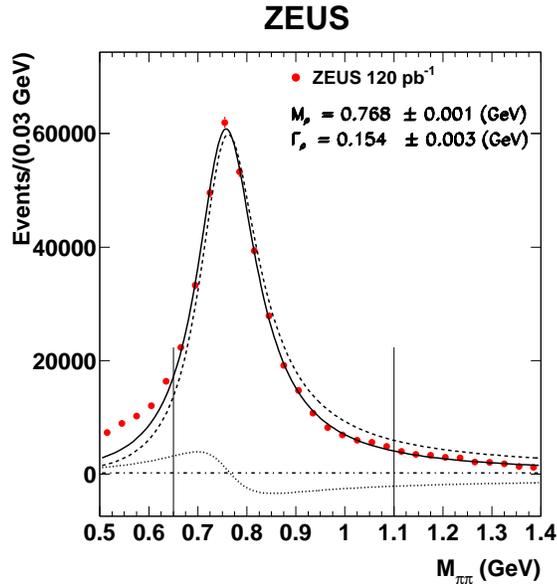,width=8cm}
\end{center} 
\caption{The $\pi^+ \pi^-$ acceptance-corrected invariant-mass distribution. The line represents a fit of the S\"oding form for $0.65<M_{\pi^+\pi^-}<1.1$\GeV. The vertical lines indicate the mass range used for the analysis, from ZEUS.}
\label{f:mpipifz}
\vfill
\end{figure}

\begin{figure}[p]
\begin{center}
\vfill
\vspace*{-1cm}
\hspace*{1cm}\epsfig{file=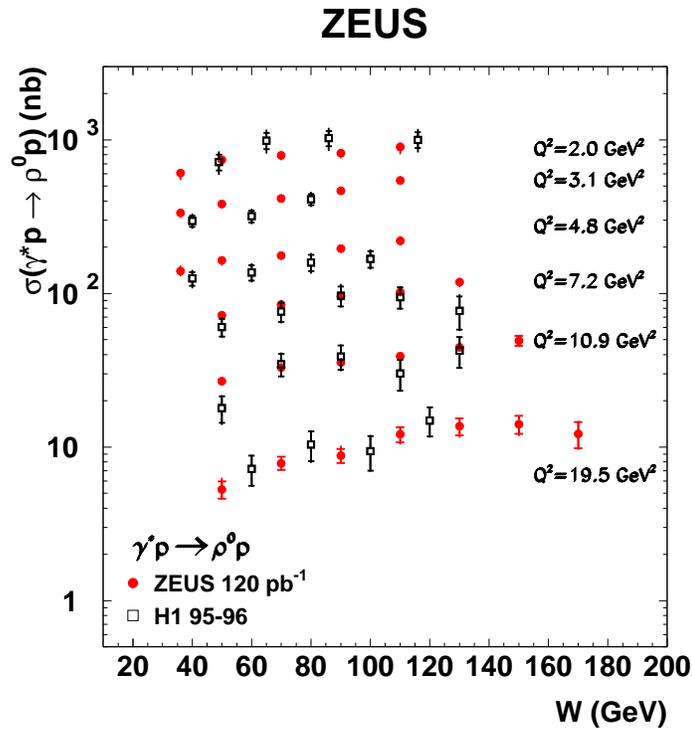,width=10cm}
\end{center} 
\caption{The cross section for $\gamma^* p \to \rho^0 p$ as a function of $W$ for different values of $Q^2$ from H1 and ZEUS.}
\label{f:sigdisrhovswhz}
\vfill
\end{figure}

An extended study of $\gamma^* p \to \rho^0 p$ was presented by ZEUS~\cite{pmc:a1:6}, based on an integrated luminosity of 120 pb$^{-1}$. The $\pi^+ \pi^-$ mass distribution (Fig.~\ref{f:mpipifz}) is dominated by $\rho^0$ production\footnote{The skewing of the mass distribution is explained by the S\"oding model, see above. The excess of events observed at masses below 0.65 GeV is due to background from $\omega \to \pi^+ \pi^- \pi^0$, where the $\pi^0$ was not detected, and from $\phi \to K^+ K^-$, where the $K's$ were misidentified as $\pi's$.}. The cross section for $\gamma^* p \to \rho^0 p$, as measured by H1 and ZEUS, is presented in Fig.~\ref{f:sigdisrhovswhz} as a function of $W$ for $Q^2$ values between 2 and 19.5 GeV$^2$. For fixed $W$, $\sigma_{\gamma^* p \to \rho^0 p}$ is falling rapidly with $Q^2$. For fixed $Q^2$, the cross section rises with $W$. A parametrisation of the form $\sigma_{\gamma^* p \to \rho^0 p} \propto W^{\delta}$ yielded the $\delta$ values shown in Fig.~\ref{f:deltavsqmhz} as a function of $(Q^2 + M^2_V)$: the power $\delta$ rises from around $0.15$ at $Q^2\approx 0$ to about $0.8$ at $(Q^2 + M^2_V) = 30$ GeV$^2$.

\begin{figure}[p]
\begin{center}
\vfill
\vspace*{-1cm}
\hspace*{1cm}\epsfig{file=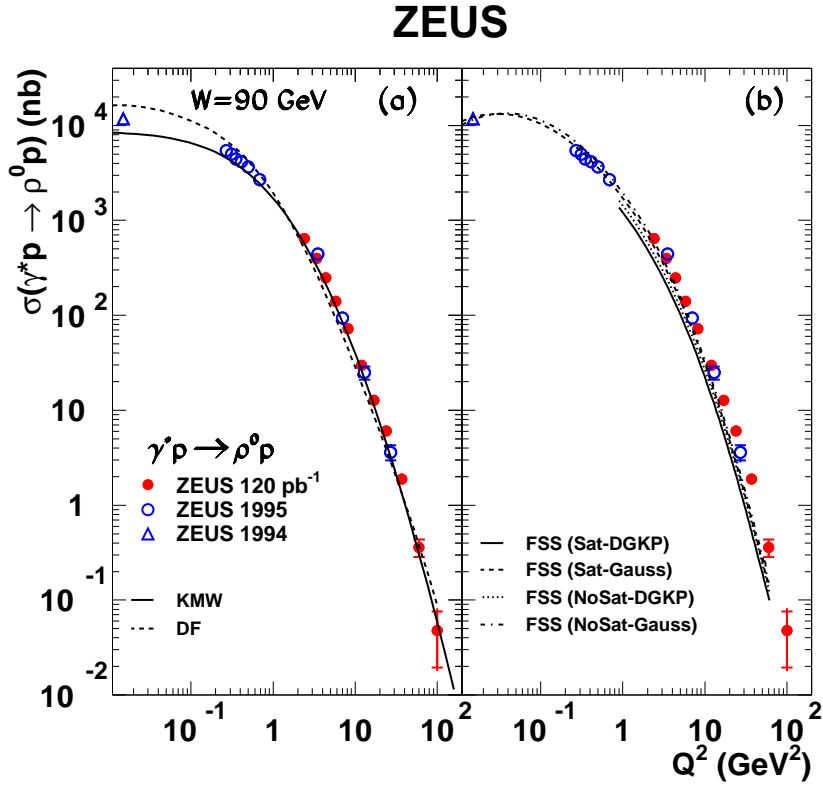,width=12cm}
\end{center} 
\caption{The cross section for $\gamma^* p \to \rho^0 p$ as a function of $Q^2$ for $W = 90$\GeV, from ZEUS; the data presented in (a) and (b) are the same. Shown are also the predictions of several models, see text.}
\label{f:sigdisrhovsq2z}
\vfill
\end{figure}

The $Q^2$ dependence of $\sigma(\gamma^* p \to \rho^0 p)$ is shown in detail~\cite{pmc:a1:6} in Fig.~\ref{f:sigdisrhovsq2z} for $W = 90$ GeV. There is an almost exponential decrease from $\sigma \approx 700$ nb at $Q^2 = 2$ GeV$^2$ to $0.04$ nb at $Q^2 = 100$ GeV$^2$. The measurements were compared with several models which are based on the dipole representation of the virtual photon: the photon fluctuates into a $q \qbar$ pair - the colour dipole - which in turn interacts with the gluon cloud of the proton to produce the $\rho^0$, see e.g.~\cite{kowtean}. For $Q^2 > 1$ GeV$^2$, the calculations of~\cite{komowa}~(KMW) and~\cite{dofe}~(DF) give a good description of the data while those of FSS~\cite{fosash04} are somewhat low for $Q^2$ above $10$ GeV$^2$.   

\begin{figure}[p]
\begin{center}
\vfill
\vspace*{-1cm}
\hspace*{1cm}\epsfig{file=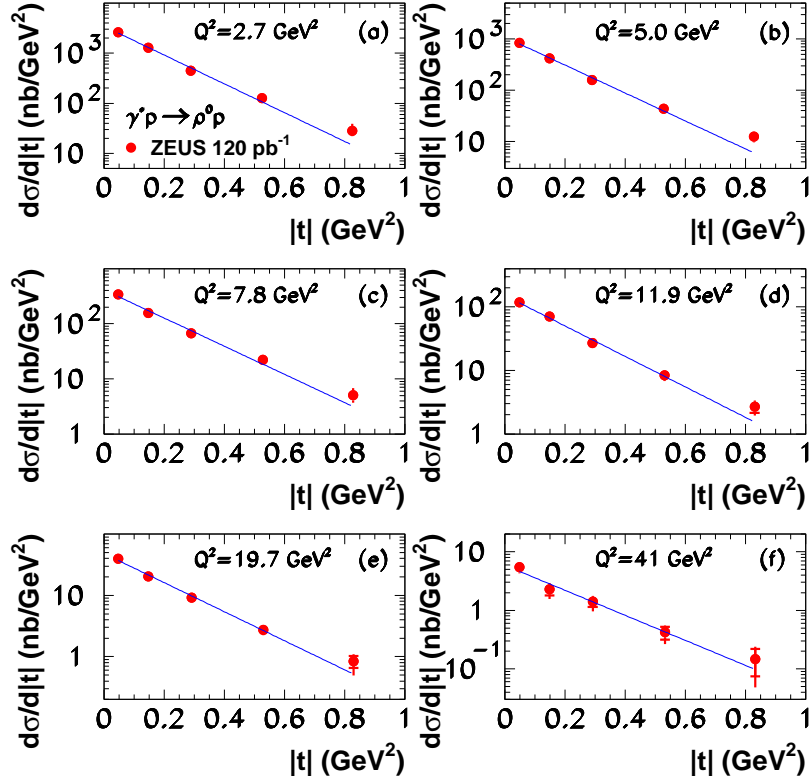,width=12cm}
\end{center} 
\caption{The differential cross section $d\sigma/d|t|$ for $\gamma^* p \to \rho^0 p$ as a function of $|t|$ for $W = 90$\GeV and fixed values of $Q^2$. The lines represent exponential fits to the data. The data stem from ZEUS.}
\label{f:dsdtvstz}
\vfill  
\end{figure}

\begin{figure}[p]
\begin{center}
\vfill
\vspace*{-0.4cm}
\hspace*{1cm}\epsfig{file=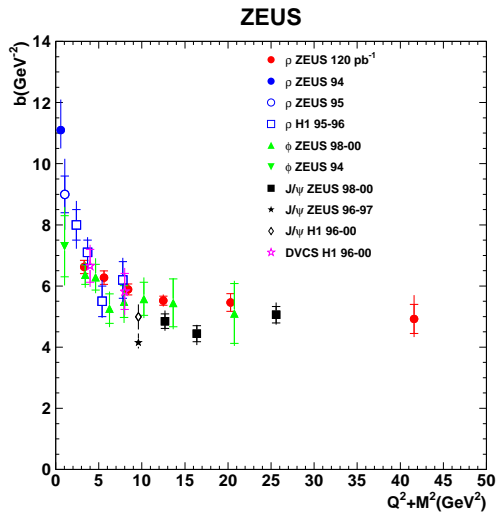,width=7cm}
\end{center} 
\caption{The slope $b$ describing the $t$ dependence of vectormeson production by the form $\gamma^* p \to Vp \propto e^{-b \cdot |t|}$ as a function of $(Q^2+M_V^2)$ where $M_V$ is the mass of the vectormeson indicated; for DVCS $M_V = 0$.  The data stem from H1 and ZEUS.}
\label{f:bvsqmhz}
\vfill
\end{figure}

As a function of $|t|$, the differential cross section $d\sigma/d|t|$ at $W = 90$ GeV is falling exponentially, $d\sigma/d|t| \propto e^{-b|t|}$~\cite{pmc:a1:6}, see Fig.~\ref{f:dsdtvstz}. The dependence of $b$ on $(Q^2 + M_V^2)$, where $M_V$ is the mass of the vectormeson, is shown in Fig.~\ref{f:bvsqmhz}. For $(Q^2 + M_V^2) \le 1$ GeV$^2$, $b$ is of the order of $10 - 12$ GeV$^{-2}$, similar to what has been measured for elastic $pp$ scattering at $W = 31 - 62$ GeV, viz. $b \approx 11$ GeV$^{-2}$~\cite{np:b248:253}. At larger values of $(Q^2 + M_V^2)$ the slope $b$ decreases rapidly approaching a constant value of $b \approx 5$ GeV$^{-2}$ for $(Q^2 + M_V^2) > 10$ GeV$^{-2}$. In an optical model, $b$ is proportional to the sum of the radii squared of the virtual photon and the proton, $b = [(R_{\gamma*})^2+(R_{p})^2]/4$. As $Q^2$ increases, the transverse extension of the virtual photon expected to be proportional to $1/Q$, goes to zero, such that $b \to (R_{p})^2/4$.

The $W$ dependence of $d\sigma/d|t|$ at fixed values of $|t|$ was used to determine the parameters of the Pomeron trajectory, $\alpha_{\pom} = \alpha_{\pom}(0) + \alpha_{\pom}^{'}*|t|$, with the following result~\cite{pmc:a1:6}:\\
\\
$<Q^2> = 3$ GeV$^2$:\\ 
$\alpha_{\pom}(0)  =  1.113\pm0.010(stat)^{+0.009}_{-0.012}(syst) $, \hspace*{0.5cm}$\alpha_{\pom}^{'} =  0.185\pm 0.042(stat)^{+0.022}_{-0.057}(syst) $ GeV$^{-2}$\\
$<Q^2> = 10$ GeV$^2$:\\ 
$\alpha_{\pom}(0)  =  1.152\pm0.011(stat)^{+0.006}_{-0.008}(syst) $, \hspace*{0.5cm}$\alpha_{\pom}^{'} =  0.114\pm0.043(stat)^{+0.026}_{-0.024}(syst) $ GeV$^{-2}$\\
\\  
The data suggest a small increase of $\alpha_{\pom}(0)$ with increasing $Q^2$.

\begin{figure}[p]
\begin{center}
\vfill
\vspace*{-1cm}
\hspace*{-0.4cm}\epsfig{file=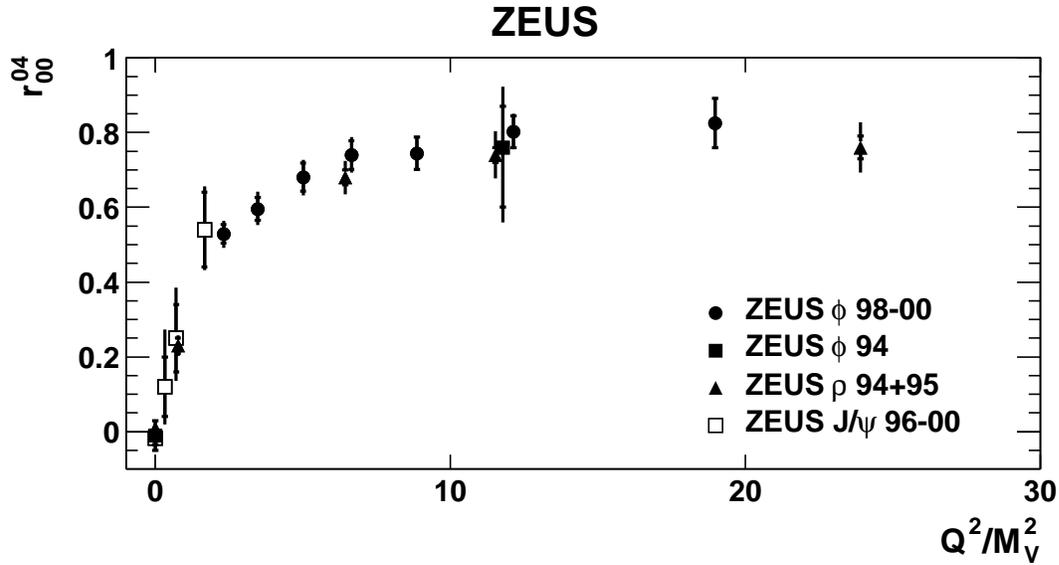,width=16cm}
\end{center} 
\caption{Reaction $\gamma^* p \to V p$: the density matrix element $r^{04}_{00}$ as a function of $Q^2/M^2_V$ for $V = \rho^0, \phi, J/\Psi$ (bottom); from ZEUS.}
\label{f:gsvpr04z}
\vfill
\end{figure}

Figure~\ref{f:gsvpr04z} summarises the measurements of $r^{04}_{00}$ for $\rho^0$, $\phi$ and $J/\Psi$ production as a function of $Q^2/M_V^2$. The same behaviour is observed for all three vectormeson species: $r^{04}_{00}$ rises sharply from zero at $Q^2 = 0$ to a value of $0.7 - 0.8$ for $Q^2 > 5$ GeV$^2$. This shows that at large $Q^2$ the vectormesons are predominantly longitudinally polarised. Under the assumption of SCHC their production is dominated by longitudinal photons.

\begin{figure}[p]
\begin{center}
\vfill
\vspace*{-0.8cm}
\hspace*{1cm}\epsfig{file=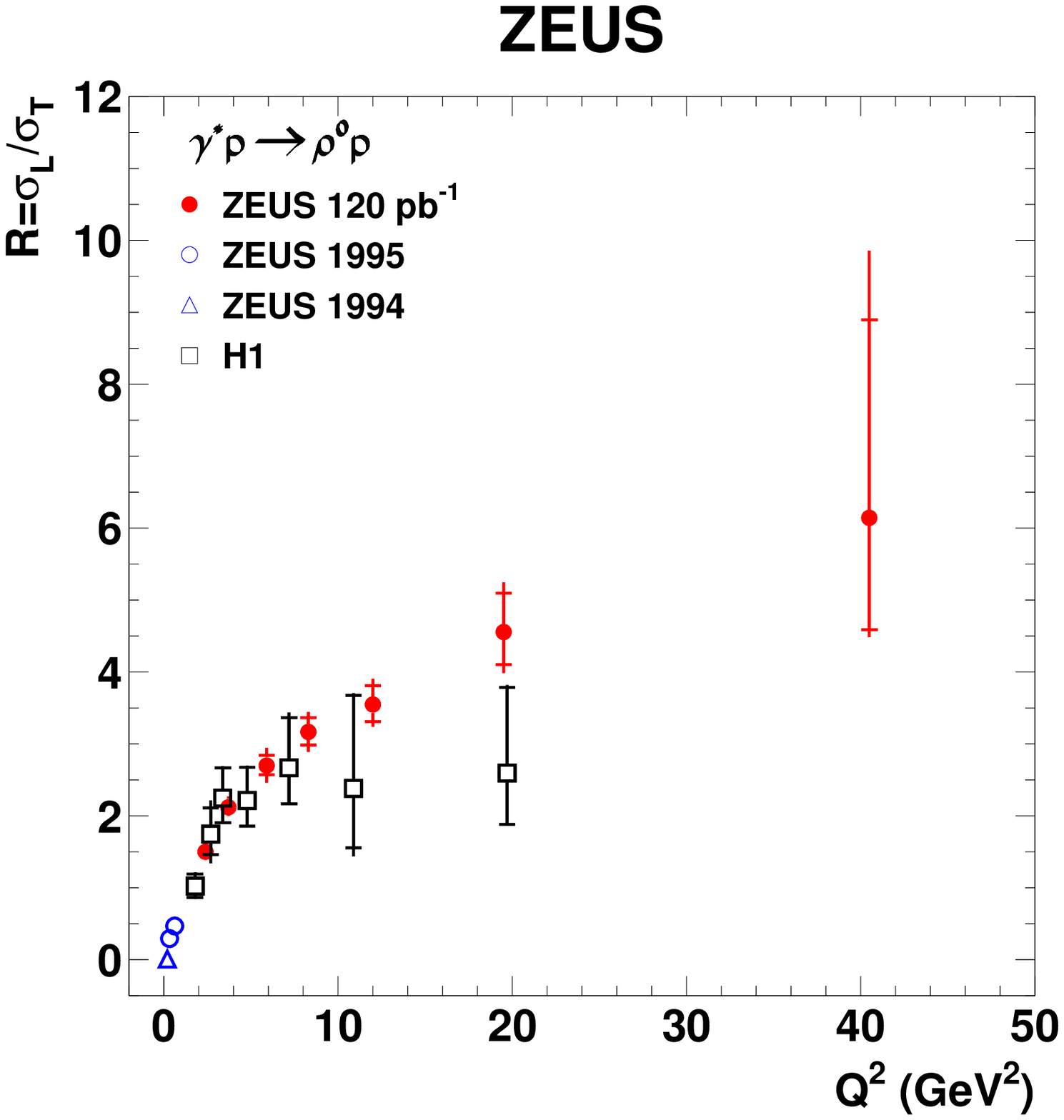,width=8cm}
\end{center} 
\caption{For $\gamma^* p \to \rho^0 p$ the ratio $R = \sigma_L/\sigma_T$ as a function of $Q^2$; the data from H1 were determined for $W = 75$\GeV, those from ZEUS for $W = 90$\GeV.}
\label{f:rsltvsqhz}
\vfill
\end{figure}

\begin{figure}[p]
\begin{center}
\vfill
\vspace*{-0.8cm}
\hspace*{1cm}\epsfig{file=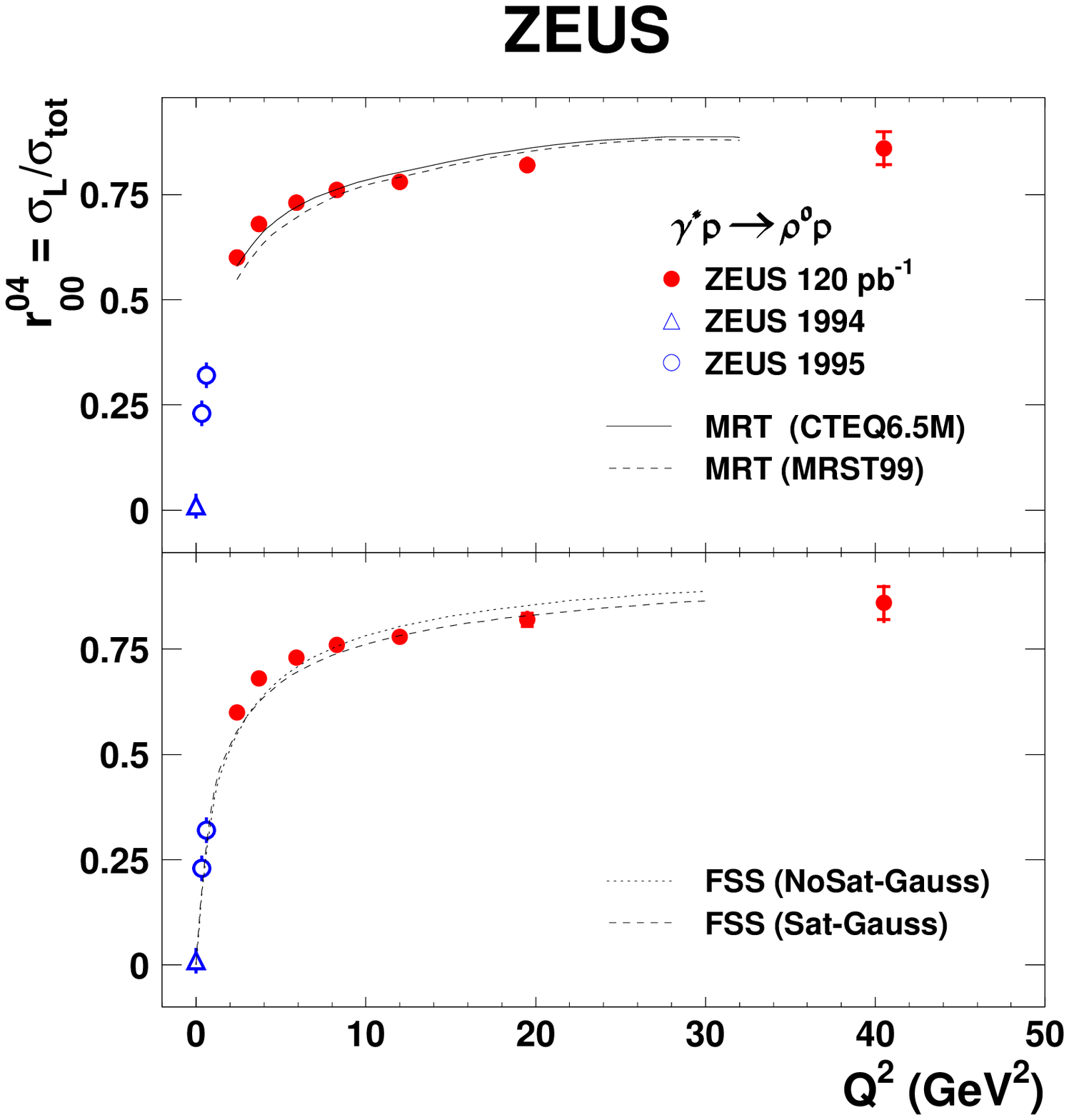,width=12cm}
\end{center} 
\caption{For $\gamma^* p \to \rho^0 p$ the ratio $R = \sigma_L/\sigma_{tot}$ as a function of $Q^2$; the data from H1 were determined for $W = 75$\GeV, those from ZEUS for $W = 90$\GeV.}
\label{f:rsltotvsqhz}
\vfill
\end{figure}

The ratio $R = \sigma_L / \sigma_T$ for $\gamma^* p \to \rho^0 p$  is shown in Fig.~\ref{f:rsltvsqhz}: $R$ rises rapidly with $Q^2$, reaching unity at $Q^2 \approx 2$ GeV$^2$ and values around $4$ at $Q^2$ = 20 GeV$^2$: for $Q^2 > 2$ GeV$^2$ the dominant contribution to $\gamma^* p \to \rho^0 p$ comes from longitudinal photons. The ratio of the contribution from longitudinal photons to the total cross section for $\rho^0$ production - as given by the density matrix element $r^{04}_{00}$ (see Fig.~\ref{f:rsltotvsqhz}) - illustrates the preponderance of longitudinal photon contributions at large $Q^2$. The measurements are well described by the models of~\cite{fosash04} and~\cite{marrysteu97}, which are based on two-gluon exchange.
 
\begin{figure}[p]
\begin{center}
\vfill
\vspace*{-0.7cm}
\hspace*{0.5cm}\epsfig{file=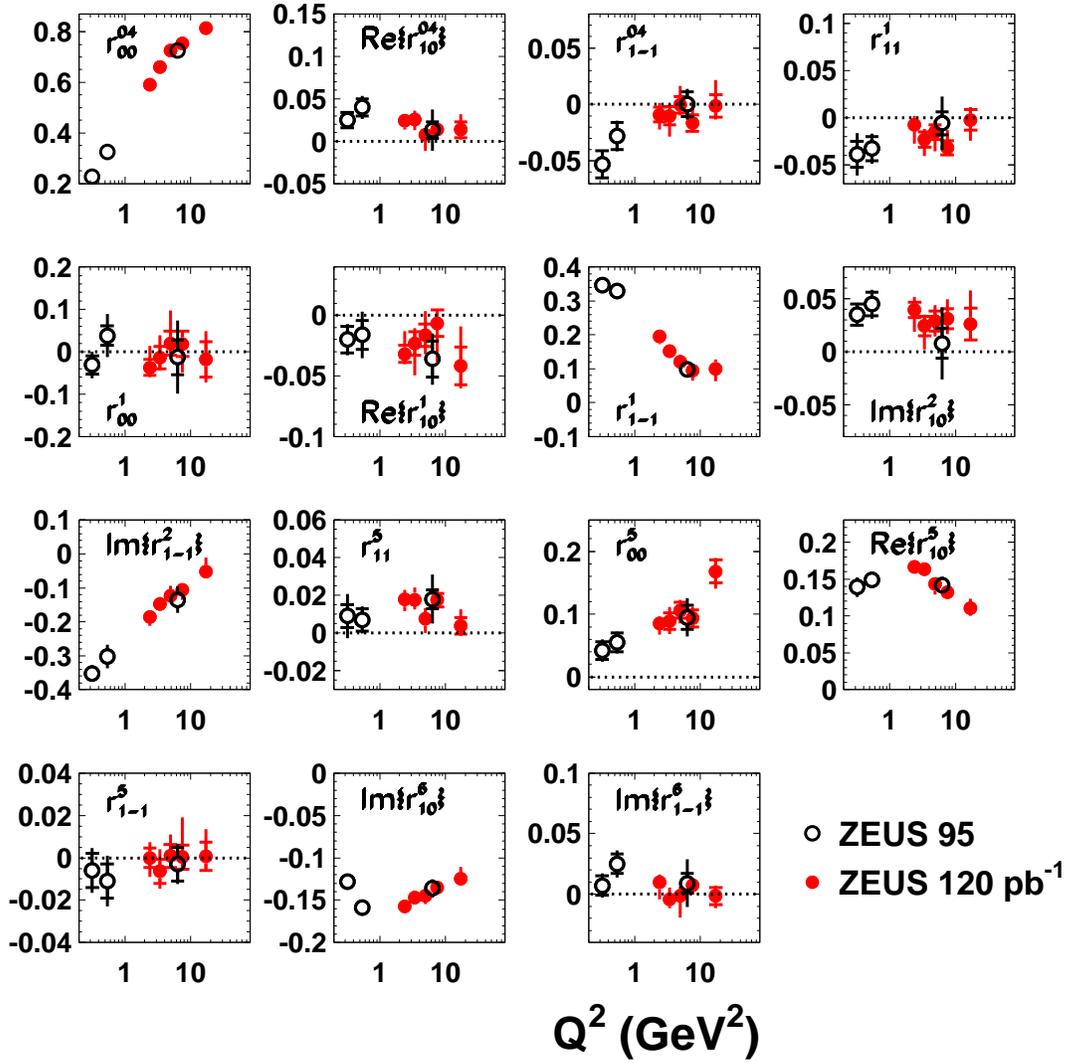,width=16cm}
\end{center} 
\caption{The density matrix elements $\rho^{\alpha}_{ik}$ for $\rho^0$ mesons produced in the reaction $\gamma^*p \to \rho^0 p$, as a function of $Q^2$; from ZEUS.}
\label{f:rhoikvsqz}
\vfill
\end{figure}

The complete set of densitiy matrix elements $\rho^{\alpha}_{ik}$ for $\gamma^* p \to \rho^0 p$ is shown in Fig.\ref{f:rhoikvsqz} as a function of $Q^2$. A strong dependence on $Q^2$ is observed for $r^{04}_{00}$, $r^{1}_{1-1}$, ${\rm Im} r^2_{1-1}$ and $r^{5}_{00}$. Under the assumption of SCHC, the dependence on $Q^2$ is driven by the dependence of $R = \sigma_L / \sigma_T$ on $Q^2$, i.e. by the rapid increase of the contribution from longitudinal relative to that from transverse photons.

\subsection{$\omega$ production} 
The data on $\omega$ production at HERA are scanty, due mainly to the difficulty in reconstructing the $\pi^0$ of the final state $\gamma p \to \omega p \to p \pi^+ \pi^- \pi^0$. Two measurements have been published, one on $\omega$ production by real photons~\cite{zfp:c73:73}, based on an integrated luminosity of 0.89pb$^{-1}$, and one on $\omega$ production by virtual photons~\cite{pl:b487:3} (integrated luminosity 37.7pb$^{-1}$).

\subsubsection{$\gamma p\to \omega p$}

\begin{figure}[p]
\begin{center}
\vfill
\vspace*{-1.2cm}
\hspace*{0.5cm}\epsfig{file=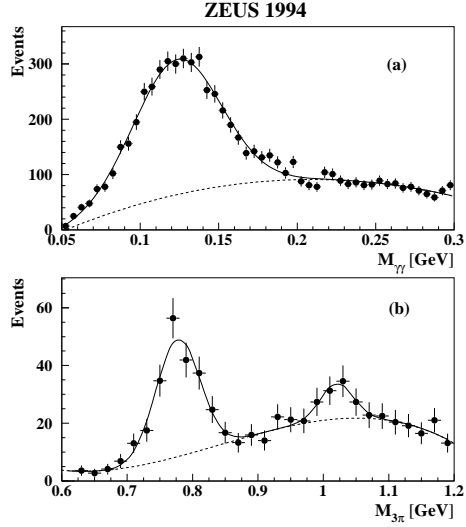,width=6cm}
\end{center} 
\caption{Reaction $\gamma p \to \pi^+\pi^- \gamma \gamma$: the $\gamma\gamma$ mass spectrum (top) and the $\pi^+\pi^-\pi^0$ mass spectrum (bottom); from ZEUS.}
\label{f:gpomegapz}
\vfill
\end{figure}

\begin{figure}[p]
\begin{center}
\vfill
\vspace*{-0.9cm}
\hspace*{0.5cm}\epsfig{file=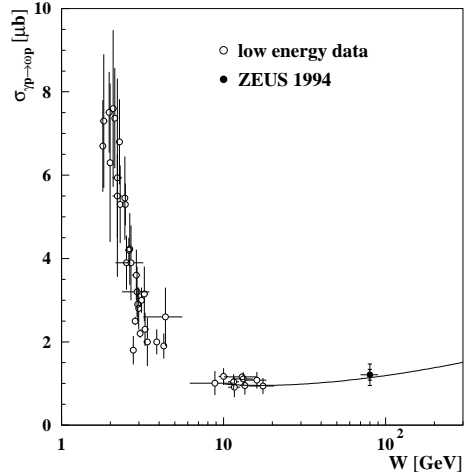,width=6cm}
\end{center} 
\caption{The cross section for $\gamma p \to \omega p$ as a function of the c.m. energy $W$; from ZEUS.}
\label{f:sigomegapz}
\vfill
\end{figure}

The $\gamma\gamma$ and $\pi^+\pi^-\pi^0$ mass spectra, measured at $Q^2 = 0, W = 70 - 90$ GeV (Fig.~\ref{f:gpomegapz}), show clear signals for $\pi^0$ (Fig.~\ref{f:gpomegapz}a), $\omega$ and $\phi$ production (Fig.~\ref{f:gpomegapz}b)~\cite{zfp:c73:73}. 
The cross section $\sigma (\gamma p\to \omega p)$ is presented in Fig.~\ref{f:sigomegapz} as a function of $W$ together with measurements by fixed target experiments from~\cite{pr:155:1468,pr:175:1669,pr:d1:790,pr:d5:15,pr:d7:3150,np:b108:45,prl:43:1545,prl:47:1782,
np:b209:56,zfp:c26:343,np:b231:15,pr:d40:1}. It is large at low c.m. energies, $W < 2 $ GeV, due to the contribution from one-pion exchange which decreases approximately proportional to $1/(W^2-m_p^2)^2$. For $W > 6$ GeV the cross section is approximately constant as a function of $W$, as expected for diffractive production. The ZEUS measurement yielded $\sigma (\gamma p\to \omega p) = 1.21 \pm 0.12(stat.)\pm 0.23(syst.) {\mu} b$ at $W=80$ GeV.  

\begin{figure}[p]
\begin{center}
\vfill
\vspace*{-0.9cm}
\hspace*{0.5cm}\epsfig{file=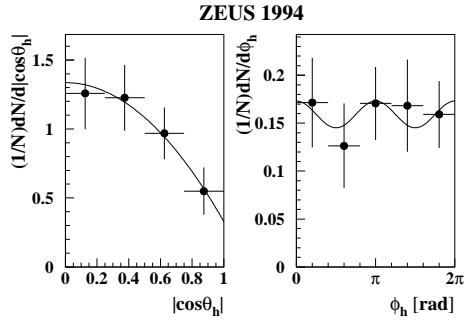,width=6cm}
\end{center} 
\caption{Reaction $\gamma p \to \omega p$: distributions of $|cos \theta_h|$ and $\phi_h$, the polar and azimuthal angles of the normal to the $\omega$ decay plane in the s-channel helicity frame; from ZEUS.}
\label{f:angleomegapz}
\vfill
\end{figure}

The distributions of the polar and azimuthal decay angles of the $\omega$ in the helicity system are shown in Fig.~\ref{f:angleomegapz}. They are of the form $dN/d{\rm cos} \theta_h \propto 1-{\rm cos}^2\theta_h$ and $dN/d\phi_h = constant$ and yield for $W \approx 80$ GeV: $r^{04}_{00} = 0.11 \pm 0.08 (stat) \pm 0.26 (syst)$ and $r^{04}_{1-1} = -0.04 \pm 0.08 (stat) \pm 0.12 (syst)$. These values are compatible with those measured for $\gamma p \to \rho^0 p$, see above, and are in agreement with s-channel helicity conservation SCHC~\cite{zfp:c73:73}.

\subsubsection{$\gamma^* p\to \omega p$}

\begin{figure}[p]
\begin{center}
\vfill
\vspace*{-0.9cm}
\hspace*{0.5cm}\epsfig{file=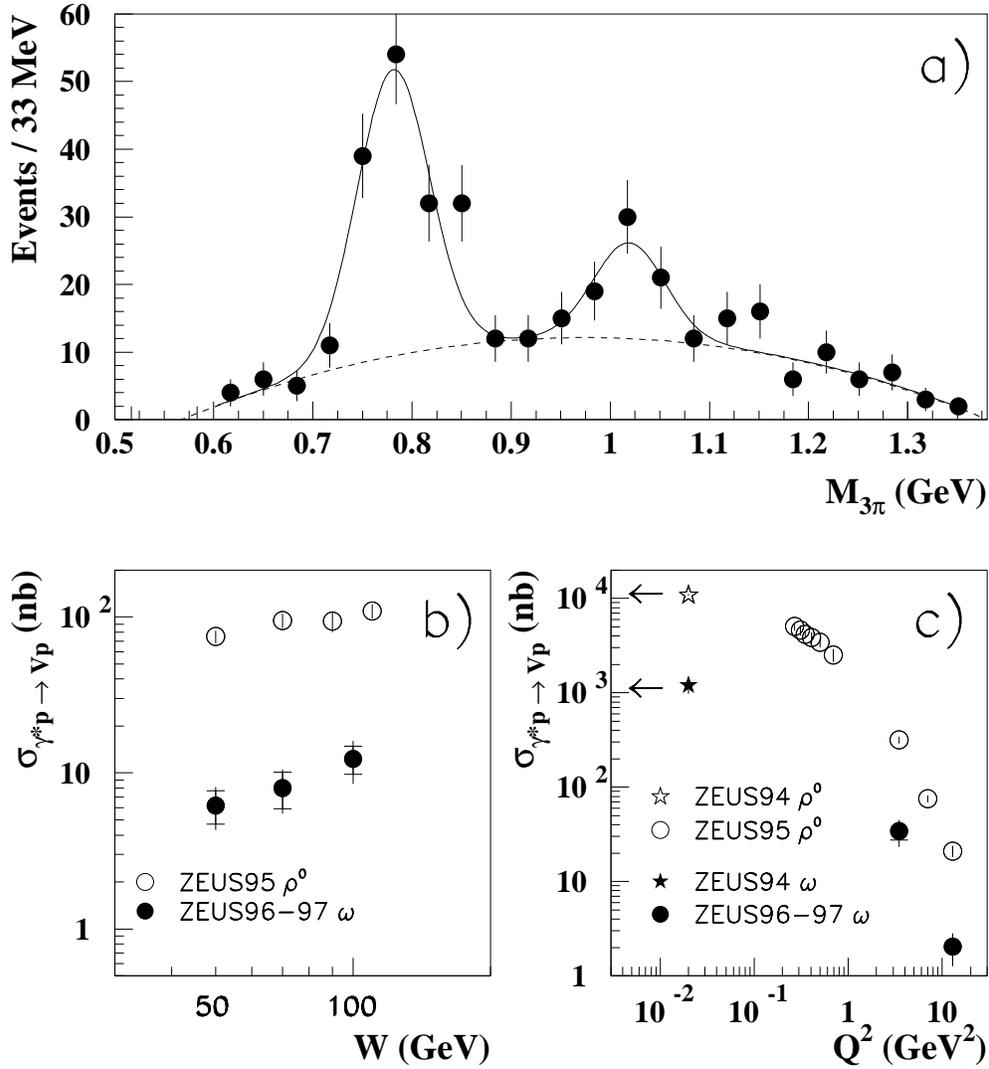,width=16cm}
\end{center} 
\caption{Reaction $\gamma^* p \to \pi^+ \pi^- \pi^0$: (a) distributions of the $\pi^+ \pi^- \pi^0$ mass spectrum; (b) cross sections for $\gamma^* p \to \rho^0 p$ and $\gamma^* p \to \omega p$ as a function of the c.m. energy $W$ at $Q^2 = 7$\GeV$^2$; (c) the same as a function of $Q^2$ at $W = 70$\GeV; from ZEUS.}
\label{f:sigsomegapz}
\vfill
\end{figure}

The production of $\omega$ mesons by virtual photons was studied~\cite{pl:b487:3} for the kinematic region: $3<Q^2<20$ GeV$^2$, $40<W<120$ GeV and $|t|<0.6$ GeV$^2$. Only events with $0.3 < M_{\pi^+ \pi^-} < 0.6$ GeV were kept. The results are presented in Fig.~\ref{f:sigsomegapz}. The $\pi^+ \pi^- \pi^0$ mass spectrum shows clear signals for the production of $\omega$ and $\phi$ mesons. The cross section for $\gamma^* p \to \omega p$ at $Q^2 = 7$ GeV$^2$ is a factor of about 10 smaller than for $\gamma^* p \to \rho^0 p$ and shows, for fixed $W = 70$ GeV, a rapid decrease with $Q^2$, similar to the $Q^2$-behaviour of $\sigma(\gamma^* p \to \rho^0 p)$.
  
\subsection{$\phi$ production}
First observations of $\phi$ production by real and virtual photons were reported by ZEUS~\cite{pl:b377:259,pl:b380:220} for integrated luminosities of 0.9 and 2.6 pb$^{-1}$, respectively. 

\subsubsection{$\gamma p \to \phi p$}
\begin{figure}[p]
\begin{center}
\vfill
\vspace*{-0.5cm}
\hspace*{0.1cm}\epsfig{file=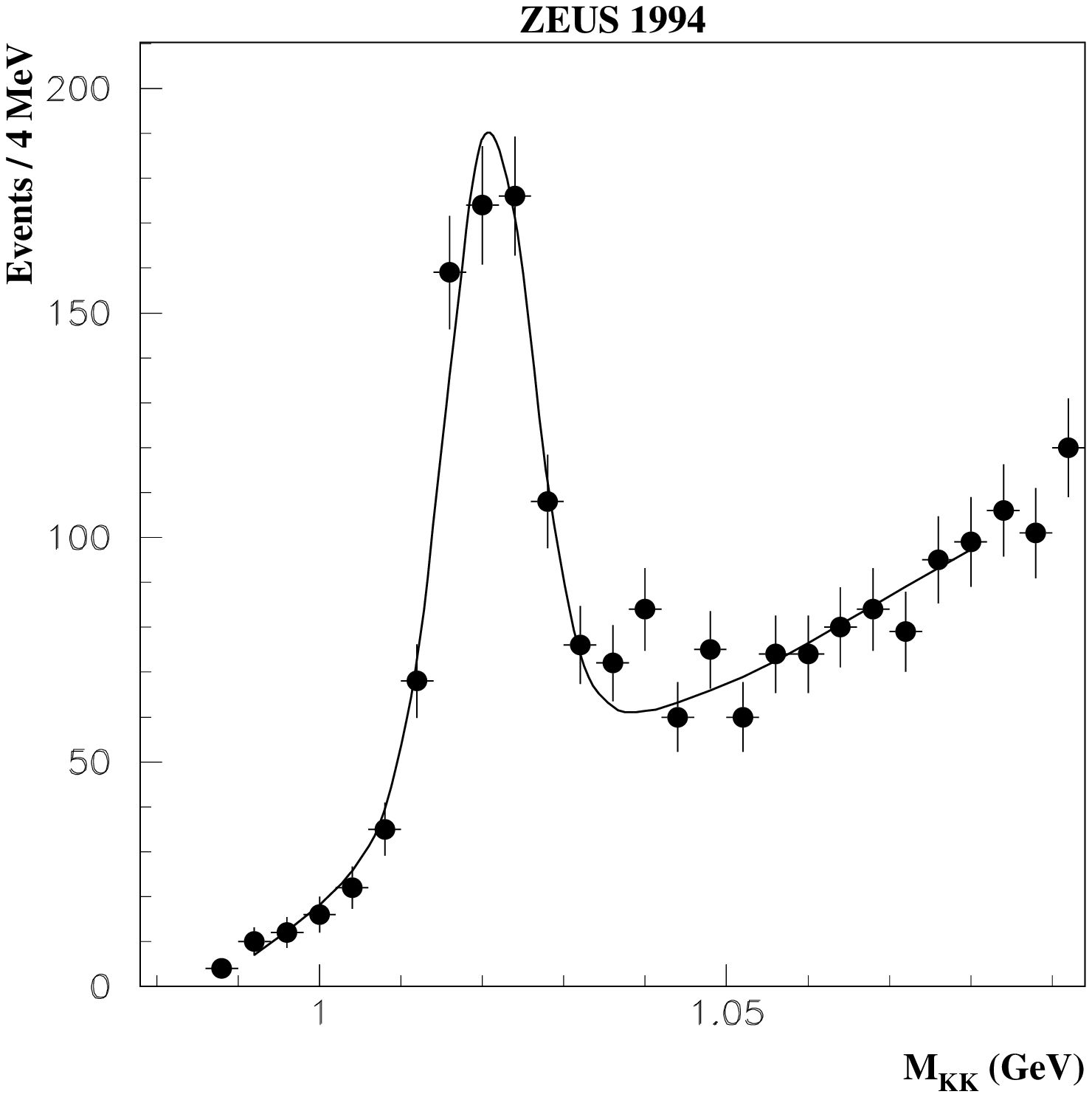,width=6.1cm}\epsfig{file=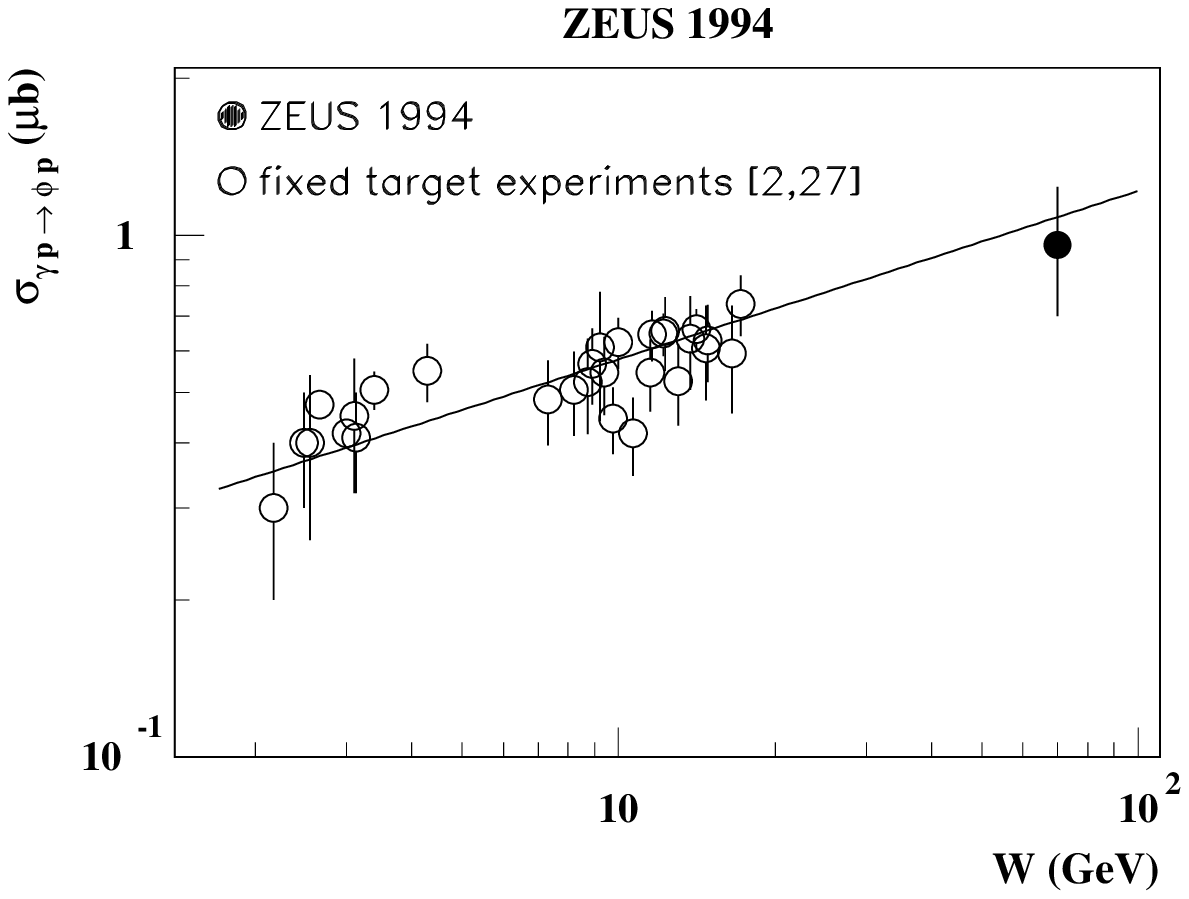,width=8.6cm}

\hspace*{1.8cm}\vspace*{-0.5cm}\epsfig{file=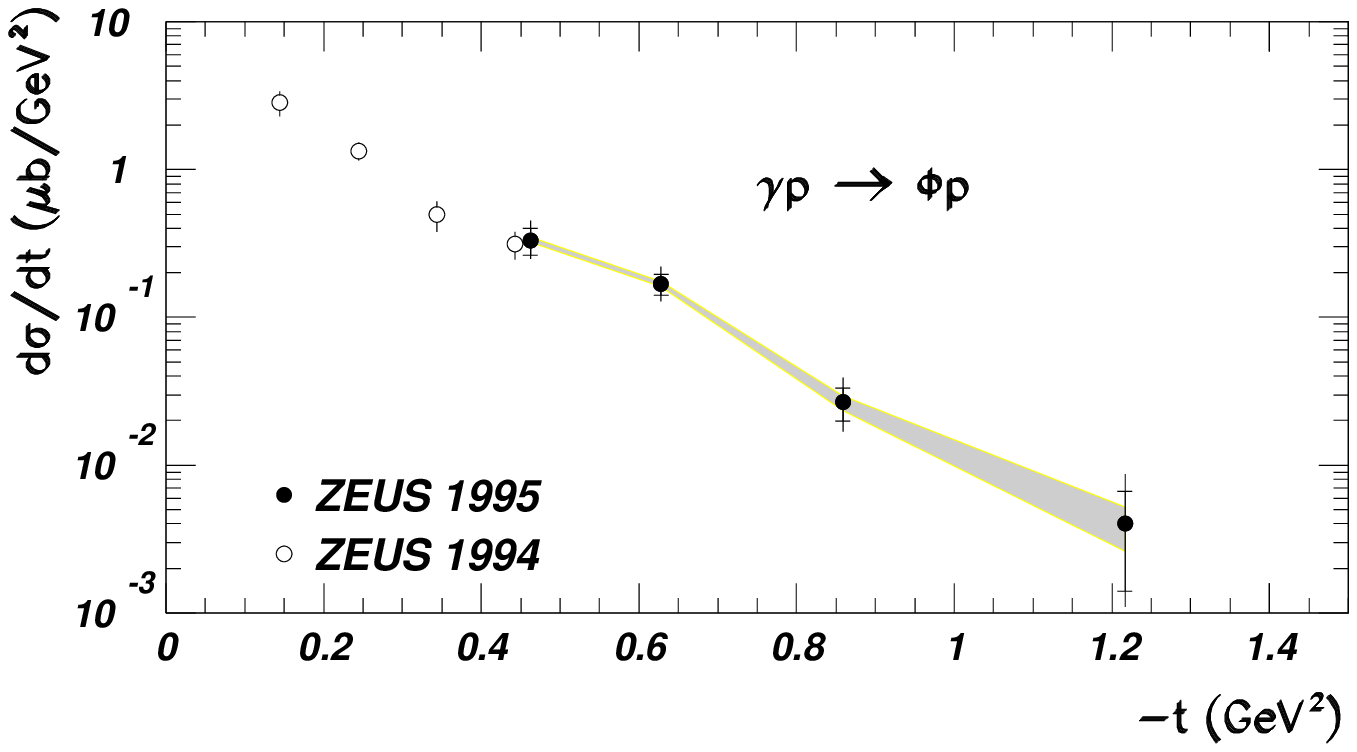,width=12cm}
\hspace*{0.7cm}\epsfig{file=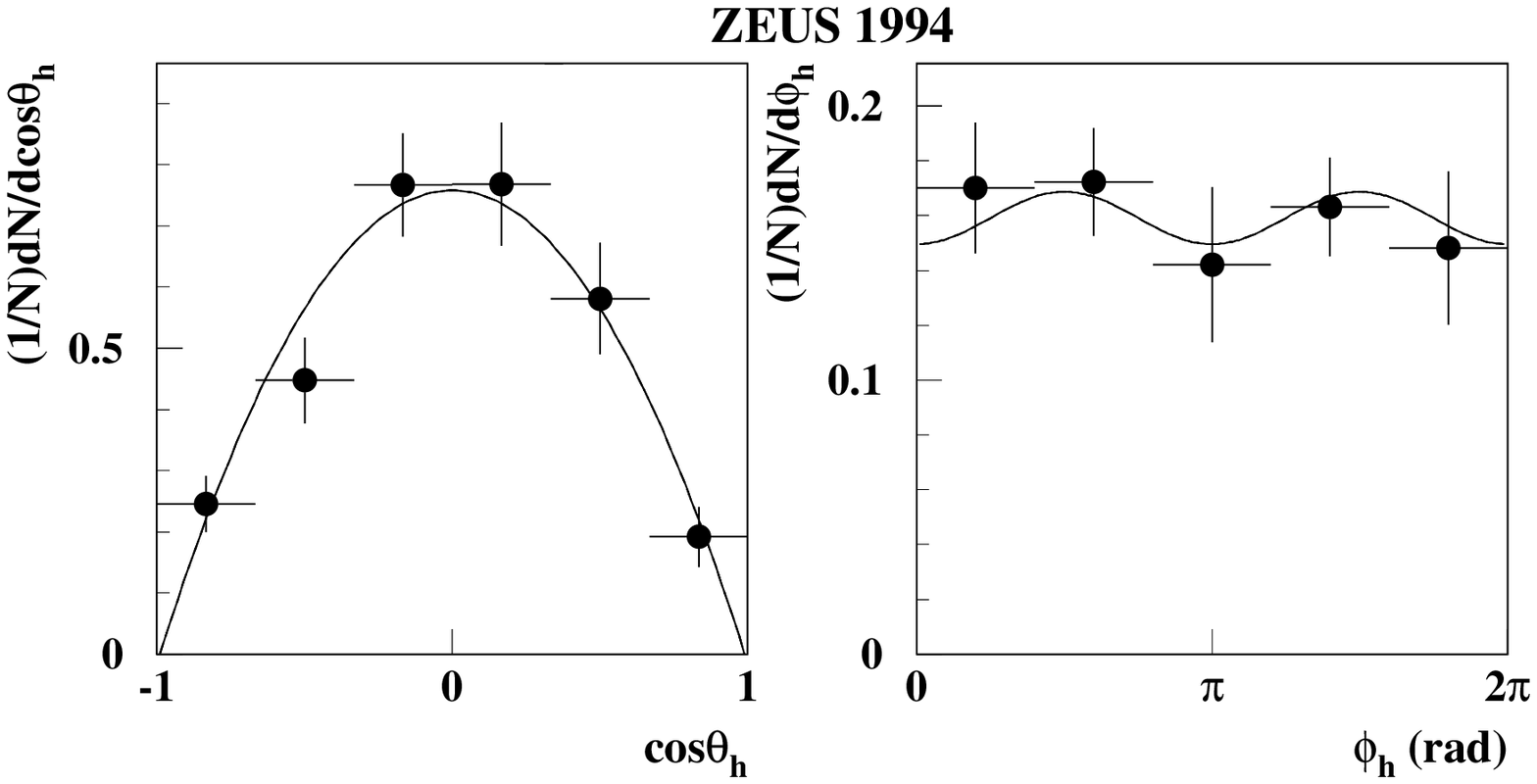,width=10.6cm}\\

\end{center} 
\caption{Reaction $\gamma p \to \phi p$: distribution of the $K^+ K^-$ mass spectrum  and the cross section as a function of the c.m. energy $W$; the cross section as a function of \hspace*{0.1cm}$t$ (center); the decay angular distributions (bottom); from ZEUS.}
\label{f:mkkz}
\vfill
\end{figure}

The reaction $\gamma p\to K^+ K^- p$ measured at a c.m. energy $W = 70$ GeV~\cite{pl:b377:259} exhibits a clean $\phi$ signal in the $M_{K+ K^-}$ mass spectrum (Fig.~\ref{f:mkkz}). The comparison with results at lower energy shows that the cross section $\sigma(\gamma p \to \phi p)$ is rising with $W$. The differential cross section, $d\sigma/d|t|$, is falling exponentially with $|t|$. The distributions of the $\phi$ meson decay angles $\theta_h$ and $\phi_h$ measured in the helicity system are shown in Fig.~\ref{f:mkkz} (bottom). They yield $r^{04}_{0 0} = -0.01 \pm 0.04$ and $r^{04}_{1 -1} = 0.03\pm 0.05$. These values are consistent with SCHC, and agree with those measured for $\gamma p \to \rho^0 p$ and $\gamma p \to \omega p$, see above.

\subsection{$\gamma^* p \to \phi p$}

\begin{figure}[p]
\begin{center}
\vfill
\vspace*{-1.0cm}
\hspace*{0.1cm}\epsfig{file=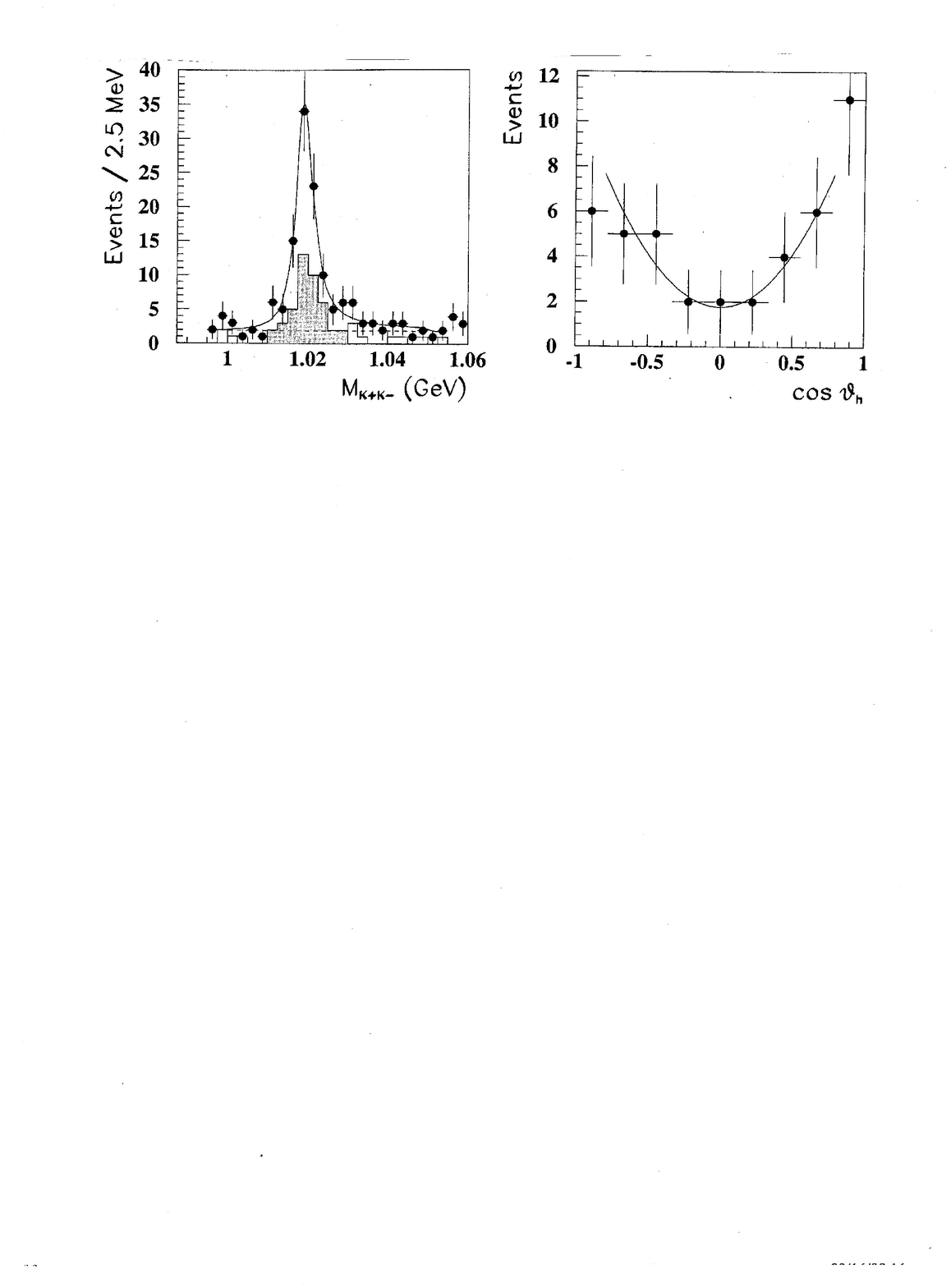,width=12cm}
\vspace*{+1cm}
\hspace*{0.5cm}\epsfig{file=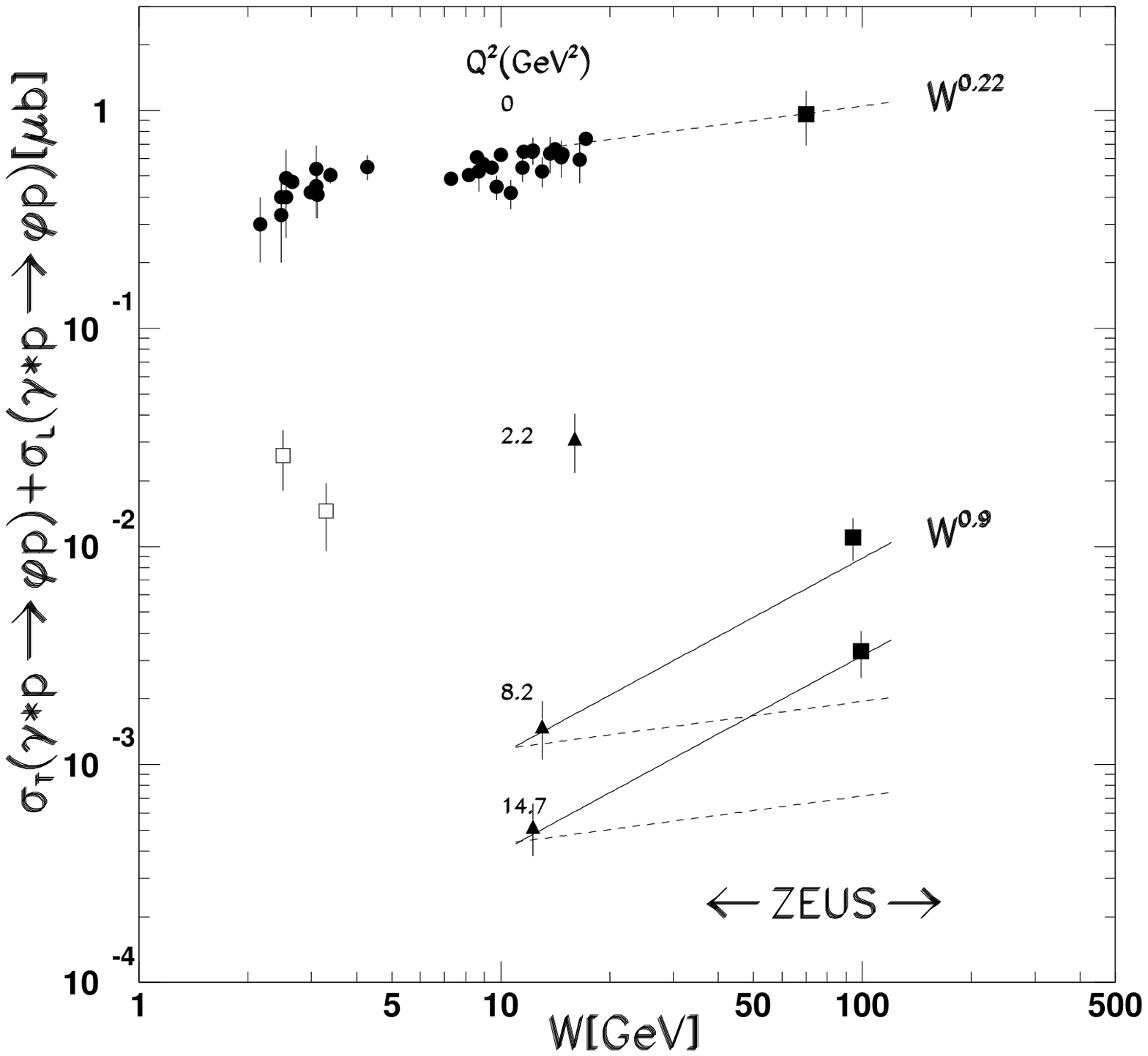,width=10cm}
\end{center} 
\caption{Reaction $\gamma^* p \to \phi p$: distributions of the $K^+ K^-$ mass and of $\cos \theta_h$ for candidate $\phi$ events (top); cross section as a function of $W$ for the $Q^2$-values indicated; from fixed target experiments and ZEUS (bottom).}
\label{f:gsphitech}
\vfill
\end{figure}

First results from HERA on $\phi$ production by $virtual$ photons, $\gamma^* p \to \phi p$, have been published by ZEUS~\cite{pl:b380:220}, based on an integrated luminosity of 2.6 $pb^{-1}$. The $K^+K^-$ mass spectrum (Fig.~\ref{f:gsphitech}) shows the $\phi$-signal. The distribution of the decay angle $\theta_h$ is approximately proportional to $\cos^2 \theta_h$ - indicating predominant production of longitudinal $\phi$'s, in contrast to $\phi$ production by real photons, see Fig.~\ref{f:mkkz}. 

\begin{figure}[p]
\begin{center}
\vfill
\vspace*{-0.5cm}
\hspace*{0.1cm}\epsfig{file=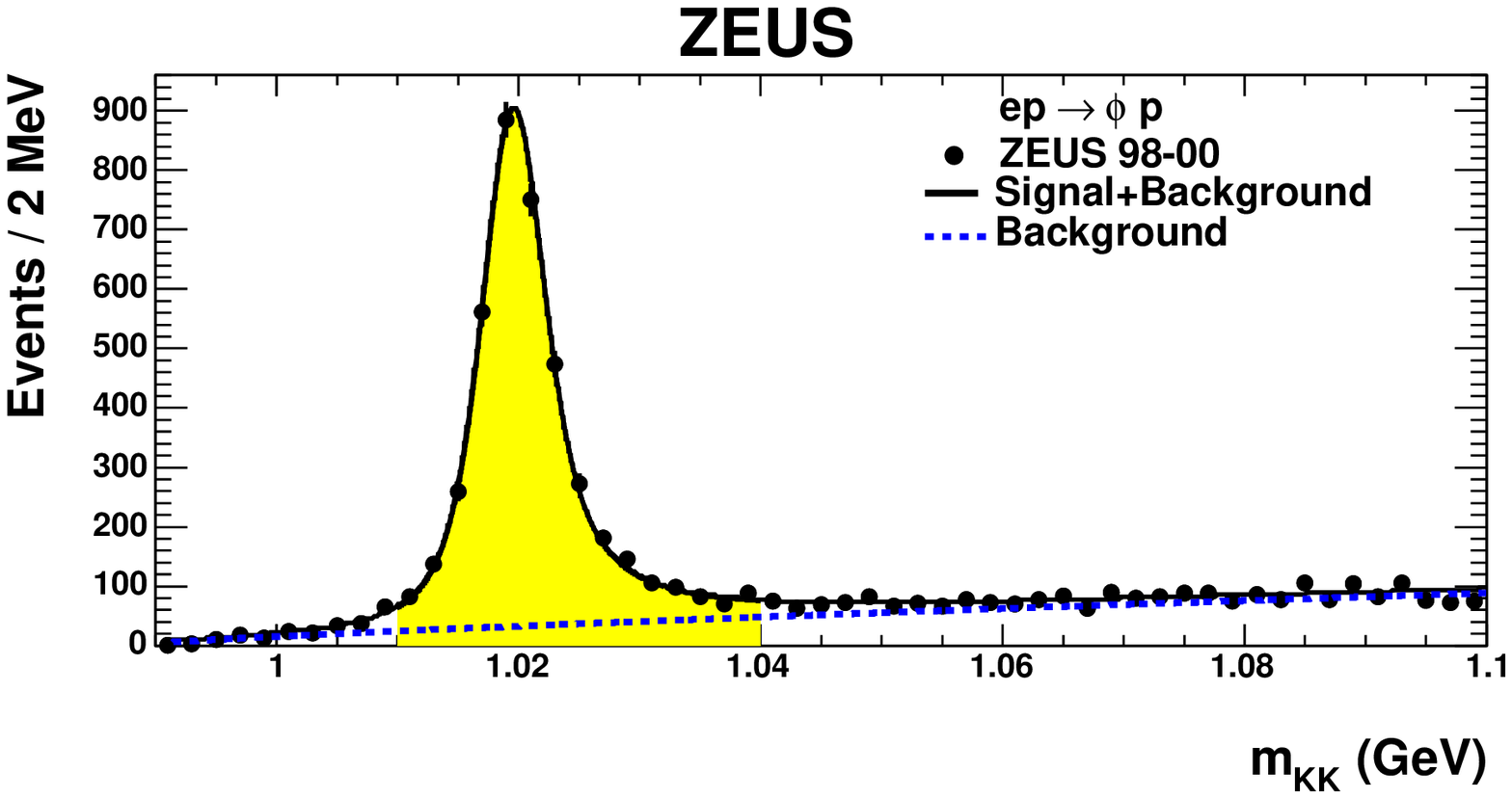,width=10cm}
\vspace*{-1cm}
\hspace*{0.1cm}\epsfig{file=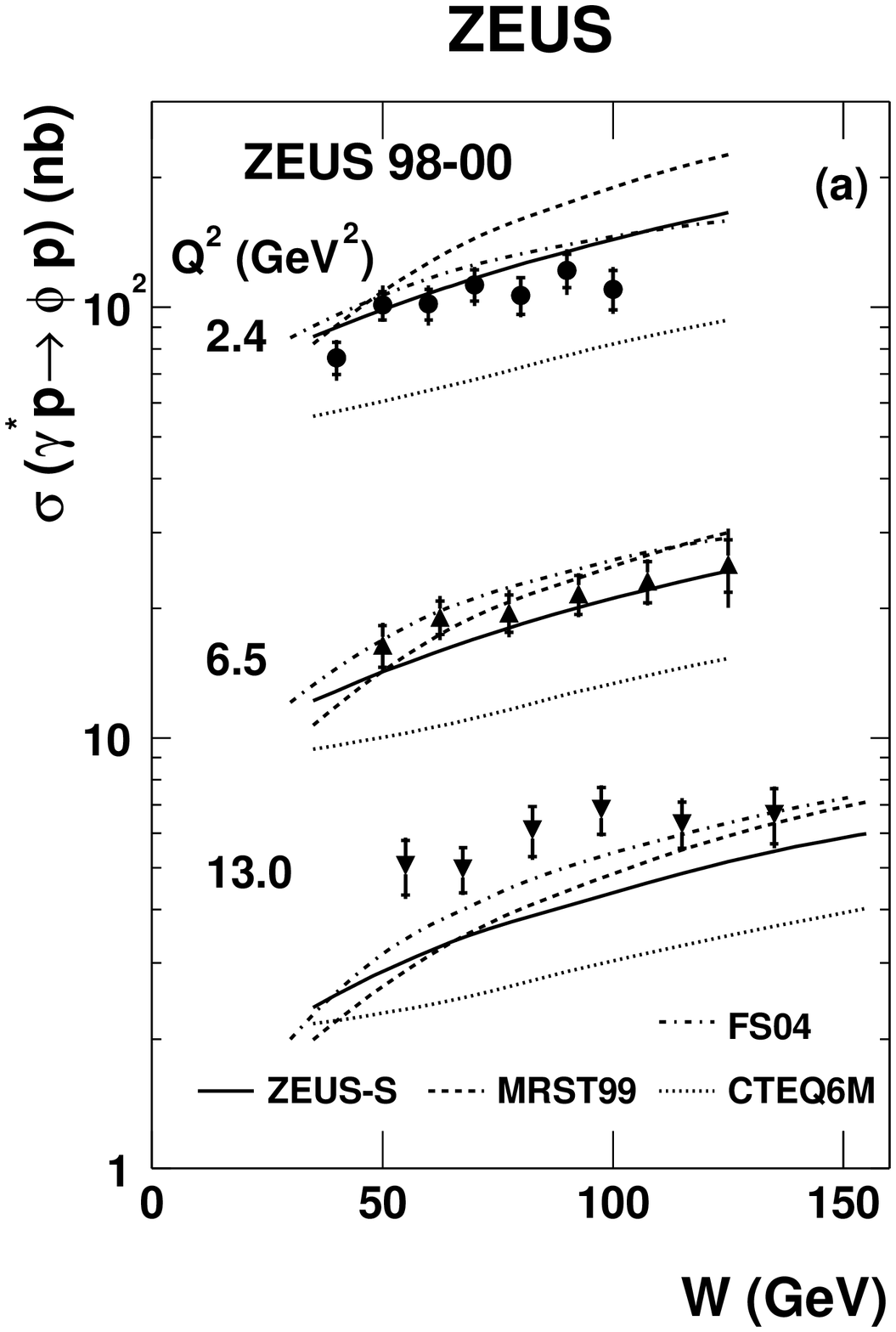,width=7.5cm}
\epsfig{file=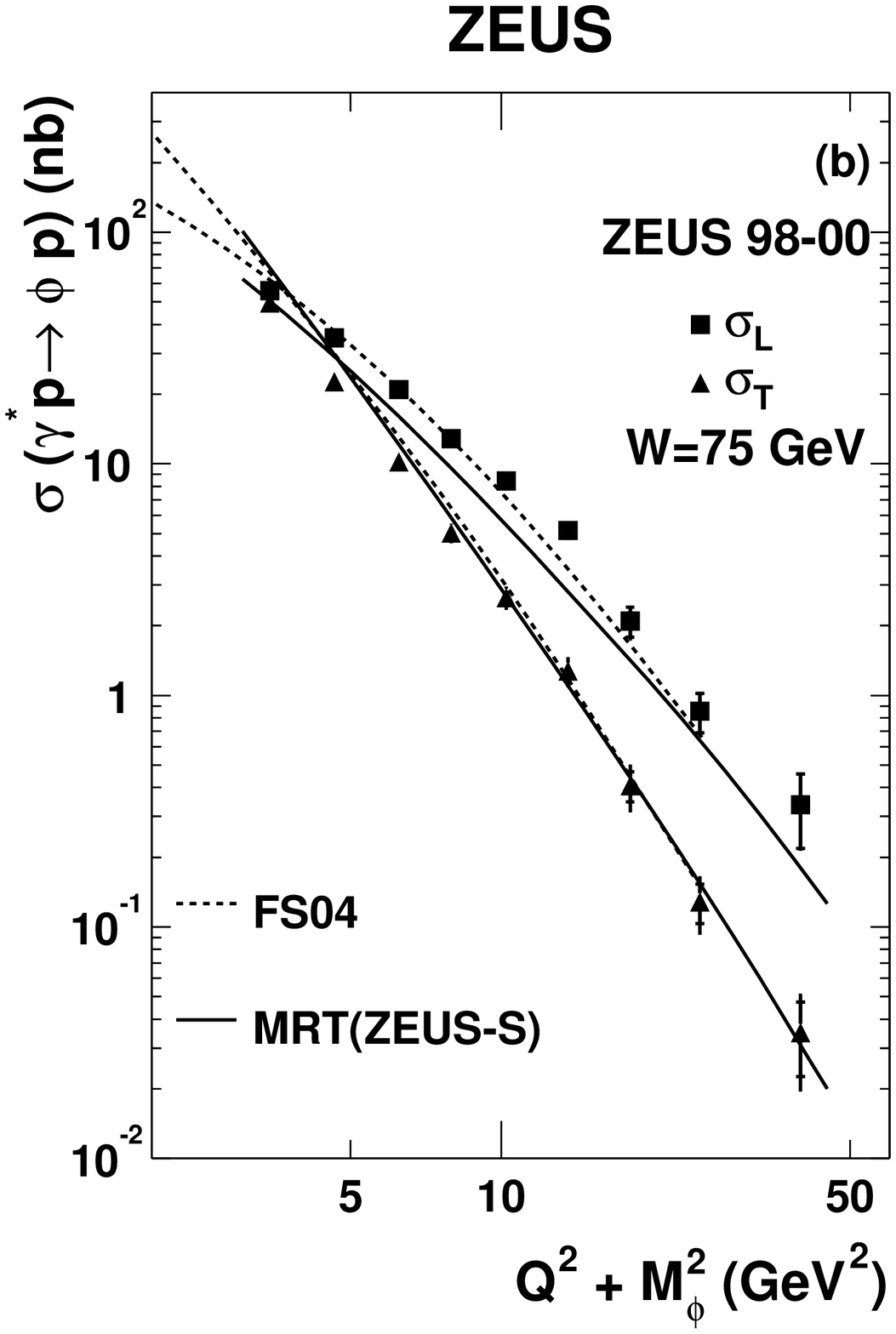,width=7.5cm}
\end{center}
\caption{Reaction $\gamma^* p \to \phi p$: 
(top) distribution of the $K^+ K^-$ mass spectrum ($L = 119$ $pb^{-1}$);
(bottom, left): cross section as a function of $W$ for $Q^2 = 2.4, 6.5, 13$\GeV$^2$(left); cross section as a function of ($Q^2 + M^2_{\phi}$) separately for the contributions from transverse (T) and longitudinal photons (L) (right); from ZEUS.}
\label{f:gsphivswvsqz}
\vfill
\end{figure}

Further results on $\phi$ production were presented by H1~\cite{pl:b483:360} (integrated luminosity L = 3.1 $pb^{-1}$) and ZEUS~\cite{pl:b553:141,np:b718:3} (L = 45 and 119 $pb^{-1}$, respectively). 
Figure~\ref{f:gsphivswvsqz} shows the results from ZEUS: the $K^+K^-$ mass spectrum is completely dominated by $\phi$ production. The cross section for $\gamma^* p \to \phi p$ at $Q^2 = 2.4, 6.5$ and $13.0$ GeV$^2$ (Fig.~\ref{f:gsphivswvsqz}a) shows a moderate rise with $W$. The dependence on $(Q^2 + M^2_{\phi})$ is shown in Fig.~\ref{f:gsphivswvsqz}b for $W = 75$ GeV, separately for the contributions from longitudinal (L) and transverse photons (T). At $(Q^2 + M^2_{\phi})= 3$ GeV$^2$, $\sigma_T$ and $\sigma_L$ are approximately equal; for larger values of $(Q^2 + M_{\phi}^2)$, $\sigma_T$ falls off more rapidly than $\sigma_L$. At $(Q^2 + M^2_{\phi}) \approx 45$ GeV$^2$ $\sigma_L$ is larger than $\sigma_T$ by about a factor of 10.

\begin{figure}[p]
\begin{center}
\vfill
\vspace*{-5cm}
\hspace*{0.1cm}\epsfig{file=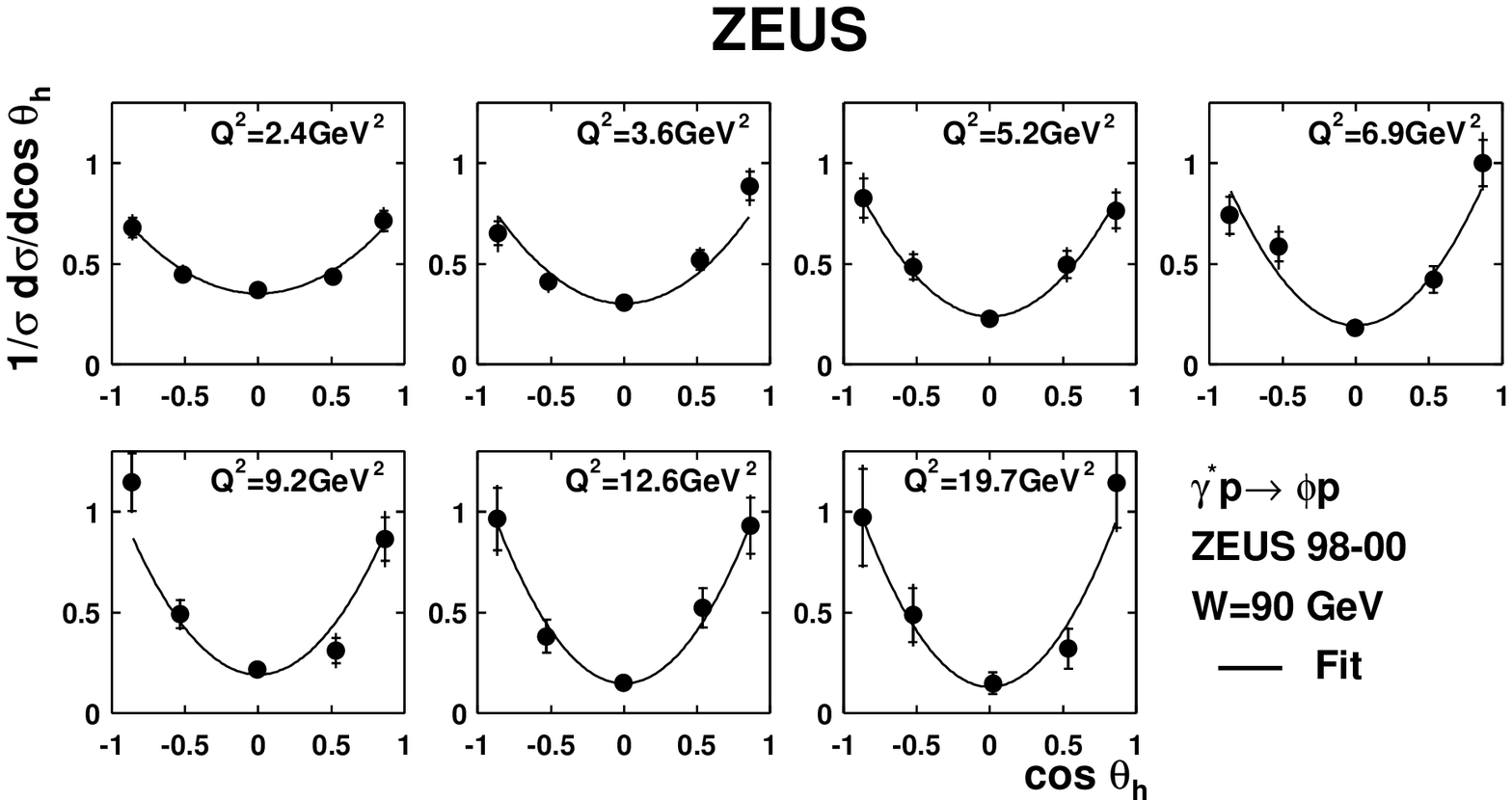,width=15cm}
\end{center} 
\caption{Reaction $\gamma^* p \to \phi p$: differential cross section $d\sigma/d|t|$ as a function of $|t|$ (top); distribution of $cos \theta_h$ (bottom); for different values of $Q^2$; from ZEUS.}
\label{f:gsphivstcosz}
\vfill
\end{figure}

The $|t|$ - dependence of $\phi$ production is displayed in Fig.~\ref{f:gsphivstcosz}(top) for different values of $Q^2$. The data yield $d\sigma/d|t| \propto e^{-b|t|}$, with $b \approx 5.5$ GeV$^{-2}$ for $(Q^2 + M_{\phi}^2) > 5$ GeV$^2$, see Fig.~\ref{f:bvsqmhz}.

The distribution of the $\phi$ polar decay angle in the helicity system, $\cos\theta_h$, is displayed in Fig.~\ref{f:gsphivstcosz} (bottom): it is of the form $dN/d\cos\theta_h \propto a + \cos^2 \theta$, where $a$ tends to zero as $Q^2$ increases from 2.4 to 19.7 GeV$^2$. This implies that at large $Q^2$, the $\phi$ mesons are predominantly longitudinally polarised - as has been observed for $\rho^0$ production (see above). Under the assumption of SCHC, $\phi$ production at $Q^2 \ge 2.4$ GeV$^2$ proceeds predominantly by longitudinal-photon proton scattering.
 
\subsection{$J/\Psi $ production}

\subsubsection{$\gamma p\to J/\Psi p$}

\begin{figure}[p]
\begin{center}
\vfill
\vspace*{-5cm}
\hspace*{3cm}\epsfig{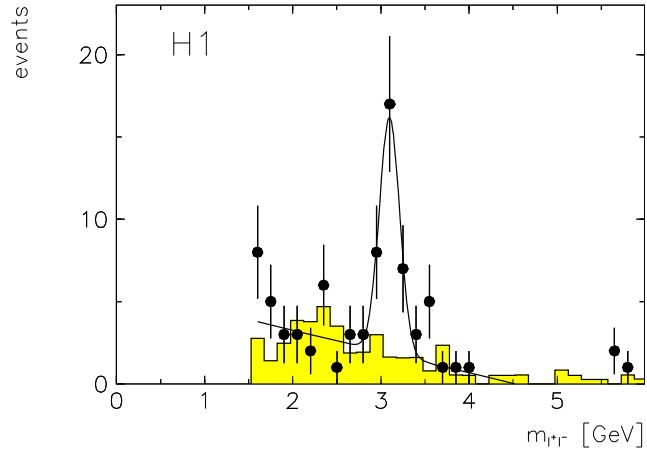}
\end{center} 
\caption{Mass distribution for $e^+ p \to l^+ l^- +X$. The curve shows a fit to $J/\Psi$ production plus background; from H1.}
\label{f:gjmllh}
\vfill
\end{figure}

\begin{figure}[p]
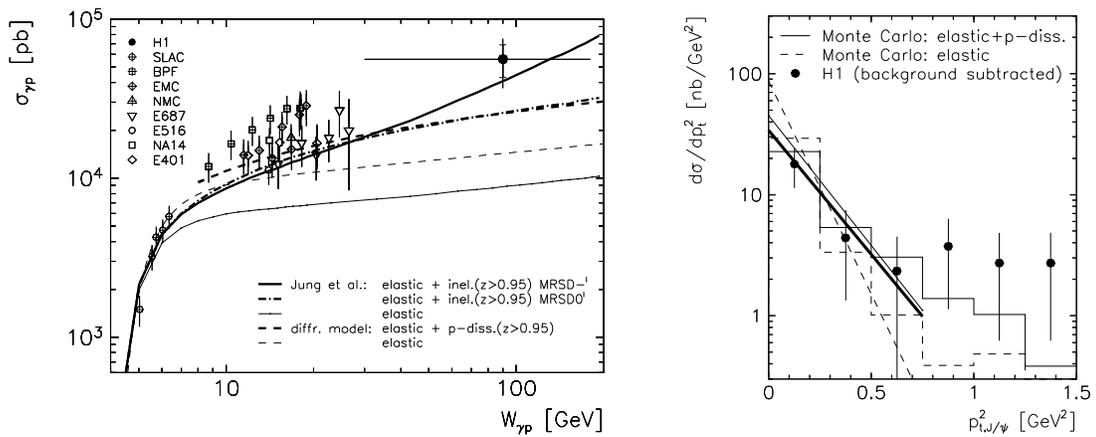

\begin{center}
\vfill
\vspace*{-5cm}
\hspace*{0.8cm}\epsfig{file=h1.DESY-94-153_4.eps,angle=90,width=14cm}\hspace*{-5.0cm}\epsfig{file=h1.DESY-94-153_5.eps,angle=90,width=14cm}
\end{center} 
\caption{Reaction $\gamma p \to J/\Psi X$: cross section as a function of $W$ (left);  differential cross section $d\sigma/dp^2_T$ as a function of $p^2_T$ (right); from H1.}
\label{f:gjwth}
\vfill
\end{figure}

First evidence at HERA for $J/\Psi$ production by $\gamma p$ scattering was obtained by H1~\cite{pl:b338:507} from the analysis of events with two leptons in the final state (40 $e^+e^-$ and 48 $\mu^+ \mu^-$  events, respectively). The $l^+ l^-$ mass spectrum (Fig.~\ref{f:gjmllh}) shows a peak at the mass of the $J/\Psi$. The cross section for $\gamma p\to J/\Psi p$ measured by H1 for $Q^2 <4$ GeV$^2$, $30<W_{\gamma p}< 180$ GeV, indicates a substantial rise with $W$ compared to the fixed target measurements at lower energies~\cite{prl:43:187,np:b213:1,pl:b332:195,pl:b316:197,prl:52:795,zfp:c33:505,prl:48:73}, and an exponential decrease with the $J/\Psi$ transverse momentum squared $p^2_T$, see Fig.~\ref{f:gjwth}.

\begin{figure}[p]
\begin{center}
\vfill
\vspace*{-3cm}
\hspace*{0.8cm}\epsfig{file=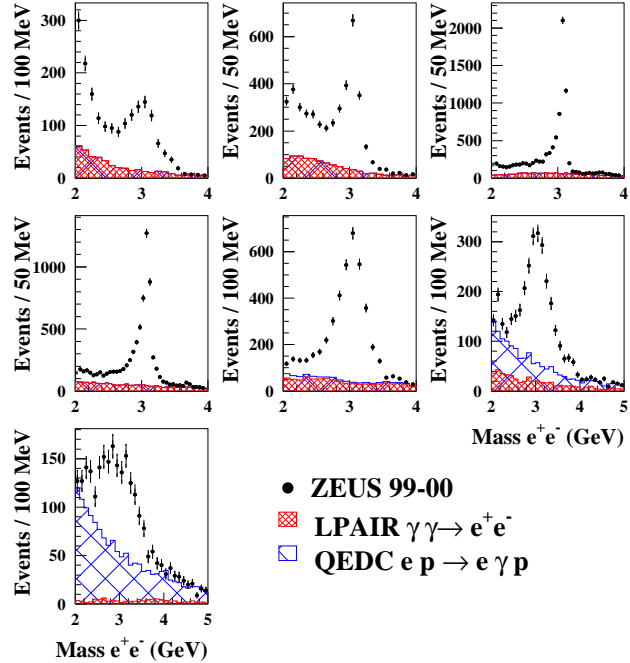,width=10cm}
\end{center} 
\caption{Reaction $\gamma p \to e^+ e^- p$: $e^+ e^-$ mass distributions for the $W$ intervals (a) 20 - 35\GeV, (b) 35 - 50\GeV, (c) 50 - 90\GeV, (d) 90 - 140\GeV, (e) 140 - 200\GeV, (f) 200 - 260\GeV and (g) 260 - 290\GeV; from ZEUS.}
\label{f:gmeez}
\vfill
\end{figure}

\begin{figure}[p]
\begin{center}
\vfill
\vspace*{-1.5cm}
\hspace*{0.8cm}\epsfig{file=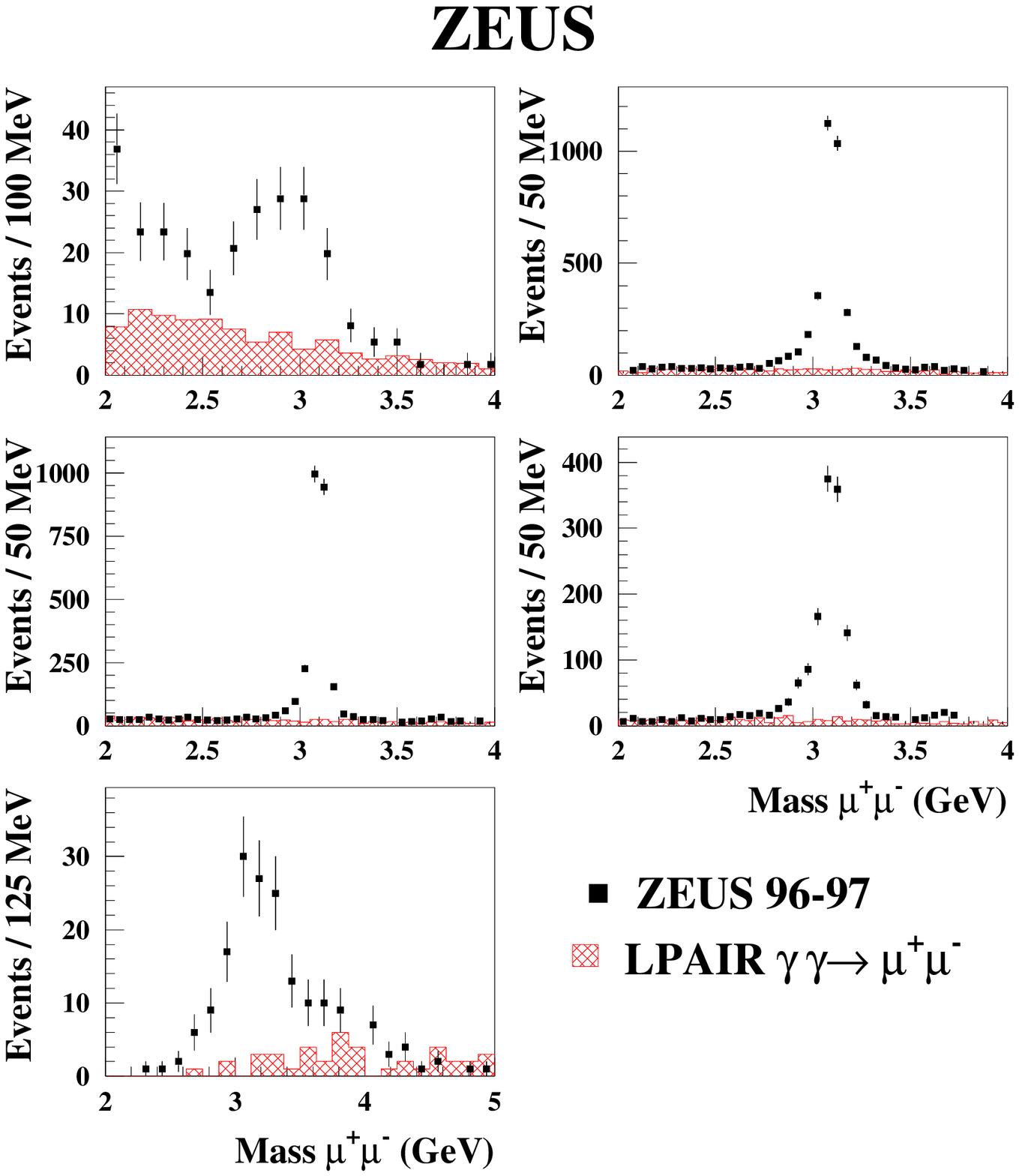,width=10cm}
\end{center} 
\caption{Reaction $\gamma p \to \mu^+ \mu^- p$: $\mu^+ \mu^-$ mass distributions for the $W$ intervals (a) 20 - 30\GeV, (b) 30 - 70\GeV, (c) 70 - 110\GeV, (d) 110 - 150\GeV and (g) 150 - 170\GeV; from ZEUS.}
\label{f:gmuuz}
\vfill
\end{figure}

The $e^+e^-$ and $\mu^+ \mu^-$ mass distributions obtained in photoproduction by ZEUS~\cite{epj:c24:345} are shown in Figs.~\ref{f:gmeez},~\ref{f:gmuuz}; they are based on integrated luminosities of 55.2 pb$^{-1}$ and 38.0 pb$^{-1}$, providing about 11000 and 6500 $J/\Psi$ events, respectively. For $W$ values between 35 and 260 GeV, $J/\Psi$ production is observed with little background.    

\begin{figure}[p]
\begin{center}
\vfill
\vspace*{-1.5cm}
\hspace*{0.8cm}\epsfig{file=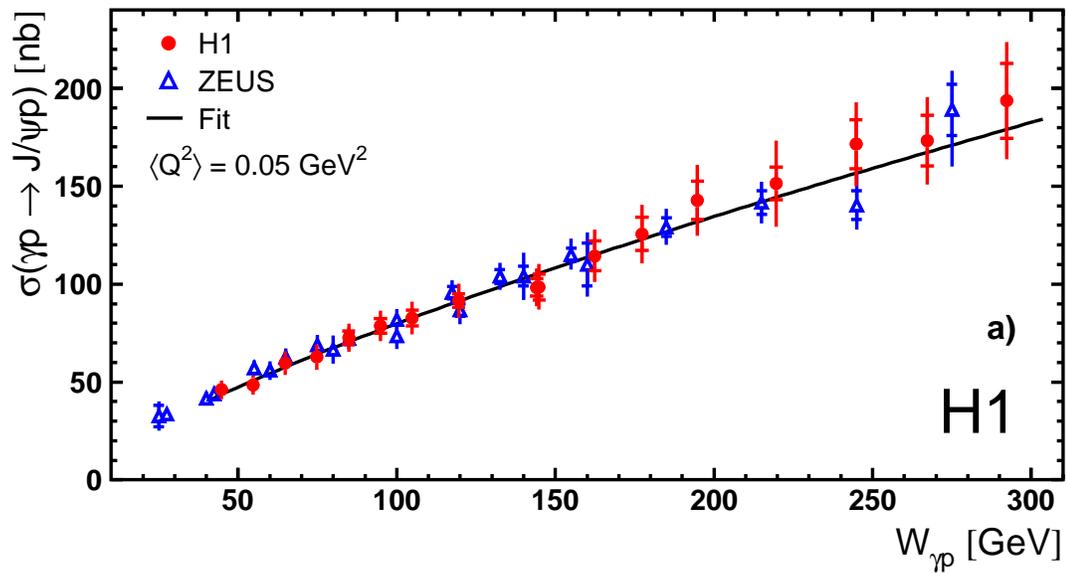,width=14cm}
\end{center} 
\caption{Reaction $\gamma p \to J/\Psi p$: the cross section $\sigma(\gamma p \to J/\Psi p)$ as a function of $W$ with measurements from H1 and ZEUS, from H1.}
\label{f:sgpjpsizh}
\vfill
\end{figure}

The $W$ dependence of the cross section $\sigma(\gamma p \to J/\Psi p)$  as obtained by ZEUS~\cite{epj:c24:345} and H1~\cite{epj:c46:585}  is shown in Fig.~\ref{f:sgpjpsizh}. There is excellent agreement between the two measurements. The $J/\Psi p$ cross section is rising proportional to $W^{\delta}$, where $\delta = 0.75 \pm 0.03(stat.) \pm 0.03(syst.)$, as determined by H1.

\begin{figure}[p]
\begin{center}
\vfill
\vspace*{-1cm}
\hspace*{-0.1cm}\epsfig{file=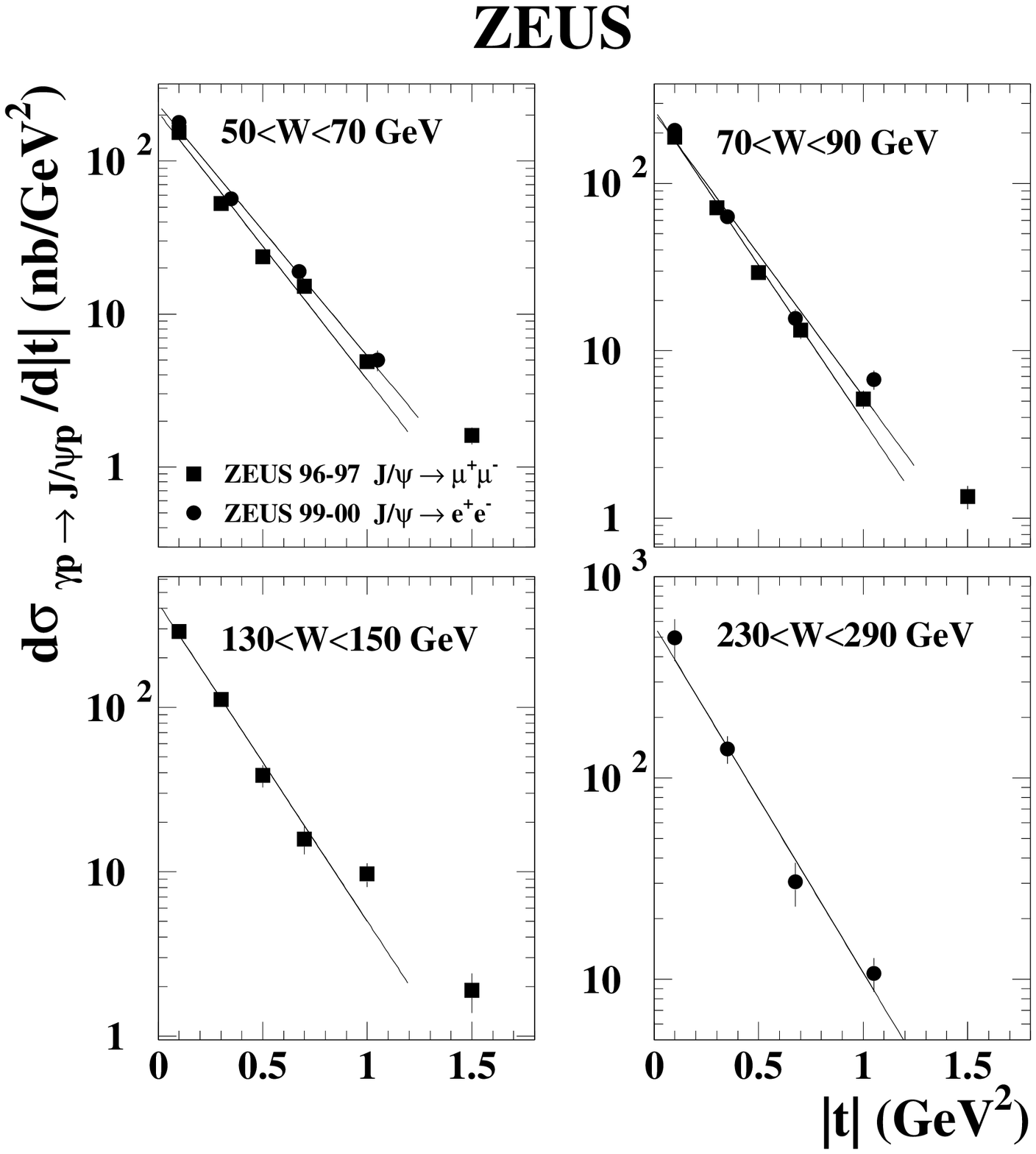,width=11cm}
\hspace*{-0.1cm}\epsfig{file=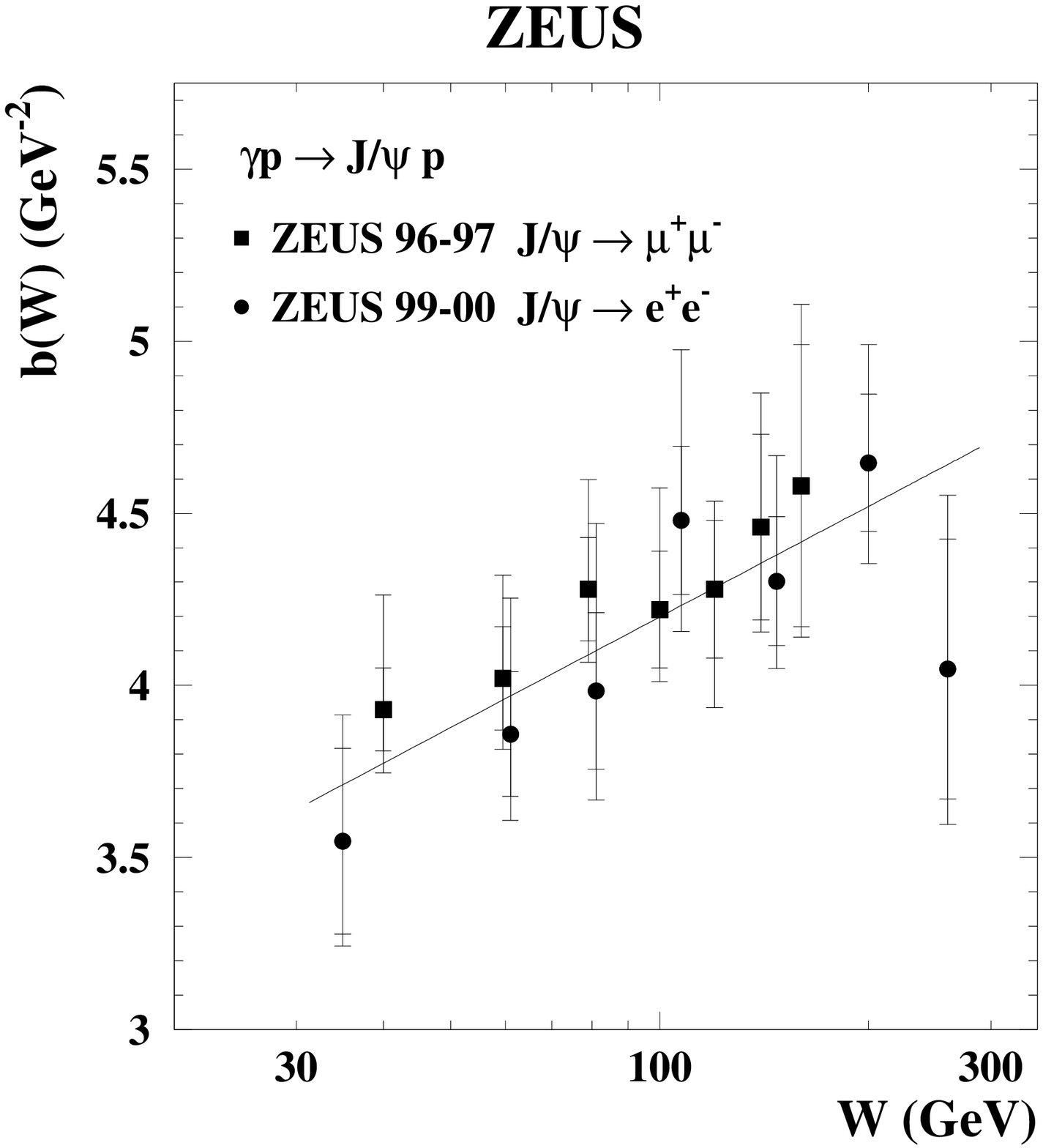,width=11cm}
\end{center} 
\caption{Reaction $\gamma p \to J/\Psi p$: the differential cross section $d\sigma/d|t|$ as a function of $|t|$ (above); the slope $b$ from a fit to $d\sigma/d|t| \propto e^{-b \cdot |t|}$ (below), from ZEUS.}
\label{f:sgpjvst}
\vfill
\end{figure}

\begin{figure}[p]
\begin{center}
\vfill
\vspace*{-1cm}
\hspace*{-0.1cm}\epsfig{file=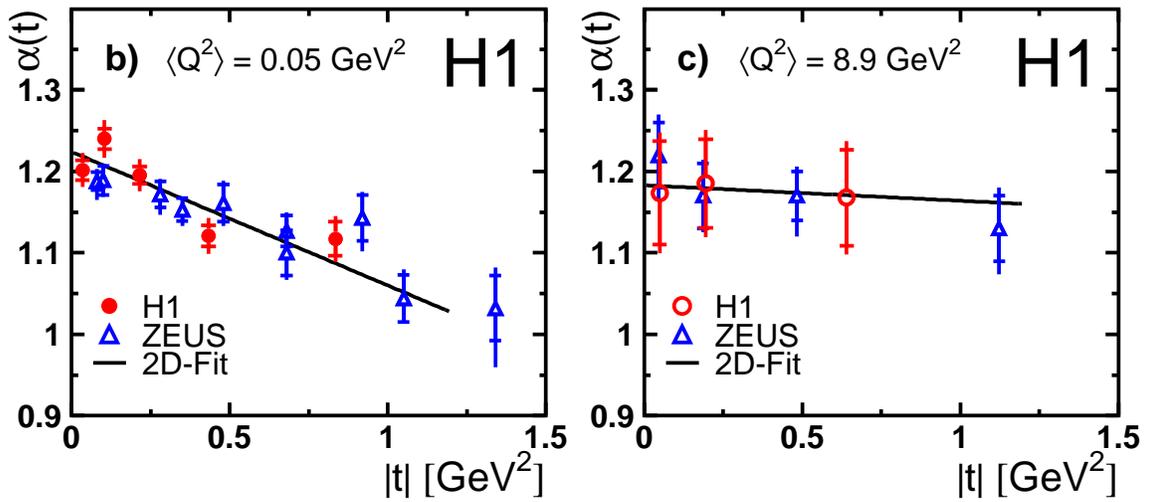,width=15cm}
\end{center}
\caption{Reaction $\gamma p \to J/\Psi p$: the effective trajectory $\alpha_{\pom}(t)$ as a function of $|t|$ for $40<W<305$\GeV, $<Q^2> = 0.05$\GeV$^2$ (left) and $<Q^2> = 8.9$\GeV$^2$ (right); measurements by H1 and ZEUS; from H1.}
\label{f:sgpjalphth}
\vfill
\end{figure}

The $|t|$ dependence of the cross section as measured by ZEUS~\cite{epj:c24:345} is shown in Fig.~\ref{f:sgpjvst} for several bins in $W$ between 50 and 290 GeV: $d\sigma/d|t|$ is of the form $c \cdot e^{-b |t|}$. The slope $b$ is rising with $W$, as shown by Fig.~\ref{f:sgpjvst}. A fit of the data~\cite{epj:c24:345} to the form
\begin{eqnarray}
 b(W) = b_0 + 4\cdot \alpha_{\pom}^{'} \cdot \ln(W/(90{\rm GeV})),
\end{eqnarray}
where $ \alpha_{\pom}^{'}$ is the slope of the Regge trajectory of the Pomeron, yielded $b_0 = 4.15 \pm 0.05 (stat.) ^{+0.30}_{-0.18} (syst.) {\rm GeV^{-2}}$ and $\alpha_{\pom}^{'}  =  0.116 \pm 0.026 (stat.)^{+0.010}_{-0.025} (syst.) {\rm GeV^{-2~}}$, see the straight line. The effective Pomeron trajectory deduced by H1~\cite{epj:c46:585} from the H1 and ZEUS data is shown in Fig.~\ref{f:sgpjalphth} as a function of $|t|$.\\
For $Q^2 = 0.05$ GeV$^2$:\\
$\alpha_{\pom} = [1.224 \pm 0.010(stat.) \pm 0.012(syst.)] - [0.164 \pm 0.028(stat.) \pm 0.030(syst.)] \cdot |t|$\\
and for $Q^2 = 8.9$ GeV$^2$:\\
$\alpha_{\pom} = [1.183 \pm 0.054(stat.) \pm 0.030(syst.)] - [0.019 \pm 0.139(stat.) \pm 0.076(syst.)] \cdot |t|$.

\begin{figure}[p]
\begin{center}
\vfill
\vspace*{-1cm}
\hspace*{-0.1cm}\epsfig{file=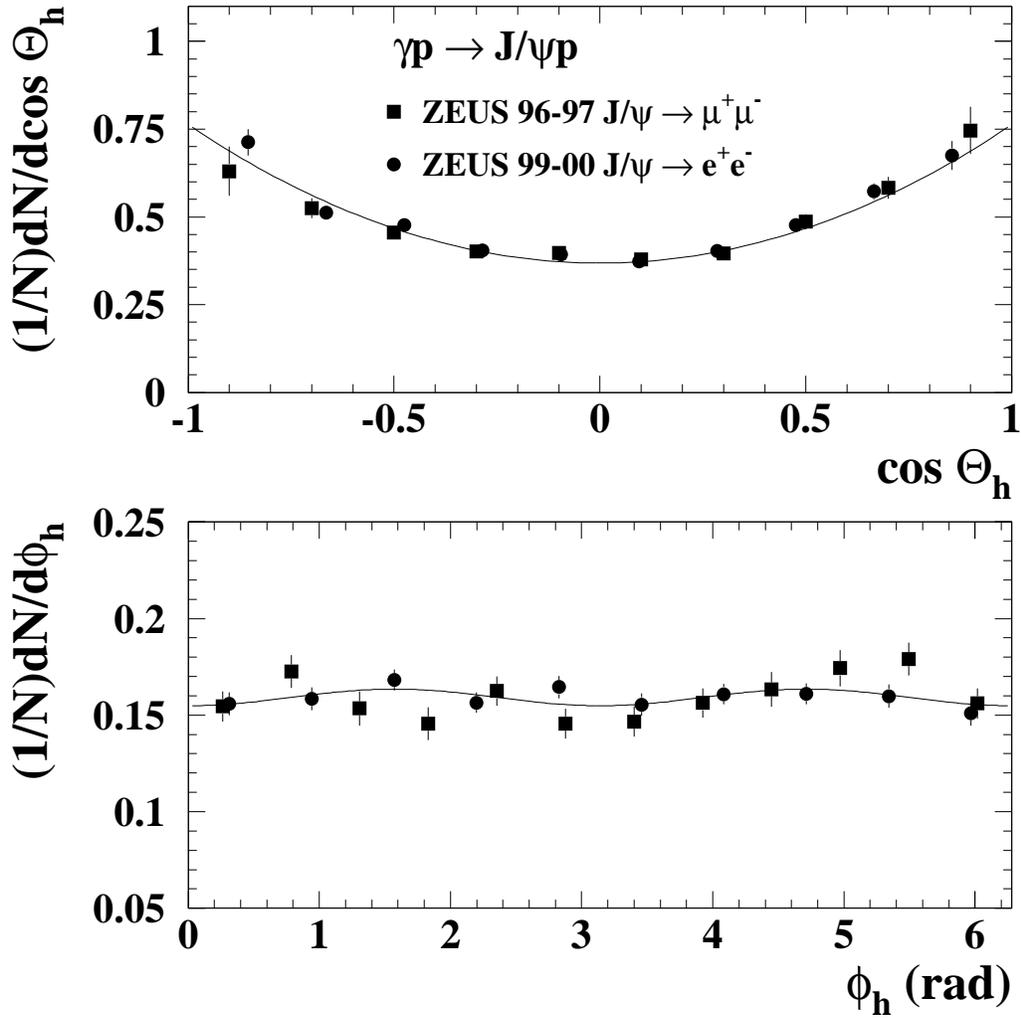,width=15cm}
\end{center}
\caption{Reaction $\gamma p \to J/\Psi p$: decay angular distributions of the $J/\Psi$ in the s-channel helicity system for $30<W<170$\GeV, $|t|<1$\GeV$^2$; from ZEUS.}
\label{f:sgpjdecay}
\vfill
\end{figure}

The $J/\Psi$ polar and azimuthal angular distributions for the $l^+l^-$ decay ($l = \mu$ or $e$) in the helicity system are shown in Fig.~\ref{f:sgpjdecay} as measured by ZEUS~\cite{epj:c24:345}. They yield $r^{04}_{0 0} = -0.017 \pm 0.015(stat.)\pm0.009(syst.)$ and $r^{04}_{1 -1} = 0.027\pm 0.013(stat.)\pm 0.005(syst.)$. If the $J/\Psi$ retains the helicity of the photon (SCHC) then $r^{04}_{0 0}$ and $r^{04}_{1 -1}$  should both be zero: the data show that both matrix elements are small.

\subsubsection{$\gamma^* p\to J/\Psi p$}
\begin{figure}[p]
\begin{center}
\vfill
\vspace*{-1cm}
\hspace*{-0.1cm}\epsfig{file=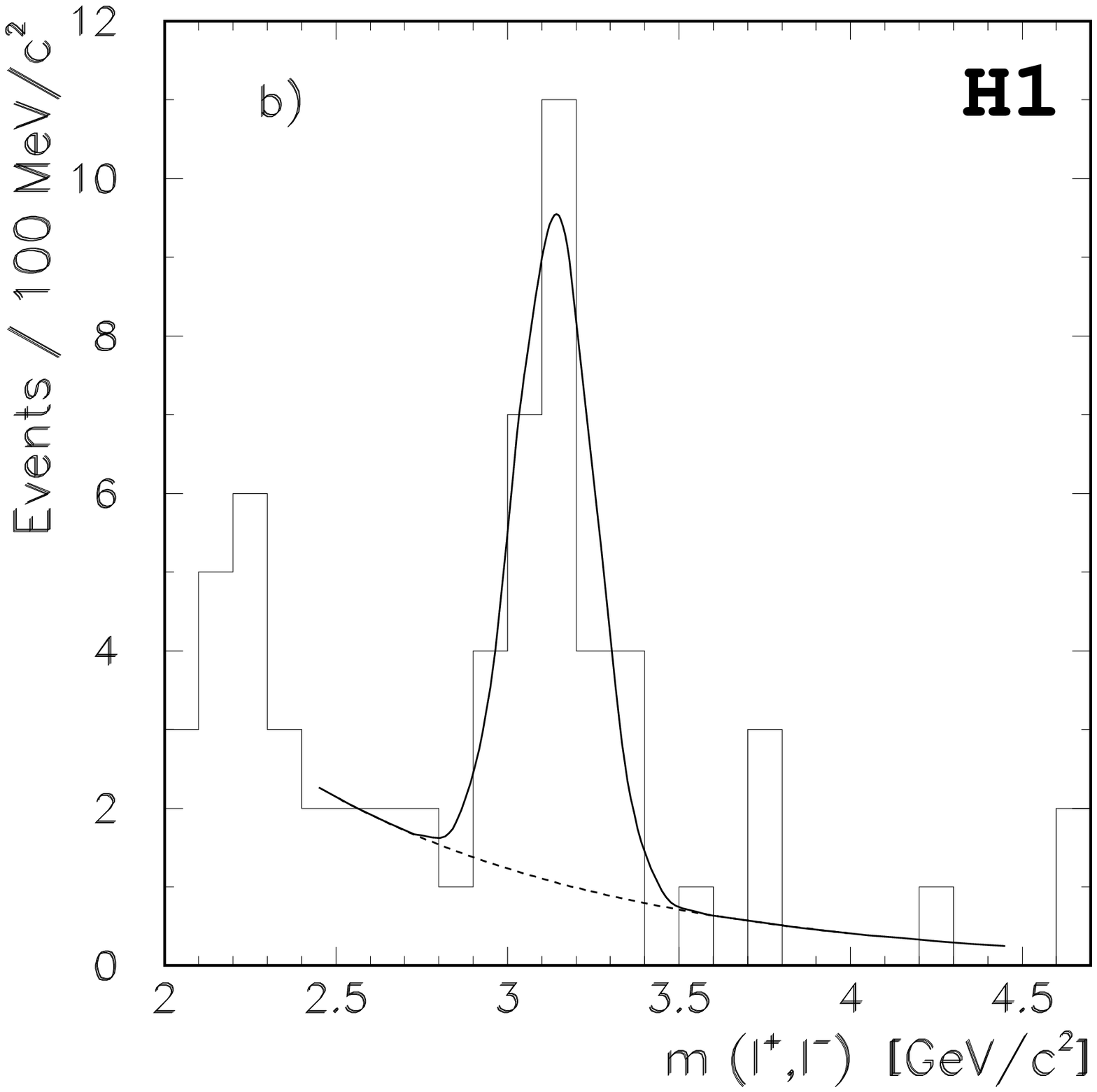,width=4.8cm}\hspace*{-0.1cm}\epsfig{file=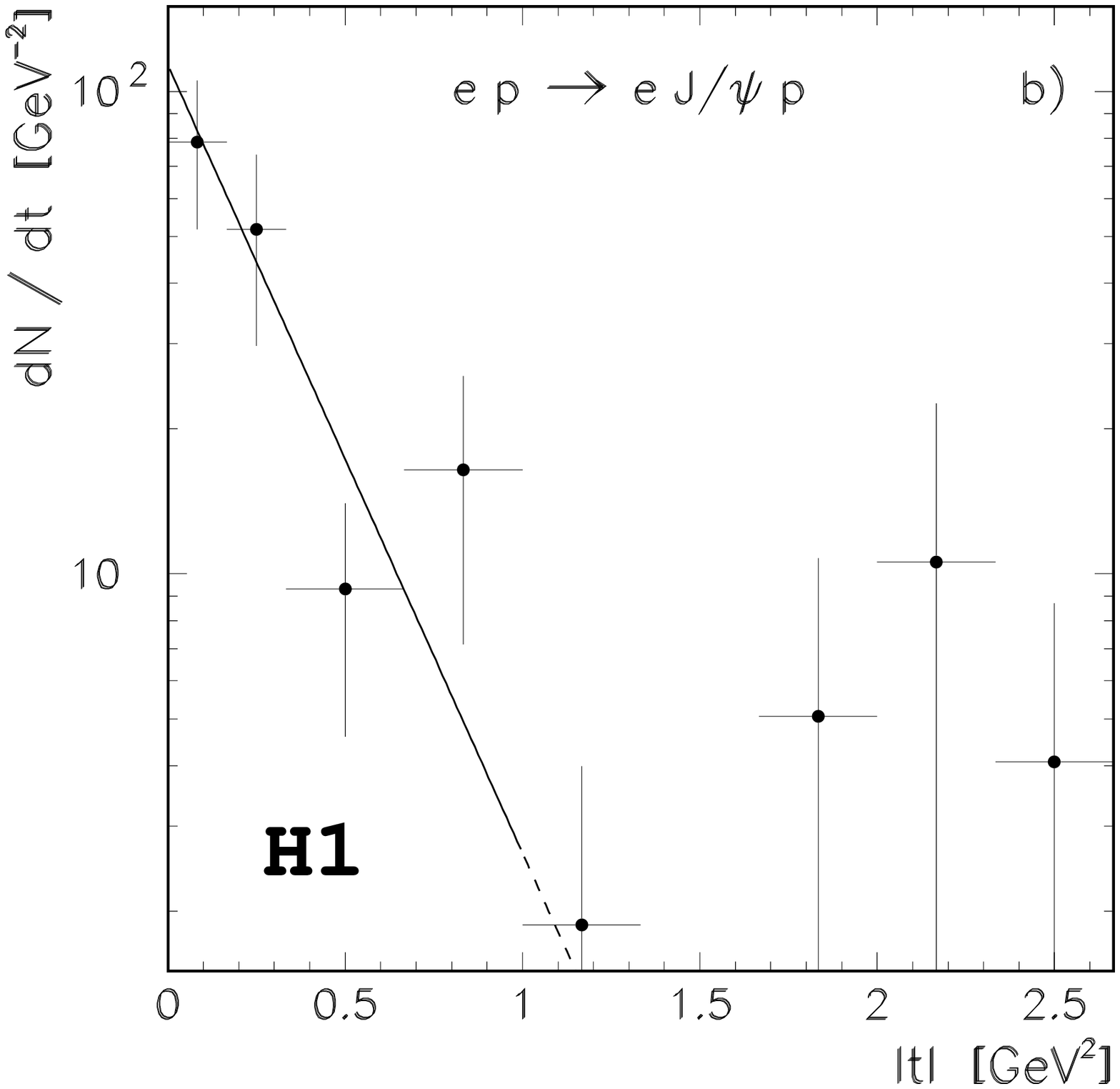,width=4.8cm}\hspace*{-0.1cm}\epsfig{file=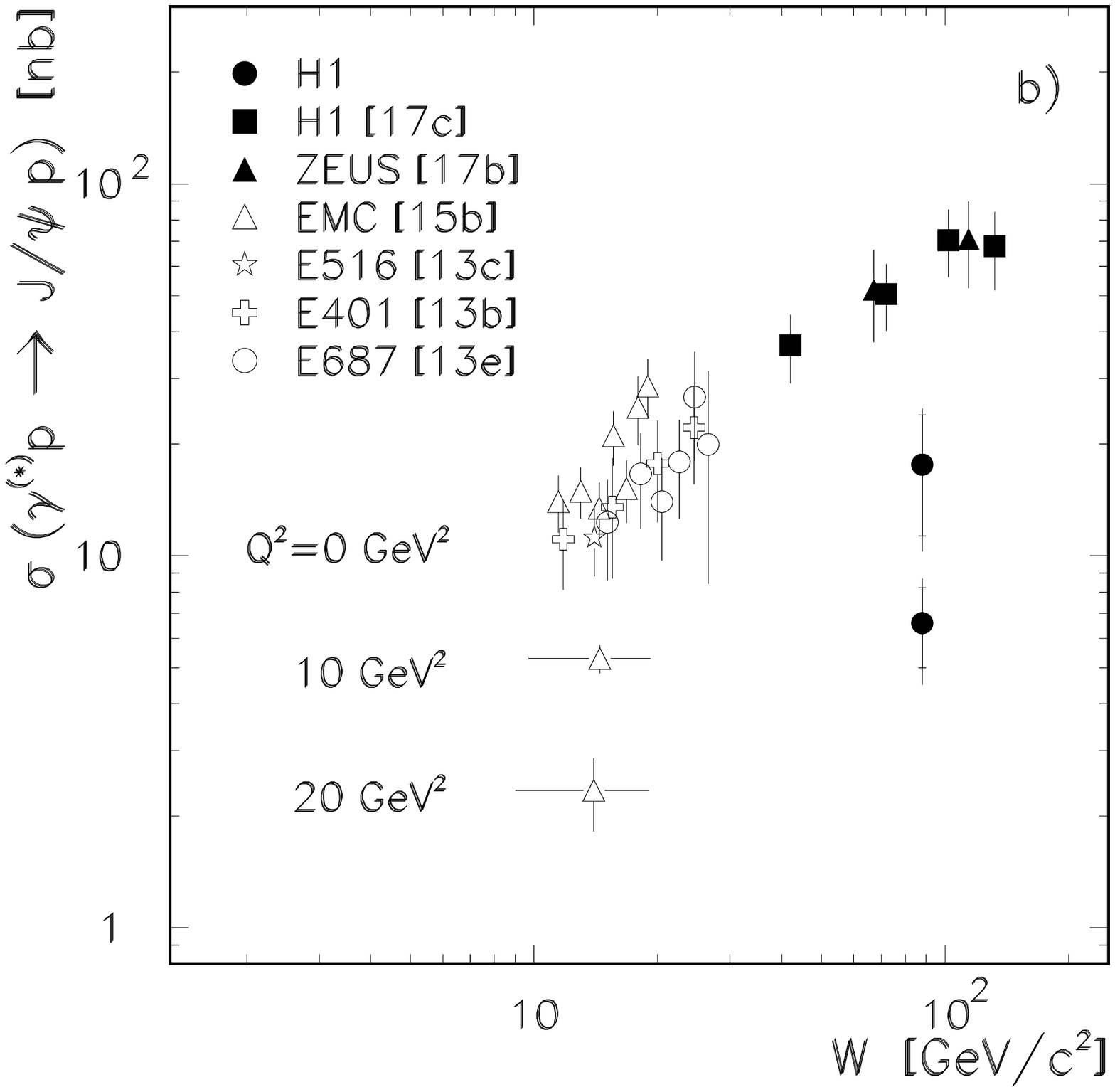,width=4.8cm}
\end{center}
\caption{Reaction $\gamma^* p \to J/\Psi p$ for $Q^2 > 8$\GeV$^2$, $W=30-150$\GeV: mass distribution of $m_{l^+l^-}$ (left), $d\sigma/d|t|$ (middle) and cross section vs $W$ (right); from H1.}
\label{f:sgsjpsih1}
\vfill
\end{figure}

The first signal for $J/\Psi$ production in deep-inelastic scattering was reported by H1~\cite{np:b468:3}: Fig.~\ref{f:sgsjpsih1} shows the lepton-lepton mass distribution, the differential cross section $d\sigma/d|t|$ and the cross section as a function of $W$. The cross sections at $Q^2 =10(20)$ GeV$^2$ are a factor of about $4(10)$  below the measurements at $Q^2 = 0$.

\begin{figure}[p]
\begin{center}
\vfill
\vspace*{-1cm}
\hspace*{-0.1cm}\epsfig{file=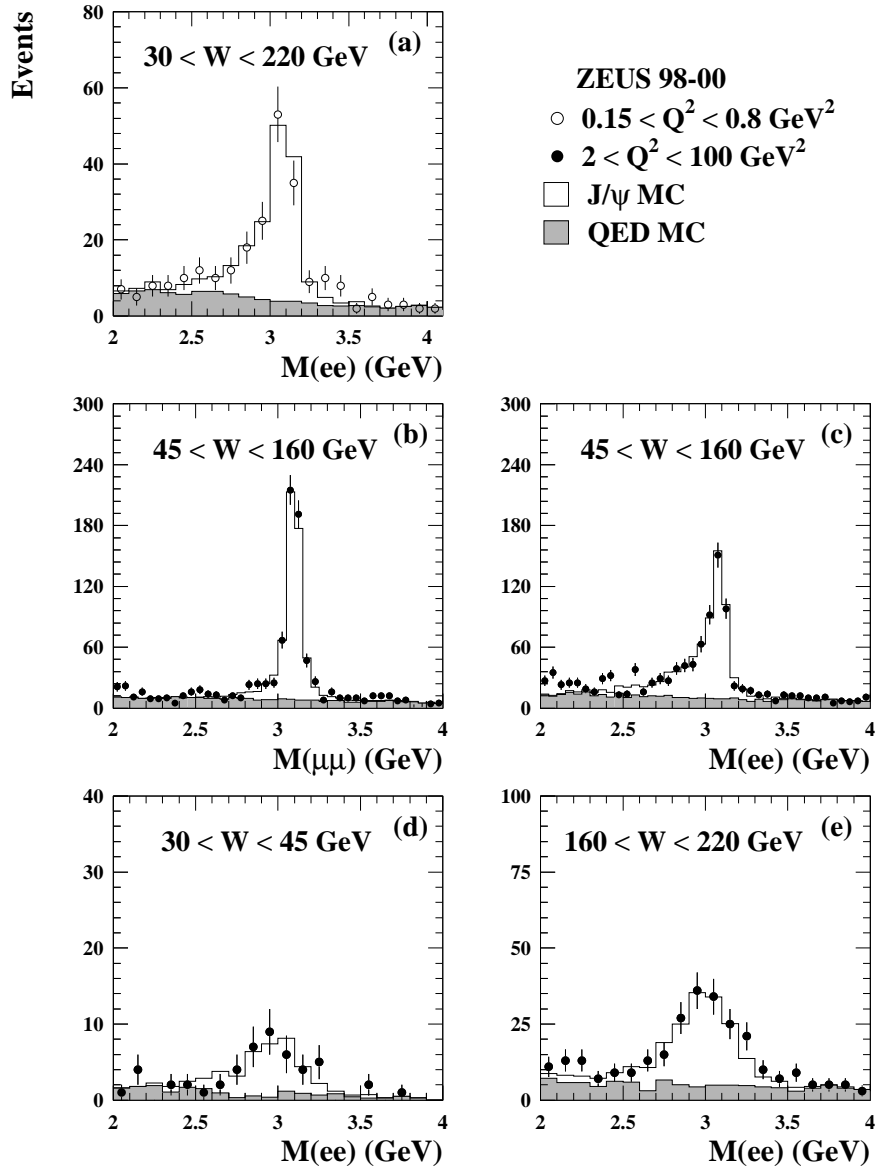,width=12cm}
\end{center}
\caption{Reaction $\gamma^* p \to J/\Psi p$: distributions of $M_{l^+l^-}$, $l=e,\mu$, for the $W,Q^2$ regions indicated; from ZEUS.}
\label{f:sgsmjpsiz}
\vfill
\end{figure}

\begin{figure}[p]
\begin{center}
\vfill
\vspace*{-1cm}
\hspace*{-0.1cm}\epsfig{file=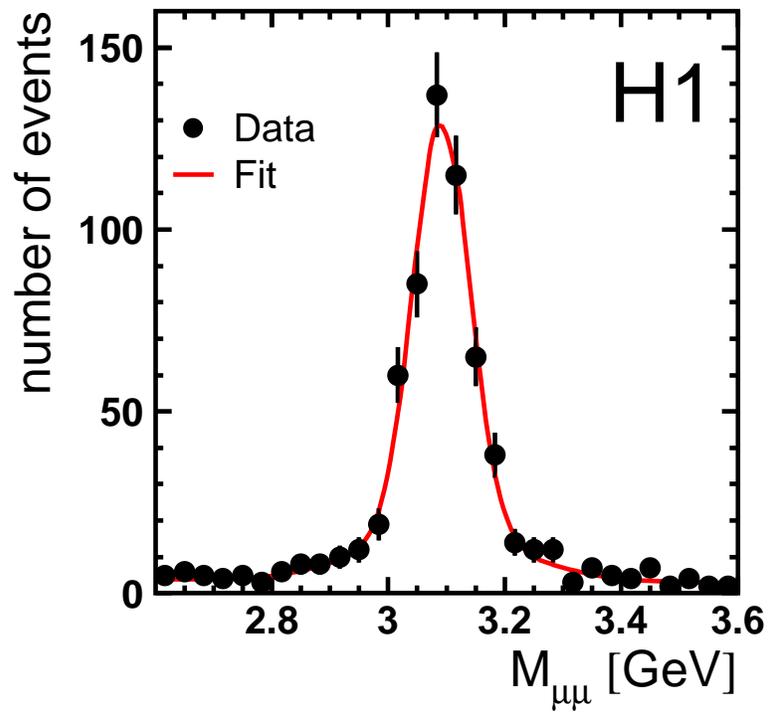,width=10cm}
\end{center}
\caption{Reaction $\gamma^* p \to J/\Psi p$: distribution of $M_{\mu^+\mu^-}$ for $Q^2 = 2-80$\GeV$^2$; from H1.}
\label{f:sgsmjpsih1}
\vfill
\end{figure}

\begin{figure}[p]
\begin{center}
\vfill
\vspace*{-1cm}
\hspace*{-0.1cm}\epsfig{file=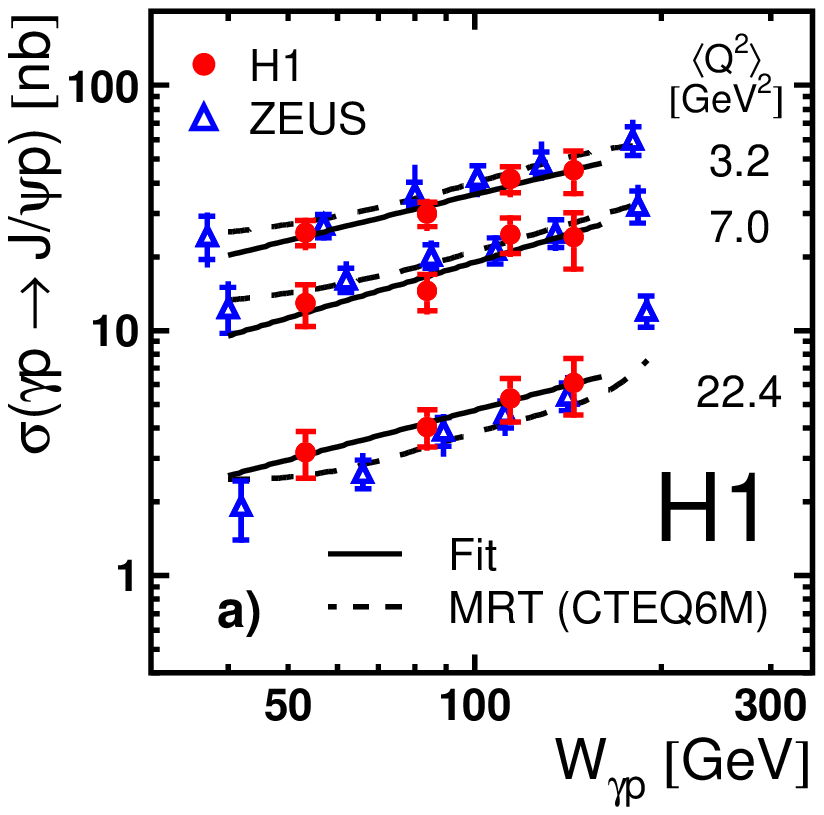,width=12cm}
\hspace*{-0.1cm}\epsfig{file=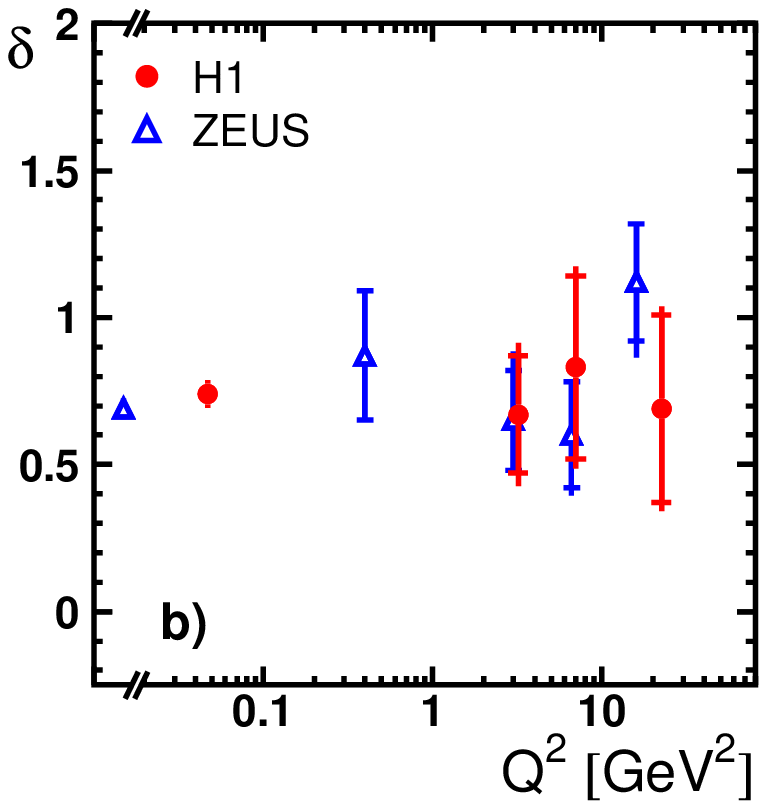,width=8cm}
\end{center}
\caption{The cross section $\gamma^* p \to J/\Psi p$ as a function of $W$ for different values of $Q^2$ as measured by H1 and ZEUS (top); parametrisation $\sigma(\gamma^* p \to J/\Psi p) \propto W^{\delta}$: the power $\delta$ as a function of $Q^2$ as determined by H1 (bottom); from H1.}
\label{f:sgsmjpcroshz}
\vfill
\end{figure}

Further results were reported by 
ZEUS~\cite{epj:c6:603,np:b695:3} and H1~\cite{epj:c10:373,epj:c46:585}.
The final results stem from about 1200 (ZEUS) and 600 (H1) $J/\Psi p$ events with $Q^2 >2$ GeV$^2$, see Figs.~\ref{f:sgsmjpsiz},~\ref{f:sgsmjpsih1}. The $W$ dependence of the cross section $\sigma (\gamma^* p\to J/\Psi)$ is shown in Fig.~\ref{f:sgsmjpcroshz} for $Q^2$ values between 3.2 and 22.4 GeV$^2$. The parametrisation $\sigma(\gamma^* p \to J/\Psi p) \propto W^{\delta}$ yields the $\delta$ values shown as a function of $Q^2$ in Fig.~\ref{f:sgsmjpcroshz}(bottom); within errors, $\delta \approx 0.7$.

\begin{figure}[p]
\begin{center}
\vfill
\vspace*{-1cm}
\hspace*{-0.1cm}\epsfig{file=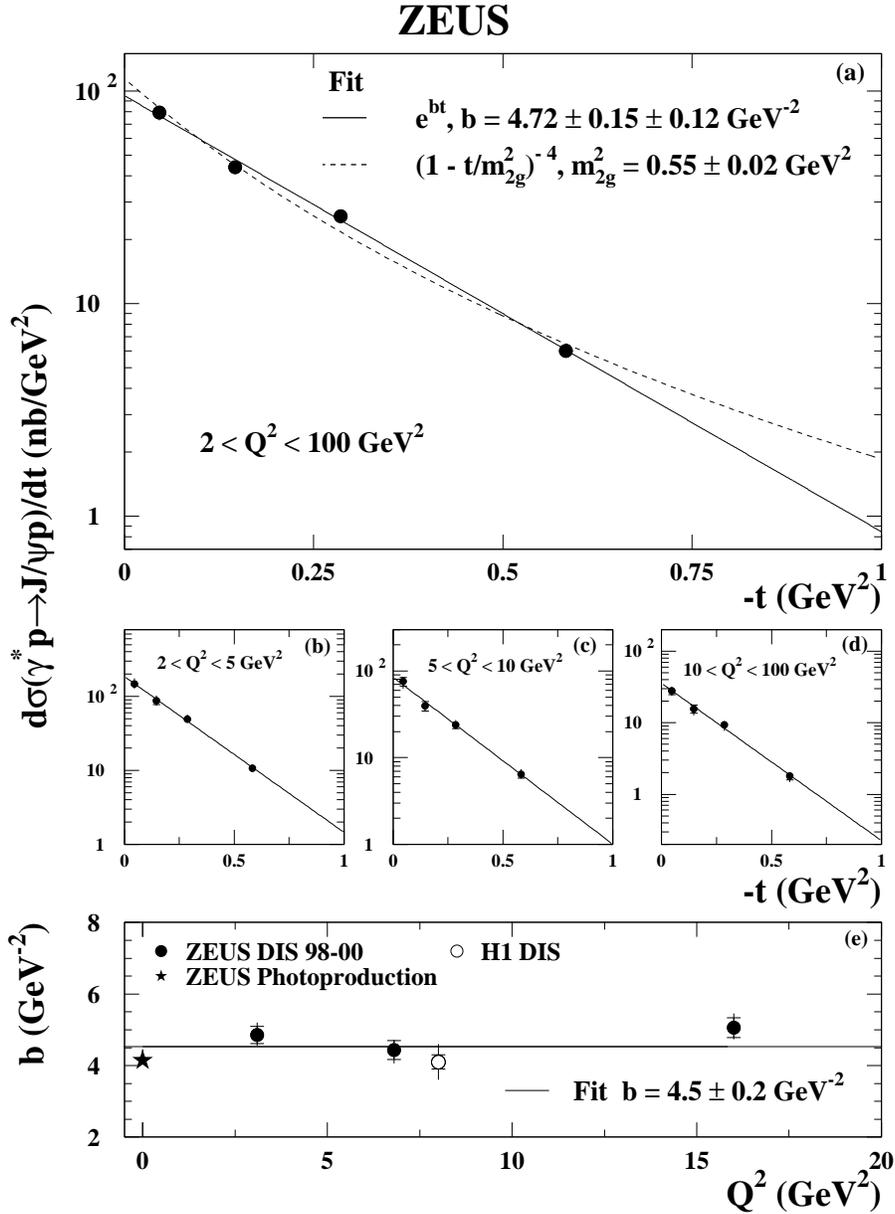,width=12cm}
\end{center}
\caption{Reaction $\gamma^* p \to J/\Psi p$: (a)-(d) differential cross section $d\sigma/dt$ as a function of $-t$ for $Q^2 = 2 -100$\GeV$^2$ and $Q^2 = 2-5, 5-10, 10-100$\GeV$^2$; (e) assuming $\sigma(\gamma^* p \to J/\Psi p) \propto W^{\delta}$: $\delta$ as a function of $Q^2$; from ZEUS.}
\label{f:dsdtsmjpsiz}
\vfill
\end{figure}

The differential cross section $d\sigma(\gamma^* p \to J/\Psi p)/d|t|$ (Fig.~\ref{f:dsdtsmjpsiz}a-d) is proportional to $e^{- b|t|}$, where $b = 4.72\pm 0.15\pm 0.12$ GeV$^{-2}$ (Fig.~\ref{f:dsdtsmjpsiz}a) independent of $Q^2$, as shown by Fig.~\ref{f:dsdtsmjpsiz}(e). The measured $b$-value of 4.7 GeV$^{-2}$ corresponds to the size of the radius of the proton alone. Hence, the radius of the virtual photon must be small.

\begin{figure}[p]
\begin{center}
\vfill
\vspace*{-1cm}
\hspace*{-0.1cm}\epsfig{file=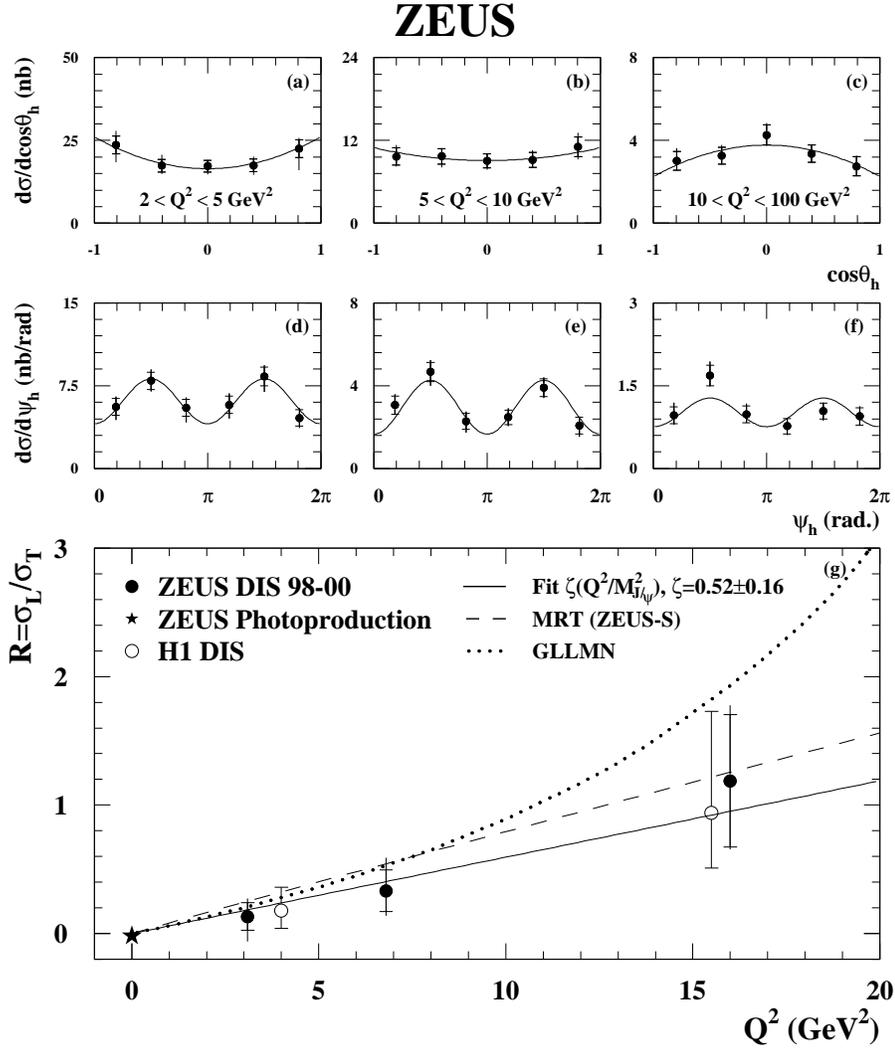,width=12cm}
\end{center}
\caption{Reaction $\gamma^* p \to J/\Psi p$: (a)-(f) decay angular distributions of the $J/\Psi$ in the s-channel helicity system for the $Q^2$ bins indicated; (g) ratio $R = \sigma_L/\sigma_T$ as a function of $Q^2$; from ZEUS.}
\label{f:decayjpsiz}
\vfill
\end{figure}

\begin{figure}[p]
\begin{center}
\vfill
\vspace*{-1cm}
\hspace*{-0.1cm}\epsfig{file=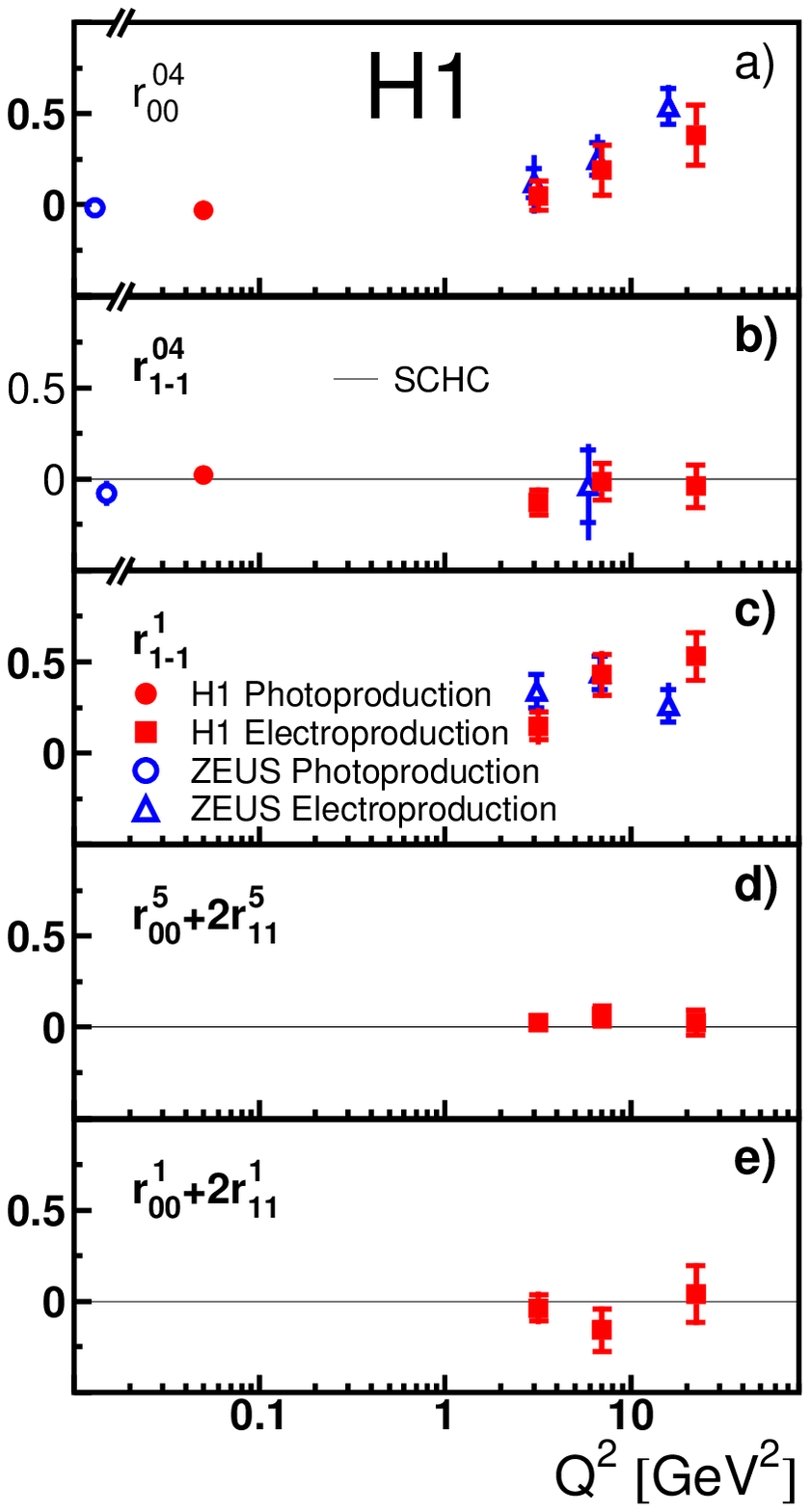,width=7cm}\hspace{-0.2cm}\epsfig{file=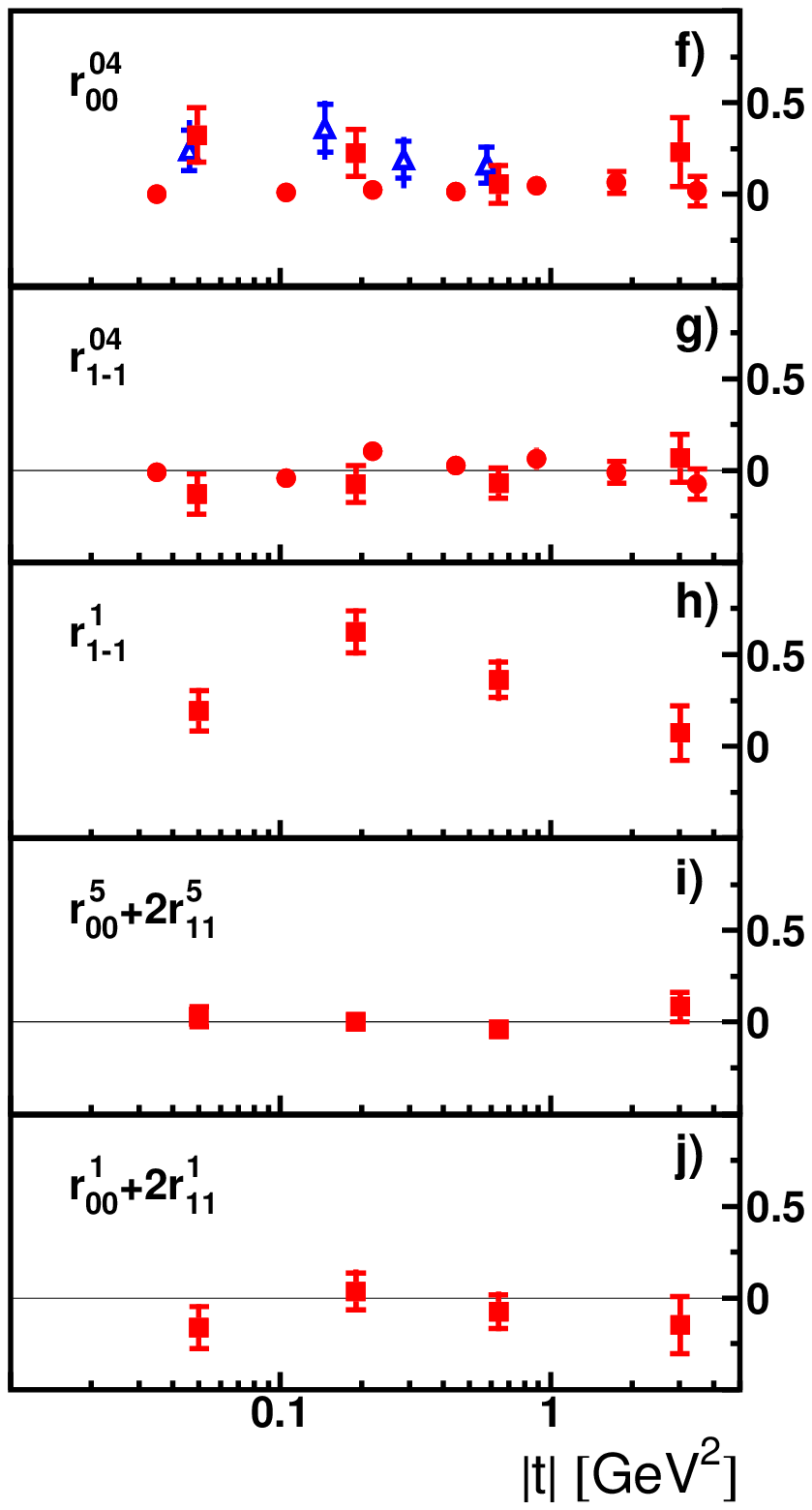,width=7cm}
\end{center}
\caption{Reaction $\gamma^* p \to J/\Psi p$: spin-density matrix elements as a function of $Q^2$ (left),$|t|$ (right), with data from H1 and ZEUS; from H1.}
\label{f:jpsirhoik}
\vfill
\end{figure}

The decay $J/\Psi \to l^+l^-$ was studied in terms of the decay angles $\theta_h$ and $\psi_h = \phi_h - \Phi_h$, where $\theta_h$ and $\phi_h$ are the polar and azimuthal angles of the positively charged decay lepton, and $\Phi_h$ is the angle between the $J/\Psi$ production plane and the lepton scattering plane. The distributions of $\cos \theta_h$ and of $\psi_h$ are shown in Figs.~\ref{f:decayjpsiz}a-f for $Q^2$ between 2 and 100 GeV$^2$.  The full set of $r^{\alpha}_{ik}$ is shown in Fig.~\ref{f:jpsirhoik} as a function of $Q^2$ (left) and $|t|$ (right). The element $r^{04}_{1-1}$ and the combinations $r^{5}_{00} +2r^{5}_{11}$ and $r^{1}_{00} +2r^{1}_{11}$ are zero, within errors, in agreement with SCHC. Assuming SCHC, the ratio of the longitudinal to transverse cross sections, $R = \sigma_L/\sigma_T$, was determined from $r^{04}_{00}$: $R = \frac{1}{\epsilon} \frac{r^{04}_{00}}{1-r^{04}_{00}}$, where $\epsilon$ is the polarisation parameter. The ratio $R$ is shown in  Fig~\ref{f:decayjpsiz}(g): it is rising proportional to $Q^2$: $R = \alpha \cdot (Q^2/M^2_{J/\Psi})$ with $\alpha = 0.52\pm0.16$. For $Q^2 = 20$ GeV$^2$, $\sigma_L$ and $\sigma_T$ are approximately equal.

\subsubsection{$\gamma p\to \Psi(2S) p$}
\begin{figure}[p]
\begin{center}
\vfill
\vspace*{-1cm}
\hspace*{-0.1cm}\epsfig{file=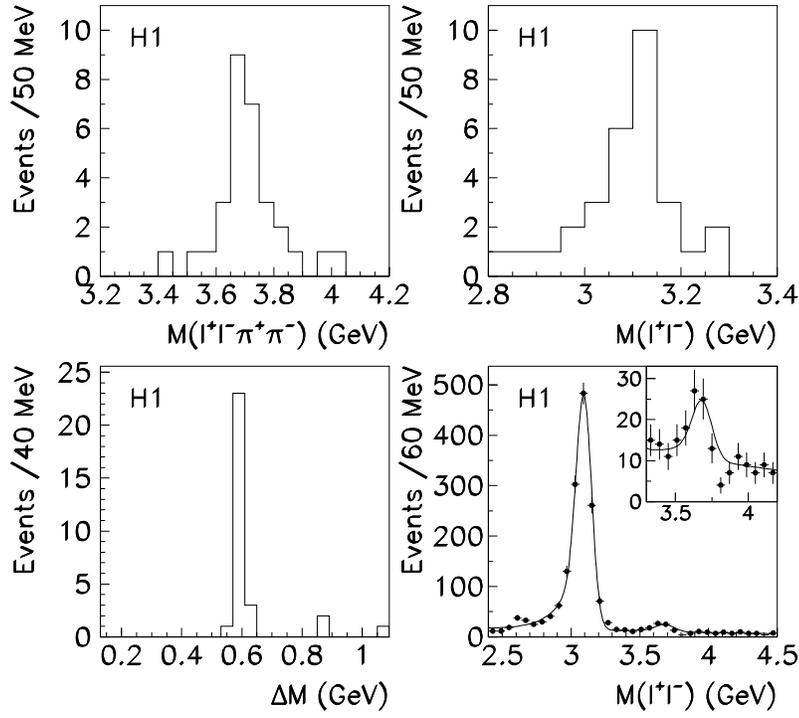,width=12cm}
\end{center}
\caption{Reaction $\gamma p \to \Psi(2S) p$: effective mass plots for the four track sample: four track effective mass; $l^+ l^-$ ($l = e,\mu$) effective mass; $\Delta M = (M_{l^+l^- \pi^+ \pi^-} - M_{l^+l^-})$; $M_{l^+l^-}$ of the two-track sample; from H1.}
\label{f:psiprimh}
\vfill
\end{figure}

\begin{figure}[p]
\begin{center}
\vfill
\vspace*{-1cm}
\hspace*{-0.1cm}\epsfig{file=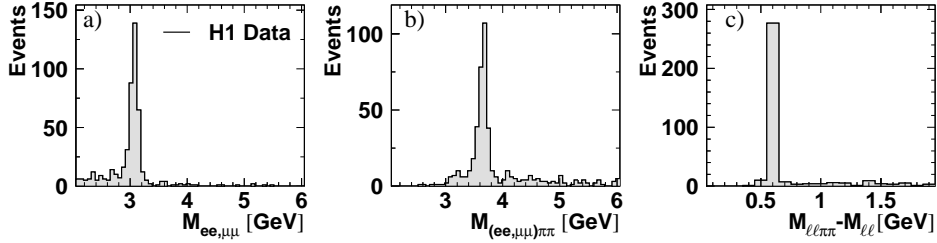,width=12cm}
\end{center}
\caption{Reaction $\gamma p \to \Psi(2S) p$, from H1.}
\label{f:psiprim2h2}
\vfill
\end{figure}

\begin{figure}[p]
\begin{center}
\vfill\vspace*{-1cm}
\hspace*{-0.1cm}\epsfig{file=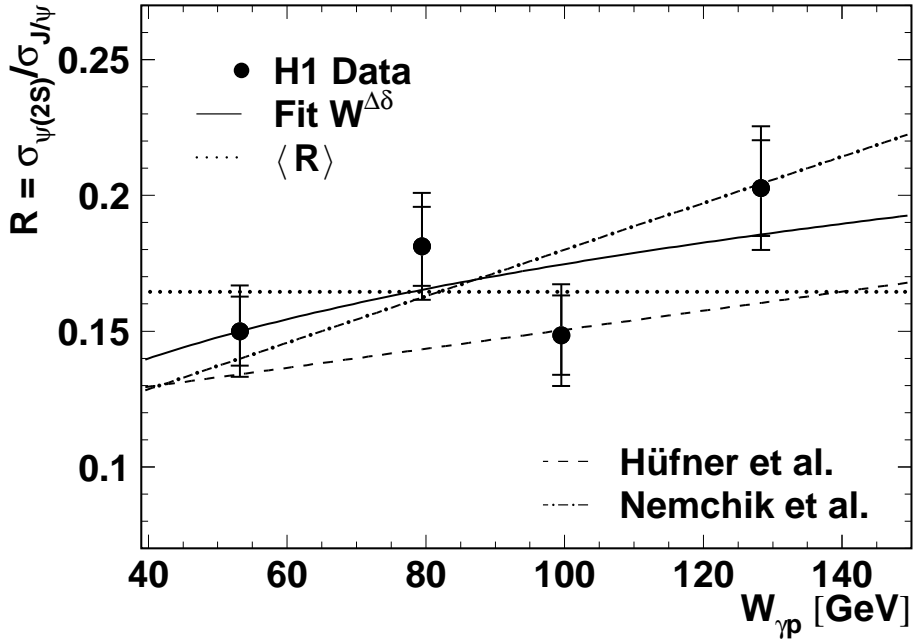,width=12cm}
\end{center}
\caption{The ratio $R(W_{\gamma p}) = \sigma(\Psi(2S))/\sigma(J/\Psi)$:
the solid line shows a fit to the data with
$R \propto W^{\Delta \delta}$ where $\Delta \delta = 0.25$. 
The dashed and dashed-dotted lines show predictions, see text; from H1.}
\label{f:rpsipsish}
\vfill
\end{figure}

The production of $\Psi(2S)$ in $\gamma p$ collisions was reported by H1~\cite{pl:b421:385,pl:b541:251}. In the first analysis, based on an integrated luminosity of 6.3 pb$^{-1}$, the $\Psi(2S)$  was identified by the decays $\Psi(2S) \to l^+l^-$, $l = e,\mu$, and $\Psi(2S) \to J/\Psi \pi^+\pi^-$, $J/\Psi \to l^+ l^-$, see Fig.~\ref{f:psiprimh}. A total of 60 $\Psi(2S)$ events were observed. In the second analysis, performed with 77 pb$^{-1}$, about 300 $\Psi(2S)$ could be isolated (Fig.~\ref{f:psiprim2h2}). Figure~\ref{f:rpsipsish} shows the cross section ratio $R = \sigma(\Psi(2S))/\sigma(J/\Psi)$. When $R$ is assumed to be independent of $W_{\gamma p}$, the data yield\\ 
\\
$R = \sigma(\Psi(2S))/\sigma(J/\Psi)= 0.166\pm 0.007(stat.)\pm 0.008(syst.) \pm 0.007(BR)$,\\
\\ 
where the last error accounts for the uncertainty of the $\Psi(2S) \to J/\Psi$ branching ratio. The dashed and dashed-dotted lines show the predictions of~\cite{jetp:86:1054},~\cite{pr:d62:094022}: in these models the photon fluctuates into a $c \cbar$ colour dipole which interacts with the proton via the exchange of two gluons. The different wave functions of $J/\Psi$ and $\Psi^{\prime}$ are taken into account. The model of Nemchik et al.~\cite{jetp:86:1054} reproduces the data somewhat better.
 
\subsubsection{$\gamma p\to \Upsilon p$}
\begin{figure}[p]
\begin{center}
\vfill\vspace*{-1cm}
\hspace*{-0.1cm}\epsfig{file=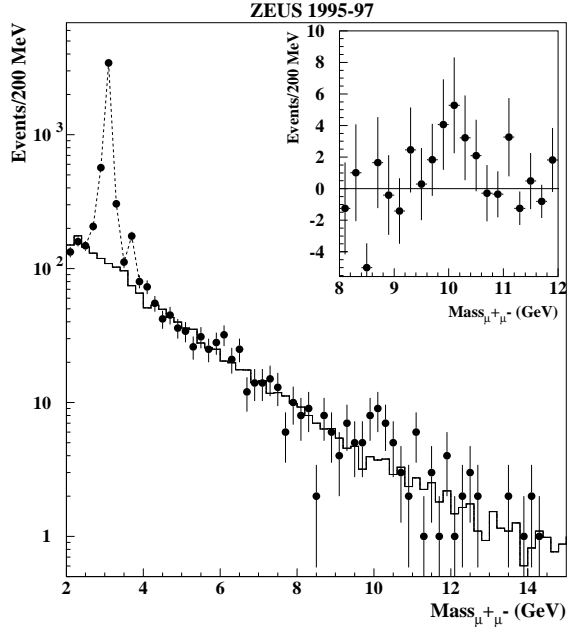,width=8cm}
\end{center}
\caption{Mass distribution of $\mu^+ \mu^-$ pairs: the histogram represents the simulated Bethe-Heitler background. Points in the $J/\Psi$ region are connected by a dotted line. The insert shows the signal in the $\Upsilon$ region after background subtraction; from ZEUS.}

\label{f:upsilonz}
\vfill
\end{figure}

The first observation of the reaction $\gamma p \to \Upsilon p$ was reported by ZEUS~\cite{pl:b437:432}, see Fig.~\ref{f:upsilonz}. Based on an integrated luminosity of 43.2 pb$^{-1}$, a total of 57 events were observed in the mass range 8.9 - 10.9 GeV of which 17 events were estimated to come from $\Upsilon$ production(assuming that the relative production cross sections for the $\Upsilon(nS)$ states are the same as those measured in $p \pbar$ collisions by CDF~\cite{prl:75:4358}). The cross section derived for a mean $W= 120$ GeV is
\begin{eqnarray}
\sigma_{\gamma p \to \Upsilon p} 
& = & 375 \pm 170(stat.)^{+75}_{-64}(syst.) \pb,
\end{eqnarray}
and the ratio
\begin{eqnarray}
\sigma_{\gamma p \to \Upsilon(1S) p}/\sigma_{\gamma p \to J/\Psi p} 
& = & [(4.8\pm 2.2 (\rm stat.)
^{+0.7}_{-0.6} (\rm syst.)] \times 10^{-3}.
\end{eqnarray}

\begin{figure}[p]
\begin{center}
\vfill\vspace*{-1cm}
\hspace*{-0.1cm}\epsfig{file=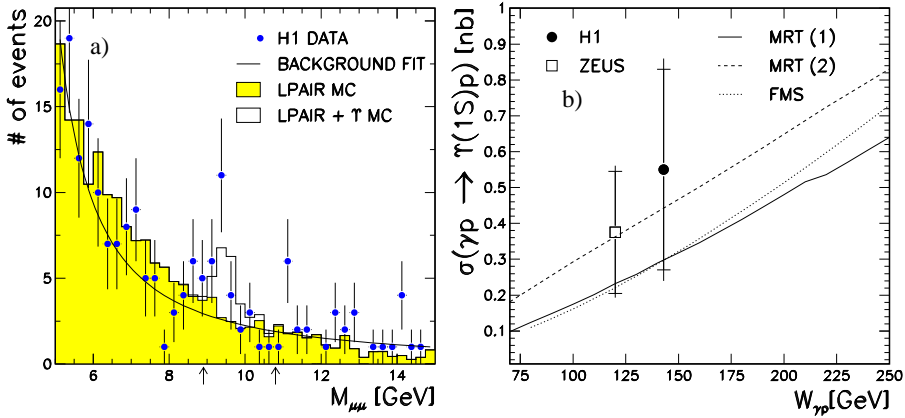,width=14cm}
\end{center}
\caption{(a) Mass distribution of $\mu^+ \mu^-$ pairs from H1: the grey histogram represents the simulated background; (b) cross section for $\gamma p \to \Upsilon (1S) p$ vs. $W$ as measured by ZEUS and H1; in both measurements, 70 \% of the signal in the $\Upsilon$ mass region is assumed to be due to $\Upsilon(1S)p$ production; from H1.}

\label{f:upsilonzh}
\vfill
\end{figure}

A further measurement of $\Upsilon$ photoproduction was presented by H1~\cite{pl:b483:23}. Based on an integrated luminosity of 27.5 pb$^{-1}$, a total of $12.2\pm 6.3$ signal events were detected, see Fig.~\ref{f:upsilonzh}(a). The cross section $\sigma_{\gamma p \to \Upsilon(1S) p}$ as measured by ZEUS and H1 is shown in Fig.~\ref{f:upsilonzh}(b): it is of the order of 0.3 - 0.6 nb for $W = 120 - 150$ GeV.

\subsection{The $W$ dependence of the photoproduction of vectormesons}
\begin{figure}[p]
\begin{center}
\vfill\vspace*{-1cm}
\hspace*{-0.1cm}\epsfig{file=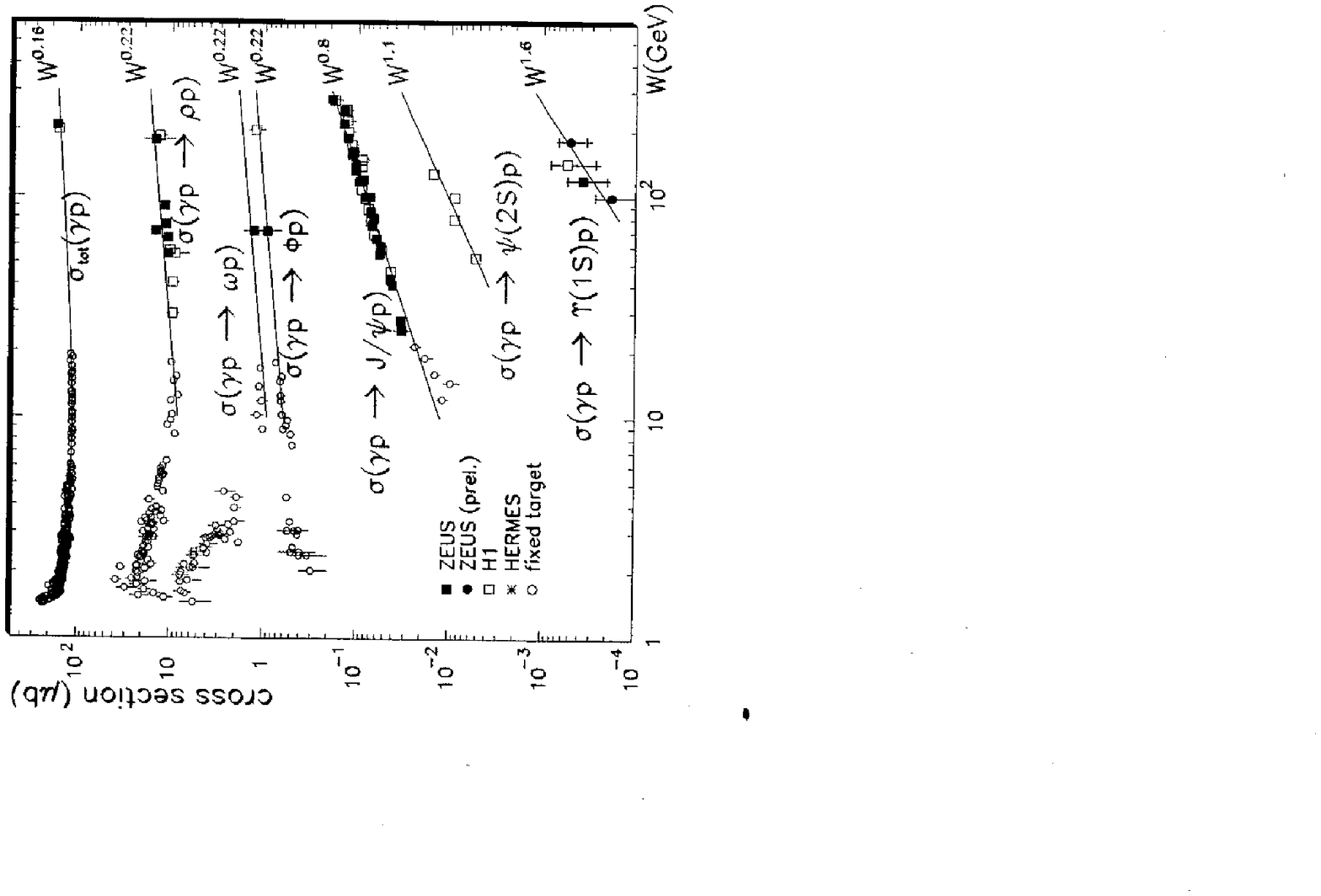,angle=270,width=14cm}
\end{center}
\caption{Photoproduction: total cross section and cross sections for the production of vectormesons, $\gamma p \to V p$, $V = \rho^0, \omega, \phi, J/\Psi, \Psi(2S), \Upsilon$, as a function of $W$; from A. Levy}

\label{f:gptovptot}
\vfill
\end{figure}

The cross sections for photoproduction of vectormesons, $\gamma p \to V^0 p$, are shown in Fig.~\ref{f:gptovptot} as a function of $W$ together with the total $\gamma p$ cross section~\cite{alevy2007}. The parametrisation of the cross sections by the form $\sigma \propto W^{\delta}$ gives the $\delta$-values indicated: for the total cross section $\delta = 0.16$; for vectormeson production $\delta$ increases with the mass of the vectormeson from $\delta = 0.22$ ($\rho^0$) to $\delta = 0.8$ ($J/\psi$) and $\delta \approx 1.6$ ($\Upsilon$).

\subsubsection{Estimate of cross sections for elastic vectormeson proton scattering}

Assuming Vector Meson Dominance (VDM), the cross sections measured for photoproduction of vectormesons, $\gamma p \to V^0 p$, can be used to estimate the cross sections for elastic scattering of transverse vectormesons  on protons, $V_T^0 p \to V_T^0 p$:
\begin{eqnarray}
\sigma(V_T^0 p \to V_T^0 p) & = & \frac{4}{\alpha}\frac{\gamma^2_V}{4 \pi} \cdot \sigma(\gamma  p \to V_T^0 p),
\end{eqnarray}
where $\frac{\gamma^2_V}{4 \pi} = \frac{\alpha^2 M_V}{12 \Gamma_{Vee}} = 0.489, 5.8, 3.56, 2.48$ for $V = \rho, \omega, \phi, J/\Psi$, respectively~\cite{PDG2006}. The resulting cross sections for $W = 70-80$ GeV are 
\begin{eqnarray}
\sigma(\rho^0_T p \to \rho^0_T p) & = & 4.0 \pm 0.5 \mb \\
\sigma(\omega_T p \to \omega_T p) & = & 3.8 \pm 1.0 \mb \\
\sigma(\phi_T p \to \phi_T p)     & = & 1.9 \pm 0.6 \mb\\
\sigma(J/\Psi_T p \to J/\Psi_T p) & = & 0.8 \pm 0.1 \mb.
\end{eqnarray} 
The elastic $\rho^0_T p$ and $\omega_T p$ cross sections are of the same magnitude as the cross sections measured for elastic $\pi^+ p$ and $\pi^- p$ scattering - albeit at a somewhat lower energy of $W \approx 26$ GeV:  $\sigma(\pi^- p \to \pi^- p) = 3.6$ mb and $\sigma(\pi^+ p \to \pi^+ p) = 3.3$ mb~\cite{RPP2006}. 

\subsection{The $Q^2$ and $t$-dependences of vectormeson production}
\begin{figure}[p]
\begin{center}
\vfill
\vspace*{-1cm}
\hspace*{0.1cm}\epsfig{file=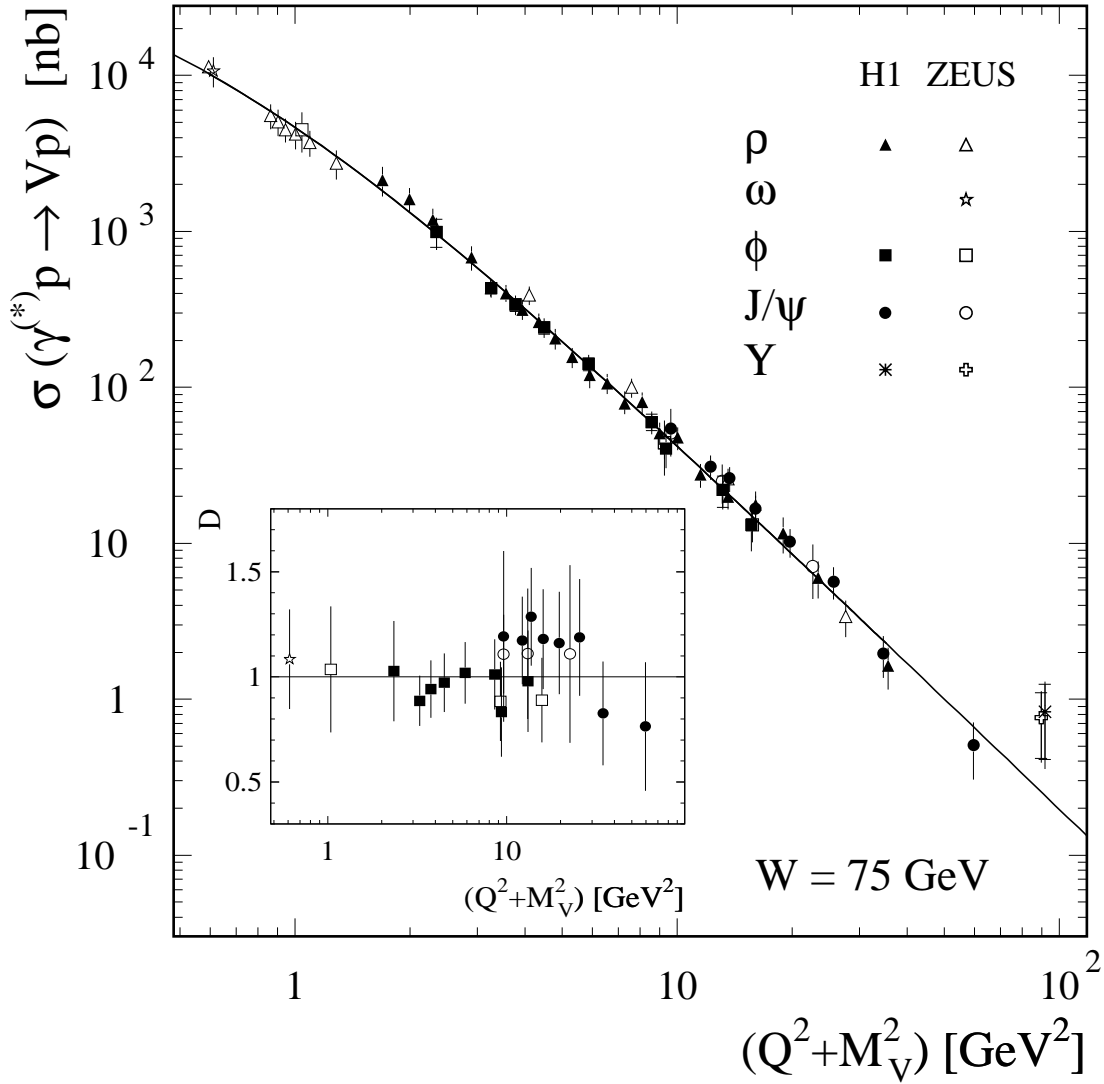,width=16cm}
\end{center} 
\caption{The cross section $\sigma(\gamma^* p \to V p)$ with $V = \rho^0, \omega, \phi, J/\Psi, Y$, as a function of $(Q^2 + M_V^2)$ for $W = 75$\GeV: results from H1 and ZEUS.}
\label{f:gsvqmvhz}
\vfill
\end{figure}

Figure~\ref{f:gsvqmvhz} shows the cross sections for $\gamma^*p \to Vp$ at $W = 75$ GeV, as a function of $(Q^2 + M^2_V)$, where $V = \rho^0, \omega, \phi, J/\Psi, \Upsilon$~\cite{pl:b483:360}. The cross sections lie on a universal curve and are falling by four orders of magnitude from $(Q^2 + M^2_V) = 1$ to $(Q^2 + M^2_V) = 100$ GeV$^2$. The curve shown has the form $\sigma = a_1(Q^2 +M^2_V + a_2)^{a_3}$, with $a_1 = 10689 \pm 165 \nb$, $a_2 = 0.42 \pm 0.09$ GeV$^2$ and $a_3 = - 2.37 \pm 0.10$~\cite{pl:b483:360}.

\begin{figure}[p]
\begin{center}
\vfill\vspace*{-1cm}
\hspace*{-0.1cm}\epsfig{file=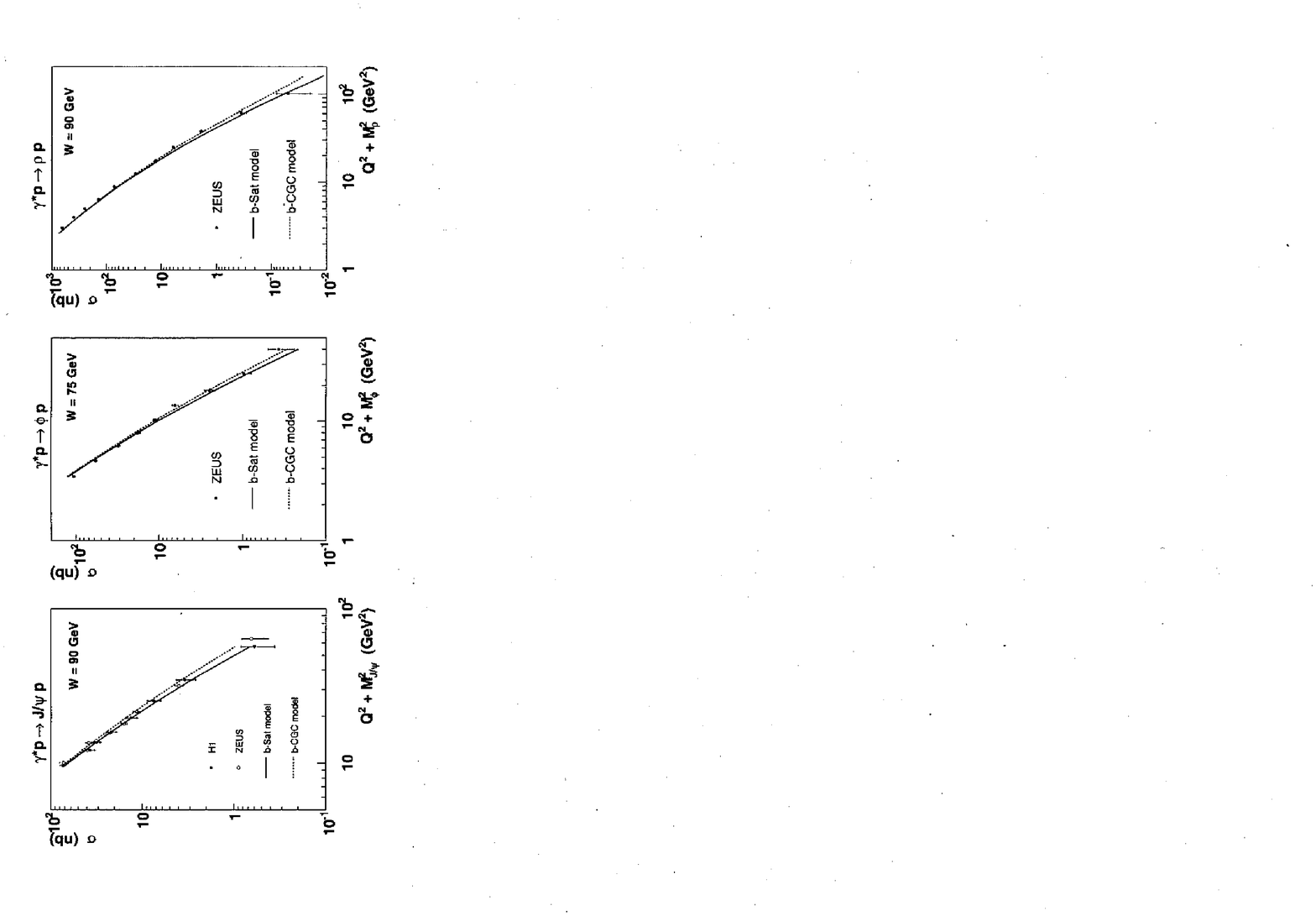,angle=270,width=15cm}
\end{center}
\caption{The cross section $\sigma(\gamma^* p \to V p)$ with $V = \rho^0, \phi, J/\Psi$, as a function of $(Q^2 + M_V^2)$ for $W = 75$\GeV: results from H1 and ZEUS. The curves have been computed with the b-Sat and b-CGC models; from Watt and Kowalski.}
\label{f:gsptovpwk}
\vfill
\end{figure}

The cross sections for $V = \rho^0, \phi, J/\Psi$ are compared in Fig.~\ref{f:gsptovpwk} with the predictions of the bSat and gGGC models~\cite{ianitamun04}~\footnote{These models are based on the socalled color glass concept~\cite{mclervenug}: the hadrons participating in the scattering process behave as a new form of matter (color glass condensate) which is made of small-x gluons. These gluons carry small fractions $x << 1$ of the total momentum and are created by slowly moving color sources - i.e. partons at large $x$ - like for a glass where, on short time scales, the constituents appear to be frozen.}, 
as provided by~\cite{wattkowal08}. A good description of the data is obtained.

\begin{figure}[p]
\begin{center}
\vfill\vspace*{-1cm}
\hspace*{-0.1cm}\epsfig{file=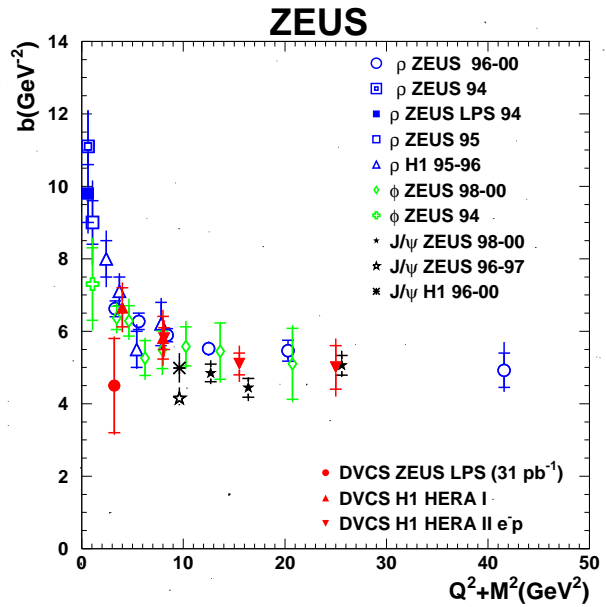,width=9cm}
\end{center}
\caption{The slope b characterizing the $t$ dependence, $d\sigma/d|t| \propto e^{-b|t|}$, of photon and vectormeson production by $\gamma,\gamma^* p$ interactions, shown for DVCS and for $V=\rho^0,\phi,J/\Psi$ production as a function of $(Q^2 + M^2_V)$, where $M_V = 0$ for DVCS and equal to the mass of the vectormeson otherwise; from ZEUS.}
\label{f:tslopegVpz}
\vfill
\end{figure}

The slope $b$ characterising the $|t|$ dependence of photon and vectormeson production, $d\sigma/d|t| \propto  e^{-b |t|}$, is shown in Fig.~\ref{f:tslopegVpz} as a function of $(Q^2 + M^2_V)$, where $M_V = 0$ for DVCS, and equal to the mass of the vectormeson produced, otherwise~\cite{desy:08:132}. Within errors, all slope values lie on a universal curve. For $(Q^2 + M^2_V) \ge 5$ GeV$^2$, $b$ reaches a constant value of about $5$ GeV$^{-2}$. As remarked before, this value of $b$ corresponds to an interaction radius of $R = \sqrt{4 b}=0.9 \cdot 10^{-13}$cm, consistent with the radius of the proton: in reactions of the type $\gamma^*p \to Vp$, the transverse extension of the virtual photon becomes negligible when $Q^2 \ge 5$ GeV$^2$.

\begin{figure}[p]
\begin{center}
\vfill\vspace*{-1cm}
\vspace*{-1cm}\hspace*{+0.1cm}\epsfig{file=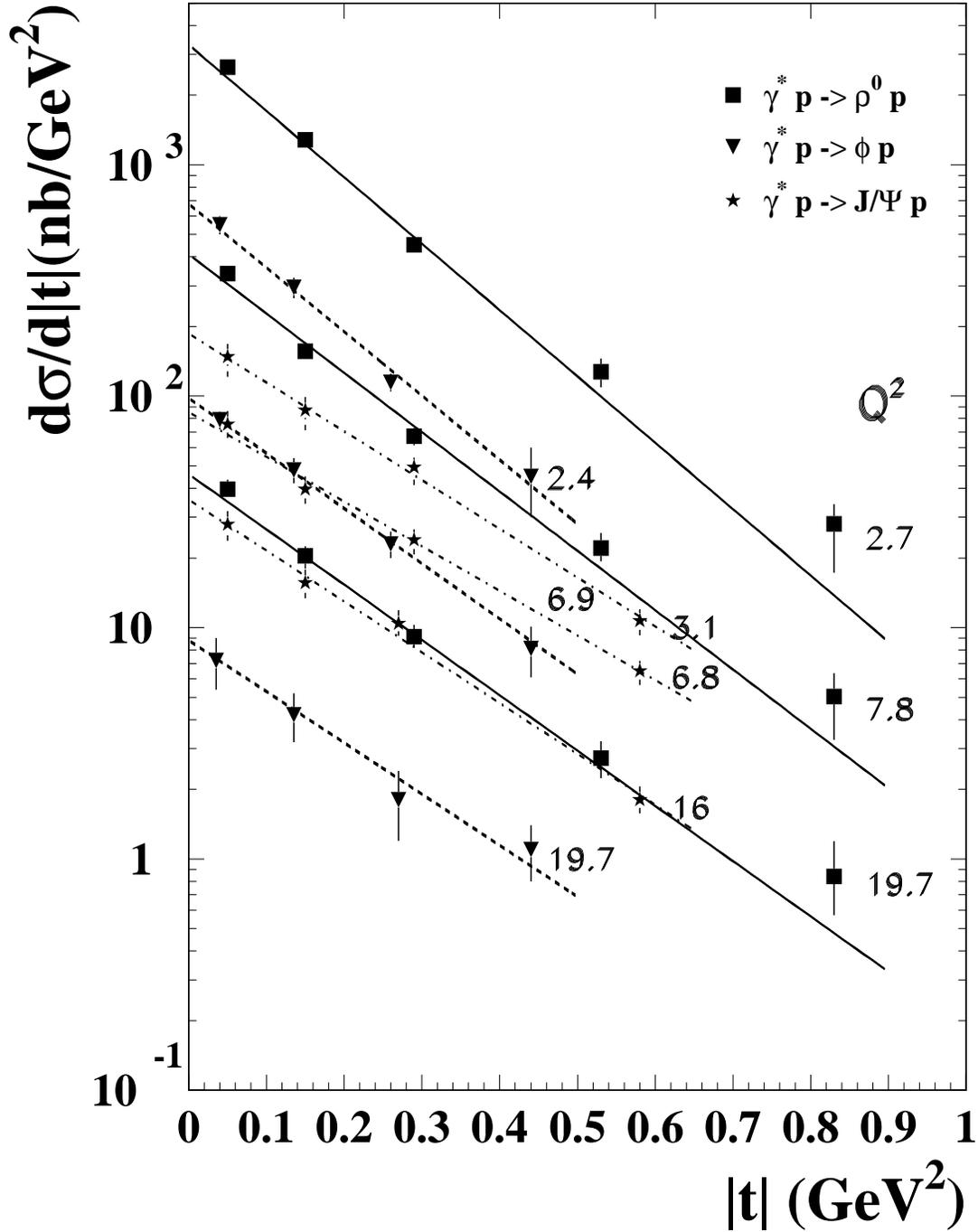,width=16cm}
\end{center}
\caption{The differential cross sections $d\sigma/d|t|$ as a function of $|t|$for vectormeson production, $\gamma^*p \to Vp$, $V = \rho^0, \phi, J/\Psi$; at $W = 90$\GeV for $\rho^0, J/\Psi$ and at $W = 75$\GeV for $\phi$; as determined by ZEUS. The lines represent the fits of the form $d\sigma/d|t| = c \cdot e^{-b|t|}$ shown above.}
\label{f:rophipsivst}
\vfill
\end{figure}

The differential cross sections $d\sigma/d|t|$ for $\gamma^*p \to Vp$ are shown in 
Fig.~\ref{f:rophipsivst} 
as a function of $|t|$, where $V = \rho^0,J/\Psi$ ($W = 90$ GeV) and $V=\phi$ ($W = 75$ GeV); $Q^2$ is fixed with values between 2.4 and 19.7 GeV$^2$. The $|t|$-dependences of the three reactions are consistent with an exponential fall-off, $d\sigma/d|t| \propto e^{-b|t|}$. 

\begin{figure}[p]
\begin{center}
\vfill\vspace*{-1cm}
\vspace*{-1cm}\hspace*{+0.1cm}\epsfig{file=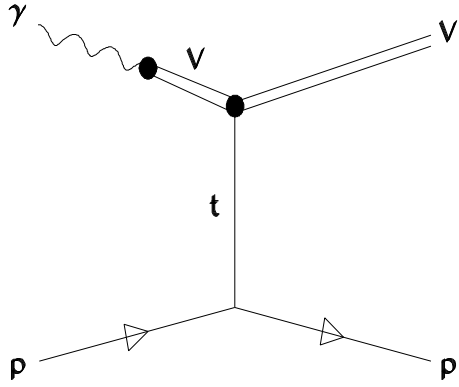,width=6cm}
\end{center}
\caption{Diagram for diffractive vectormeson production, $\gamma^* p \to V p$.}
\label{f:diagprophipsi}
\vfill
\end{figure}

The relative size of these cross sections at large $Q^2$ can be understood by assuming that the virtual - predominantly longitudinal - photon couples to a vectormeson $V$, which scatters elastically on the proton, see Fig~\ref{f:diagprophipsi}. The cross section for $\gamma_L^* p \to V_L p$ is given by:
\begin{eqnarray}
\sigma (\gamma_L^* p \to V^0_L p) = \frac{\alpha}{4} \cdot \frac{1}{\gamma_V^2/{4 \pi}} \cdot \frac{M_V^4}{(Q^2 + M_V^2)^2} \cdot \frac{Q^2}{M^2_V} \sigma (V^0_L p \to V^0_L p).
\end{eqnarray}

From Fig.~\ref{f:rophipsivst} one finds for $|t|$ = 0.5 GeV$^2$:\\
\begin{eqnarray}
\frac{d\sigma(\phi_L p \to \phi_L p,W=75,Q^2=19.7)/d|t|}{d\sigma(\rho^0_L p \to \rho^0_L p,W=90,Q^2=19.7)/d|t|} = 1.03\pm0.05\\
\frac{d\sigma(J/\Psi_L p \to J/\Psi_L p, W=90, Q^2 =19.7)/d|t|}{d\sigma(\rho^0_L p \to \rho^0_L p, W=90, Q^2=19.7)/d|t|} = 0.59 \pm 0.15 
\end{eqnarray}
Both ratios are about a factor of two larger than those obtained with transverse photons at $Q^2 = 0$, see above.

\subsection{$\gamma^* p \to D^{*\pm}(2010) X p$ and $\gamma^* p \to c \bar{c} p$}

\begin{figure}[p]
\begin{center}
\vfill
\vspace*{-1cm}
\hspace*{-0.1cm}\epsfig{file=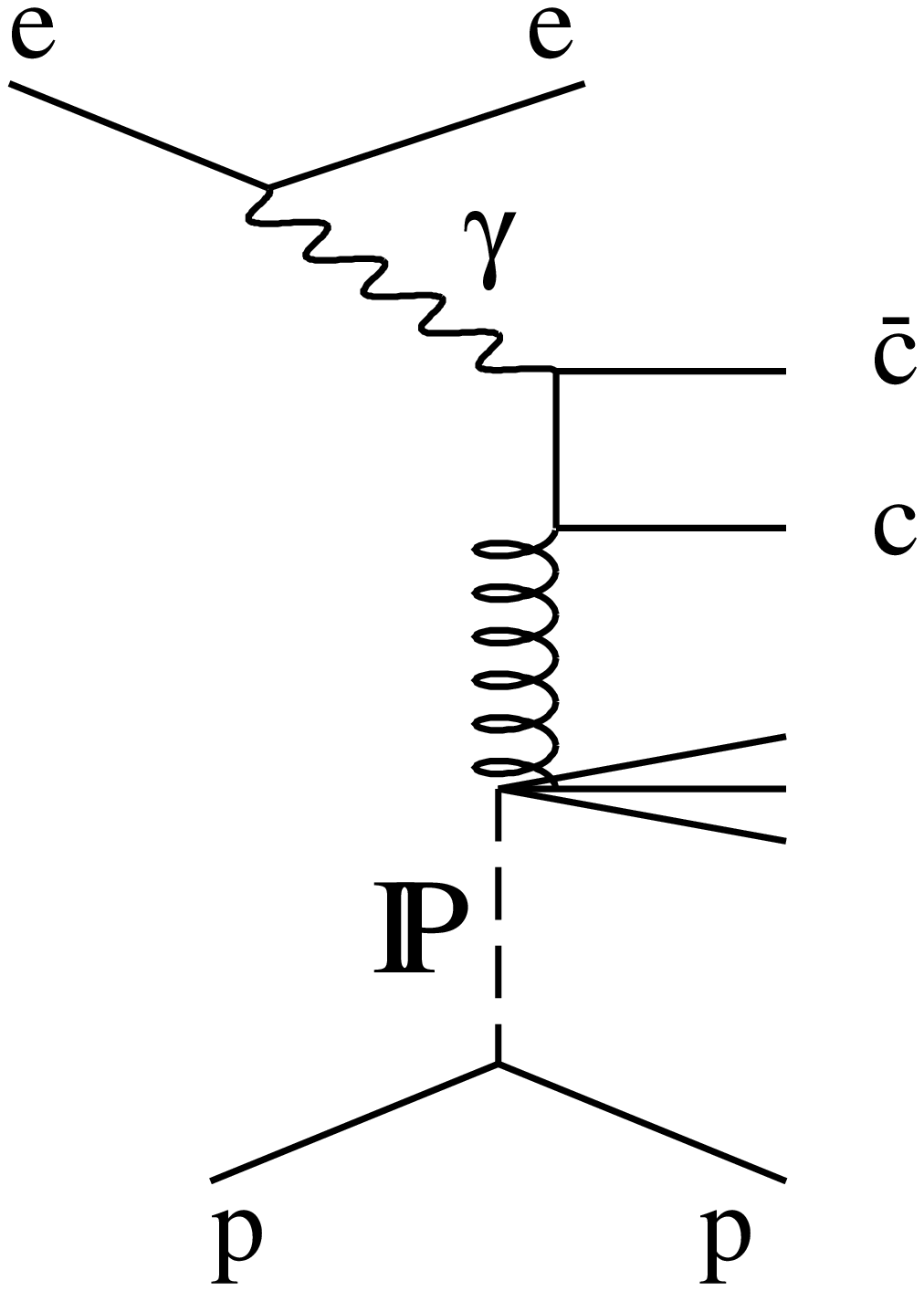,width=4.8cm}\hspace*{-0.1cm}\epsfig{file=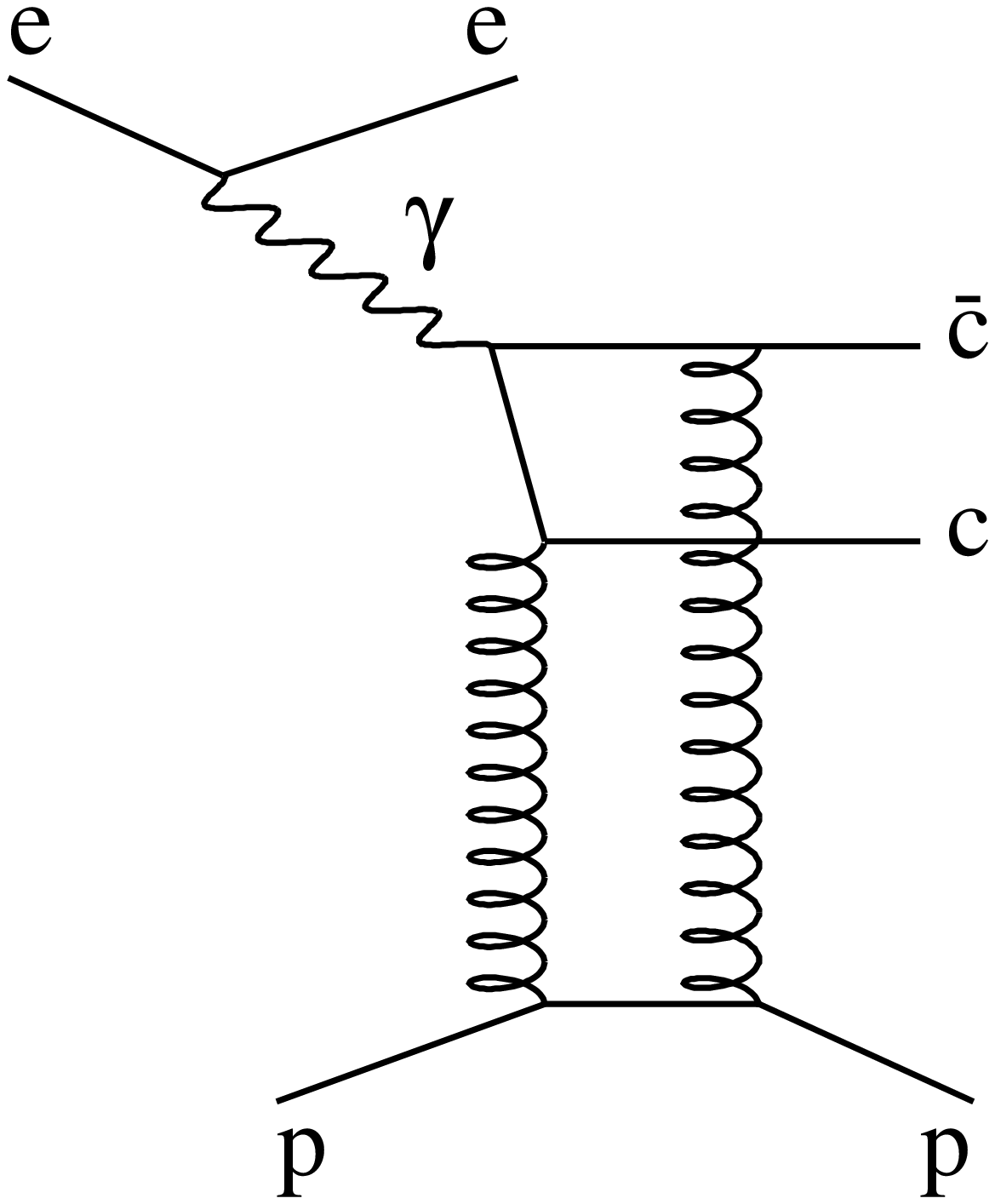,width=4.8cm}\hspace*{-0.1cm}\epsfig{file=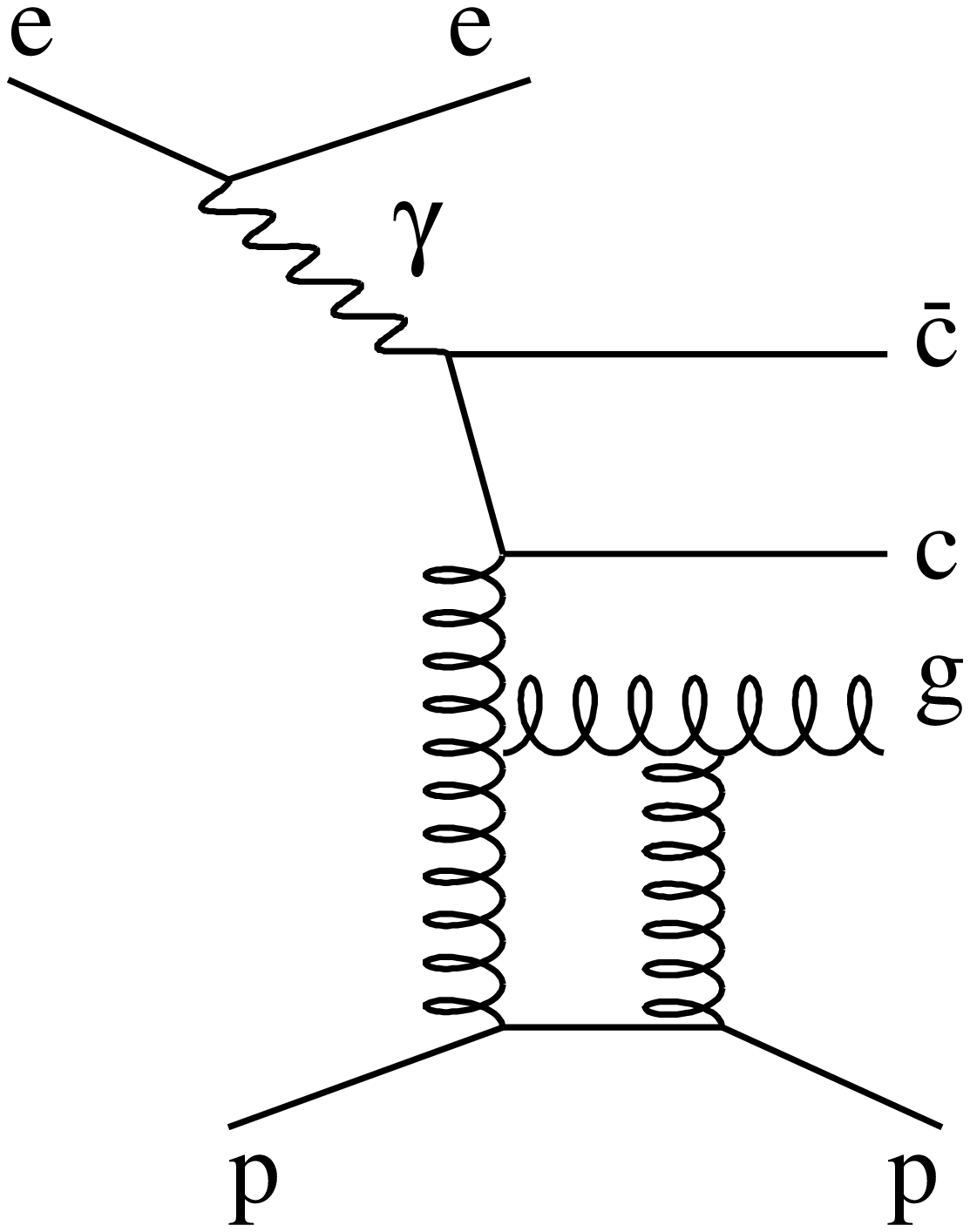,width=4.8cm}
\end{center}
\caption{Diagrams for charm production in diffractive $ep$ scattering: boson-gluon fusion and $c {\bar c}$, $c {\bar c} g$ states in the two-gluon exchange model; from ZEUS.}
\label{f:ccbardiaz}
\vfill
\end{figure}

\begin{figure}[p]
\begin{center}
\vfill\vspace*{-1cm}
\hspace*{-0.1cm}\epsfig{file=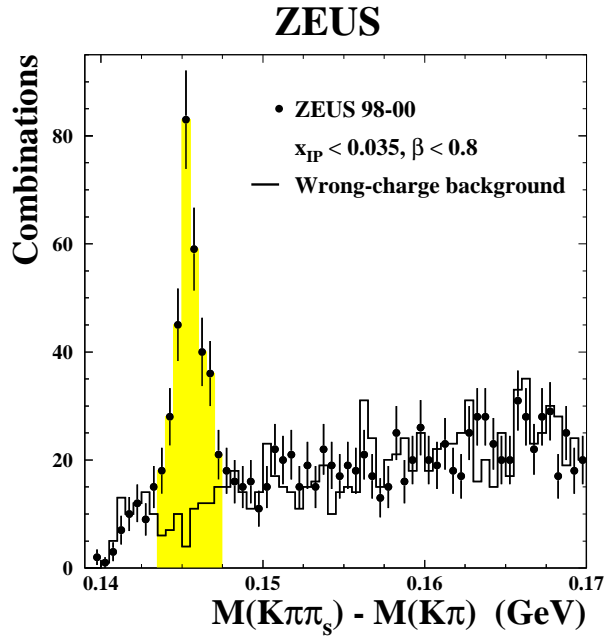,width=9cm}
\end{center}
\caption{Distribution of the mass difference, $\Delta M = m(K\pi\pi_s) - M(k\pi)$; from ZEUS.}
\label{f:dstarccz}
\vfill
\end{figure}

The contribution from diffractive production of $c \bar{c}$ pairs (see diagrams in Fig.~\ref{f:ccbardiaz}) was determined by ZEUS~\cite{epj:c51:301}, studying the reaction $\gamma^* p \to D^*(2010) X p$ with an integrated luminosity of 81.8 pb$^{-1}$. Events with a $D^{*{+(-)}} \to D^0 \pi^{+(-)}$ decay were isolated by making use of the small mass difference between $D^{*}$ and $D^0$, see Fig.~\ref{f:dstarccz}. For $Q^2 > 1$ GeV$^2$, in total $253 \pm 21$ events with an identified $D^*$ meson were obtained. 

\begin{figure}[p]
\begin{center}
\vfill\vspace*{-1cm}
\hspace*{-0.1cm}\epsfig{file=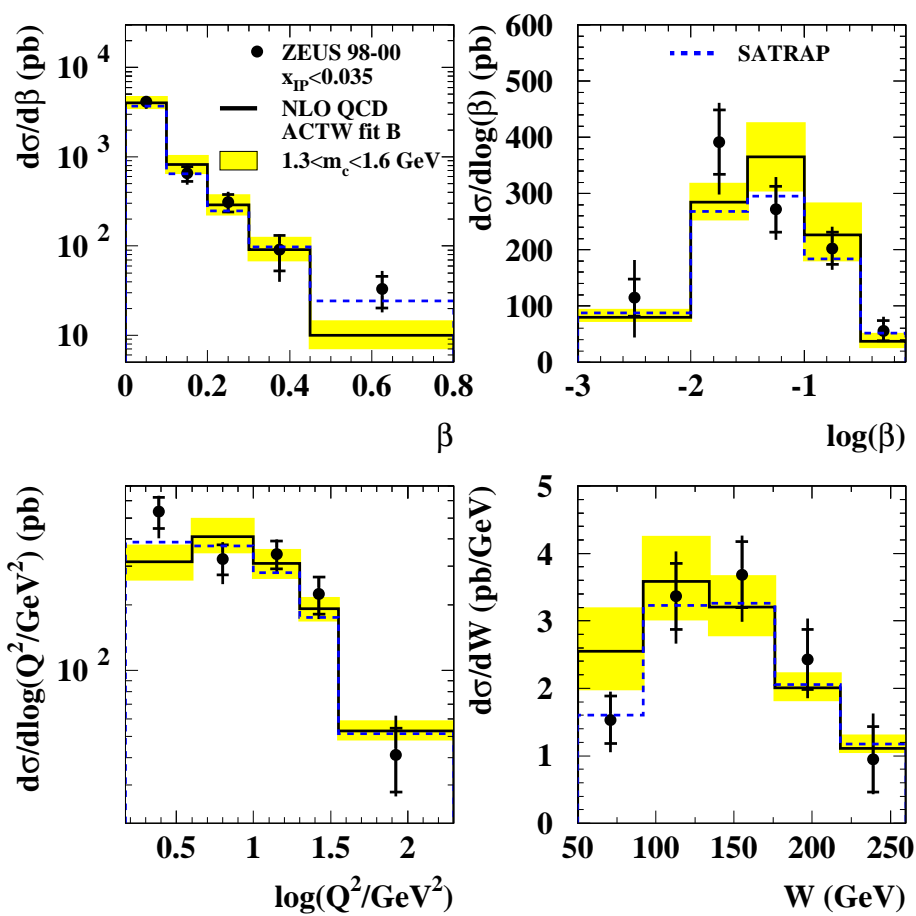,width=14cm}
\end{center}
\caption{Reaction $\gamma^* p \to D^{*+-} X p$: Differential cross sections $d\sigma/d\beta$, $d\sigma/dlog(\beta)$, $d\sigma/dlog(Q^2)$ and $d\sigma/dW$; from ZEUS.}
\label{f:sigdstarz}
\vfill
\end{figure}

\begin{figure}[p]
\begin{center}
\vfill\vspace*{-1cm}
\hspace*{-0.1cm}\epsfig{file=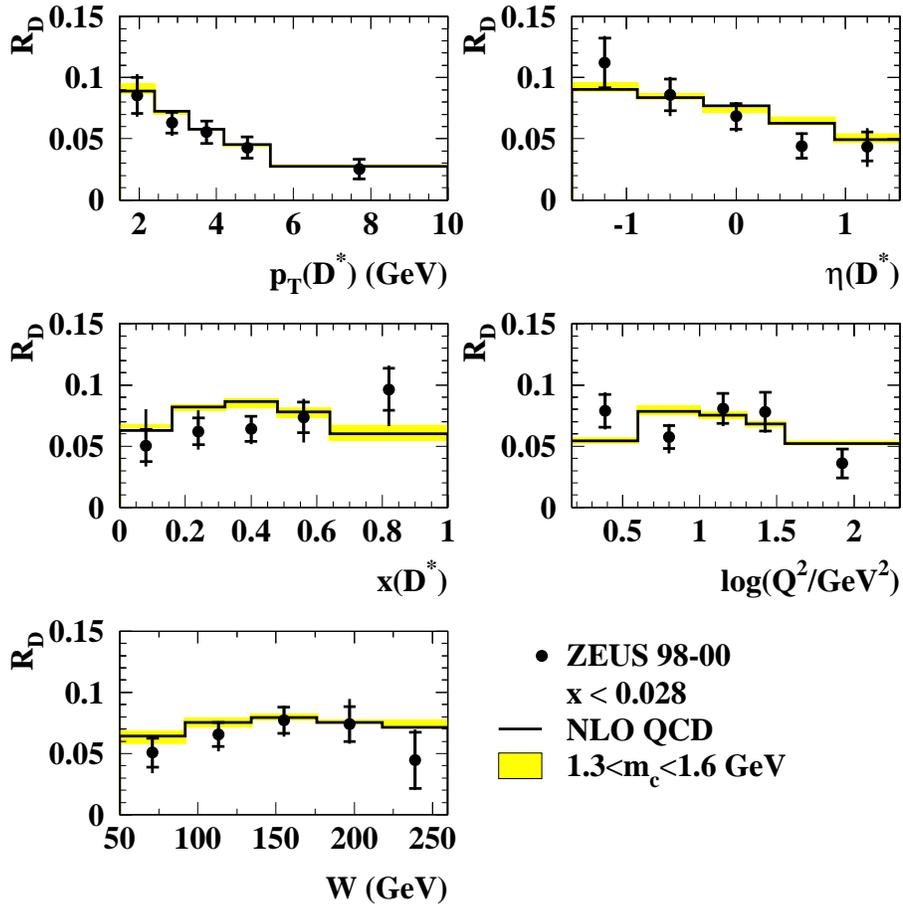,width=14cm}
\end{center}
\caption{Reaction $\gamma^* p \to D^* + {\rm anything}$: ratio of diffractively produced $D^{*\pm}$ mesons to inclusive $D^{*\pm}$ production as a function of $p_T, \eta, x$ of the $D^*$, and as a function of $log Q^2$ and $W$; from ZEUS.}
\label{f:ratioddiftot}
\vfill
\end{figure}

\begin{figure}[p]
\begin{center}
\vfill\vspace*{-1cm}
\hspace*{-0.1cm}\epsfig{file=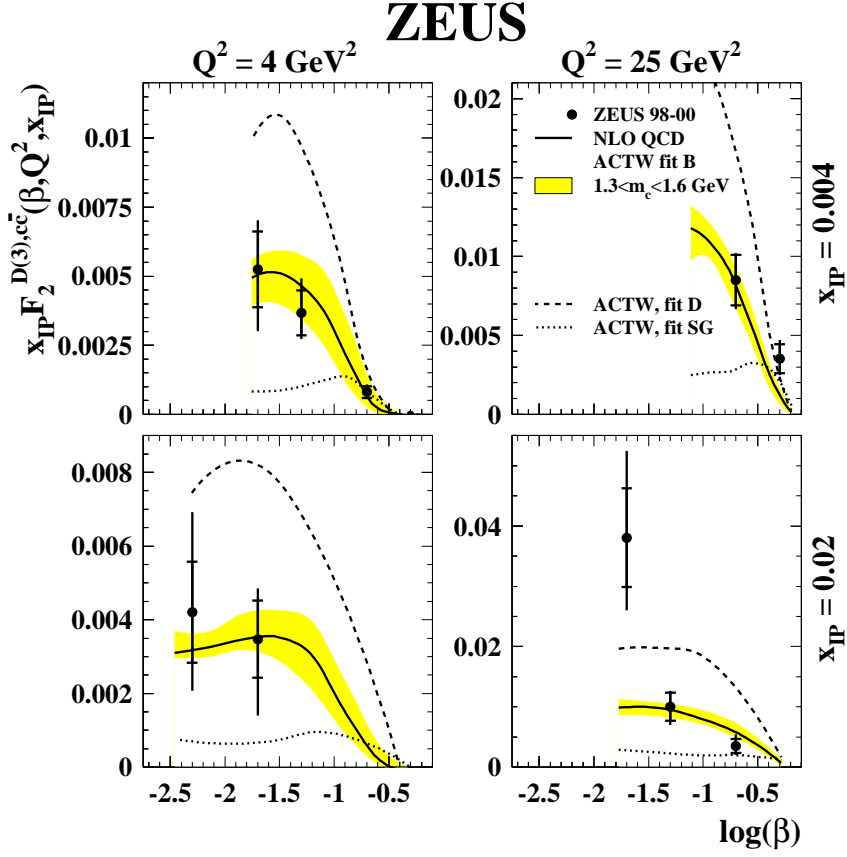,width=12cm}
\end{center}
\caption{Reaction $\gamma^* p \to c {\bar c}p$: diffractive structure function $\xpom F_2^{D(3),c \bar{c}}(\beta,Q^2,\xpom)$ versus $\log (\beta)$ for the $Q^2, \beta$ values indicated; from ZEUS.}
\label{f:xpfd3ccz}
\vfill
\end{figure}

\begin{figure}[p]
\begin{center}
\vfill
\vspace*{-1cm}
\hspace*{-0.1cm}\epsfig{file=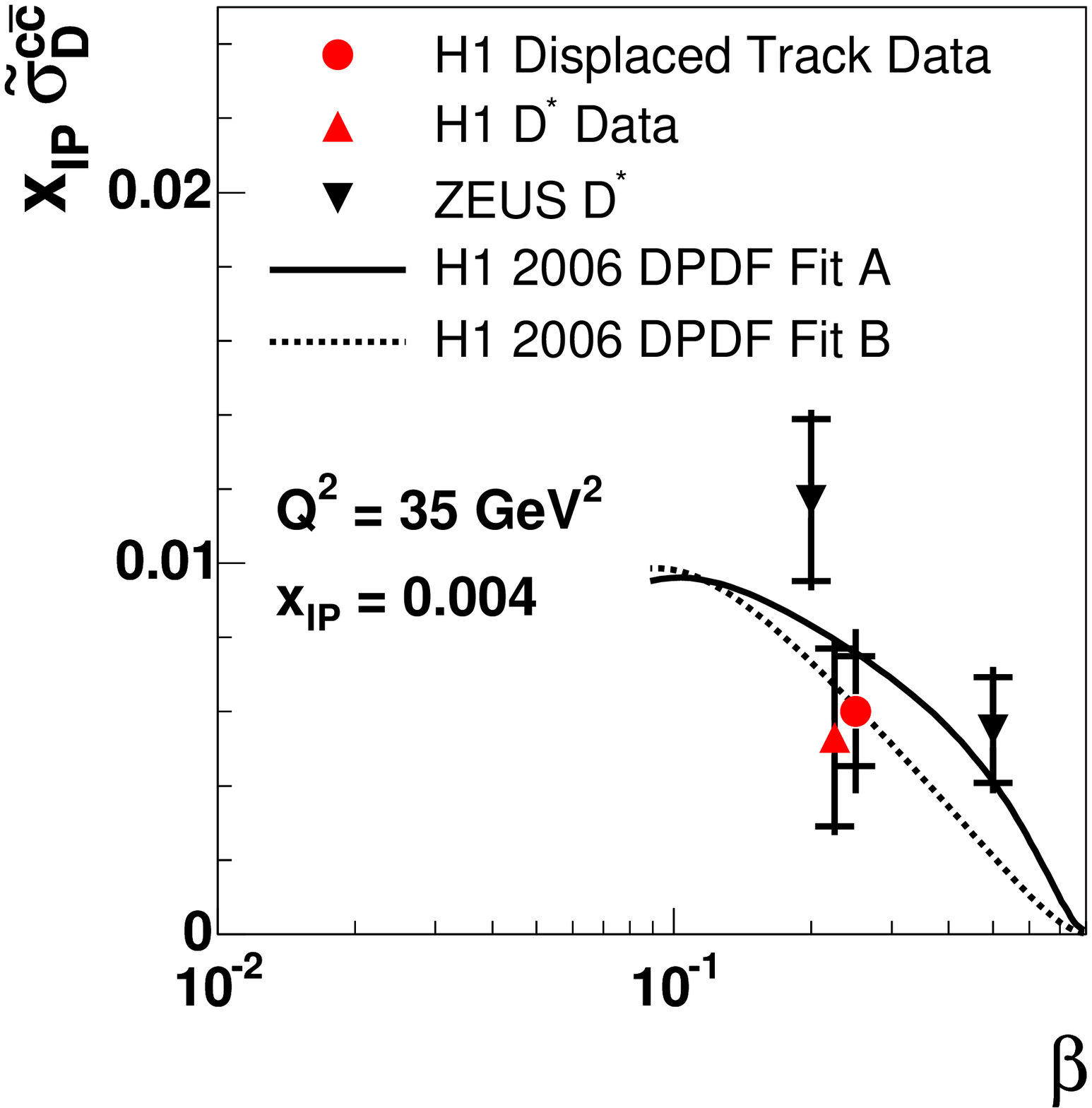,width=6.0cm}\hspace*{-0.1cm}\epsfig{file=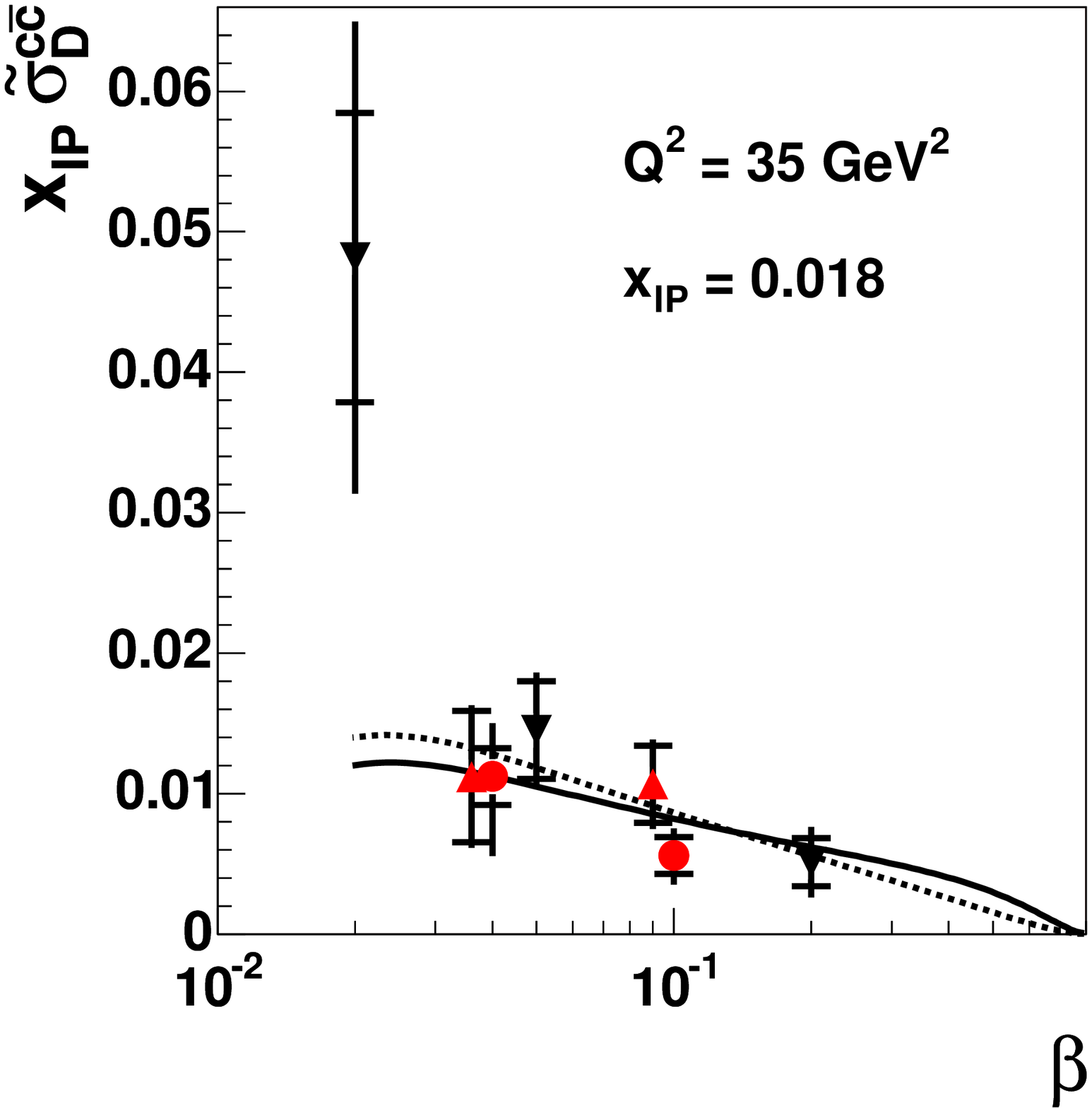,width=6.0cm}
\end{center}
\caption{Reaction $\gamma^* p \to c {\bar c}p$: Reduced cross section $\xpom {\sigma}^{c\bar c}_D \approx \xpom F_2^{D(3),c \bar{c}}$ versus $\beta$ for the $Q^2,\xpom$ values indicated, as measured by H1 and ZEUS; from H1.}
\label{f:xpfd3cchz}
\vfill
\end{figure}

The differential cross sections for $\gamma^* p \to D^{*\pm}(2010) X p$ as function of $\beta = Q^2/(Q^2+M_X^2)$, $Q^2$ and $W$, are shown in Fig.~\ref{f:sigdstarz}. The ratio of diffractively produced $D^{*\pm}$ mesons to inclusive $D^{*\pm}$ production is of the order of 0.05 to 0.1, see Fig.~\ref{f:ratioddiftot}. The charm contribution to the diffractive structure function of the proton, $\xpom F^{D(3),c \bar{c}}(\beta,Q^2,\xpom)$, is presented in Figs.~\ref{f:xpfd3ccz},~\ref{f:xpfd3cchz}
as a function of $\log (\beta)$ for $Q^2 = 4$ and $25$ GeV$^2$ and $\xpom = 0.004, 0.02$: the structure function rises rapidly as $\beta$ decreases. The ACTW calculation~\cite{pr:d59:074022} with the gluon-dominated fit B reproduces the data.

\begin{figure}[p]
\begin{center}
\vfill
\vspace*{-1cm}
\hspace*{-0.1cm}\epsfig{file=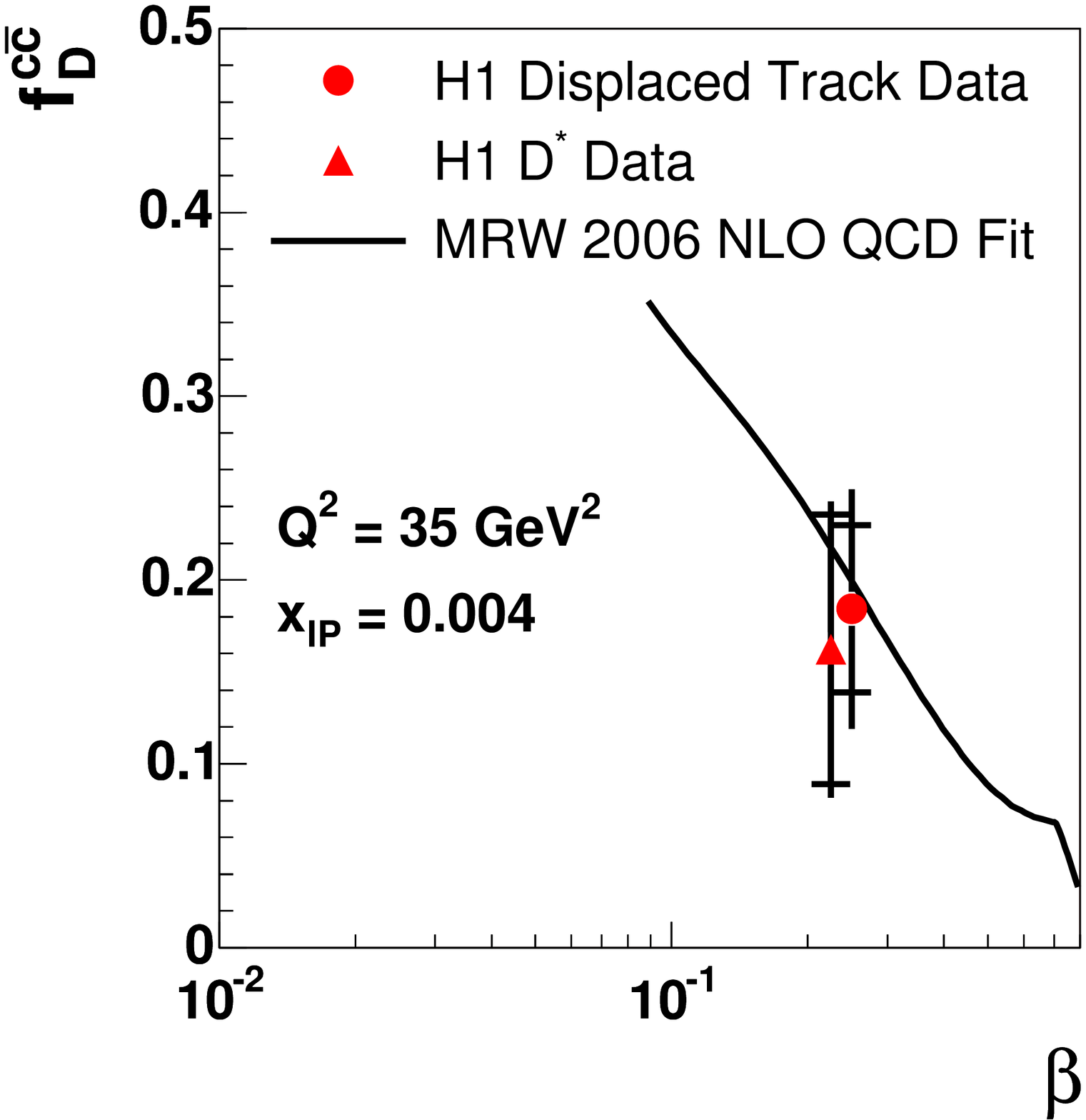,width=7cm}\hspace{-0.2cm}\epsfig{file=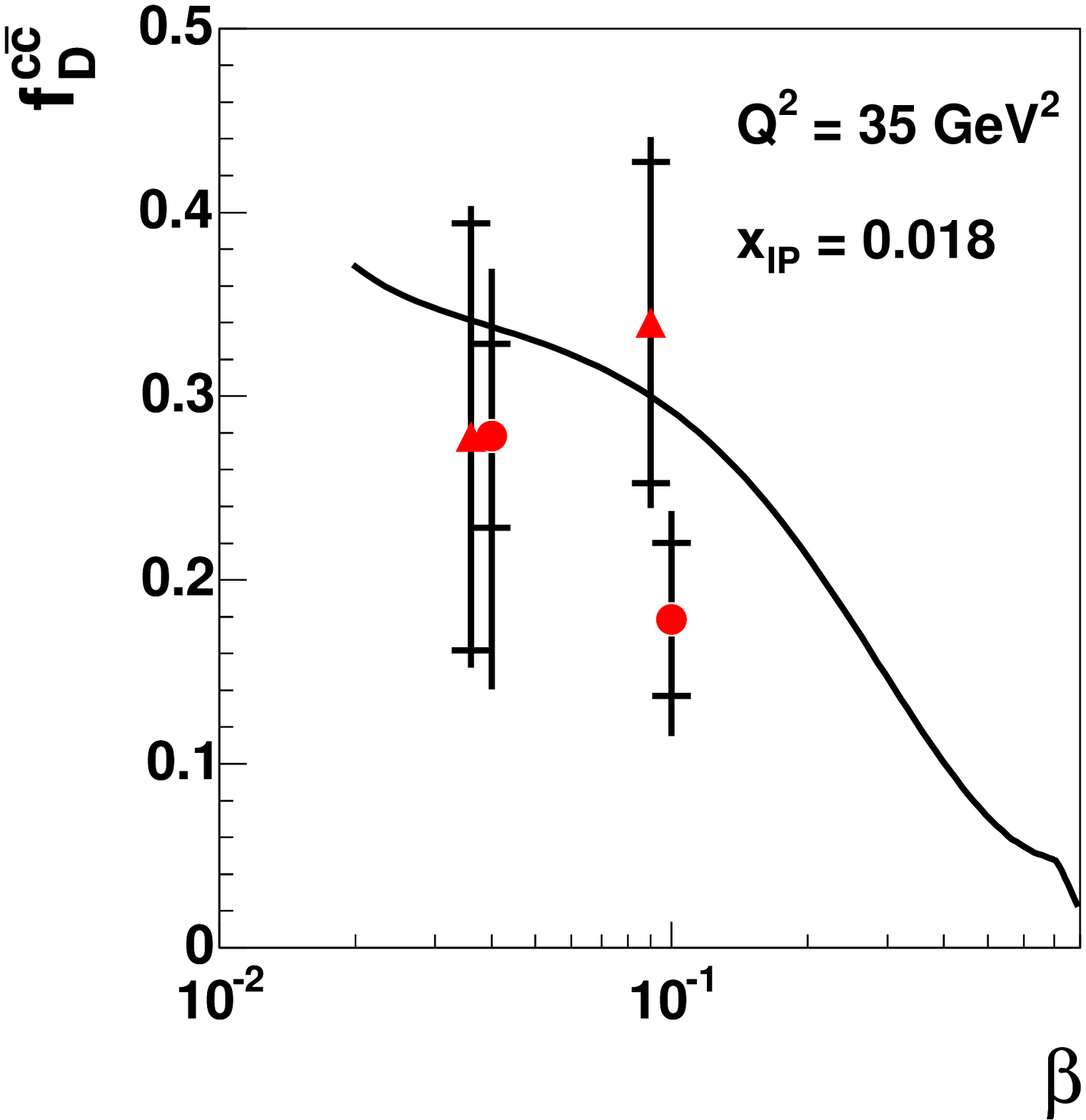,width=7cm}
\end{center}
\caption{Contribution of $c \cbar$ production to the total diffractive cross section; from H1.}
\label{f:cctodifh}
\vfill
\end{figure}

A similar measurement was performed by H1~\cite{epj:xx8:yy8}. Based on an integrated luminosity of 47 pb$^{-1}$, $70 \pm 13$ ($124 \pm 15$) events with $D^*(2010)$ production were detected at $Q^2 < 0.01$ GeV$^2$ ($Q^2>2$ GeV$^2$). Figure~\ref{f:xpfd3cchz} shows for $Q^2 = 35$ GeV$^2$ the combined results from H1 and ZEUS for $\xpom {\sigma}^{c\bar c}_D = \xpom F_2^{D(3),c \bar{c}}$: within the limited statistics there is agreement between the two data sets. The fraction $f^{c\cbar}_D$ of the total diffractive cross section contributed by the production of charm quarks is shown in Fig.~\ref{f:cctodifh} as a function of $\beta$ for $Q^2 = 35$ GeV$^2$, $\xpom = 0.004$ and $0.018$. For $\beta$ around 0.03 - 0.1, $f^{c\cbar}_D$  is of the order of 20-30\%.

\section{Jet production by $\gamma p$ and $\gamma^* p$ diffractive scattering}

\begin{figure}[p]
\begin{center}
\vfill
\vspace*{-2cm}
\hspace*{+0.5cm}\epsfig{file=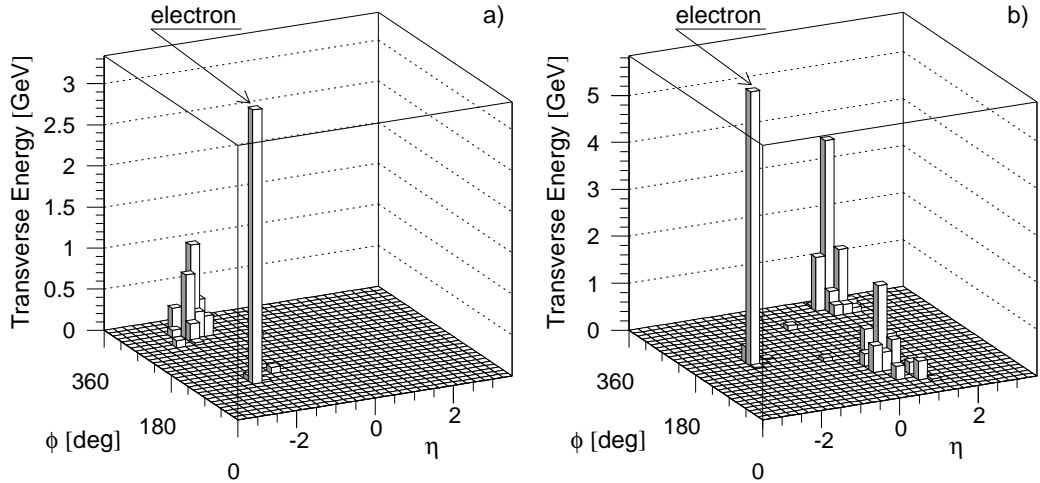,width=14.5cm}
\end{center}
\caption{Transverse energy deposition in $\eta -\phi$ space for a large rapidity gap event with one hadronic jet balancing the momentum of the scattered electron (left); (b) for a large rapidity gap event with two jets (right); from ZEUS.}
\label{f:lrgevtz}
\vfill
\end{figure}

\begin{figure}[p]
\begin{center}
\vfill
\vspace*{-3cm}
\hspace*{-0.5cm}\epsfig{file=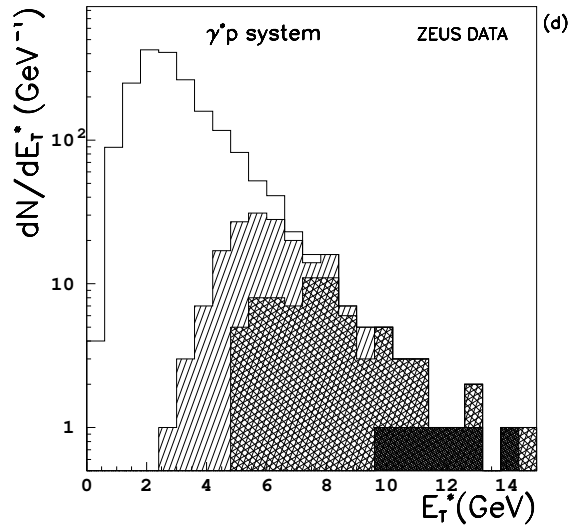,width=7cm}
\end{center}
\caption{Total hadronic energy transverse to the direction of the virtual photon in $\eta -\phi$ space for DIS events with a large rapidity gap and zero, $\ge 1$(hashed), $\ge2$(cross-hashed) or 3 jets(solid); from ZEUS.}
\label{f:etlrgetaz}
\vfill
\end{figure}

The study of jet production by virtual-photon proton scattering led ZEUS to the observation of events with a large rapidity gap between the jet - or system of jets - and the direction of the outgoing proton~\cite{pl:b332:228}. With an integrated luminosity of 0.55 pb$^{-1}$ a total of $39000$ events were collected for $Q^2 > 10$ GeV$^2$. A fraction of these events showed a large gap between the direction of the outgoing proton and the direction of particle(s) emitted closest to the outgoing proton. The size of the gap was quantified by the variable $\eta_{max} = -ln \tan \theta_{min}/2$, where $\theta_{min}$ is the polar angle of the most forward going particle measured w.r.t the proton beam direction. Figure~\ref{f:lrgevtz} shows two events with a large rapidity gap in $\eta -\phi$ space ($\phi$ = azimuthal angle), (a) with one jet and (b) with two jets (b). In both events there is a large space in $\phi$ (a) or $\eta$ (b) without energy deposition. The distribution of the total hadronic transverse energy, $E^*_T$, as determined in the $\gamma^* p$ c.m. system, is shown in Fig.~\ref{f:etlrgetaz} for events with 0,1,2 or 3 jets and a large rapidity gap. A fraction of these events have one ($\approx 15\%$) or two jets ($\approx 0.3 \%$). For $E^*_T>7$ GeV, the dominant production mechanism is jet production. In the case of two-jet production the two jets are back-to-back in the $\gamma^* p$ system (not shown).

\subsection{$\gamma p \to \hspace*{0.2cm}two \hspace*{0.2cm} jets \hspace*{0.2cm} + \hspace*{0.1cm}N$}

\begin{figure}[p]
\begin{center}
\vfill
\vspace*{-1cm}
\hspace*{+1cm}\epsfig{file=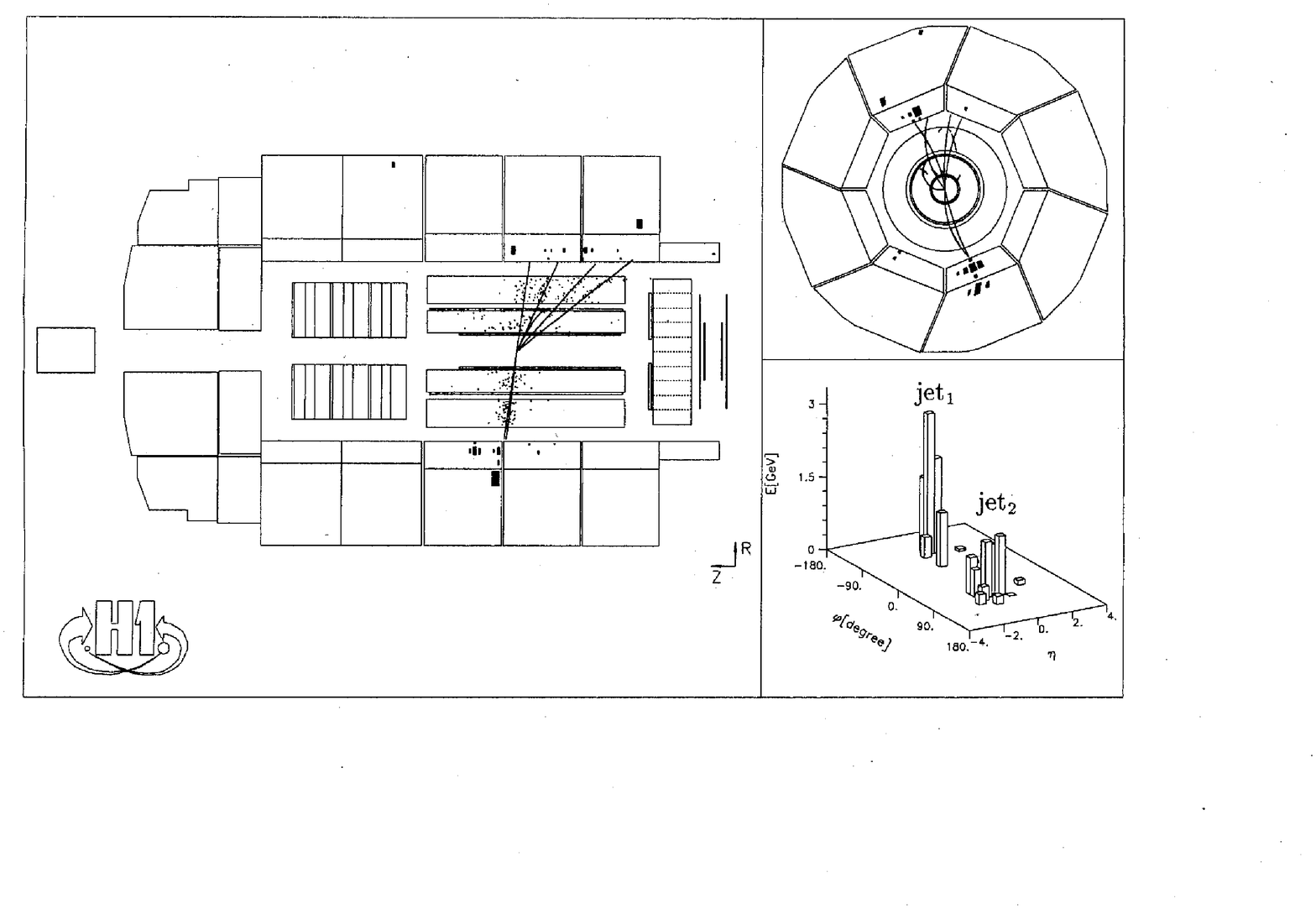,width=14cm}
\end{center}
\vspace*{-1cm}
\caption{Photoproduction of a two-jet event with a large rapidity gap; from H1.}
\label{f:twojetlrh}
\vfill
\end{figure}

\begin{figure}[p]
\begin{center}
\vfill
\vspace*{-1cm}
\hspace*{-4cm}\epsfig{file=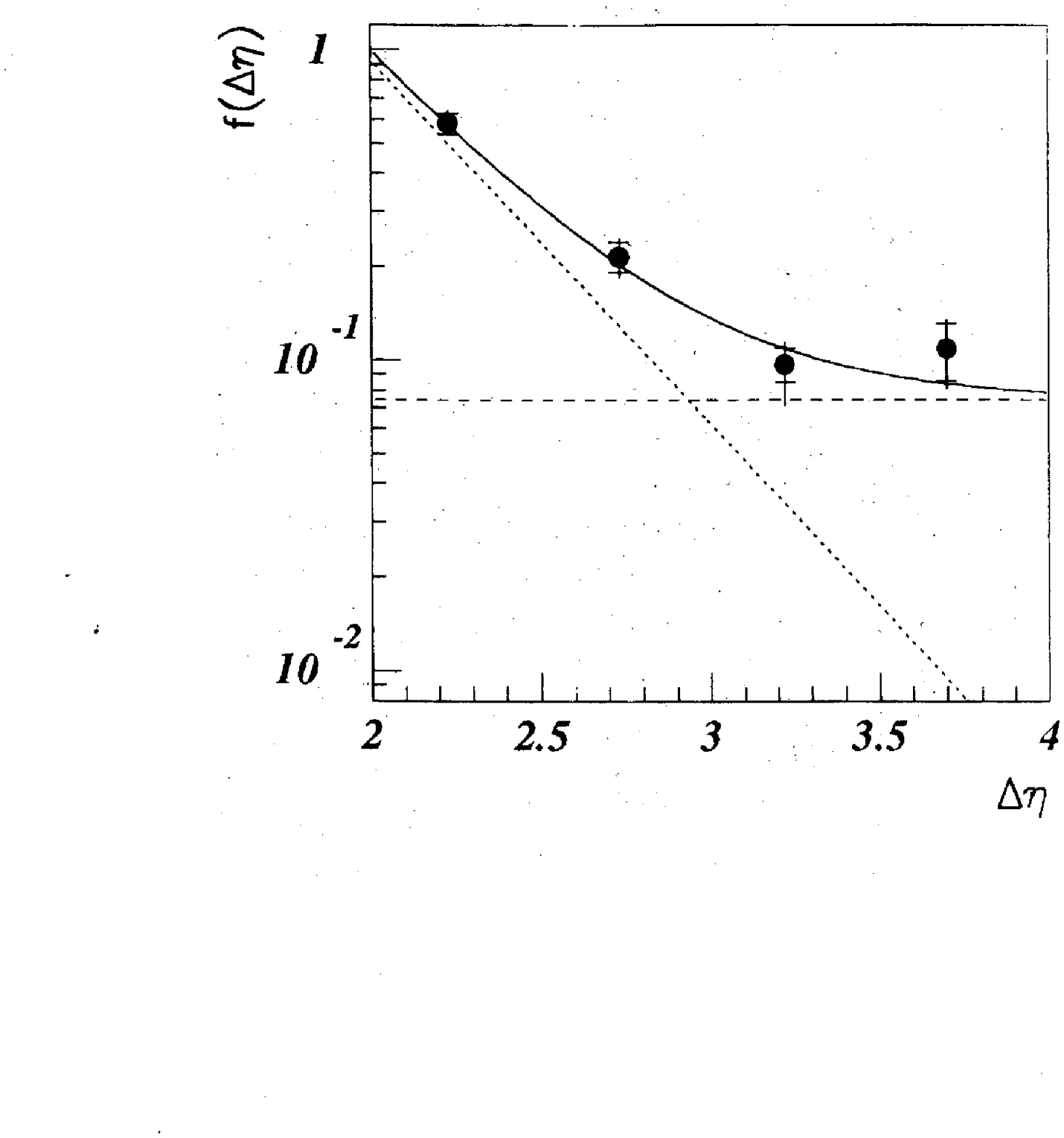,angle=270,width=12cm}
\end{center}
\vspace*{-3cm}
\caption{Photoproduction of two jets which are back-to-back in azimuth and separated  by a pseudorapidity gap $\Delta \eta$: gap fraction $f(\Delta \eta)$ as a function of $\Delta \eta$ (dots). Shown are also the expectations for colour non-singlet exchange (dotted line) and colour singlet exchange (solid line); from ZEUS.}
\label{f:deleta2jz}
\vfill
\end{figure}

An early observation of {\rm photo-}produced events with a large rapidity gap and two back-to-back jets was reported by H1~\cite{np:b435:3}, see Fig.~\ref{f:twojetlrh}. An analysis by ZEUS~\cite{pl:b356:129} indicated that between 30\% and 80\% of the Pomeron momentum carried by partons is due to hard gluons. Photoproduction of two or more jets with jet-transverse energies above 6 GeV was studied by ZEUS~\cite{pl:b369:55}. For a fraction of events with two jets, the jets are separated by a large rapidity gap $\Delta \eta$ of up to four units, with little hadronic energy in between the jets. The fraction of such events decreases exponentially with $\Delta \eta$ up to $\Delta \eta \approx 3$ where it reaches a constant value of about 0.08, see Fig.~\ref{f:deleta2jz}. The excess of events above the exponential fall-off gave evidence for hard diffractive scattering in photoproduction.

\begin{figure}[p]
\begin{center}
\vfill
\vspace*{-2cm}
\hspace*{+1cm}\epsfig{file=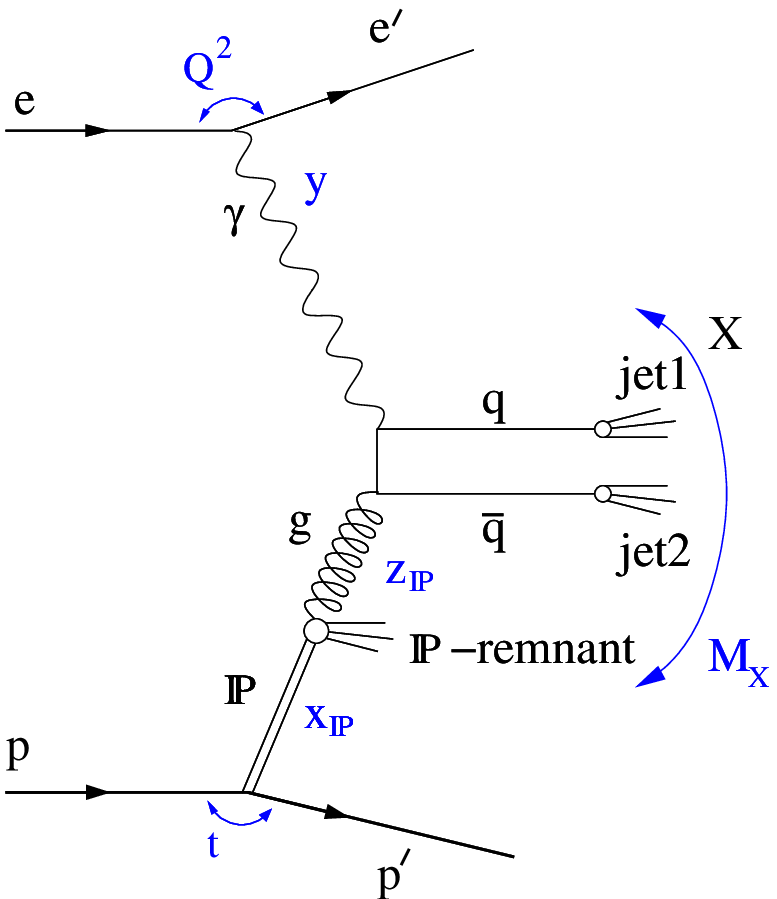,width=6cm}\hspace*{+1cm}\epsfig{file=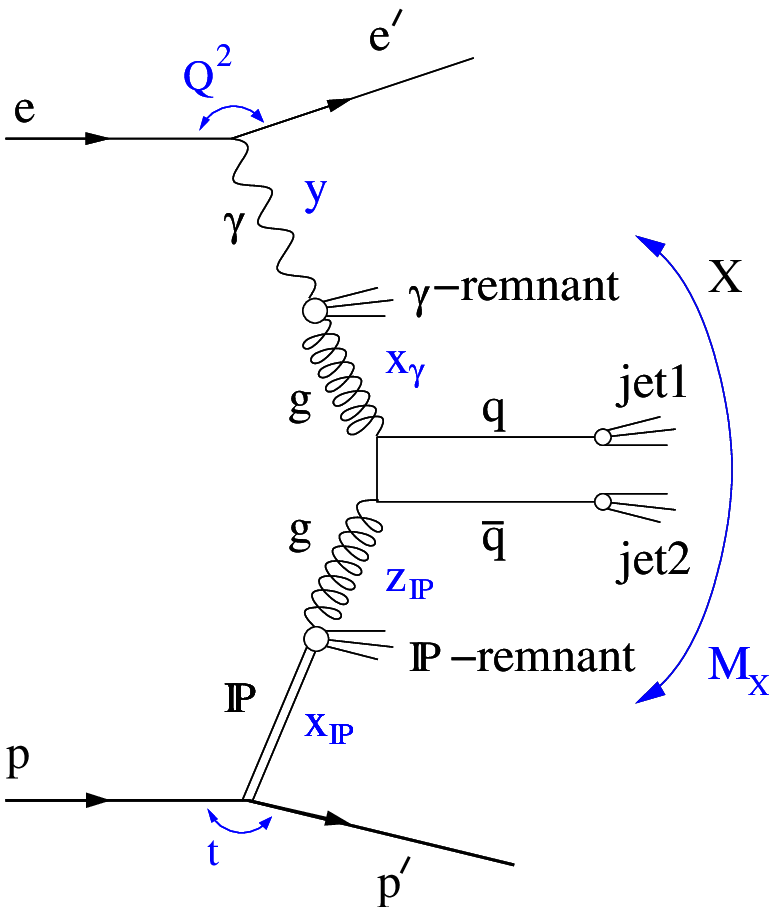,width=6cm}
\end{center}
\caption{Leading-order diagrams for direct (left) and resolved (right) processes in diffractive photoproduction of dijets via the exchange of a Pomeron ($\pom$).}
\label{f:diadifdirresgp}
\end{figure}

Further measurements on diffractive photoproduction of two jets have been reported by ZEUS~\cite{epj:c5:41,epj:c55:177}
and H1~\cite{epj:c6:421,epj:c24:517,epj:c51:549}. Figure~\ref{f:diadifdirresgp} shows leading-order digrams for direct and resolved diffractive photoproduction, 
\begin{eqnarray}
e + p \to e^{\prime} + X + p^{\prime}.
\end{eqnarray}
The process is described in terms of the four-momenta $e,e^{\prime}$ of the incoming and scattered electron, the incoming and scattered proton $p, p^{\prime}$, and the outgoing system $X$. Defining the four-momentum of the virtual photon, $q = e -e^{\prime}$, and the square of the photon-proton c.m. energy, $W^2  =  (p+q)^2$, the fraction of the energy of the incoming electron transferred to the proton is given by $y = \frac{p \cdot q}{p \cdot e} \approx \frac{W^2}{2p\cdot e}$. Under the assumption that the virtual photon interacts with the Pomeron ($\pom$) and produces a hadronic system $X$ of mass $M_X$, the fraction of the proton momentum carried by the Pomeron is $\xpom = \frac{(p-p^{\prime})\cdot q}{p \cdot q}$. The partons from the resolved photon and the diffractive exchange have fractional momenta 
$x_{\gamma} = \frac{(p \cdot u)}{(p \cdot q)}$
(where $u$ is the four-momentum of the parton in the resolved photon) and 
$z_{\pom} = \frac{v \cdot q}{(p -p^{\prime})\cdot q}$
(where $v$ is the four-momentum of the parton in the diffractive exchange). The variables $x_{\gamma}$ and $z_{\pom}$ are approximately given by the energies and pseudorapidities of the two jets with the highest transverse energies $E^{jet1,2}_T$ in the laboratory system:

\begin{eqnarray}
x^{obs}_{\gamma} = \frac{\sum_{jet1,2}E^{jet}_T e^{-\eta^{jet}}}{2 y E_e},\hspace{1cm}z^{obs}_{\pom} = \frac{\sum_{jet1,2}E^{jet}_T e^{\eta^{jet}}}{2 x_{\pom} E_p}
\end{eqnarray}  

\begin{figure}[p]
\begin{center}
\vfill
\vspace*{-2cm}
\hspace*{+0.5cm}\epsfig{file=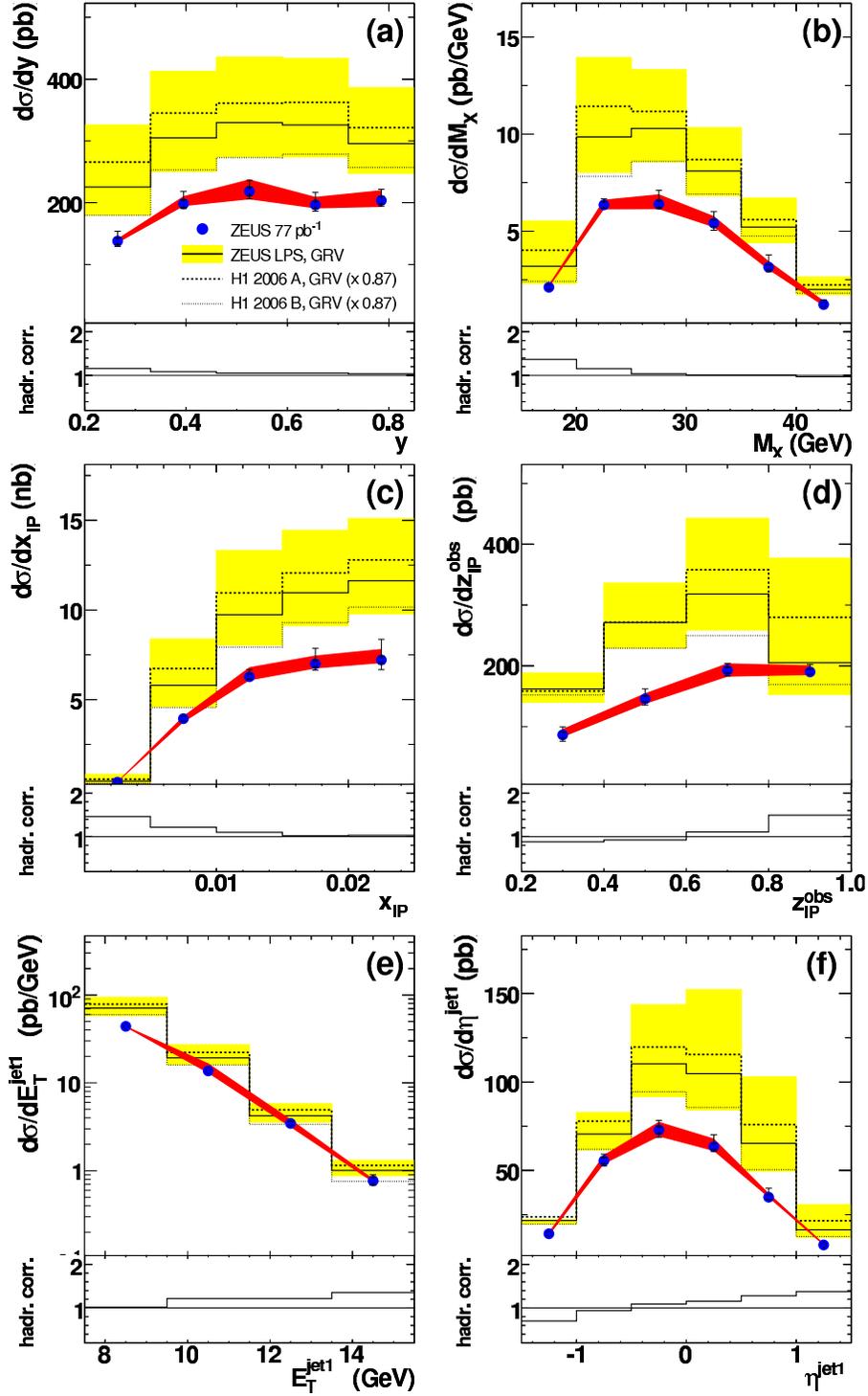,width=12cm}
\end{center}
\caption{Diffractive photoproduction of dijets as function of $y$, $M_X$, $\xpom$, $z^{obs}_{\pom}$, $E^{jet1}_T$ and $\eta^{jet1}$, from ZEUS; also shown are NLO QCD predictions based on the H1 2006 DPDFs.}
\label{f:gpdiff2jetsy}
\vfill
\end{figure}

\begin{figure}[p]
\begin{center}
\vfill
\vspace*{-2cm}
\hspace*{+0.5cm}\epsfig{file=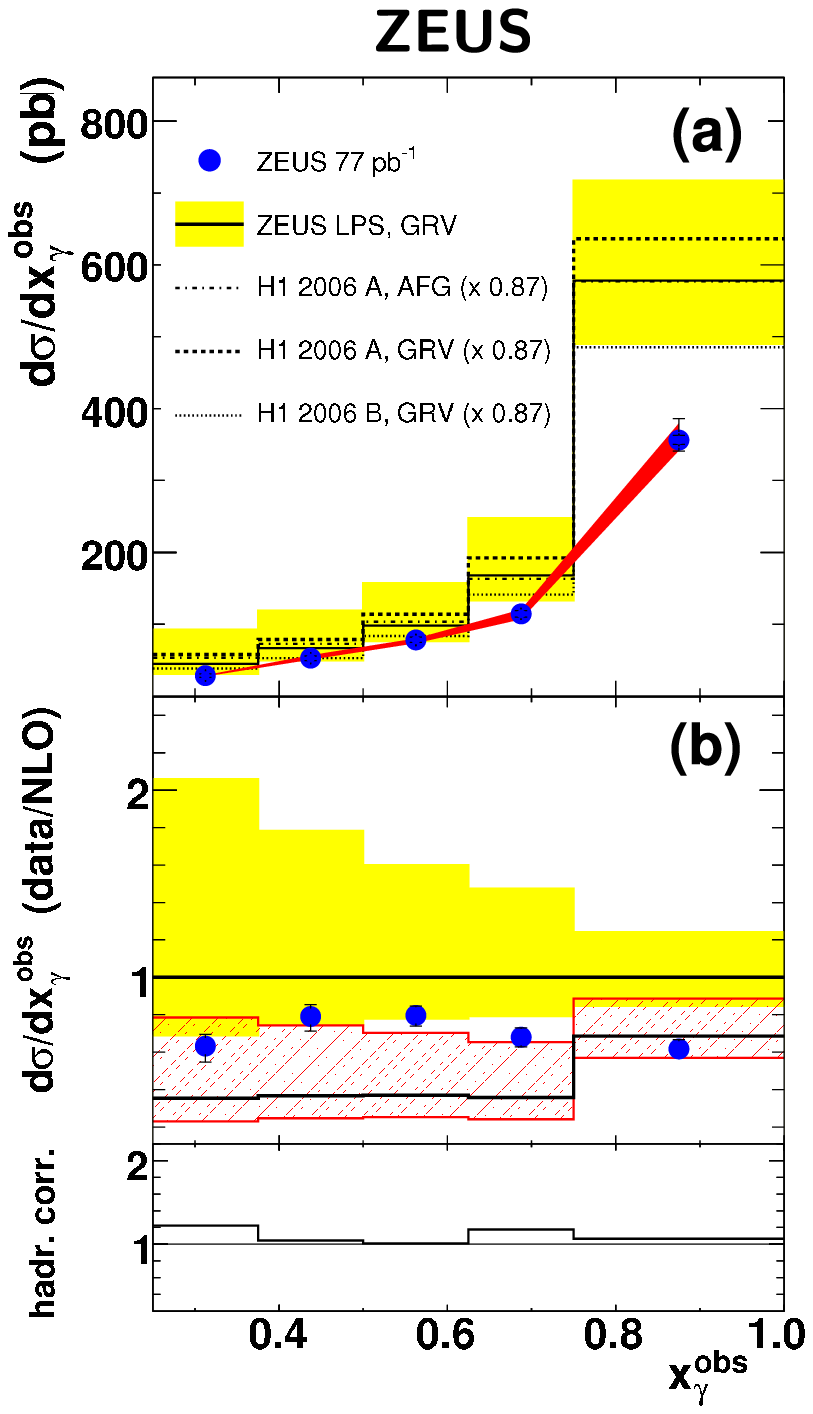,width=12cm}
\end{center}
\caption{Diffractive photoproduction of dijets as function of $x^{obs}_{\gamma}$, from ZEUS; also shown are NLO QCD predictions based on the H1 2006 DPDFs.}
\label{f:gpdiff2jetsxg}
\vfill
\end{figure}

The differential cross sections as measured by ZEUS~\cite{epj:c55:177} are shown in Figs.~\ref{f:gpdiff2jetsy},~\ref{f:gpdiff2jetsxg} together with NLO QCD predictions~\cite{epj:c38:93}: they reproduce the data to within a factor of about 1.5.

\subsection{$\gamma^* p \to \hspace*{0.2cm}two \hspace*{0.2cm} jets + \hspace*{0.1cm}N \hspace*{0.2cm}{\rm and} \hspace{0.2cm} three \hspace*{0.2cm} jets \hspace*{0.2cm} + \hspace*{0.1cm}N$} 

\begin{figure}[p]
\begin{center}
\vfill
\vspace*{-2cm}
\epsfig{file=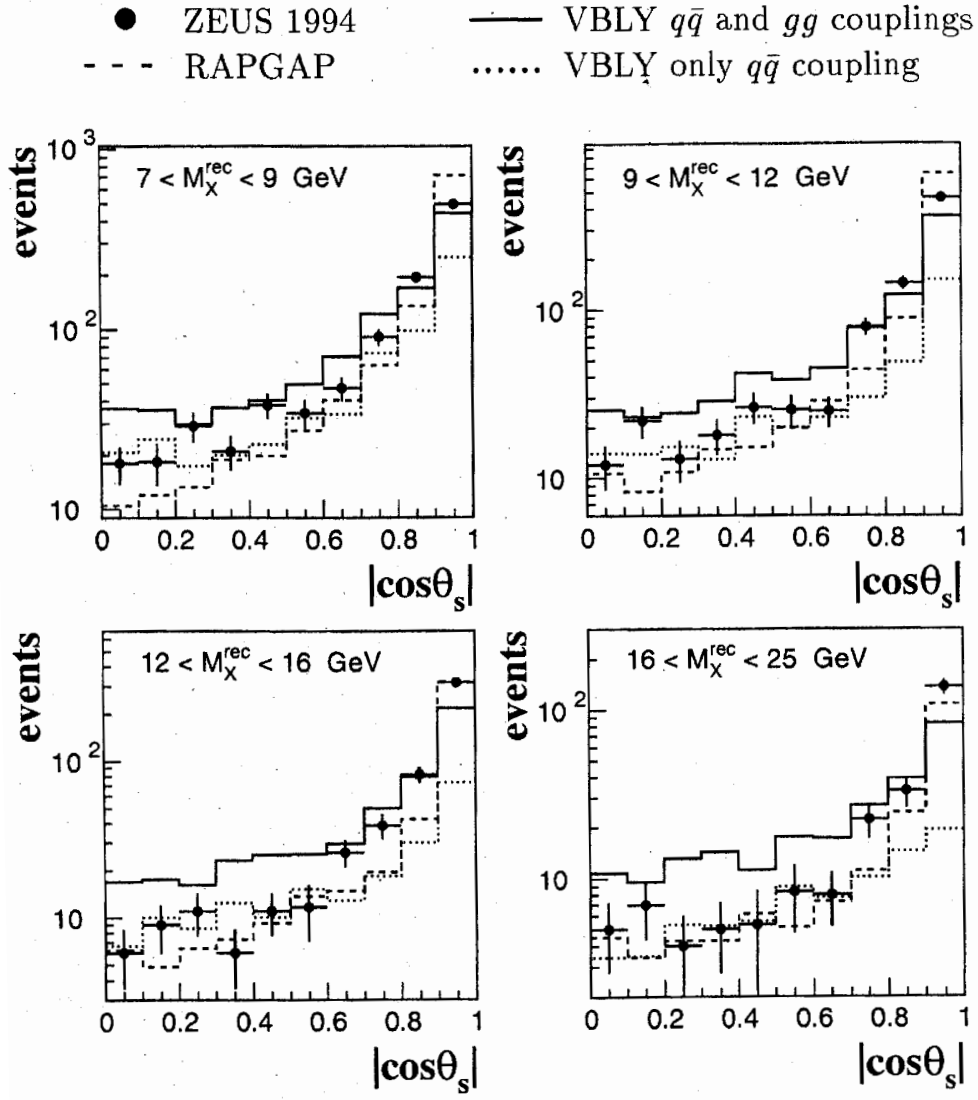,angle=271,width=14cm}
\end{center}
\vspace*{-3cm}
\caption{Multihadron final states of mass $M_X$ produced by deep-inelastic scattering with $160 < W < 250$\GeV, $5< Q^2 < 185$\GeV$^2$, where the most forward going (in proton direction) condensate has pseudorapidity $\eta^{rec}_{max} \le 1.8$: distribution of $\cos \theta_S$, where $\theta_S$ is the angle between the sphericity axis and the direction of the virtual photon in the $\gamma^* - \pom$ rest frame, for the range of $M_X$ values indicated; from ZEUS.}
\label{f:g*sphercosz}
\vfill
\end{figure}

A first in-depth analysis of deep-inelastic scattering leading to the production of two jets combined with a large rapidity gap has been presented by ZEUS~\cite{pl:b421:368} for $160<W<250$ GeV, $5<Q^2<185$ GeV$^2$ and $\eta_{max} \le 1.8$. For this event sample the value of $\xpom$ is below 0.01; thus, dominance by Pomeron exchange can be expected. The observed multihadron final state $X$ of mass $M_X$ is studied in its rest system which is interpreted as the $\gamma^*$-Pomeron rest system. As shown by Fig.~\ref{f:g*sphercosz}, the system $X$ is collimated around the direction of the virtual photon. The observed features are close to those observed in $e^+ e^-$ annihilation at a total c.m. energy $\sqrt{s} \approx M_X$~\cite{zfp:c12:297,zfp:c26:157}. The solid and dotted histograms show the comparison with the VBLY model~\cite{verbalayn} which is based on QCD. In this model, the Pomeron is assumed to have a pointlike coupling to $q \qbar$, and an additional gluon may be radiated. The prediction for $q \qbar$ alone agrees with the data for $\cos \theta_S <0.7$, but is too low for $\cos \theta_S >0.8$, while the prediction for $q \qbar$ plus $q \qbar g$ agrees with the data at $\cos \theta_S >0.8$ and overshoots the data when $\cos \theta_S < 0.7$. RAPGAP~\cite{jung95}, which also considers the sum of the contributions from $q \qbar$ plus $q \qbar g$, gives a fair representation of the data for the full $\cos \theta_S$ range.   

\begin{figure}[p]
\begin{center}
\vfill
\vspace*{-1cm}
\hspace*{-0.1cm}\epsfig{file=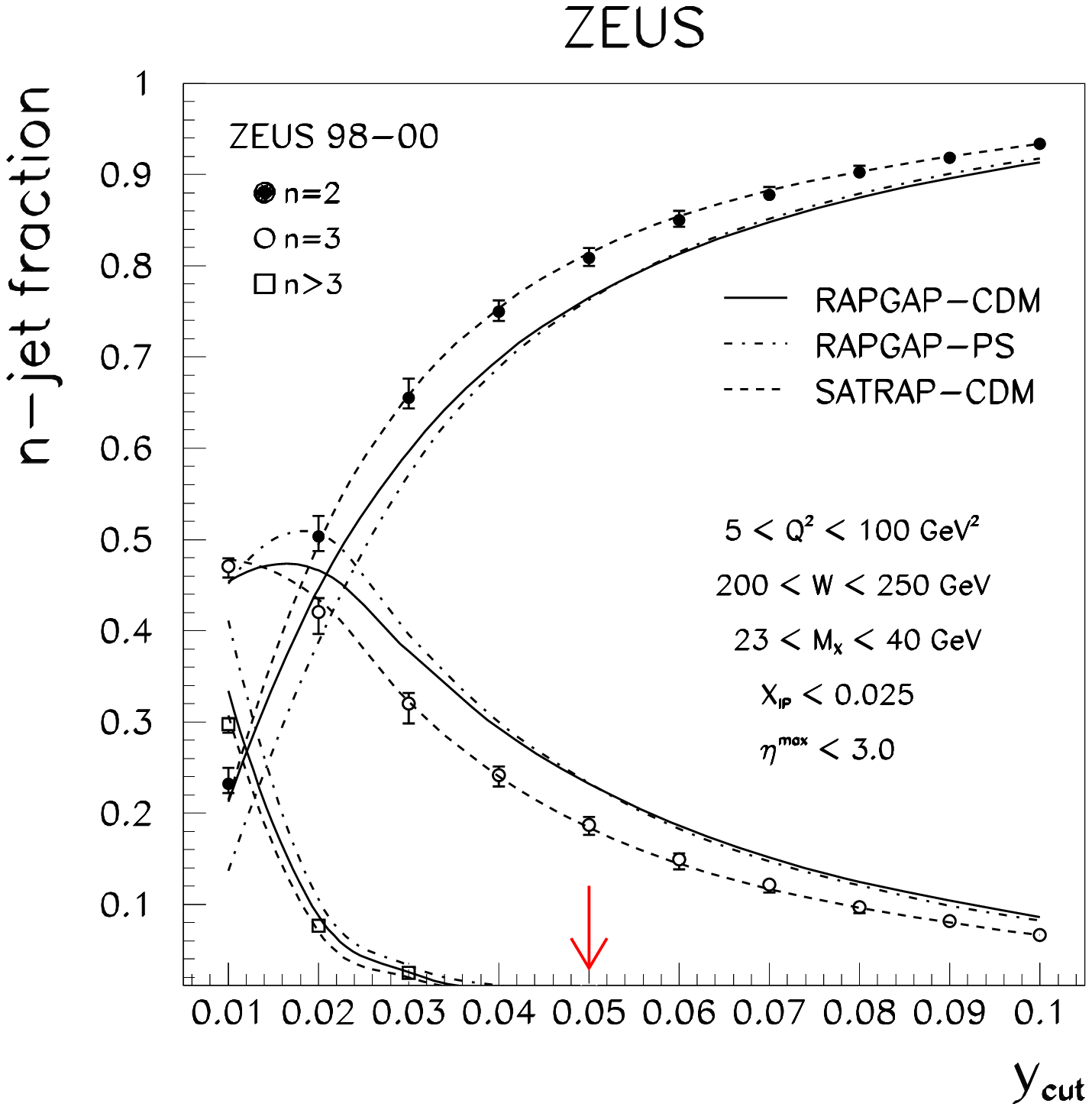,width=9cm}\hspace*{-1cm}\epsfig{file=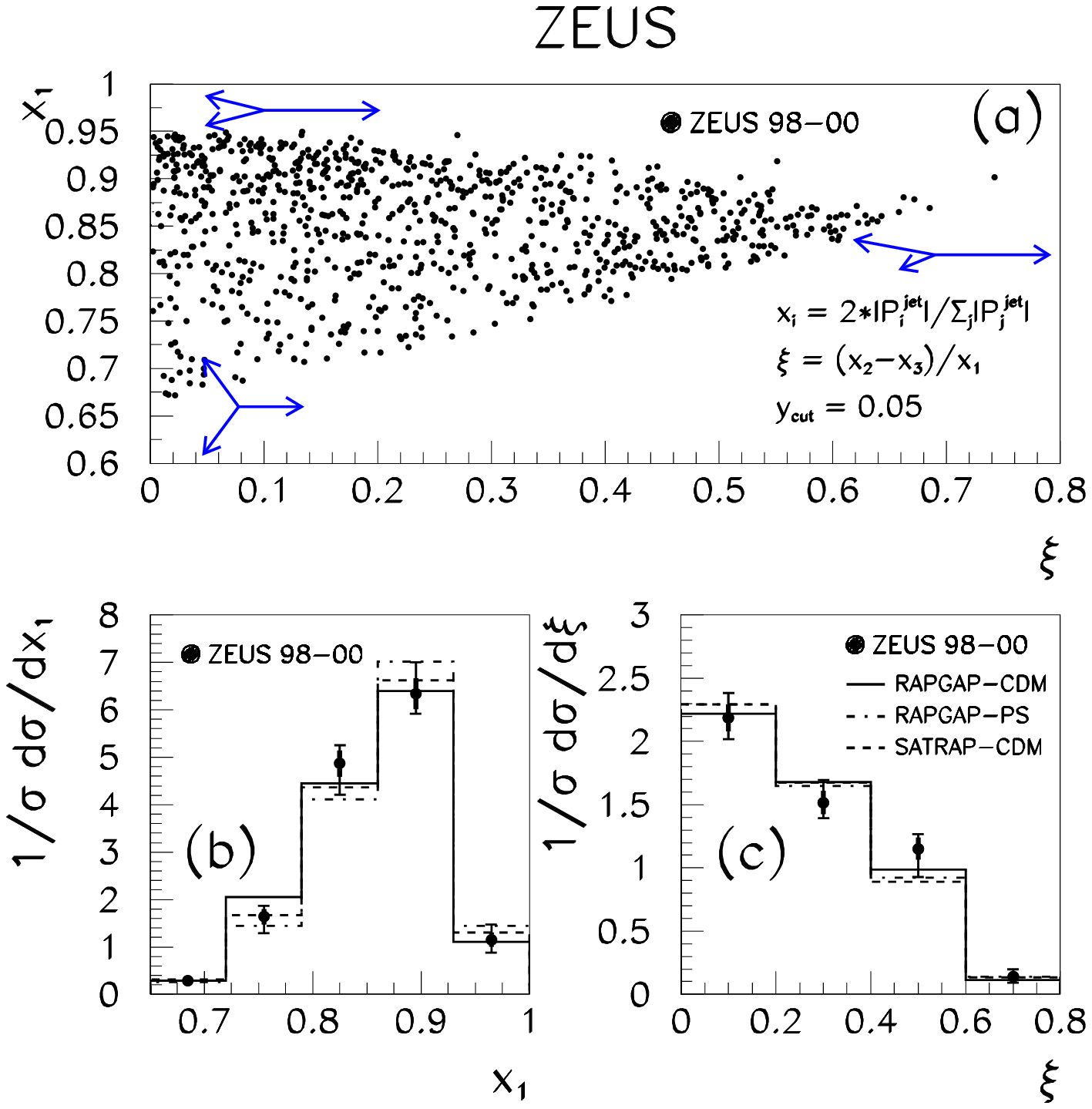,width=9cm}
\end{center}
\caption{Diffractive production of jets by deep-inelastic scattering: (left) jet fraction as a function of the jet resolution parameter $y_{cut}$ for $n_{jet}=2,3$ and $>3$, for the kinematic range indicated; (right) distribution of the 3-jet sample in the $\xi, x_1$ plane. The MC expectations were calculated with RAPGAP; from ZEUS.}
\label{f:jetfraction}
\vfill
\end{figure}

\begin{figure}[p]
\begin{center}
\vfill
\vspace*{+0.5cm}
\hspace*{+0.5cm}\epsfig{file=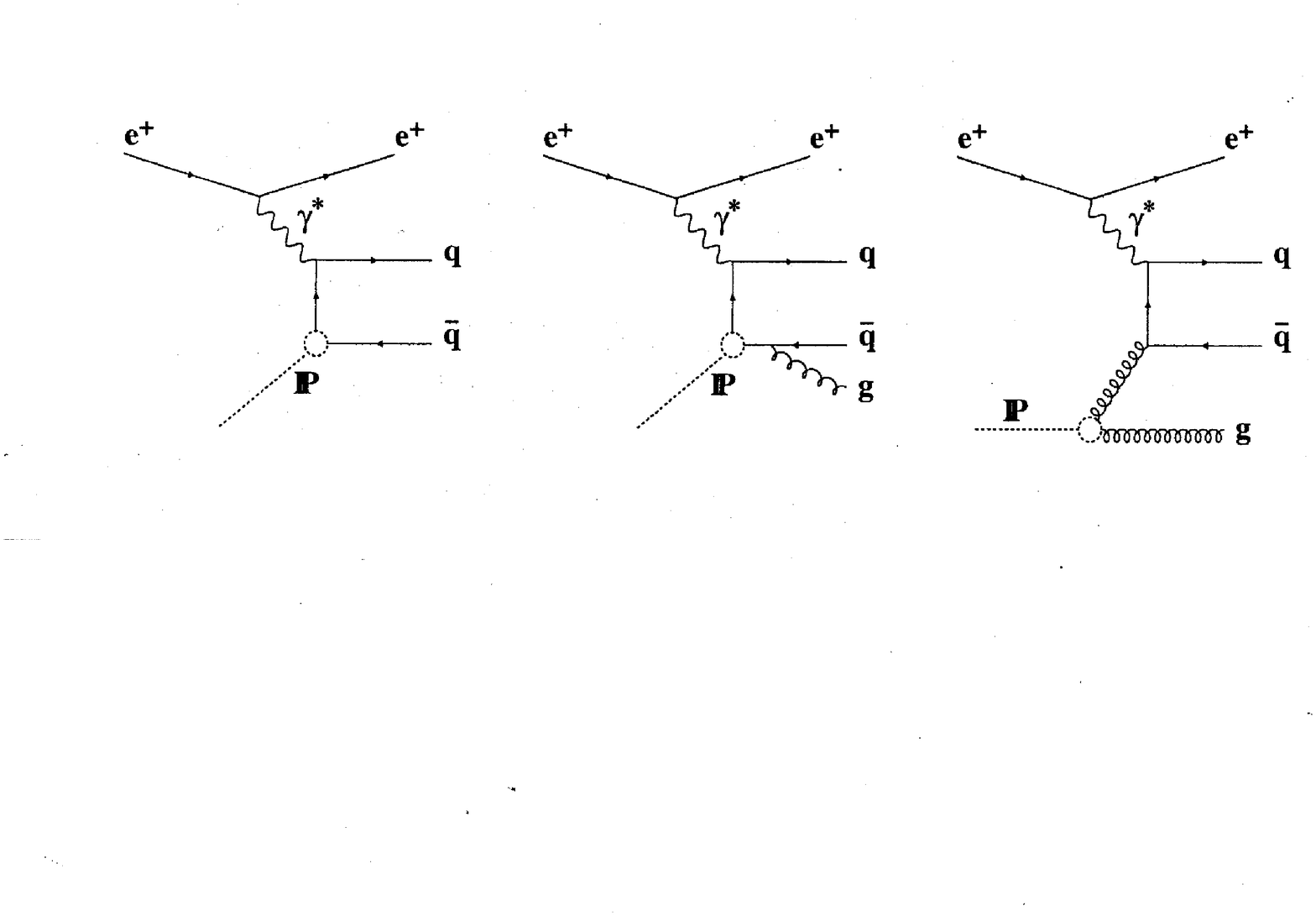,width=15cm}
\end{center}
\vspace*{-2.5cm}
\hspace*{1.5cm}\caption{Diagrams for two- and three-jet production in deep-inelastic diffractive scattering.}
\label{f:g*diaeqq}
\vfill
\end{figure}

The importance of multi-jet production by diffraction is also illustrated by Fig.~\ref{f:jetfraction}~\cite{pl:b516:3} which shows the fraction of events with $\eta_{max} < 3$ and two or more jets, as a function of the jet resolution parameter $y_{cut}$. For $y_{cut} \ge 0.02$, the dominant contributions are of the two- and three-jet type. QCD diagrams leading to two- and three-jet configurations by diffractive processes are shown in Fig.~\ref{f:g*diaeqq}. The curves in Fig.~\ref{f:jetfraction} show predictions of several models: the Monte Carlo generator SATRAP~\cite{kowalski99} combined with higher order QCD contributions~\cite{yamashita,jung} describes the data best.

\begin{figure}[p]
\begin{center}
\vfill
\vspace*{-1cm}
\hspace*{+0.5cm}\epsfig{file=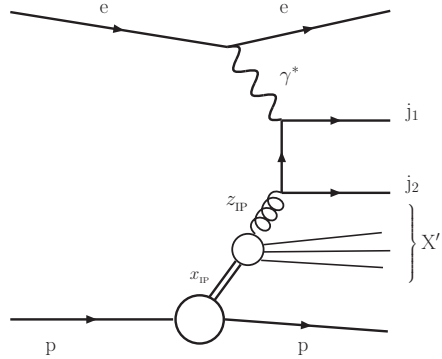,width=6cm}
\end{center}
\vspace*{-1cm}
\hspace*{1.5cm}\caption{Boson-gluon fusion for LO dijet production in diffractive DIS.}
\label{f:bosglufusion}
\vfill
\end{figure}

\begin{figure}[p]
\begin{center}
\vfill
\vspace*{-1cm}
\hspace*{-0.1cm}\epsfig{file=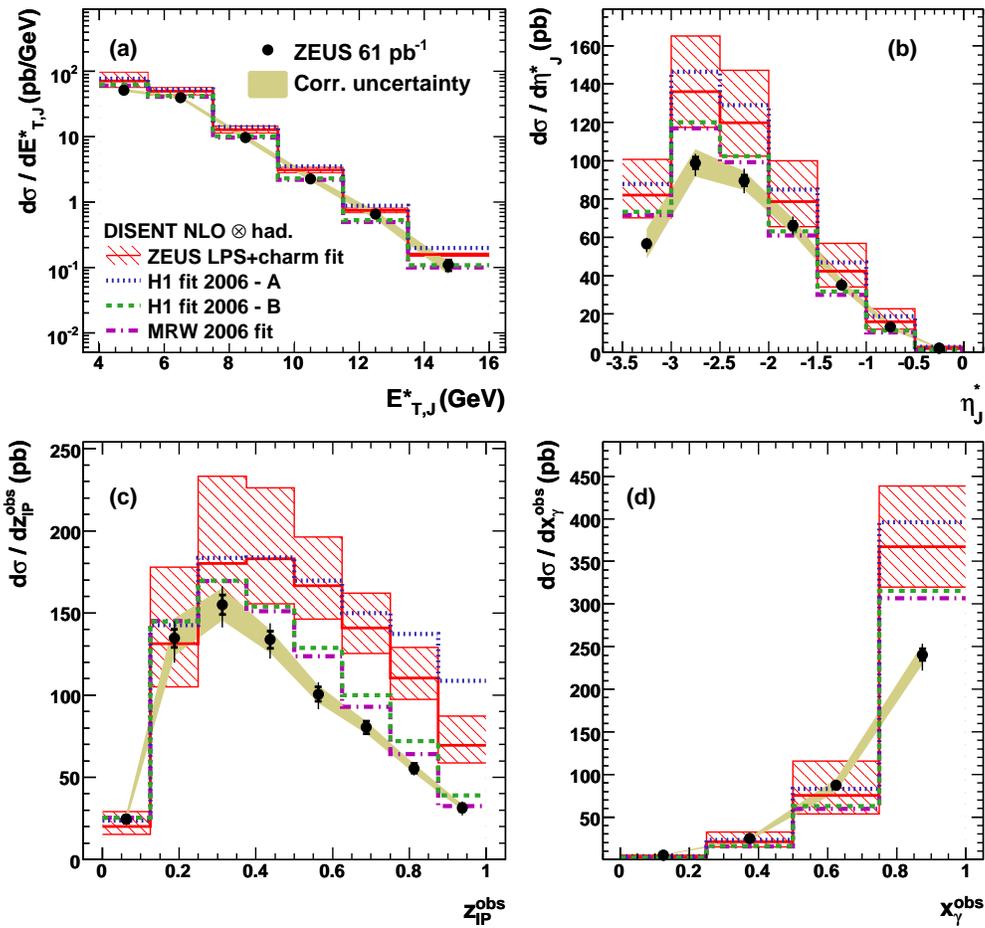,width=14cm}
\end{center}
\caption{Cross section for diffractive dijet production by deep-inelastic scattering with $Q^2 < 100$\GeV$^2$, $100 < W <250$\GeV. Also shown are NLO predictions using the dPDFs indicated; from ZEUS.}
\label{f:diffdis2jet}
\vfill
\end{figure}

In Ref.~\cite{epj:c52:813} diffractive dijet production (see Fig.~\ref{f:bosglufusion}) for $100<W<250$ GeV, $5<Q^2<100$ GeV$^2$ has been studied by ZEUS with an integrated luminosity of 61 pb$^{-1}$. The jet transverse energies in the $\gamma^* p$ rest system were required to be $E^*_{T,j1} > 5$ GeV and $E^*_{T,j2} > 4$ GeV, respectively; their pseudorapidities had to satisfy $-3.5 < \eta^*_{j} < 0$ and their pseudorapidities in the laboratory frame were restricted to $|\eta^{LAB}_{j}| < 2$. Diffractive events were selected by requiring less than 1 GeV energy deposition in a small forward calorimeter (FPC), which ensures a large rapidity gap, and by demanding $x^{obs}_{\pom} = \frac{Q^2 + M^2_X}{Q^2 + W^2} < 0.03$. Figure~\ref{f:diffdis2jet} shows the differential cross section as functions of $E^{*}_{T,J}$, $\eta^*_J$, $z^{obs} = \frac{Q^2 + M^2_{jj}}{Q^2 + M^2_X}$ and of $x_{\gamma}$, the fractional momentum of the virtual photon. For $E^{*}_{T,J}> 6$ GeV, the jet cross section is falling exponentially with $E^{*}_{T,J}$. The requirement of two high $E_T$ jets suppresses the contribution at low $\xpom$. The low value of the peak position in $z^{obs}_{\pom}$ indicates additional production of hadrons. 
    
The data were compared with NLO calculations which include direct and resolved photon contributions and use diffractive PDFs (DPDF) of Refs.~\cite{epj:c48:715,pl:b644:131}. They determine the probability to find in the proton a parton of type i carrying a fraction $\xpom \beta$ of the proton momentum with a probe resolution $Q^2$ under the condition that the proton stays intact. They have been determined by fits to inclusive diffractive DIS data. The predictions from the ZEUS LPS+charm fit, as well as those from the H1-2006 fits A and B, lie in general above the data. The MRW fit~\cite{pl:b644:131} is in broad agreement with the data with the exception of the high $x^{obs}_{\gamma}$ region, where the predictions are too high. 

\begin{figure}[p]
\begin{center}
\vfill
\vspace*{-1cm}
\hspace*{-0.1cm}\epsfig{file=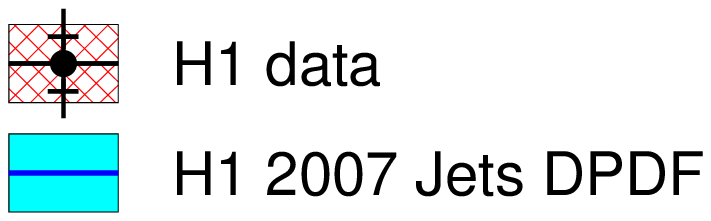,width=5cm}
\hspace*{-0.1cm}\epsfig{file=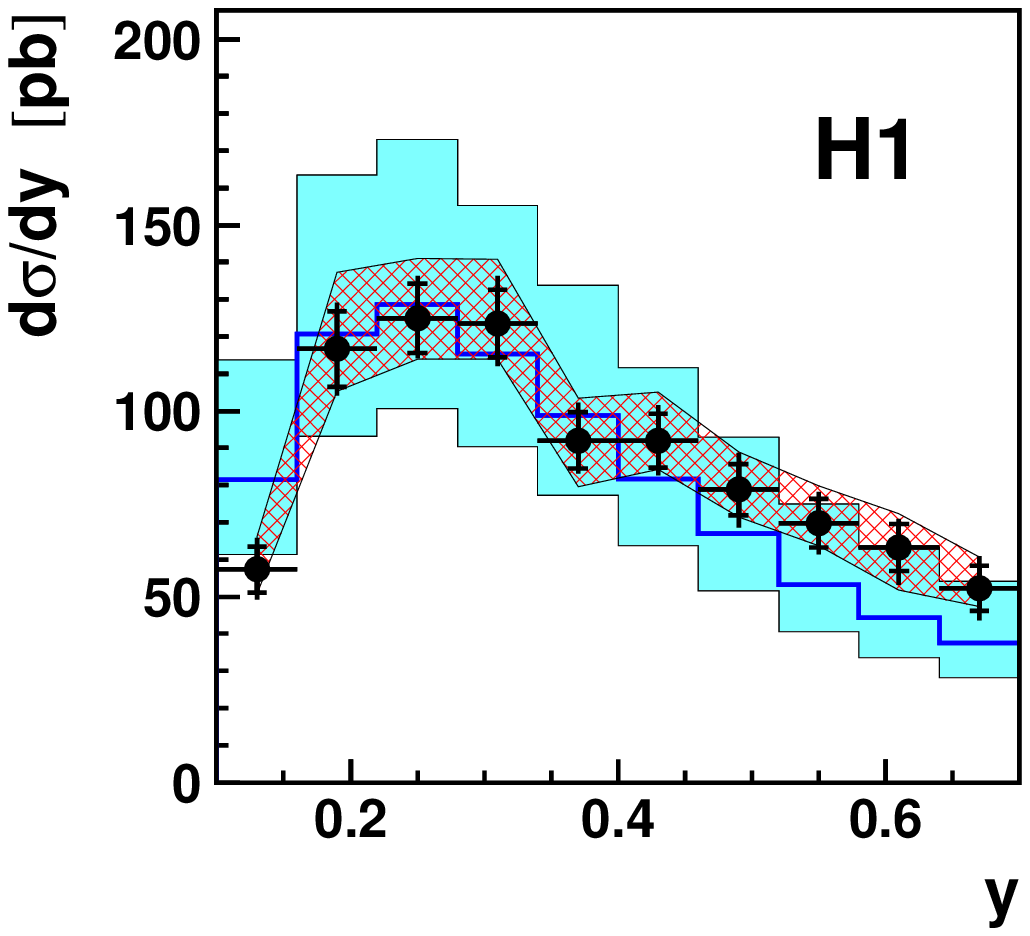,width=7cm}\hspace*{-0.1cm}\epsfig{file=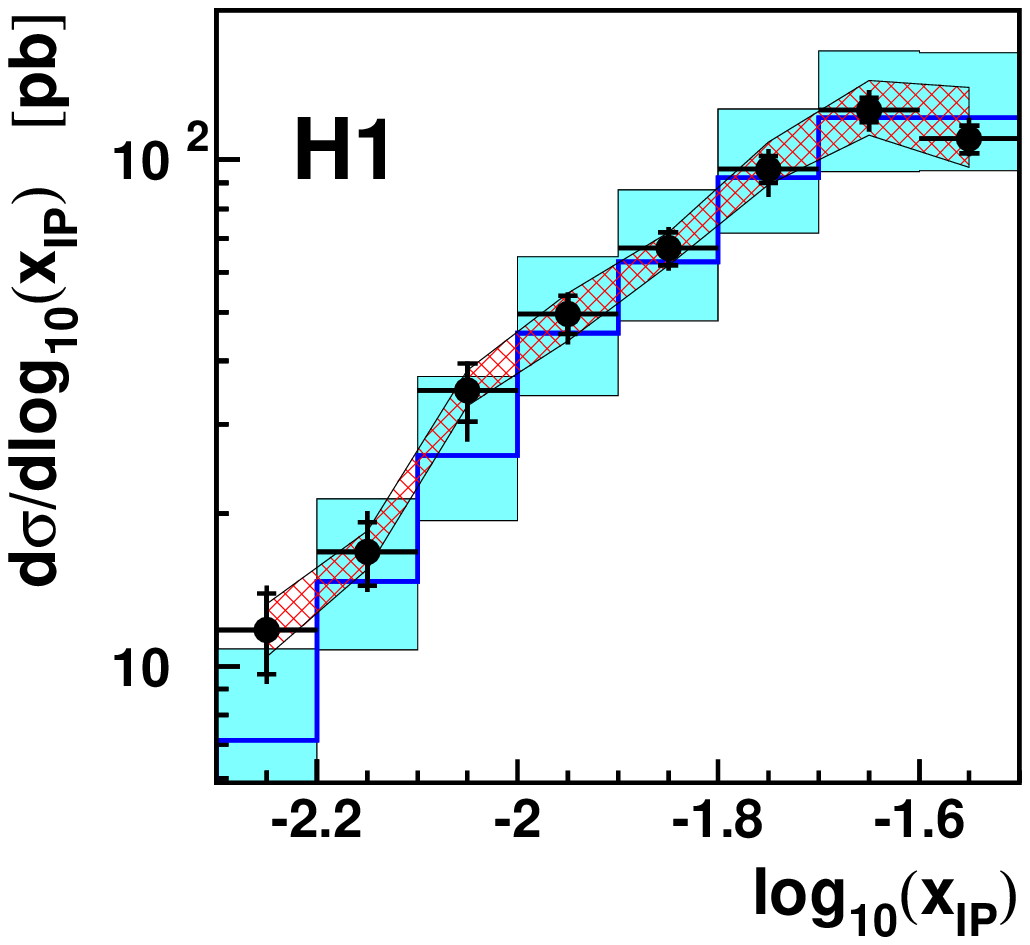,width=7cm}
\hspace*{-0.2cm}\epsfig{file=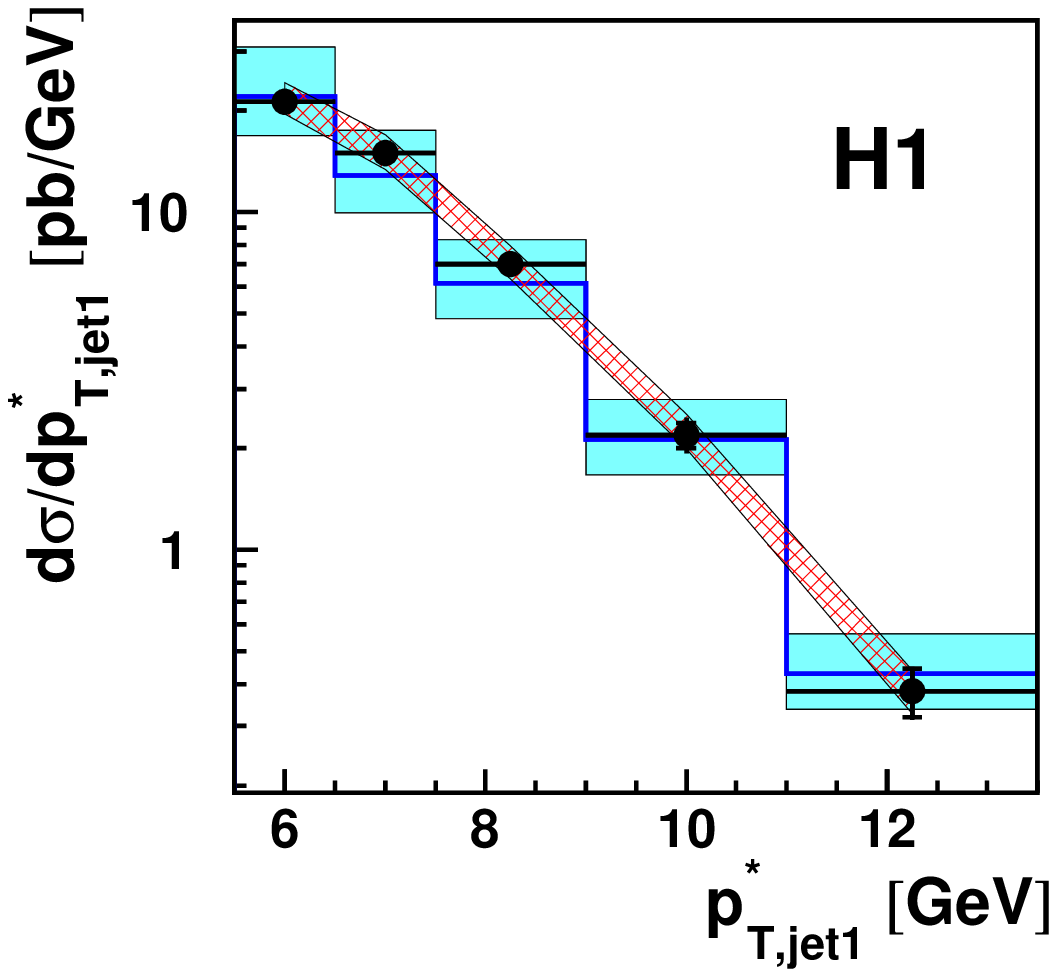,width=7cm}\hspace*{-0.1cm}\epsfig{file=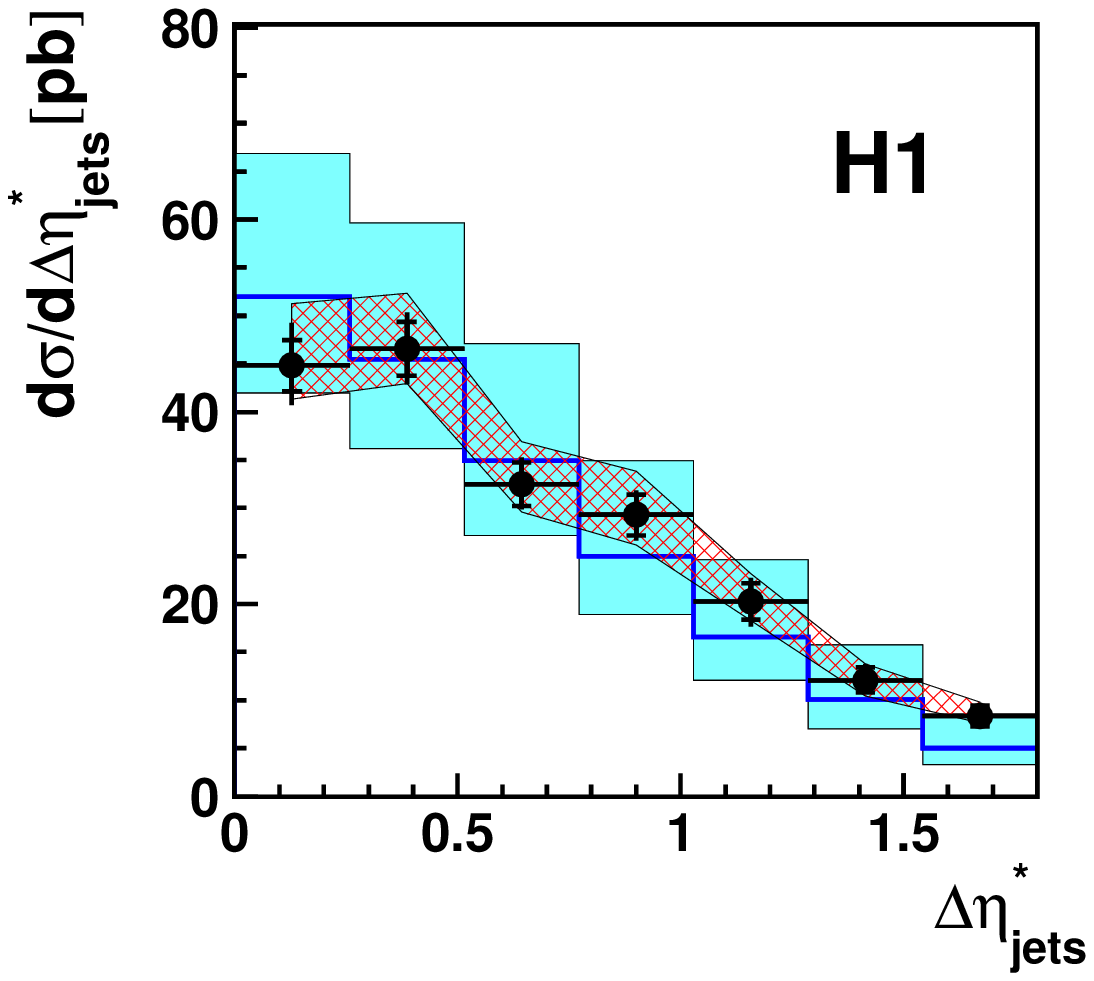,width=7cm}

\end{center}
\caption{Cross section for diffractive production of dijets by deep-inelastic scattering with $Q^2 = 4 - 80$\GeV$^2$, $101 < W <266$\GeV 
. Also shown are NLO predictions based on the H1 2006 PDFD fits: H1 2006 DPDF fit A (dotted) and fit B (dashed) with error band; from H1.}
\label{f:diffdis2jeth}
\vfill
\end{figure}

\begin{figure}[p]
\begin{center}
\vfill
\vspace*{-1cm}
\hspace*{-2.9cm}\epsfig{file=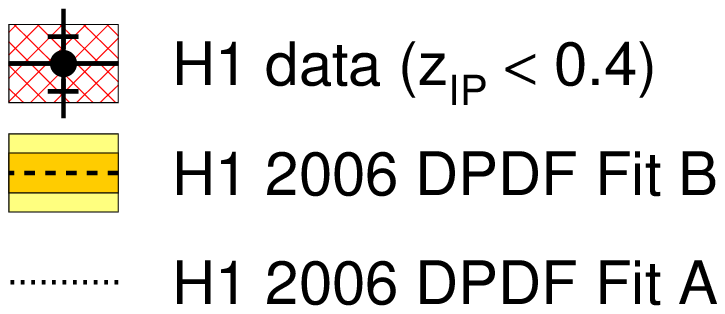,width=5cm}
\hspace*{-0.1cm}\epsfig{file=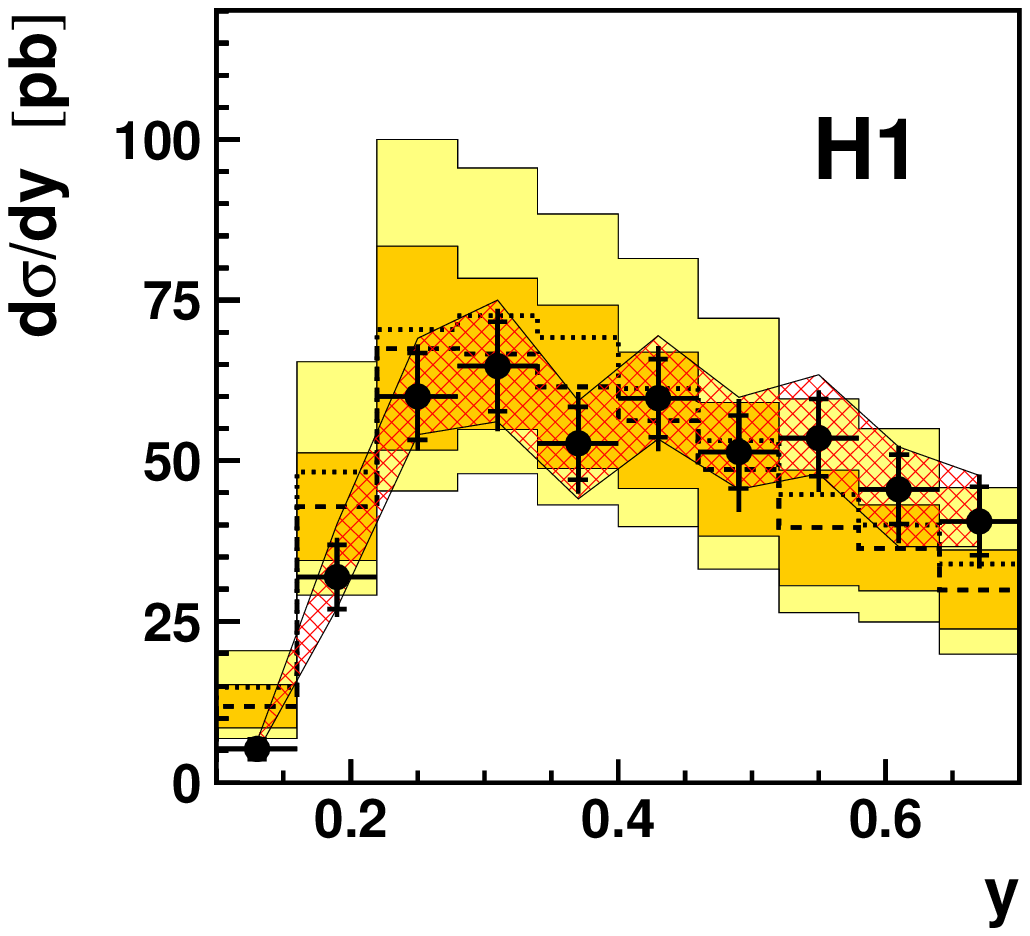,width=7cm}\hspace*{-0.1cm}\epsfig{file=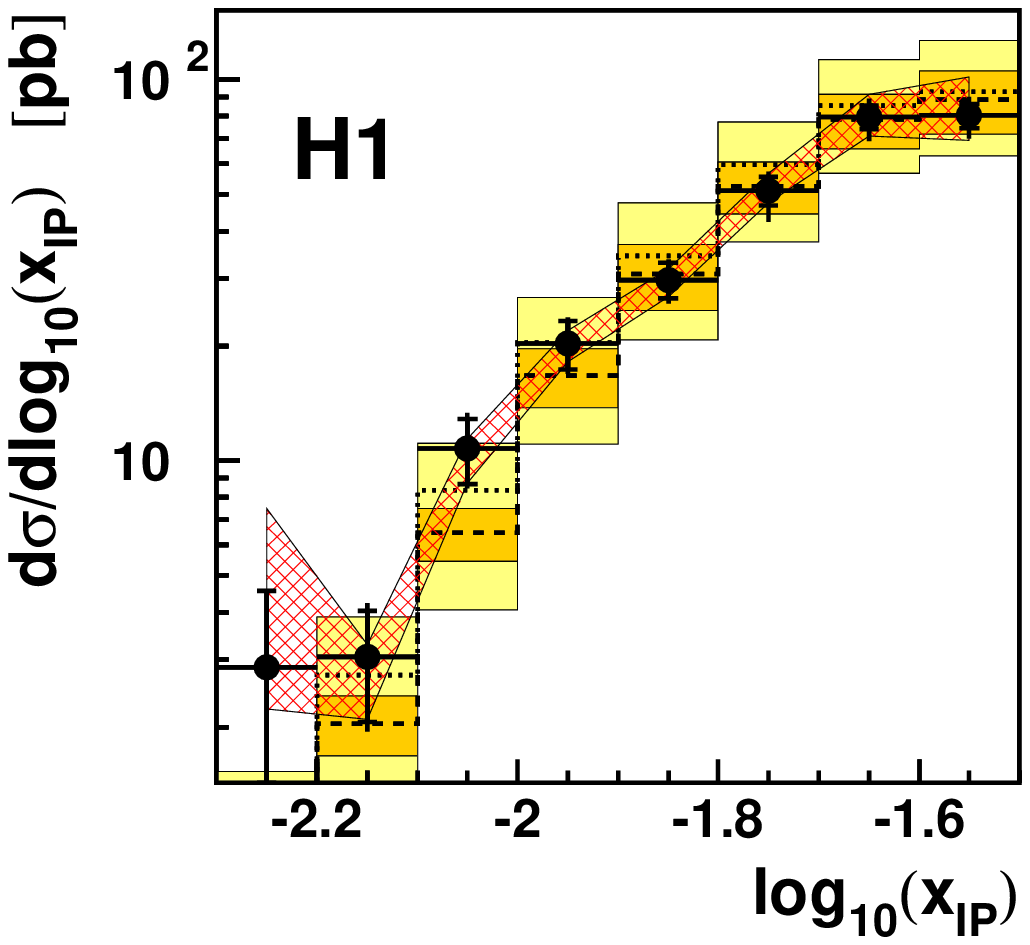,width=7cm}
\hspace*{-0.2cm}\epsfig{file=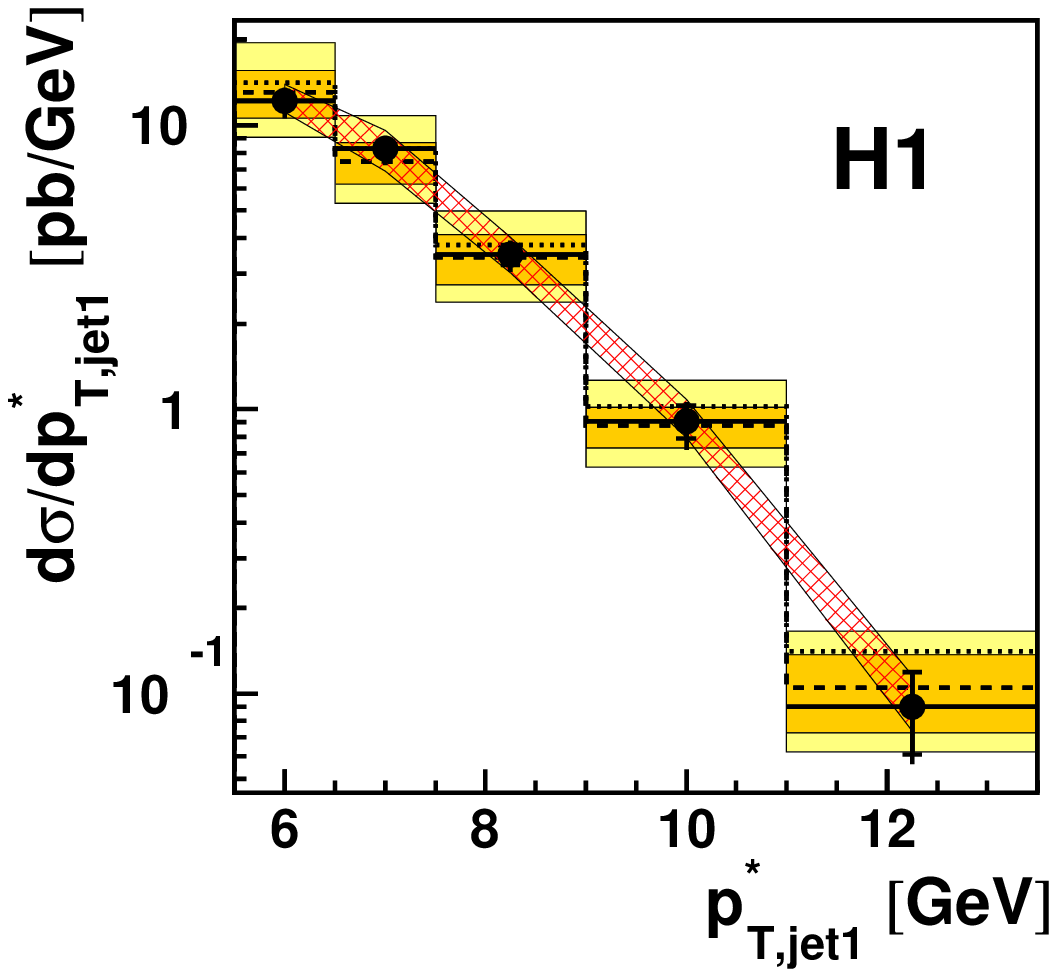,width=7cm}\hspace*{-0.1cm}\epsfig{file=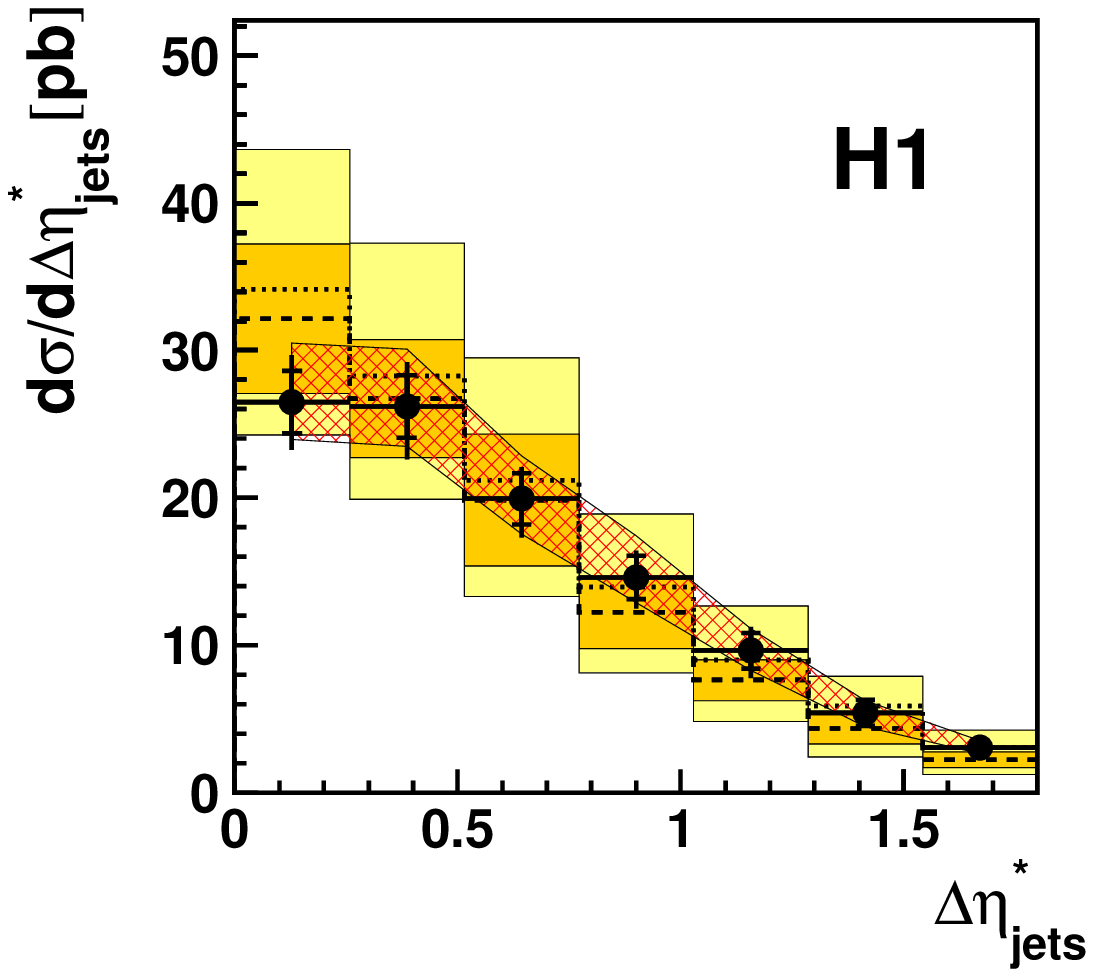,width=7cm}

\end{center}
\caption{Cross section for diffractive production of dijets by deep-inelastic scattering with $Q^2 = 4 - 80$\GeV$^2$, $101 < W <266$\GeV restricted to $z_{\pom} < 0.4$. Also shown are NLO predictions based on the H1 2006 PDFD fits: H1 2006 DPDF fit A (dotted) and fit B (dashed) with error band; from H1.}
\label{f:diffdis2jetlt04h}
\vfill
\end{figure}

Diffractive production of dijets has also been reported by H1~\cite{JHEP:0710:042}, using data from an integrated luminosity of 51.5 $pb^{-1}$ with $100<W<250$ GeV and $5<Q^2<100$ GeV$^2$. Diffractive events were required to have $\xpom < 0.03$, $M_N < 1.6$ GeV, where $M_N$ is the mass of the nucleonic system produced in the forward (= proton direction), and $|t| < 1$ GeV$^2$. The jet selection required $p^*_{T,jet1} >5.5$ GeV, $p^*_{T,jet2} >4$ GeV and $-3 < \eta^*_{jet} <0$, where the index $*$ indicates evaluation relative to the collision axis in the $\gamma^* p$ centre of mass frame. Figures~\ref{f:diffdis2jeth},~\ref{f:diffdis2jetlt04h} show the differential cross section as functions of $y, \xpom, p^*_{T,jet1}$ and $\Delta^*\eta_{jets} = |\eta^*_{jet1}-\eta^*_{jet2}|$, for all $z_{\pom}$ ($z_{\pom}$ is the fractional longitudinal momentum of the diffractive exchange carried by the parton entering the hard interaction), and for $z_{\pom} < 0.4$, respectively. 

\begin{figure}[p]
\begin{center}
\vfill
\epsfig{file=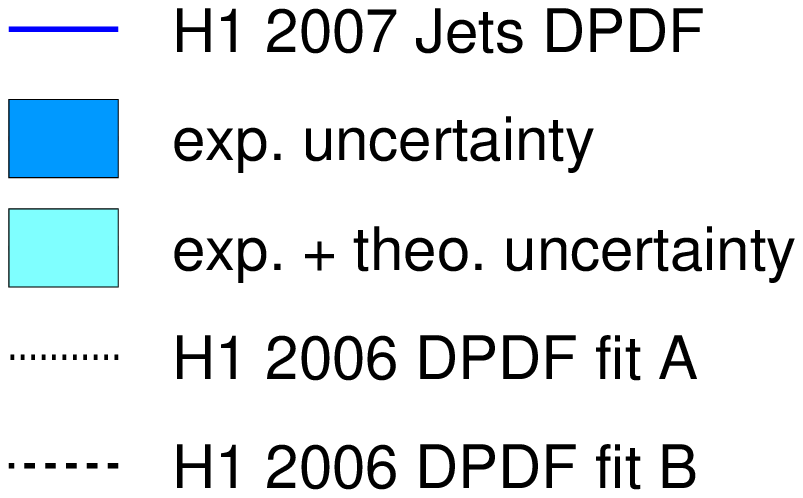,width=5cm}
\vspace*{1cm}\hspace*{-0.1cm}\epsfig{file=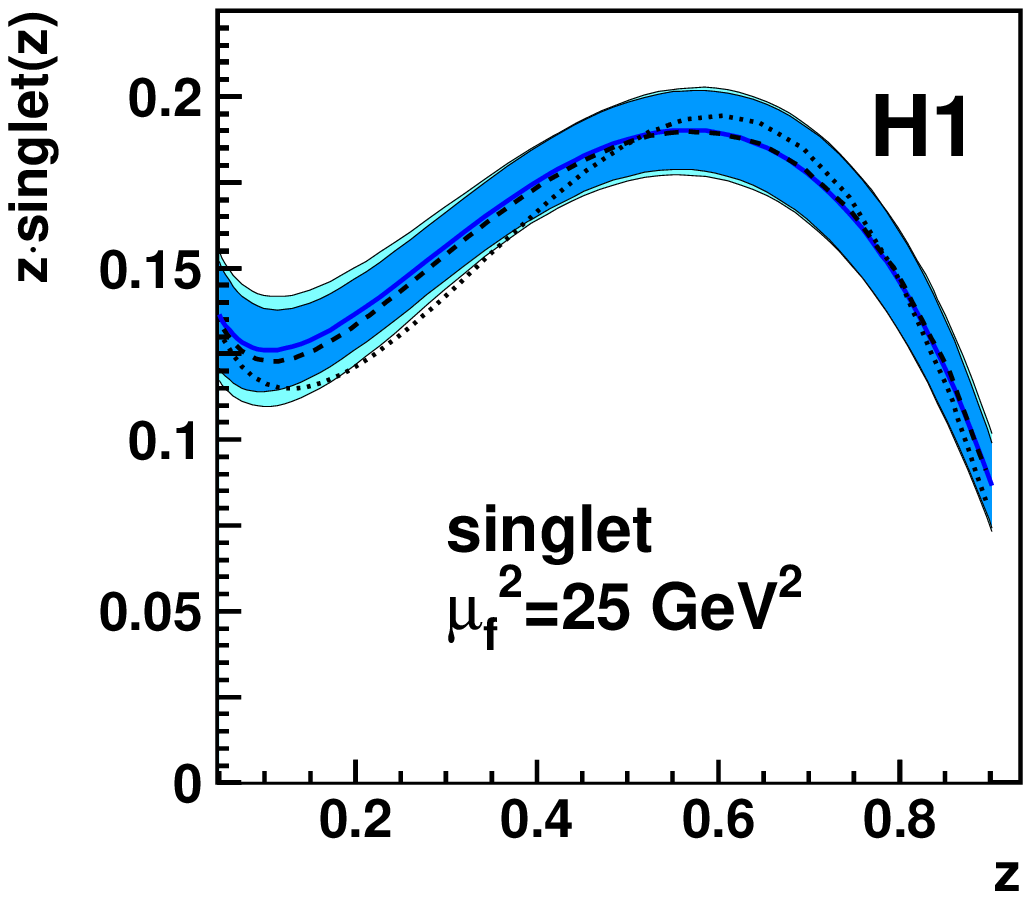,width=7cm}\hspace*{+0.2cm}\epsfig{file=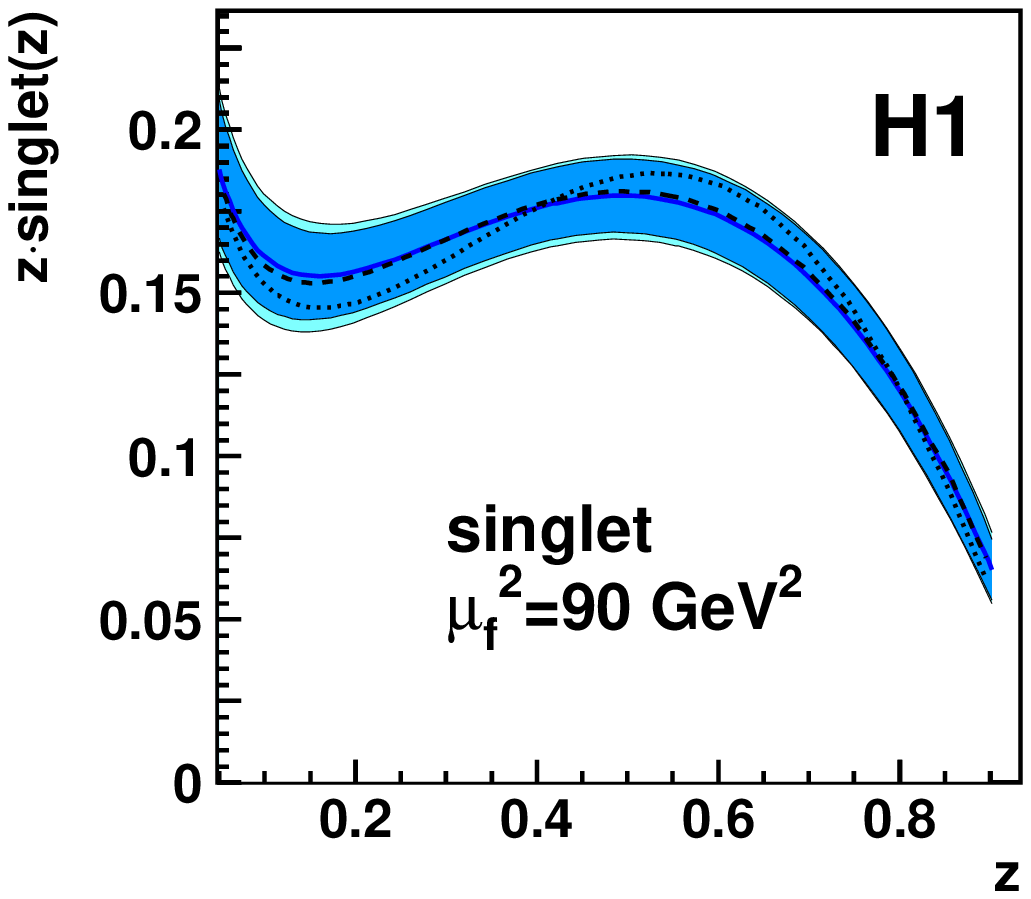,width=7cm}
\hspace*{-0.2cm}\epsfig{file=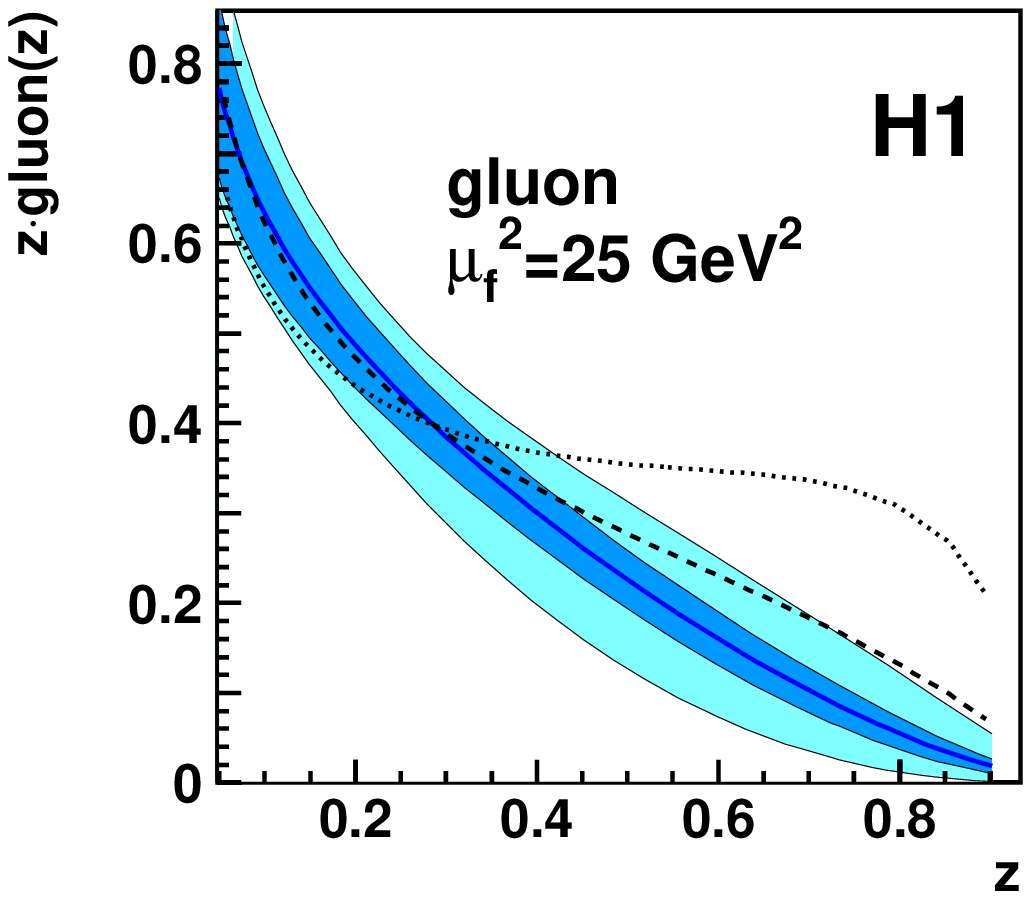,width=7cm}\hspace*{+0.2cm}\epsfig{file=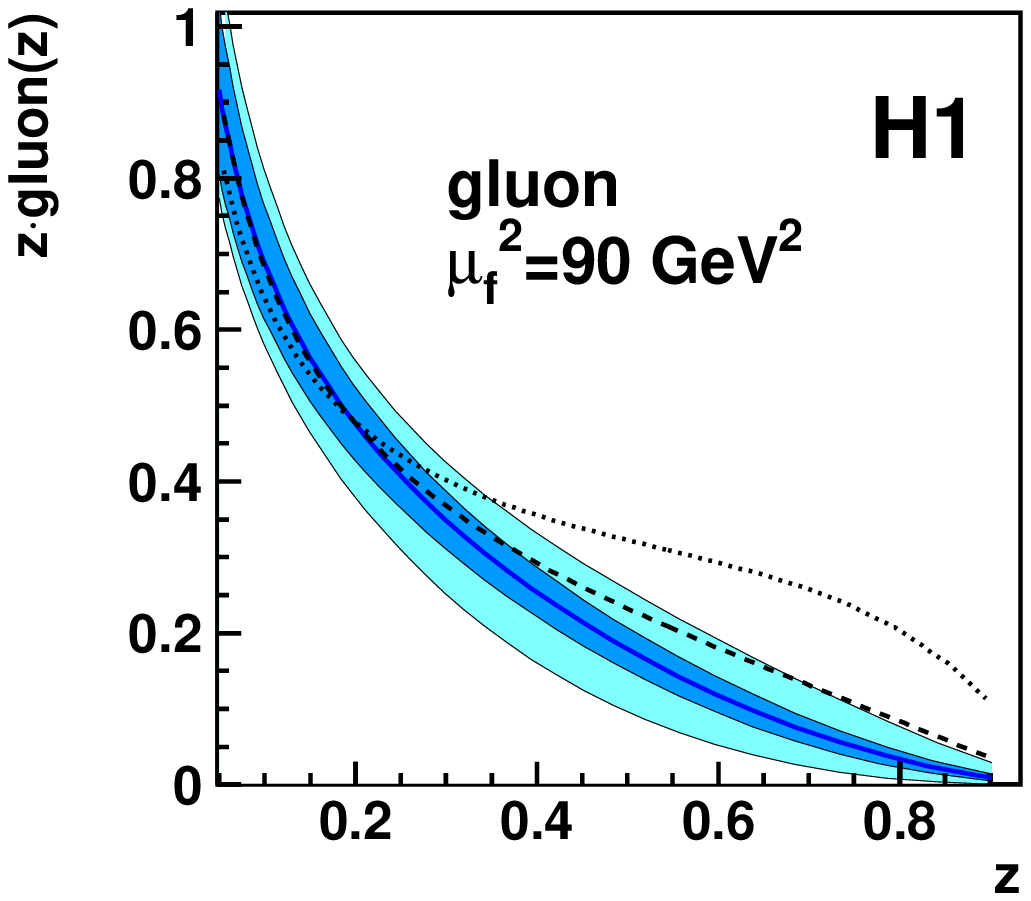,width=7cm}

\end{center}
\caption{Diffractive quark and gluon densities as a function $z$, the fraction of the longitudinal momentum of the diffractive exchange carried by quark (gluon), for the factorisation scales $\mu^2_f = 25$ and 90\GeV$^2$, respectively; from H1.}
\label{f:diffqandgh}
\vfill
\end{figure}

The dijet data are compared with QCD predictions at next-to-leading order based on diffractive parton distribution functions obtained from a fit to these data (H1 2007 Jets DPDF fit B, based on Refs.~\cite{List99,Nagy01}), and to H1 data on inclusive diffraction in deep-inelastic scattering. As a function of $z_{\pom}$ the dijet data at low and intermediate $z_{\pom}$ are well described. The resulting diffractive quark and gluon densities are shown in Fig.~\ref{f:diffqandgh} (for factorisation scales $\mu^2$ = 25 and 90 GeV$^2$, respectively) as a function of $z = z_{\pom} $. The singlet quark densities show a (small) rise towards $z = 0$ and a broad maximum around $z = 0.5-0.6$. In contrast, the gluon density is falling rapidly with $z$.

\section{Inclusive diffraction}
Following the first observation of events with a large rapidity gap~\cite{pl:b315:481}, ZEUS presented a quantitative study of the diffractive contribution to deep-inelastic scattering~\cite{pl:b338:483}. The energy flow was analyzed in the Breit frame, where the virtual photon carries only a space-like momentum component, $q = (0,0,-Q,0)$, and the momentum vector of the incoming proton is given by $P = (0,0,Q/2x,Q/2x)$. In the Quark-Parton Model, the interacting quark from the proton carries a momentum $p=xP$; the $z$ component of the incoming quark is $Q/2$ before, and $-Q/2$ after interaction with the virtual photon. 

\begin{figure}[p]
\begin{center}
\vfill
\vspace*{-1cm}
\hspace*{-0.1cm}\epsfig{file=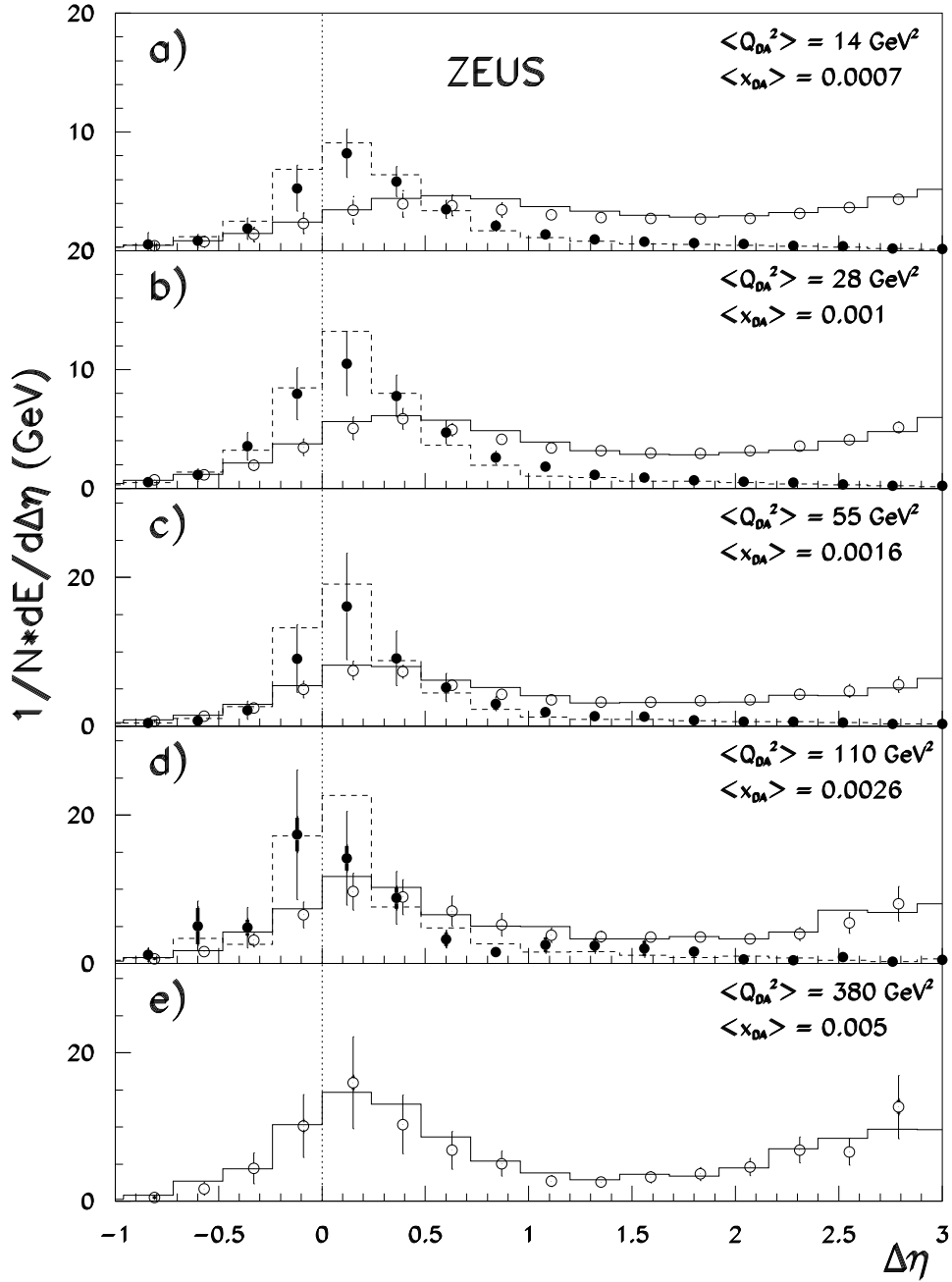,width=14cm}
\end{center}
\caption{Energy flow as a function of the rapidity gap $\Delta \eta = \eta_{cell} -\eta_{\gamma H}$. The open circles show the distributions of events without a rapidity gap ($\eta_{max} > 1.5$); the solid circles show events with a large rapidity-gap (LRG: $\eta_{cell} \le 2.5$). The full histograms represent the expectation of the Colour Dipole Model (CDMBGF) for nondiffractive production where large rapidity gaps are exponentially suppressed. The dotted histograms show the expectation of POMPYT for diffractive production; from ZEUS.}
\label{f:eflowlrgz}
\vfill
\end{figure}

Figure~\ref{f:eflowlrgz} shows for $Q^2$ between 14 and 380 GeV$^2$ the energy flow distributions, $1/N\cdot(dE/d(\Delta \eta)$, as a function of $\Delta \eta = \eta_{cell} -\eta_{(\gamma_H)}$, where $\eta_{cell}$ is the pseudorapidity of a calorimeter cell with energy deposit above 60 MeV (110 MeV) depending on the calorimeter section, and $\gamma_H$ corresponds to the scattering angle of the massless struck quark emerging as the current jet. Here, $\eta_{cell}$ and $\eta_{\gamma_H}$ were measured in the laboratory frame. For non-rapidity gap events, $\eta_{max}>1.5$ (open circles), the energy flow rises towards large $\Delta \eta$ in the direction of the incoming proton. For events with a large rapidity gap, $\eta_{cell} \le 2.5$, the energy flow peaks around the direction of the virtual photon, $\Delta \eta = 0$, and is small at large $\Delta \eta$. Note, the outgoing proton or low-mass nucleonic system arising from diffractive production was not detected in this analysis.

The colour dipole model (CDMBGF~\cite{np:b301:554,cpc:39:347,cpc:43:367}) which decribes standard deep-inelastic scattering without diffraction, considers the production of arbitrarily many jets in leading-log approximation for parton shower development. It gives an excellent representation of the non-rapidity gap events (solid histograms) but fails for events with a large rapidity gap. Diffractive scattering is modelled by POMPYT~\cite{desy:93:187} where the beam proton emits a pomeron, whose constituents take part in a hard scattering process with the virtual photon. POMPYT reproduces the distribution of events with a large rapidity gap (dashed histograms).
    
\subsection{Inclusive diffraction by $\gamma p$ interactions}
\begin{figure}[p]
\begin{center}
\vfill
\vspace*{-1cm}
\hspace*{-0.1cm}\epsfig{file=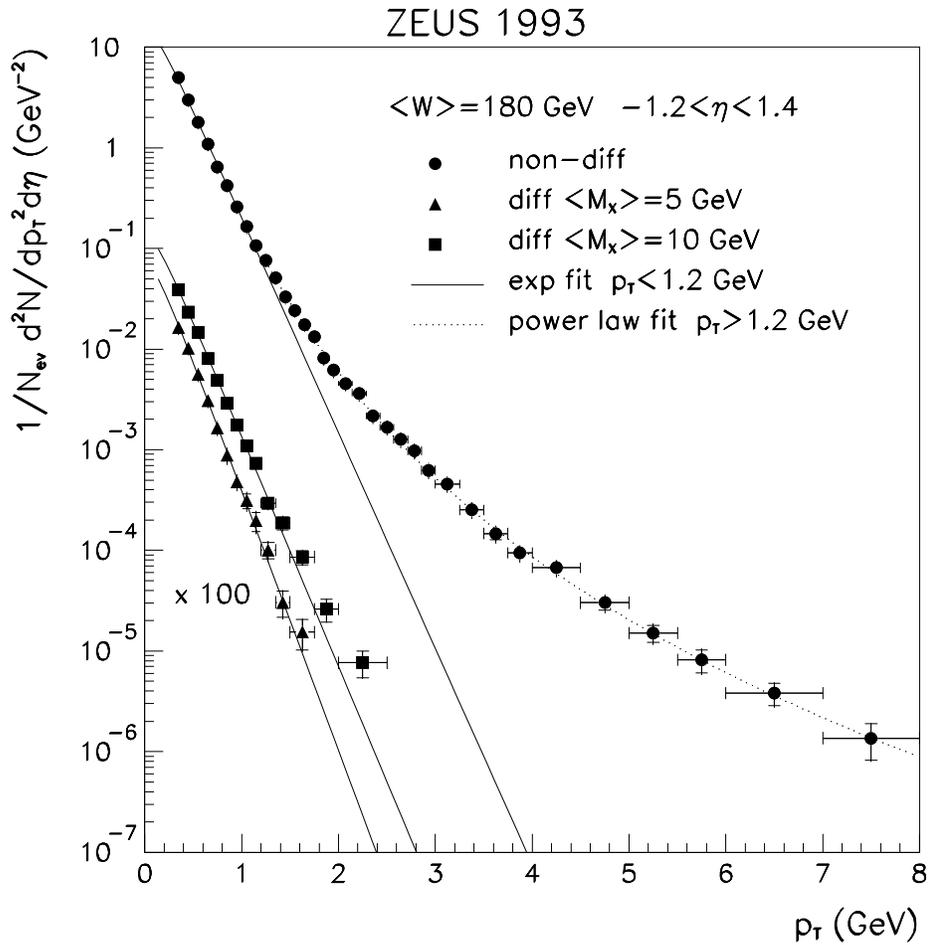,width=14cm}
\end{center}
\caption{Inclusive transverse momentum distributions of charged particles in photoproduction at $<W> = 180$\GeV for the interval $-1.2 < \eta <1.4$. Solid lines indicate a fit to the data with $p_T < 1.2$\GeV; the dotted line shows a power law fit to the nondiffractive data for $p_T > 1.2$\GeV. {\rm Note, for the sake of clarity the diffractive data are shifted downwards by two orders of magnitude}; from ZEUS.}
\label{f:ptdinodiz}
\vfill
\end{figure}

First results on inclusive photoproduction by diffraction were presented by ZEUS~\cite{zfp:c67:227}. The selection of events with $W \approx 180$ GeV, $\eta_{max} < 2$ and $\eta_{max} >2$, respectively, provided separate samples of diffractive and nondiffractive events. For diffraction, events with average hadronic masses reconstructed in the detector, $M_X$ = 5 and 10 GeV, respectively, were selected. The $p_T$ spectra for charged particles with $-1.2 < \eta < 1.4$ from the nondiffractive and diffractive event samples are displayed in Fig.~\ref{f:ptdinodiz}: for the diffractive samples the $p_T$-spectra fall off $\propto e^{-b\cdot p_T}$ with $b = 5.9 (5.3)$ GeV$^{-1}$ (average $M_X = 5 (10)$ GeV), while the $p_T$ spectra for the nondiffractive sample show a long tail towards large $p_T$.

\subsection{Inclusive diffraction by $\gamma^* p$ interactions}
The early observations of inclusive diffractive production in deep-inelastic scattering by ZEUS~\cite{pl:b315:481,pl:b338:483} (see Figs.\ref{f:zeusdifevt},\ref{f:zetamax},\ref{f:eflowlrgz}), and subsequently by H1~\cite{np:b429:477}, were followed by detailed studies of this subject. 

The first measurement of the diffractive structure function was presented by H1~\cite{pl:b348:681}, based on an integrated luminosity of 0.27 pb$^{-1}$. For unpolarised beams, the differential cross section for $\gamma^*p \to X p$, can be described in terms of the diffractive structure functions, $F^{D(4)}_{2,L} (\beta, Q^2,\xpom,t)$:
\begin{eqnarray}
\frac{d^4 \sigma_{diff}}{d\beta dQ^2 d_{\xpom} dt} = \frac{2 \pi \alpha^2}{\beta Q^4} [(1+(1-y)^2) F_2^{D(4)}x - y^2 F^{D(4)}_L],
\end{eqnarray}
where $y=Q^2/(x \cdot s)$, $s$ is the square of the $ep$ c.m. energy, and the relation $x = \beta \cdot \xpom$ has been used. A possible contribution from longitudinal photons, as measured by $F^{D(4)}_L$, will be neglected in the following: it would increase the value of $F^{D(4)}$ extracted from the differential cross section by at most $17 \%$. 

\begin{figure}[p]
\begin{center}
\vfill
\vspace*{-1cm}
\hspace*{+0.4cm}\epsfig{file=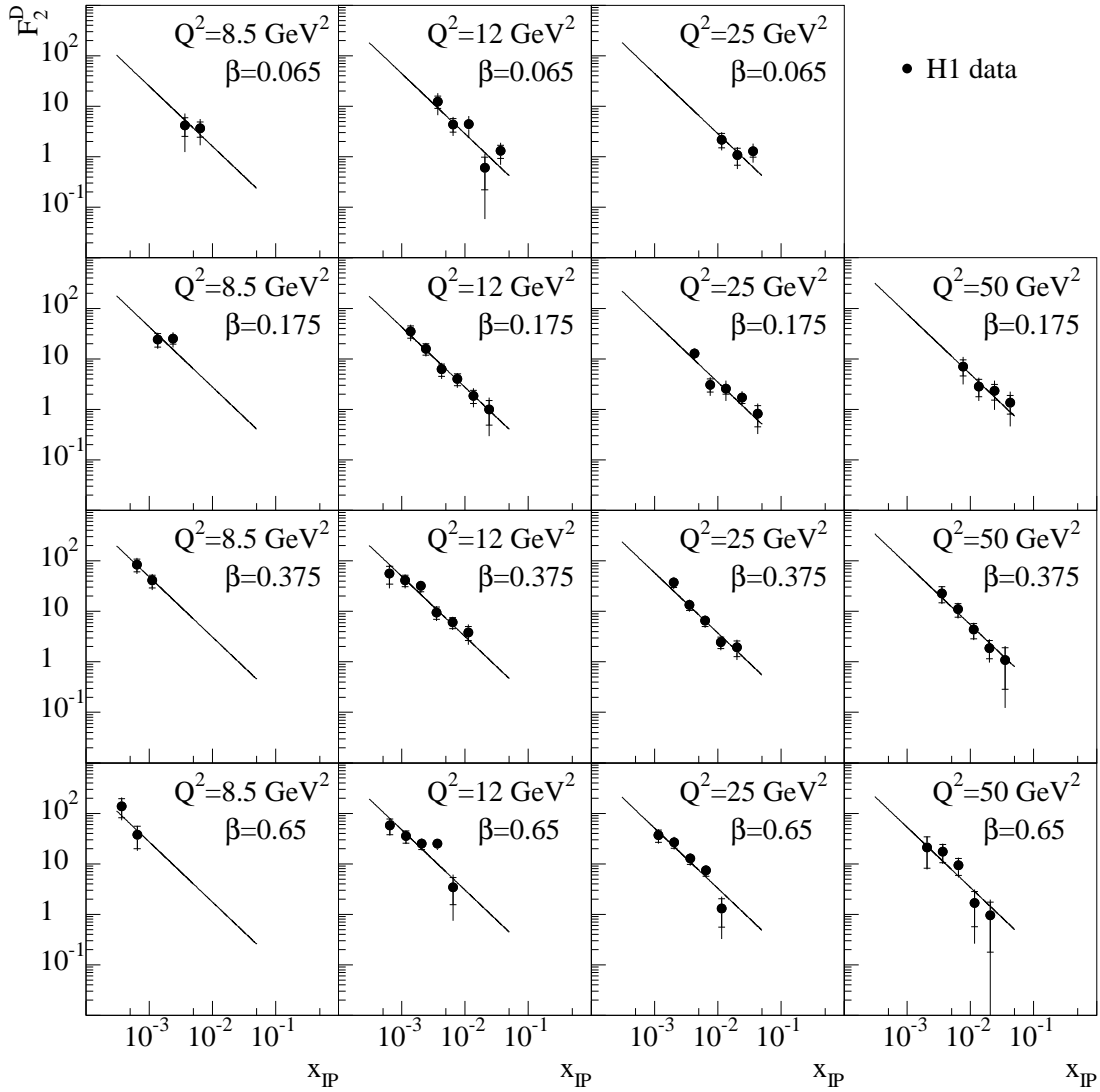,width=16cm}
\end{center}
\caption{The diffractive function $F^{D}_2(\beta,Q^2,\xpom)$ (denoted by $F^{D(3)}_2$ in the text) relative to the proton structure function $F_2$ as a function of $\xpom$ for different $\beta$ and $Q^2$; from H1.}
\label{f:f2d3vsxpomh}
\vfill
\end{figure}

Figure~\ref{f:f2d3vsxpomh} shows the diffractive structure function measured by H1~\cite{pl:b348:681} as a function of $\xpom$ for different values of $\beta$ and $Q^2$. For fixed $Q^2$ and $\beta$, $F_2^{D(4)}$ decreases with $\xpom$ proportional to $(\xpom)^{(-n)}$ with $n = 1.19 \pm 0.06(stat.)\pm 0.07(syst.)$, independent of $Q^2$ and $\beta$. This behaviour suggests a colourless target ($T$) in the proton, which carries only a small fraction of the proton momentum. In this case $F_2^{D(4)}\propto f_{T/p}(\xpom)$, where $f_{T/p}$ describes the flux of $T$ in the proton, and the cross section factorises into the flux $f$ and the cross section for virtual photon - proton scattering on target $T$. The observed $\xpom$ dependence corresponds to an effective Regge trajectory of $\alpha(t=0) = 1.10\pm 0.03(stat.)\pm 0.04(syst.)$ which agrees with that of the Pomeron. Note, for $\rho$ or $\omega$ exchange one would expect $\alpha(t=0) \approx 0.5$. Therefore, the large rapidity gap events observed in deep-inelastic $ep$ scattering result predominantly from diffractive scattering.

A similar conclusion was reached by ZEUS in Ref.~\cite{zfp:c68:569}. Furthermore, the ratio of diffractive to total $\gamma^* p$ cross section was found to be of the order of 10 to 20\%, independent of $x$ and $W$, for $Q^2 = 13$ and 39 GeV$^2$. Further results were reported by H1~\cite{zfp:c70:609}.

\begin{figure}[p]
\begin{center}
\vfill
\vspace*{-1cm}
\hspace*{+0.4cm}\epsfig{file=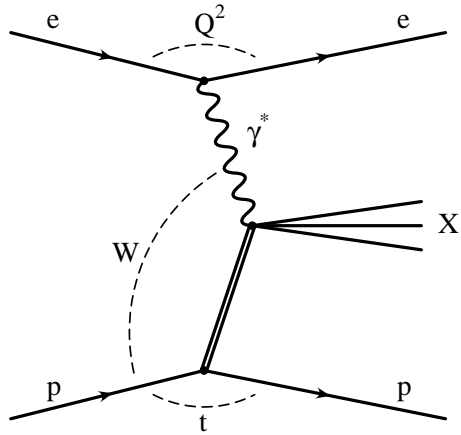,width=6cm}
\end{center}
\caption{Diagram for the reaction $ep \to eXp$.}
\label{f:diadift}
\vfill
\end{figure}

A novel method ($M_X$-method) in the analysis of inclusive diffractive production was introduced by ZEUS~\cite{zfp:c70:391}. Until then, the diffractive contribution had been extracted by selecting events with a large rapidity gap ($\eta_{max}$ cut), or with a leading proton. The drawback of both procedures is that not only diffractive production (by the exchange of the Pomeron), but any $t$-channel exchange (see Fig.~\ref{f:diadift}) of a colourless and electrical neutral particle such as the Reggeons $\rho^0$, $\omega$ etc. can produce events with a large rapidity gap and/or with a leading proton. The $W$ dependence of the cross section in the case of Reggeon exchange is proportional to $W^{2\alpha_R -2}$ at $t = 0$: for instance $\rho$-exchange, with $\alpha_R \approx 0.5$ at $t = 0$, leads to $\sigma \propto W^{-1}$. 

\begin{figure}[p]
\begin{center}
\vfill
\vspace*{-1cm}
\hspace*{+0.4cm}\epsfig{file=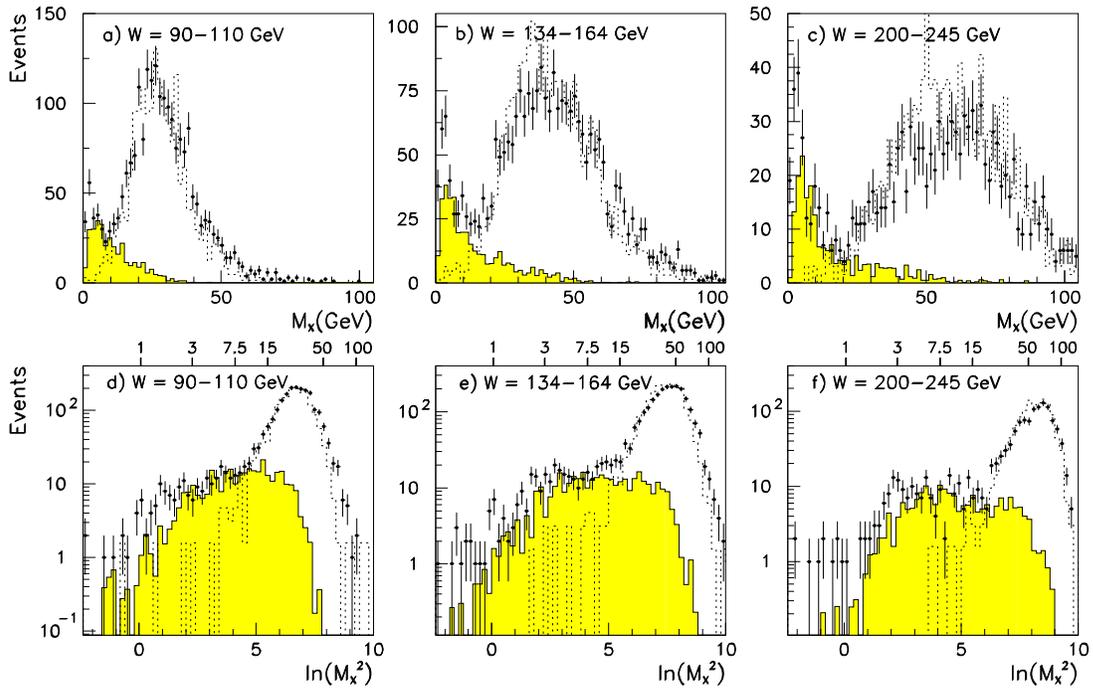,width=15cm}
\end{center}
\caption{Reaction $ep \to eX$: distribution of $M_X$(top) and $\ln M^2_X$ (bottom) for different regions of $W$ at $Q^2 = 14$\GeV$^2$. The shaded distributions show the prediction of the NZ model for diffractive production; the dashed histograms show the expectations for nondiffractive production predicted by the CDMBGF model; from ZEUS.}
\label{f:lnmx2z}
\vfill
\end{figure}

Such background is avoided by using the $M_X$-method\cite{kowrocwlf} for the extraction of the diffractive part since contributions from Reggeon exchange are exponentially suppressed. Here, $M_X$ is the mass of the hadronic system observed in the detector. The first extraction of the diffractive cross section with the $M_X$-method was performed by ZEUS~\cite{zfp:c70:391}, using data from an integrated luminosity of 0.54 pb$^{-1}$. The $M_X$ and $\ln M^2_X$ distributions are shown in Fig.~\ref{f:lnmx2z} at $Q^2 = 14$ GeV$^2$ for several bins of $W$. In the $M_X$ spectra (top) the diffractive contribution is falling rapidly with increasing $M_X$, while the $\ln M^2_X$ spectra (bottom) show a plateau for the diffractive component once $M_X$ is sufficiently above the kinematical threshold. This allows the extraction of the diffractive contribution by a fit to the $\ln M^2_X$ spectrum of the form:
\begin{eqnarray}
\frac{dN}{d\ln M^2_X} = D + c \cdot exp(b \cdot \ln M^2_X)
\end{eqnarray}   
The fit is applied in the interval $\ln M^2_X \le \ln W^2 - \eta_0$, where $\ln W^2 - \eta_0$ is the maximum value of $\ln M^2_X$ up to which the exponential behaviour of the nondiffractive part holds. The diffractive contribution is not taken from the fitted value of $D$; rather it is obtained by subtracting the nondiffractive contribution determined by the fit ($c \cdot exp(b \cdot \ln M^2_X)$) from the observed number of events.

\begin{figure}[p]
\begin{center}
\vfill
\vspace*{-1cm}
\hspace*{+0.4cm}\epsfig{file=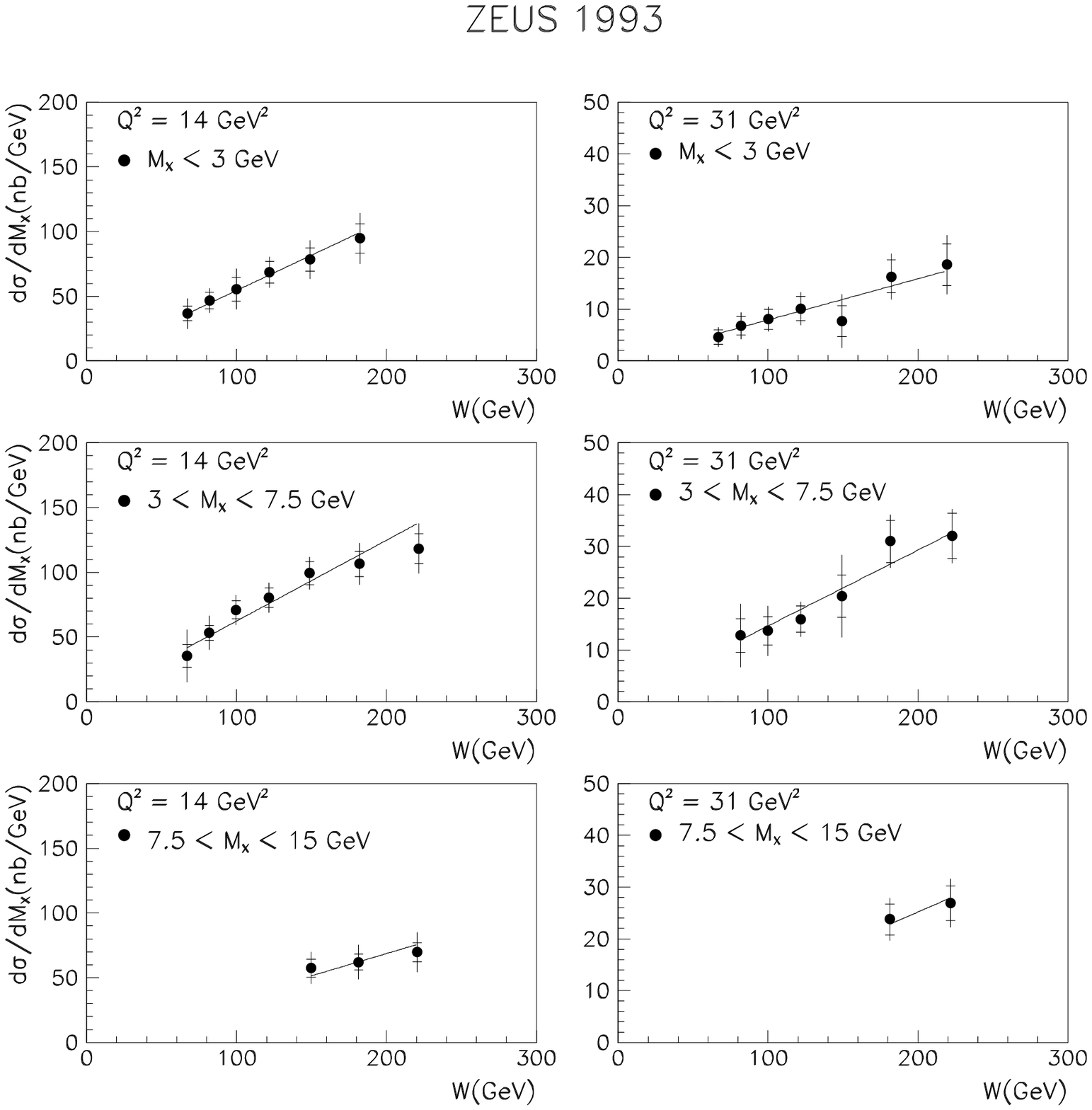,width=15cm}
\end{center}
\caption{The cross section $d\sigma^{diff}(\gamma^*p \to XN)/dM_X$ as a function of $W$ averaged over the $M_X$ intervals indicated, for $Q^2 = 14$ and $31$\GeV$^2$. 
The lines show the result of a fit to all data with the form 
$d\sigma^{diff}/dM_X \propto (W^2)^{(2 \bar{\alpha_{\pom}}-2)}$, see text; from ZEUS.}

\label{f:dsigdmx1993z}
\vfill
\end{figure}

The cross section for diffractive scattering, $d\sigma^{diff}_{\gamma^* p \to XN}/dM_X$, is shown in Fig.~\ref{f:dsigdmx1993z} as a function of $W$ for different values of $Q^2$ and $M_X$. The diffractive cross section is rising with $W$ for all $Q^2$, $M_X$ values. Assuming that only Pomeron exchange is contributing, the $W$ dependence for fixed $Q^2$ was parametrised in terms of the Regge trajectory of the Pomeron,
\begin{eqnarray}
\frac{d\sigma^{diff}_{\gamma^*p \to XN}}{dM_X}\propto (W^2)^{(2 \bar{\alpha_{\pom}}-2)}
\end{eqnarray}  
The data yielded for the Regge trajectory of the Pomeron~\cite{zfp:c70:391},
\begin{eqnarray}
\bar{ \alpha_{\pom}} = 1.23 \pm 0.02(stat) \pm 0.04 (syst),
\end{eqnarray}
in agreement with the result obtained by H1, see above.

\begin{figure}[p]
\begin{center}
\vfill
\vspace*{-1cm}
\hspace*{+0.1cm}\epsfig{file=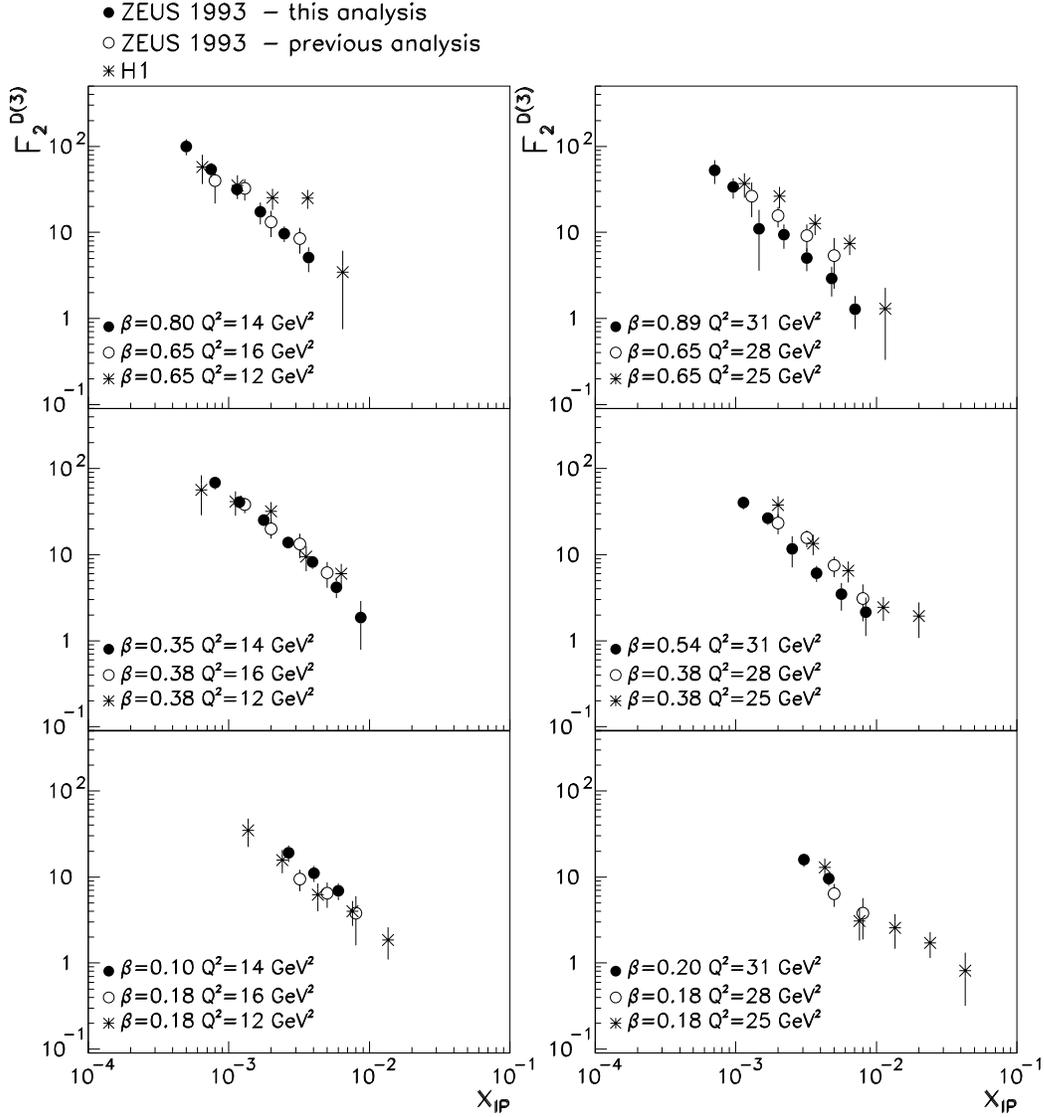,width=14cm}
\end{center}
\caption{The diffractive structure function $F^{D(3)}_2$ as a function of $\xpom$: solid points from ZEUS obtained with the $M_X$ - analysis; open points from a previous ZEUS analysis using the $\eta_{max}$-method; stars from H1, obtained with the $\eta_{max}$-method; from ZEUS.}
\label{f:f2d393z}
\vfill
\end{figure}

The diffractive structure function determined with the $M_X$-method is shown in Fig.~\ref{f:f2d393z} together with the results from previous measurements of H1~\cite{pl:b348:681} and ZEUS~\cite{zfp:c68:569}. As indicated, the three measurements were performed at slightly different values of $\beta$ and $Q^2$. Broad agreement is observed between the H1 and ZEUS measurements. 

\begin{figure}[p]
\begin{center}
\vfill
\vspace*{-1cm}
\end{center}
\vspace*{-3cm}
\hspace*{+0.1cm}\epsfig{file=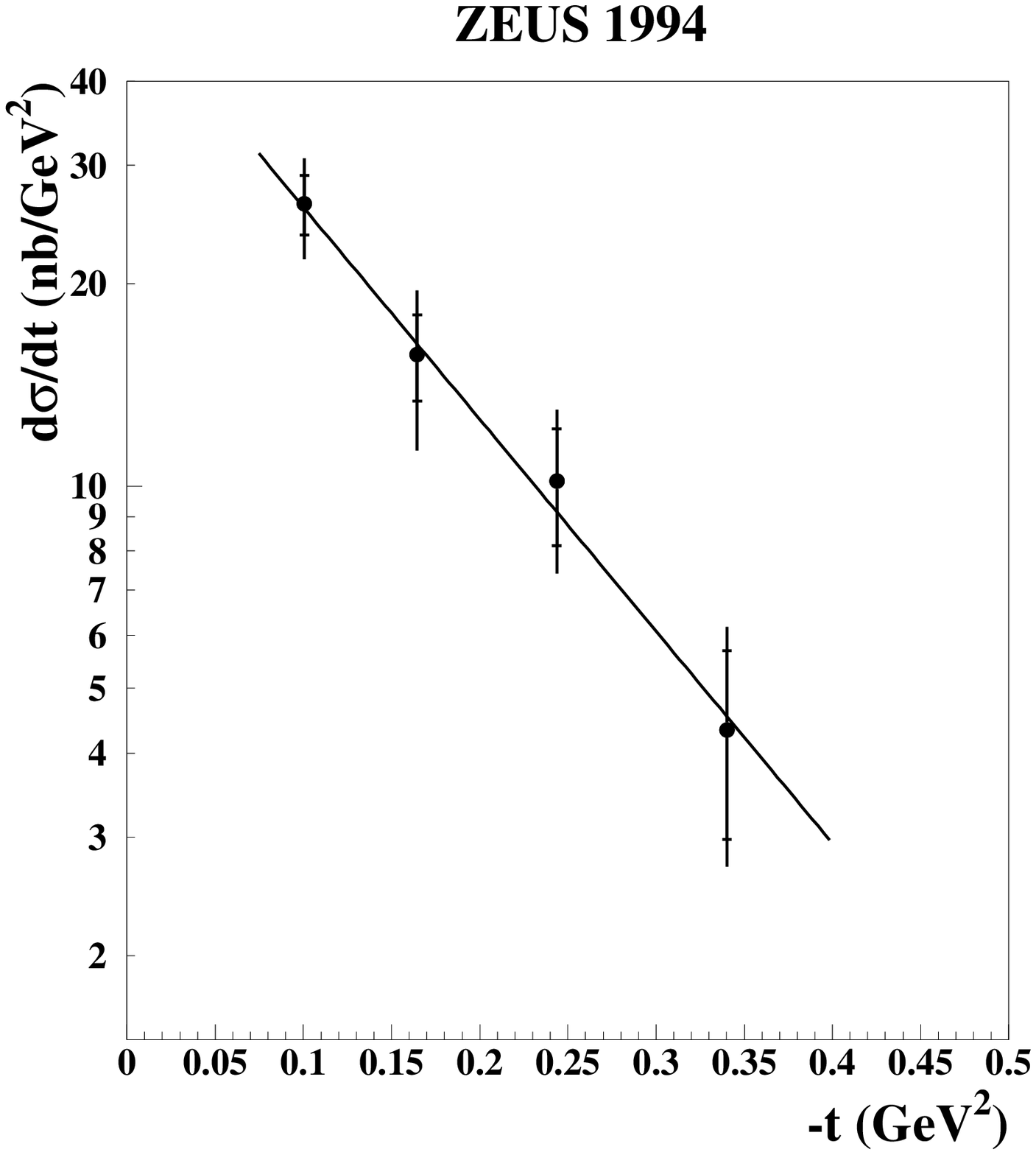,width=7cm}\hspace*{+1cm}\epsfig{file=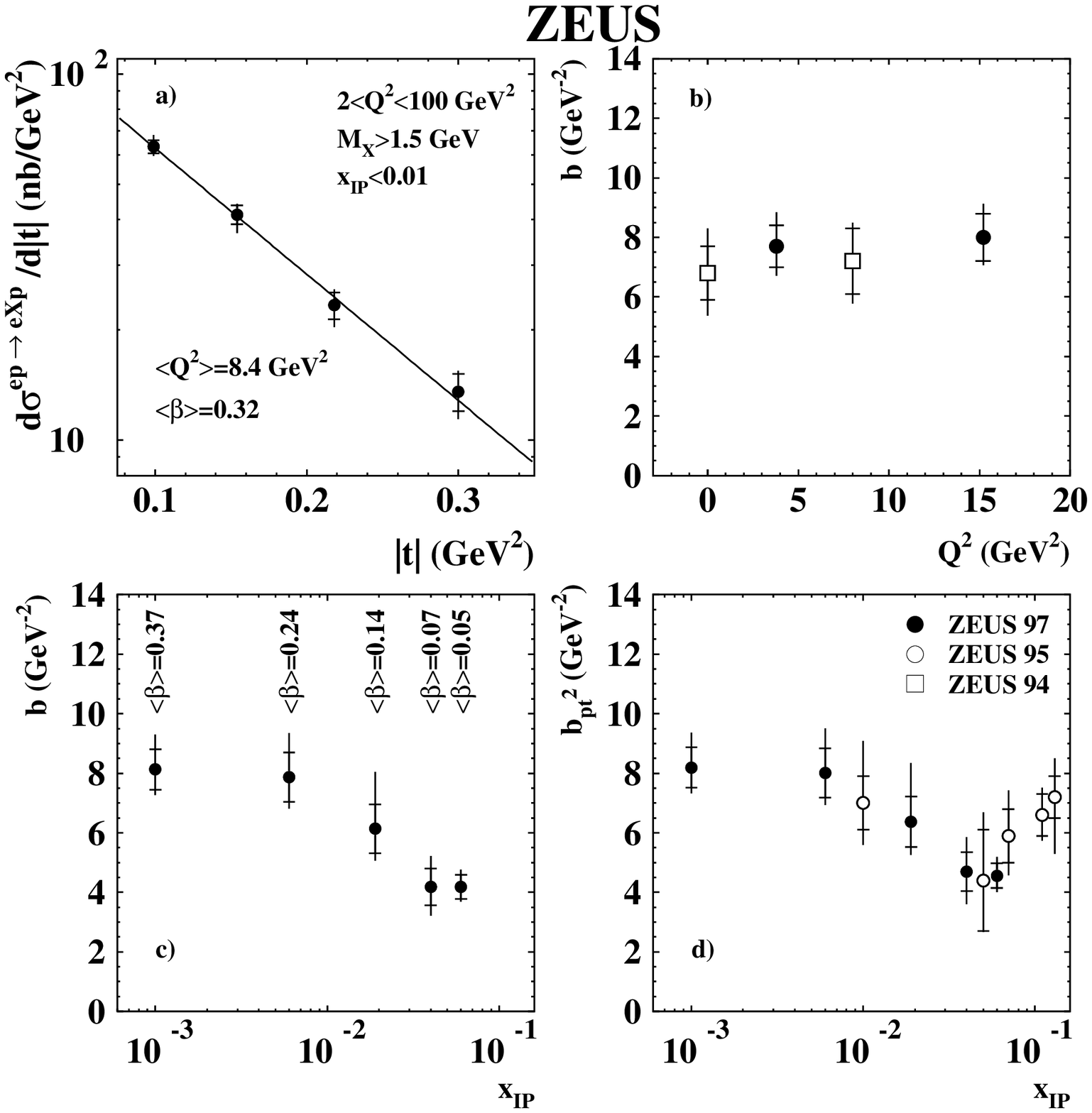,width=8cm}
\hspace*{+4cm}\epsfig{file=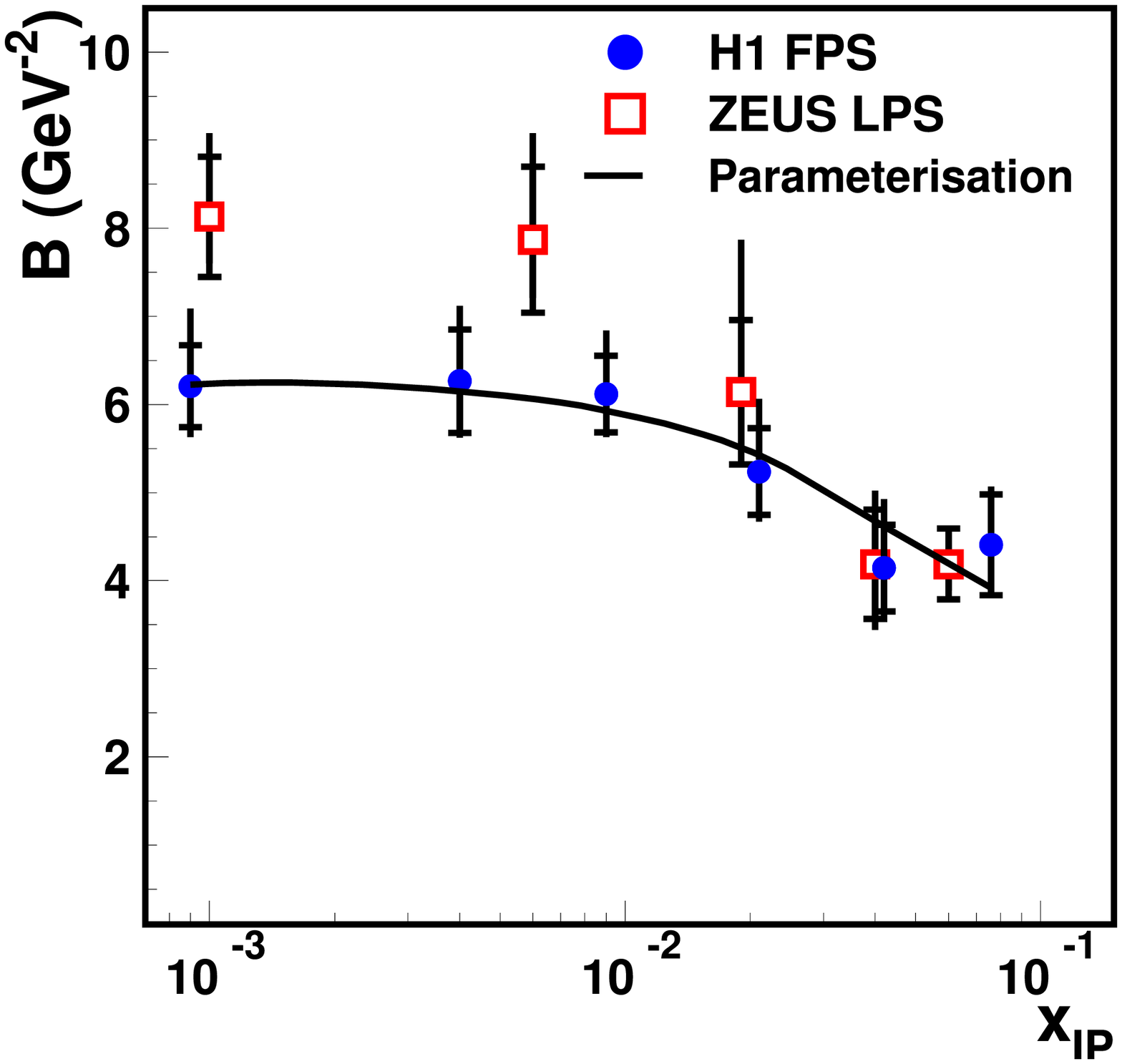,width=8cm}
\caption{Top left: The differential cross section $d\sigma/d|t|$ for diffractive events with a leading proton of $x_L > 0.7$, determined in the range $5<Q^2<20$\GeV$^2$, $50<W<270$\GeV and $0.15 < \beta < 0.5$. Top right: a) The differential cross section $d\sigma^{ep \to Xp}/d|t|$ for $\xpom < 0.01$, $Q^2 = 2-100$\GeV$^2$, $M_X > 1.5$\GeV; b) the slope $b$ as a function of $Q^2$, and c) as a function of $\xpom$; d) the slope $b_{pT^2}$ as a function of $\xpom$; from ZEUS. Bottom: The slope $b$ of the $|t|$ distribution - here denoted by $B$ - as a function of $\xpom$ for $Q^2 = 2 - 50$\GeV$^2$ and $y= 0.02 - 0.6$; from H1.}
\label{f:disdsdtz}
\vfill
\end{figure}

The first measurement of $\gamma^* p \to Xp$, where the scattered proton was momentum analyzed in a magnetic spectrometer, has been reported by ZEUS~\cite{epj:c1:81}. Figure~\ref{f:disdsdtz} shows $d\sigma/d|t|$ as a function of the four-momentum transfer squared $t$, for the range $5<Q^2<20$ GeV$^2$, $50<W<270$ GeV and $0.15 < \beta < 0.5$. A fit of the form $d\sigma/d|t| = A\cdot e^{-b|t|}$ yielded for the slope $b = 7.2 \pm 1.1({\rm{stat}})^{\rm{+0.7}}_{\rm{-0.9}}({\rm syst})$ GeV$^{-2}$. Further results on $b$ from~\cite{epj:c38:43,epj:c48:749} are also presented in Fig.~\ref{f:disdsdtz}. Since for these $Q^2$-values the transverse extension of the virtual photon is negligible compared to the radius of the proton $R_p$ then $b = (R_p^2)/4$. Elastic $\bar{p}p$ scattering at a c.m. energy of 546 GeV yields $b = 15.35$ GeV$^{-2}$\cite{pl:b147:385} which is expected to be equal to $(R_p^2)/2$. This result is in good agreement with the data from HERA.

In the following, the high statistics results on inclusive diffraction as obtained by H1 and ZEUS are presented.
 
\begin{figure}[p]
\begin{center}
\vfill
\epsfig{file=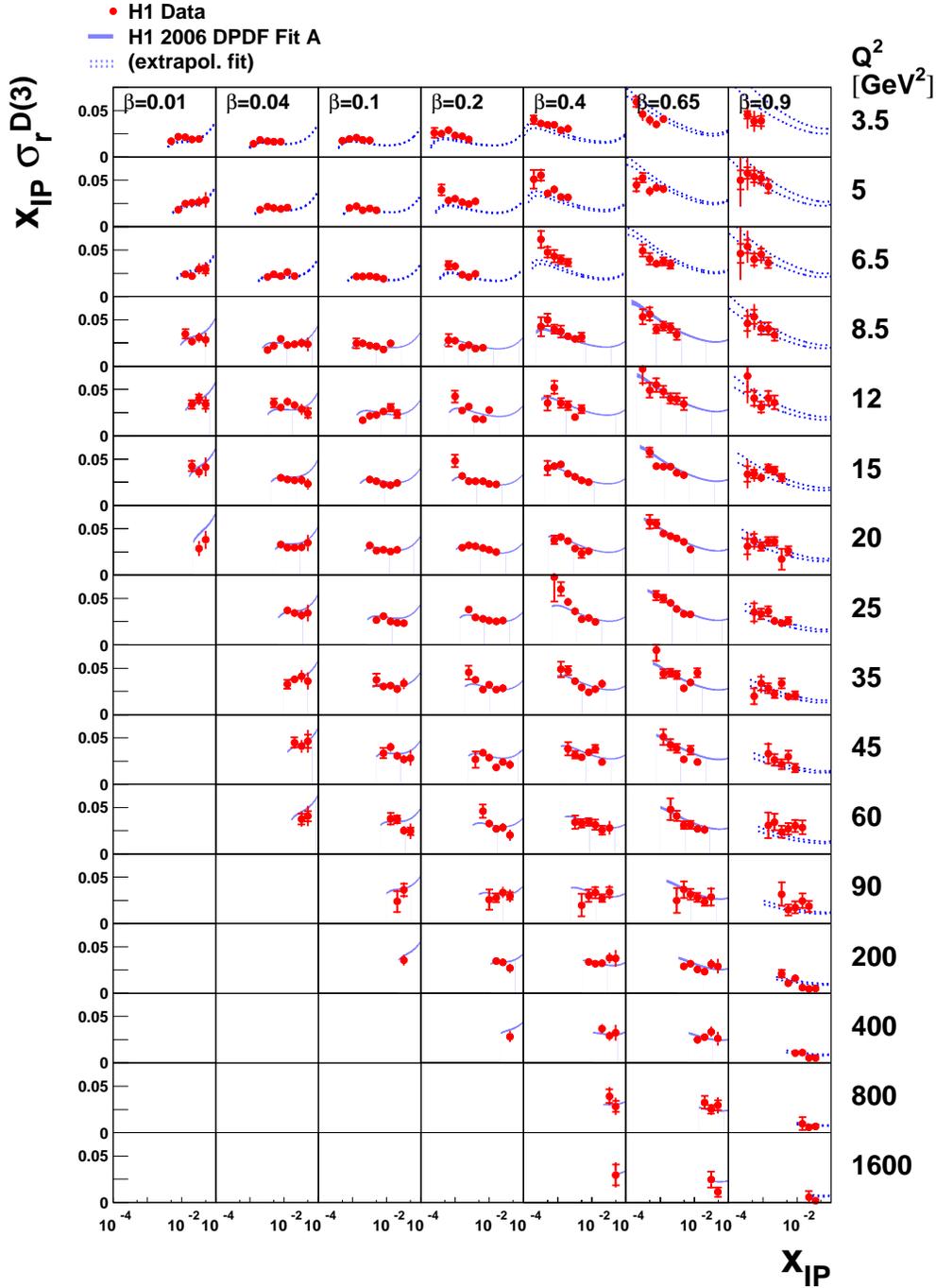,width=13cm}
\caption{The diffractive structure function $\xpom F^{D(3)}_2$ (here denoted by $\xpom \sigma^{D(3)}_r$), as a function of $\xpom$ for the $\beta,Q^2$ values indicated. The curves show the H1 2006 DPDF Fit A to these data as a shaded error band in kinematic regions included in the fit, and as a pair of dashed lines in regions excluded from the fit; from H1.}
\label{f:fd3allbetq2h1}
\end{center}
\end{figure}

The diffractive structure function measured by H1 for $\beta$ values between 0.01 and 0.9, and $Q^2$ values from 3.5 to 1600 GeV$^2$ is shown in Fig.~\ref{f:fd3allbetq2h1} as a function of $\xpom$. (From here on the term ``diffractive structure function'' is used for the function $\xpom F^{(D(3)}_2$ which H1 denotes by $\xpom \sigma_r^{D(3)}$ while ZEUS uses the notation $\xpom F^{D(3)}_2$; multiplication by $\xpom$ takes out a trivial dependence on $\xpom$ and elucidates better the $\xpom$ dependence of the diffractive structure function.). In the H1 measurement, the mass of the nucleonic system $Y$ escaping through the forward beam hole is limited to $M_Y < 1.6$ GeV.  

Figure~\ref{f:fd3allbetq2h1} shows that $\xpom F^{D(3)}_2$ increases as $\xpom \to 0$, provided $\beta \ge 0.2$. A fit of the data by a QCD motivated model, which includes the contributions from Pomeron plus Reggeon exchanges in the $t$-channel, gives a good account of the data (see solid curves): it indicates that a large fraction of the momentum of the Pomeron exchanged in the $t$-channel is carried by gluons (see dashed curves). In the model, the rise of $\xpom F^{D(3)}_2$ for $\xpom > 0.1$ is due to Pomeron-Reggeon interference.

\begin{figure}[p]
\begin{center}
\vfill
\epsfig{file=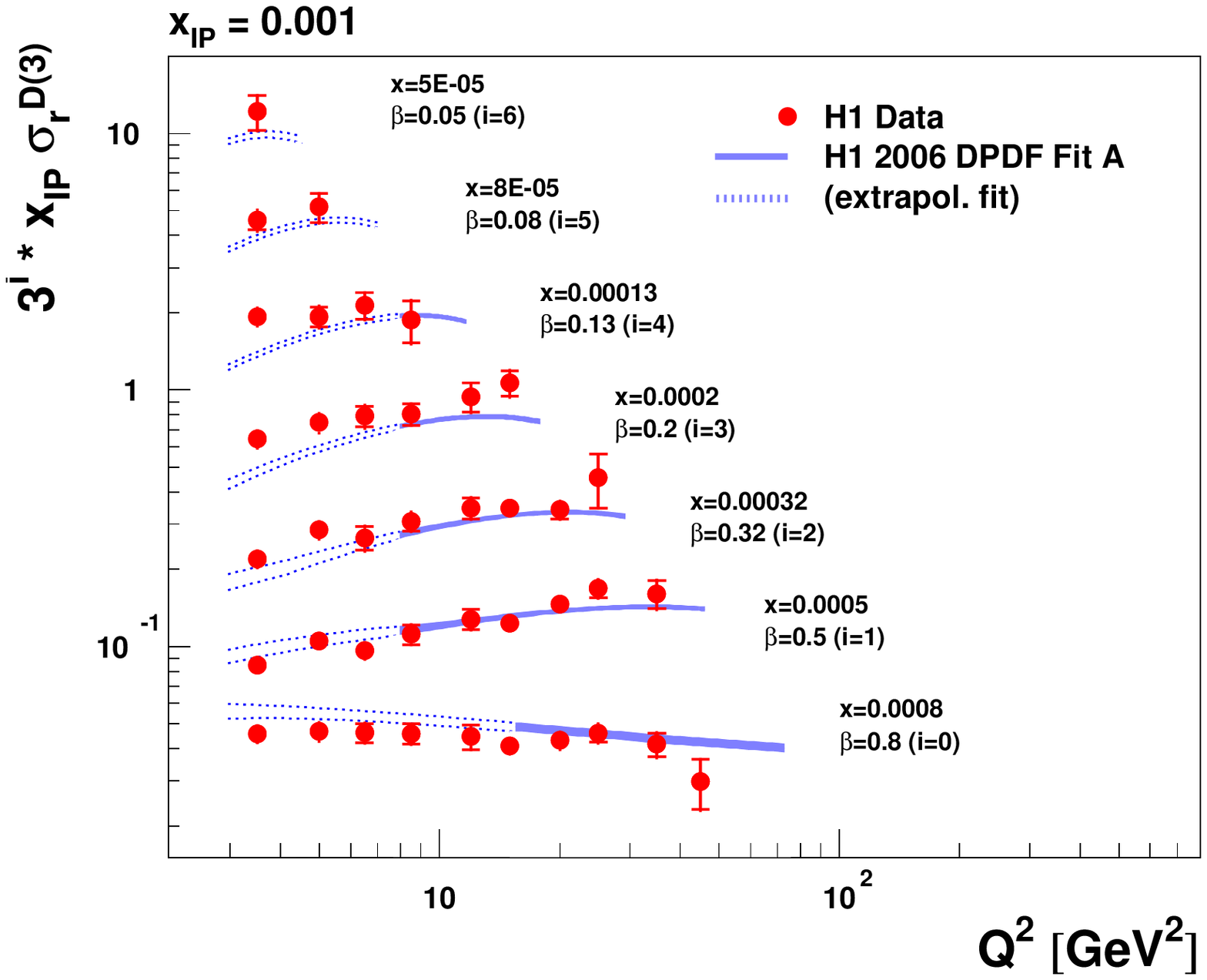,angle=90,width=8cm}
\epsfig{file=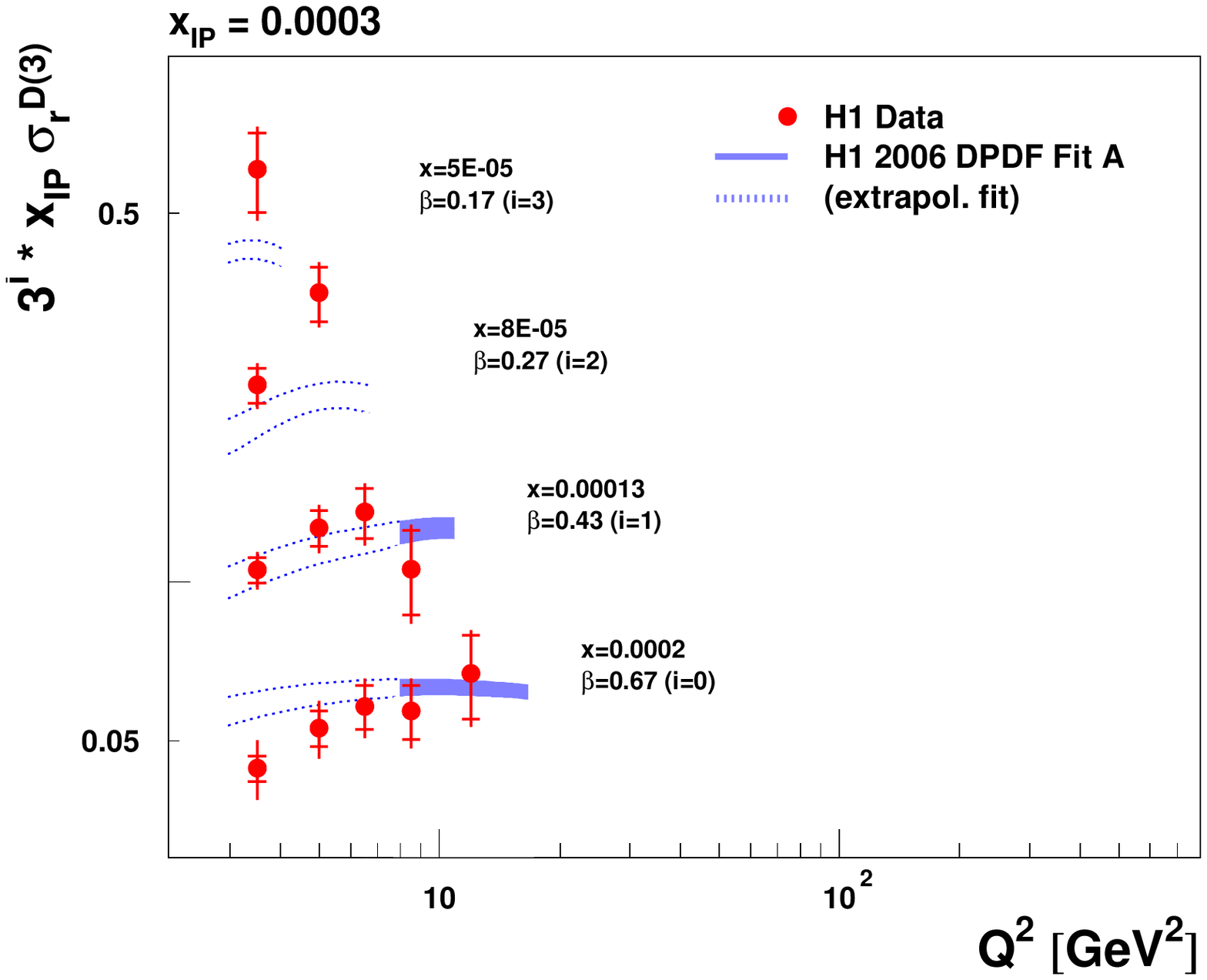,angle=90,width=8cm}
\caption{The diffractive structure function $\xpom F^{D(3)}_2$ (here denoted by $\xpom \sigma^{D(3)}_r$) at $\xpom = 0.0003$ and $0.001$ as a function of $Q^2$, for the $x$ values indicated; from H1.}
\label{f:fd3xp0003h1}
\end{center}
\end{figure}

\begin{figure}[p]
\begin{center}
\vfill
\epsfig{file=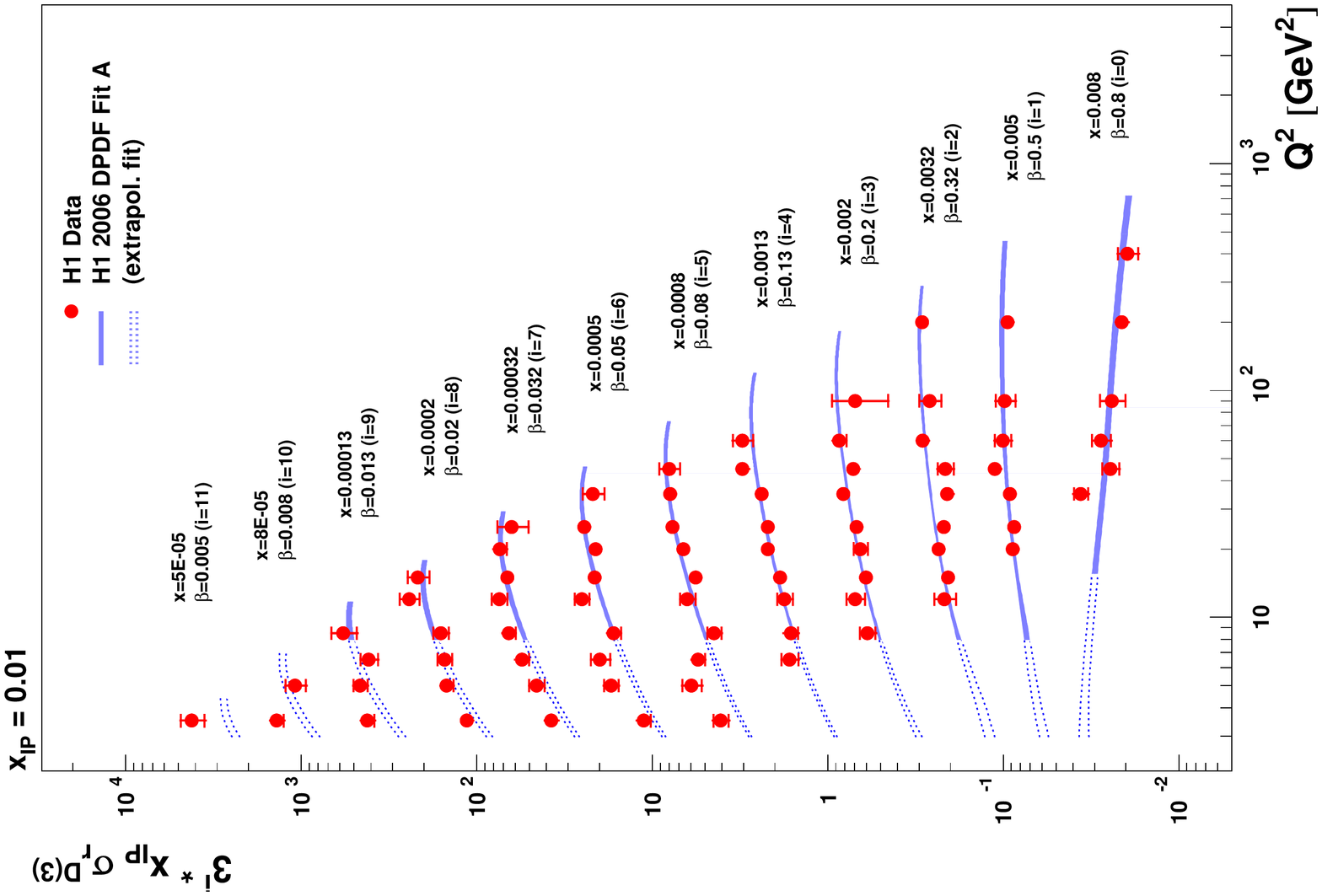,angle=90,width=10cm}
\epsfig{file=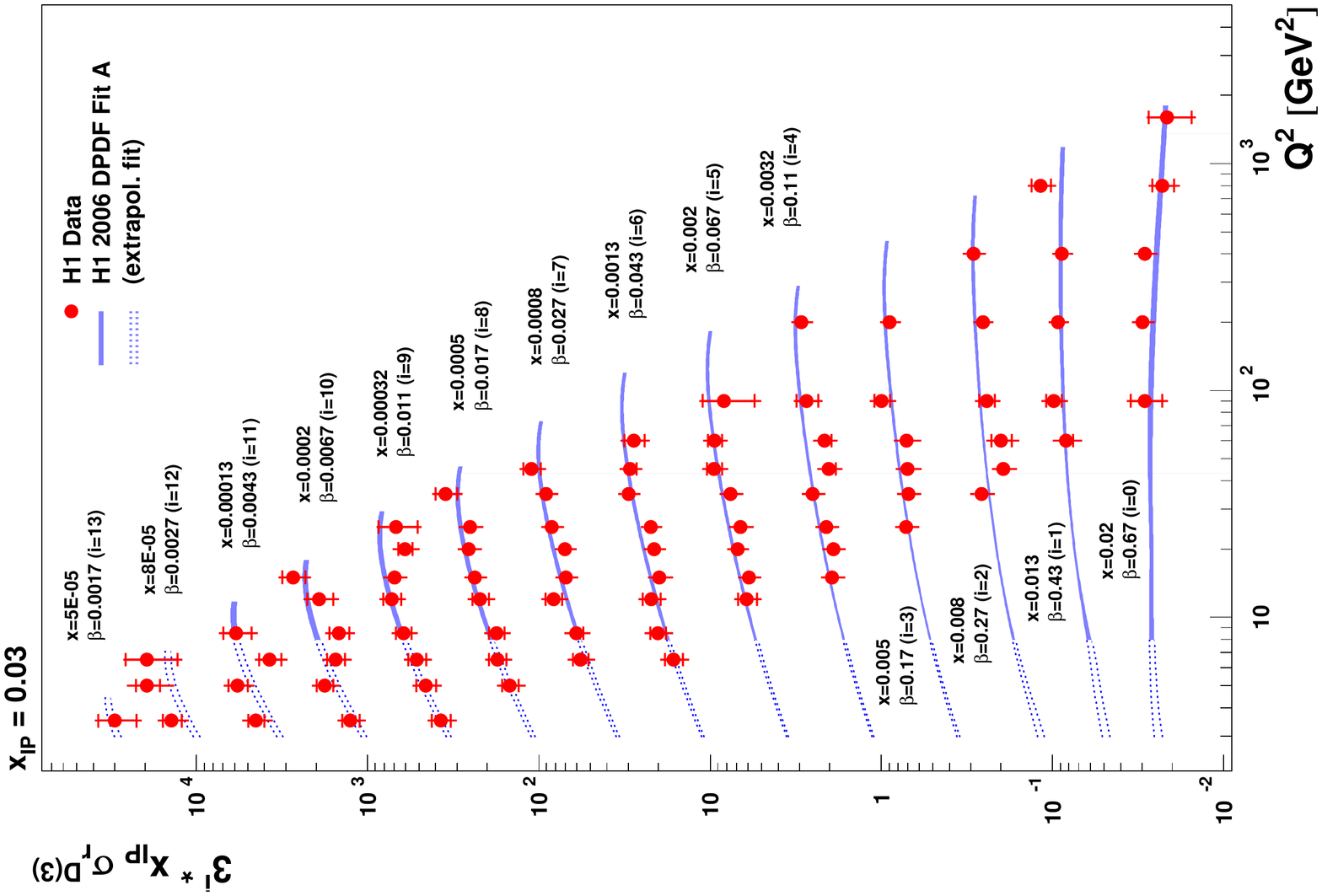,angle=90,width=10cm}
\epsfig{file=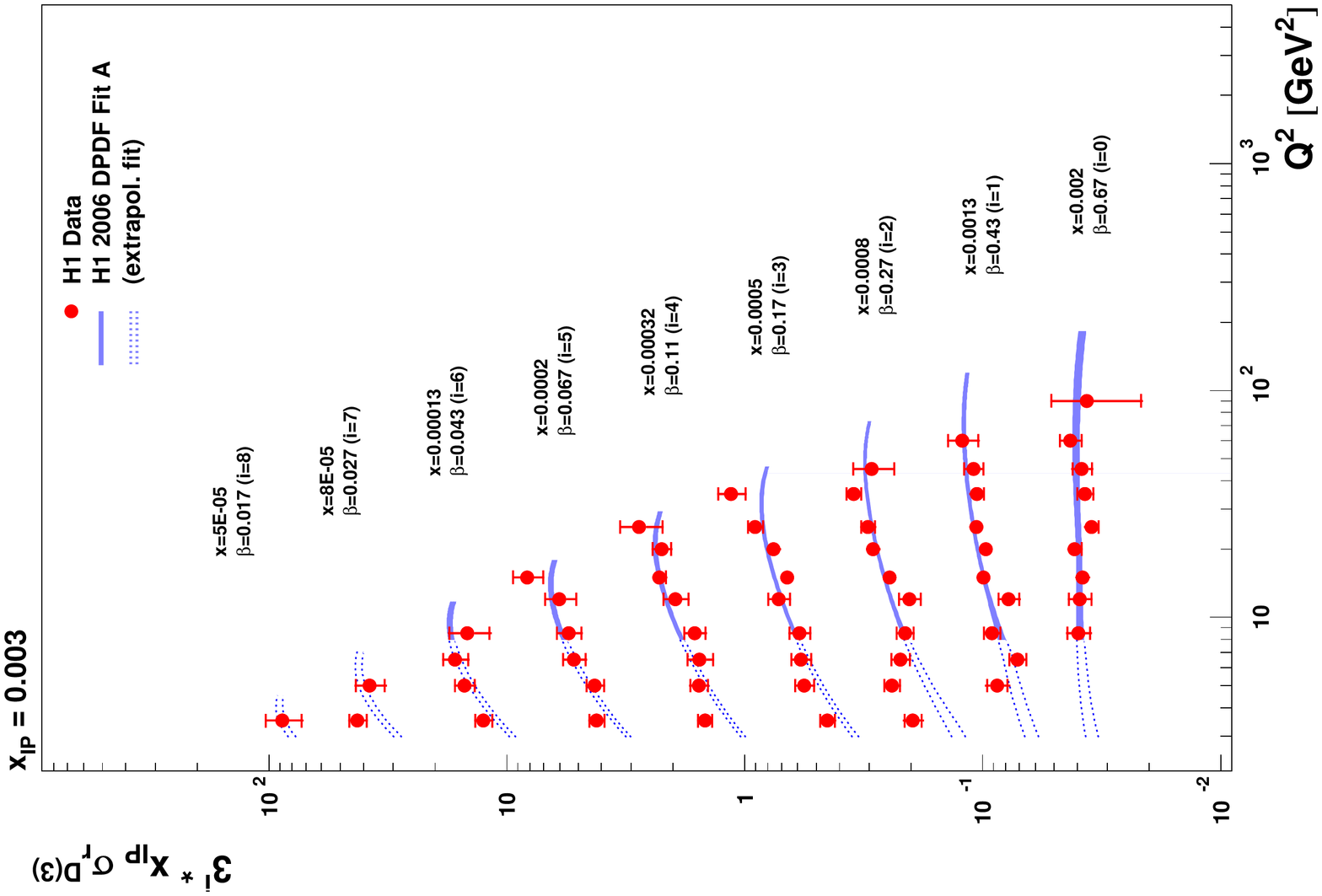,angle=90,width=10cm}
\caption{The diffractive structure function $\xpom F^{D(3)}_2$ (here denoted by $\xpom \sigma^{D(3)}_r$) at $\xpom = 0.003$, $0.01$ and $0.03$ as a function of $Q^2$, for the $x$ values indicated; from H1.}
\label{f:fd3xp003h1}
\end{center}
\end{figure}

In Figs.~\ref{f:fd3xp0003h1},~\ref{f:fd3xp003h1} 
the H1 measurements for $\xpom F^{D(3)}_2$ are shown as a function of $Q^2$ at fixed values of $\xpom$ and $\beta$. The dependence on $Q^2$ can be summarised as follows: for $\beta \le 0.5$ the diffractive structure function is rising with increasing $Q^2$, while for $\beta >0.8$ it is decreasing with increasing $Q^2$. The data are well reproduced by the QCD fit.

\subsection{Inclusive diffraction at $W = 45-220 $ GeV, $Q^2 = 2.7 - 320$ GeV$^2$}

Inclusive diffraction in deep-inelastic scattering has been measured by ZEUS over a wide space in $Q^2$, $W$ and $M_X$, making use of the forward-plug calorimeter (FPC). The FPC had a beamhole of only 3.15 cm radius for the passage of the beams. This limited the mass of the nucleonic system escaping undetected in the forward (= proton) direction to $M_N < 2.3$ GeV, on average.

\begin{figure}[p]
\begin{center}
\vfill
\includegraphics[totalheight=12cm]{{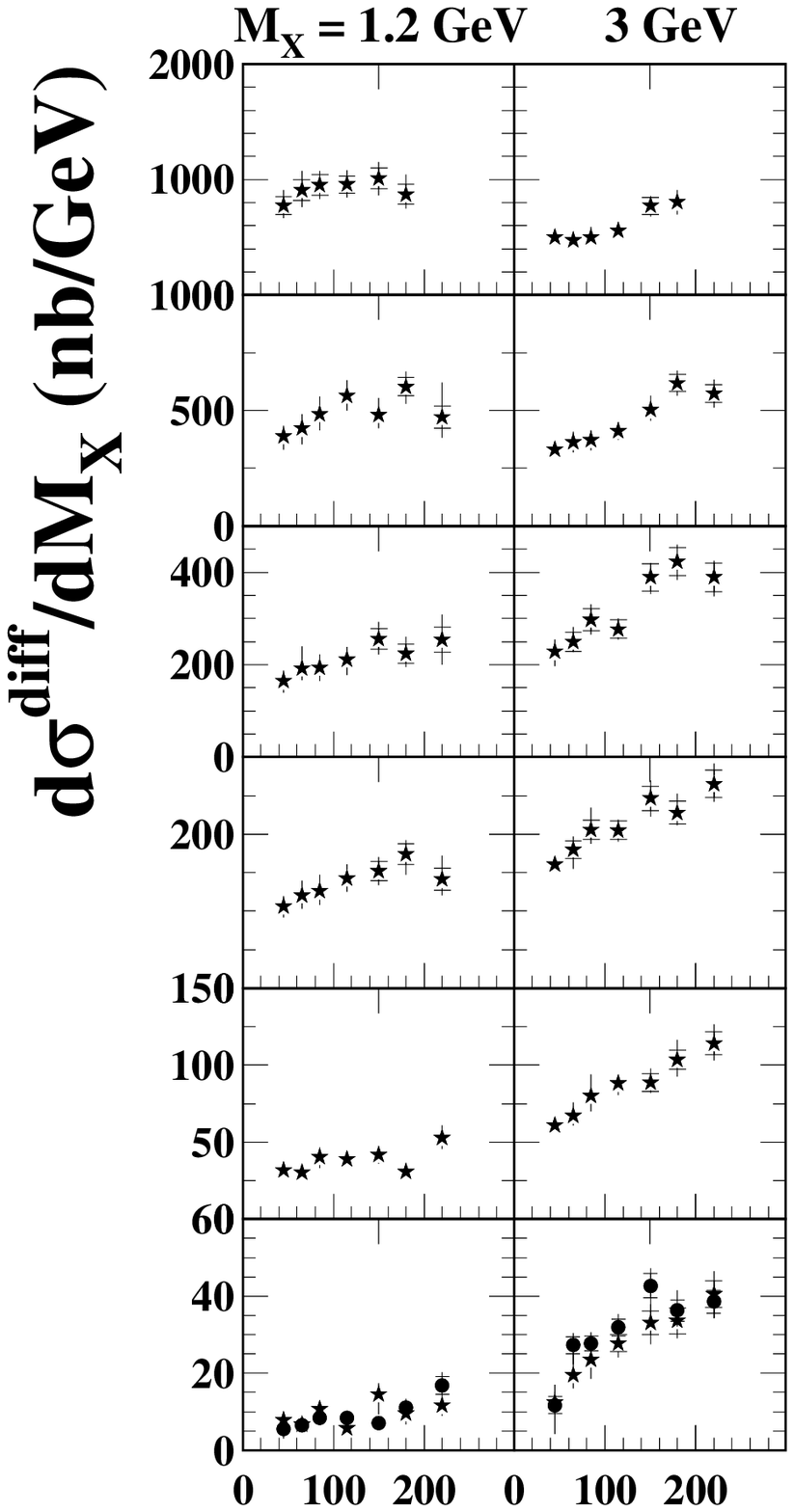}}
\includegraphics[totalheight=12cm]{{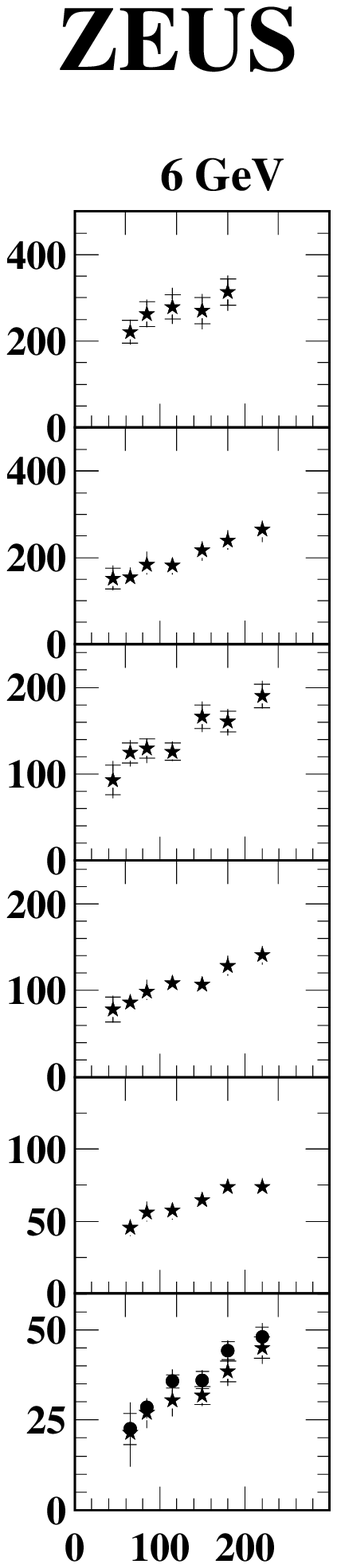}}
\includegraphics[totalheight=12cm]{{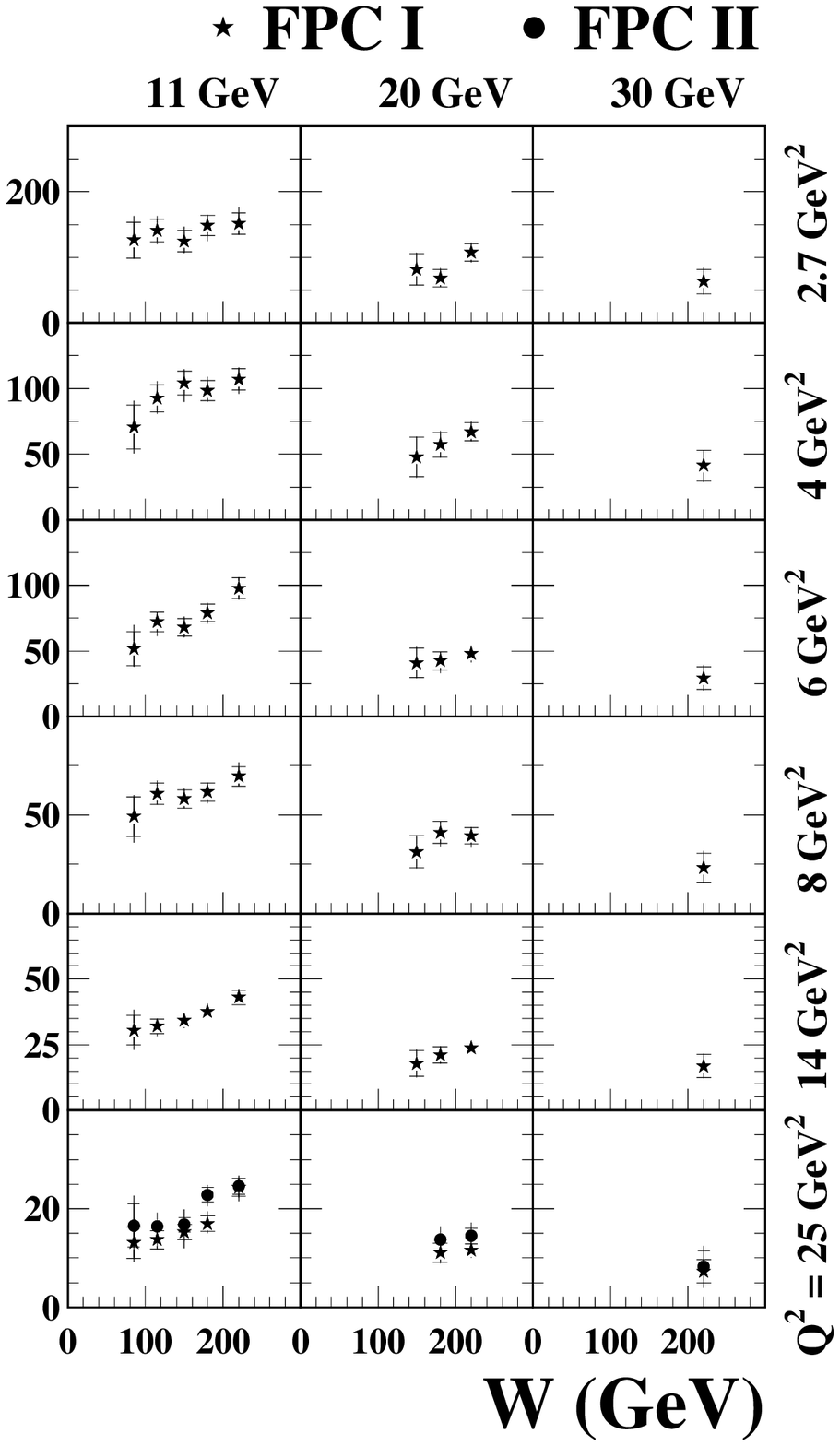}}
\caption{The differential cross section, $d\sigma^{\rm diff}_{\gamma^*p \to XN}/dM_X$, $M_N < 2.3$\GeV, as a function of $W$ for bins of $Q^2$, $M_X$, for FPC~I data (stars) and FPC~II data (dots), for $Q^2$ between 2.7 and 25\GeV$^2$; from ZEUS.}
\label{f:dsigdmxlh2.7.25}
\end{center}
\end{figure}

\begin{figure}[p]
\begin{center}
\vfill
\includegraphics[totalheight=14cm]{{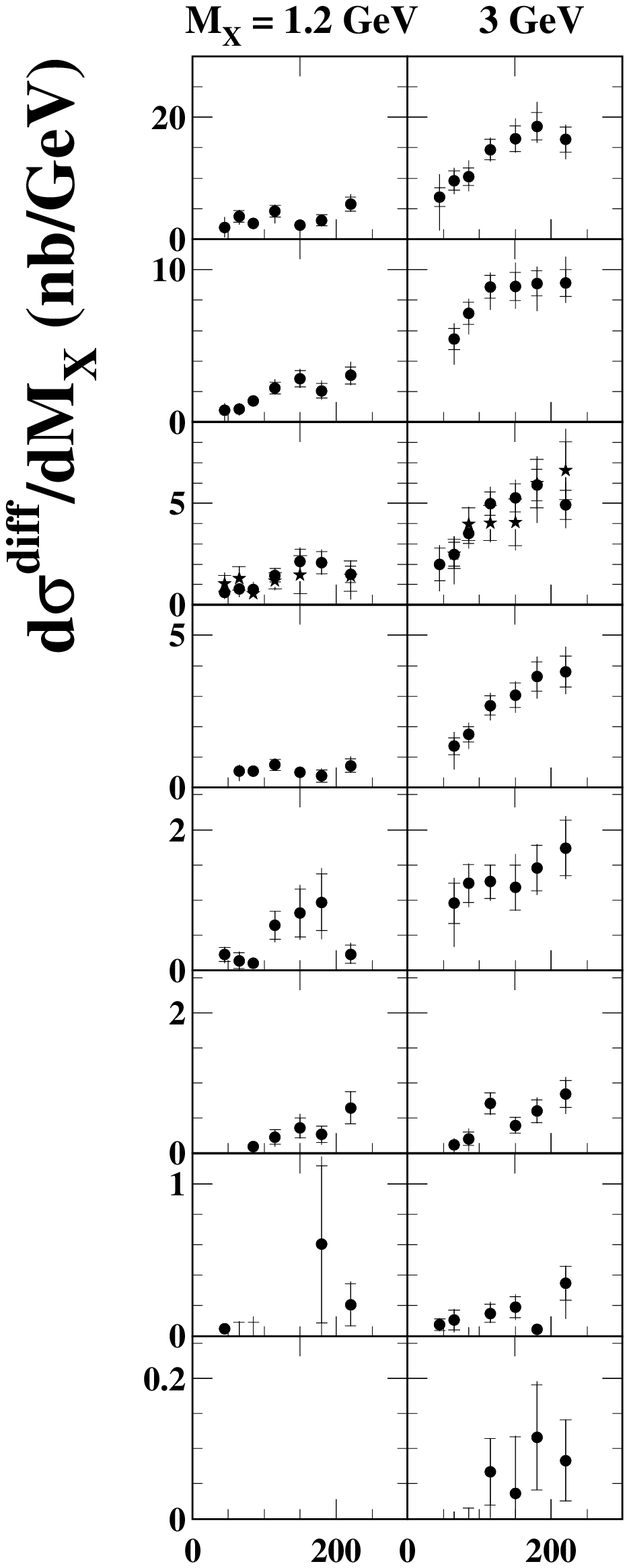}}
\includegraphics[totalheight=14cm]{{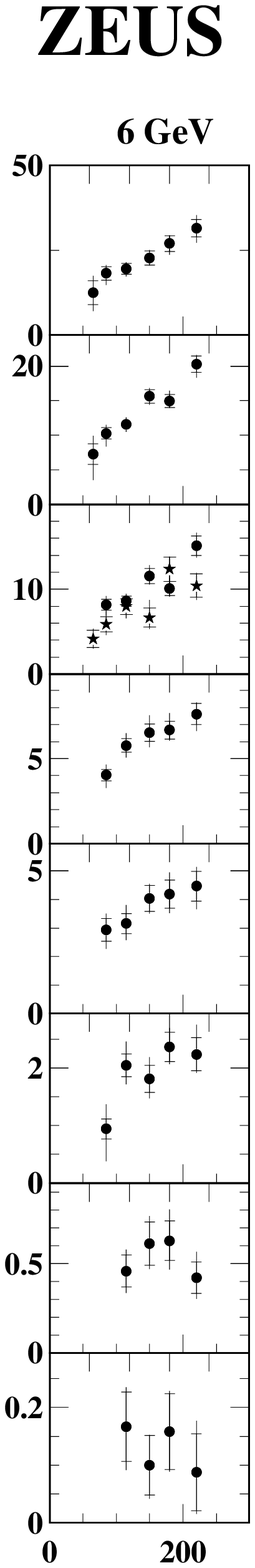}}
\includegraphics[totalheight=14cm]{{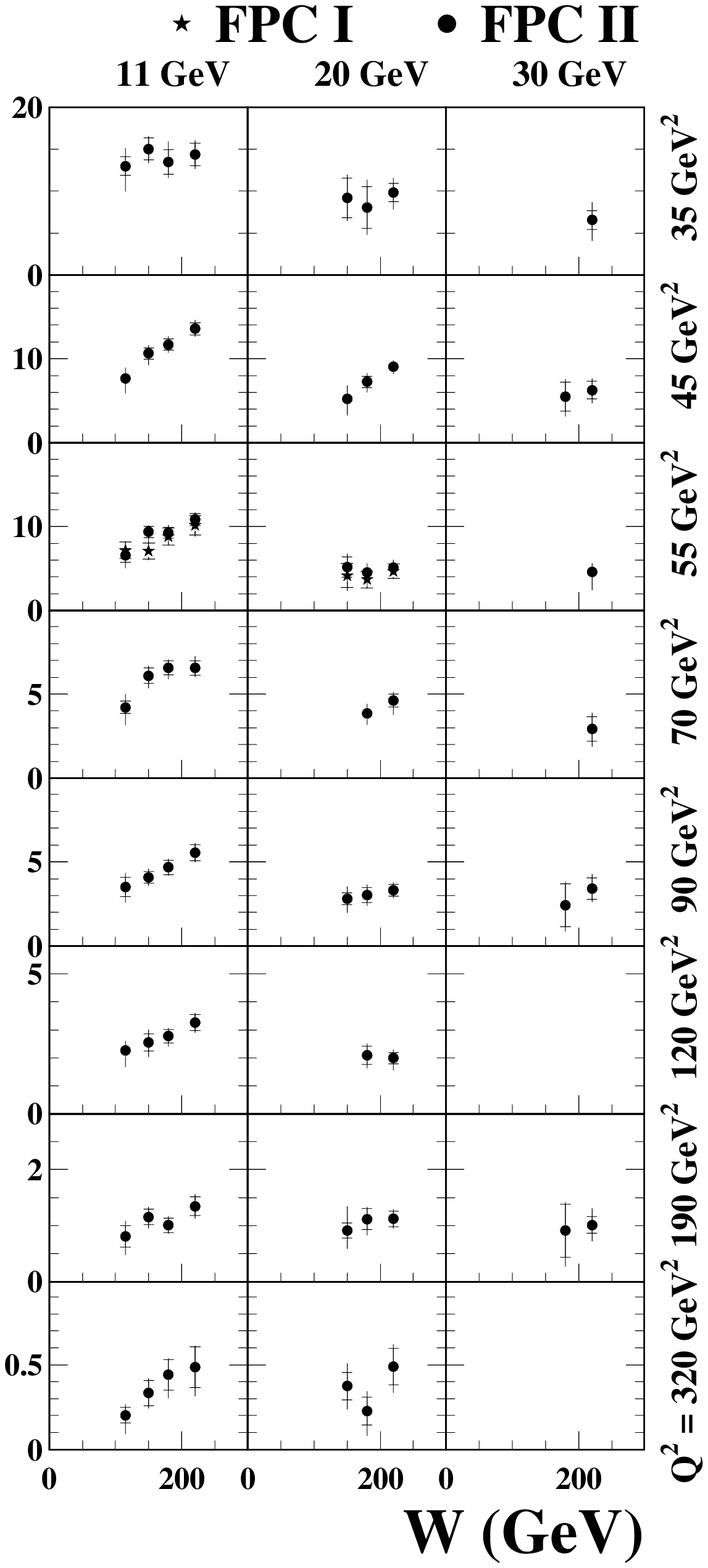}}
\caption{The differential cross section, $d\sigma^{\rm diff}_{\gamma^*p \to XN}/dM_X$, $M_N < 2.3$\GeV, as a function of $W$ for bins of $Q^2$, $M_X$, for FPC~I data (stars) and FPC~II data (dots), for $Q^2$ between 35 and 320\GeV$^2$; from ZEUS.}
\label{f:dsigdmxlh2.7.320}
\end{center}
\end{figure}

Data were collected in two periods for $Q^2 = 2.7 - 80$ GeV$^2$ and $Q^2 = 25 - 320$ GeV$^2$ with integrated luminosities of 4.2~\cite{np:b713:3} and 52.4 pb$^{-1}$~\cite{np:b800:1}, respectively. The diffractive cross section $d\sigma^{diff}/dM_X$ is shown in Figs.~\ref{f:dsigdmxlh2.7.25},~\ref{f:dsigdmxlh2.7.320} as a function of $W$ for the bins of $Q^2$ and $M_X$ indicated. In all $Q^2$, $M_X$ bins with sufficient $W$ coverage the diffractive cross section shows a strong rise with $W$. 

\begin{figure}[p]
\vfill
{\epsfig{file=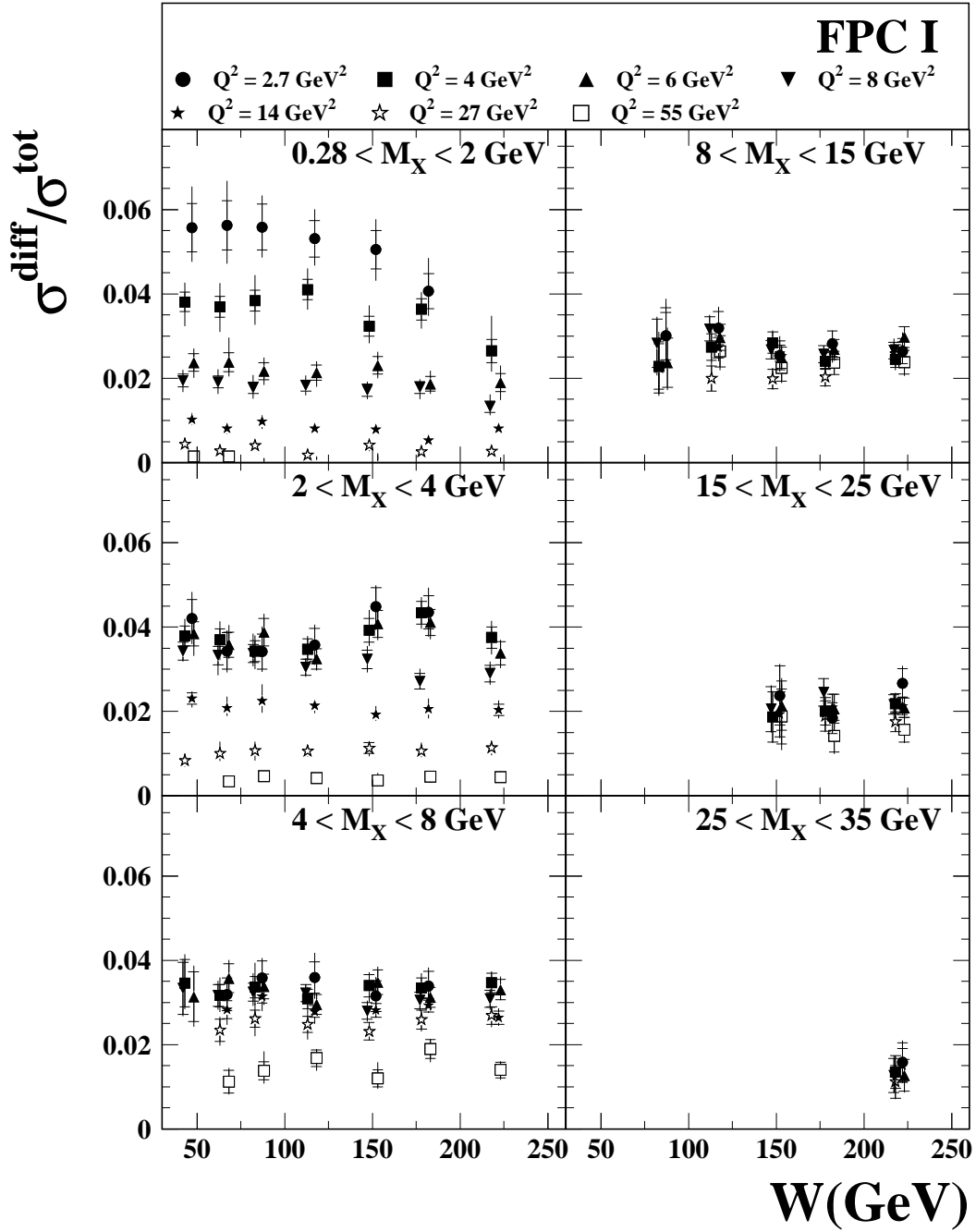,width=14.7cm,clip=}}
\caption{The ratio of the diffractive cross section, integrated over the $M_X$ intervals indicated, $\int^{M_b}_{M_a} dM_X d\sigma^{\rm diff}_{\gamma^* p \to XN, M_N < 2.3 {\rm GeV}}/dM_X$, to the total $\gamma ^{\ast}p$ cross section, $r^{\rm diff}_{\rm tot} = \sigma^{\rm diff}/\sigma^{\rm tot}_{\gamma^*p}$, as a function of $W$ for $Q^2$ between 2.7 and 55\GeV$^2$. The inner error bars show the statistical uncertainties and the full bars the statistical and systematic uncertainties added in quadrature; from ZEUS.}
\label{f:rdiftotl}
\vfill
\end{figure}

\begin{figure}[p]
\vfill
{\epsfig{file=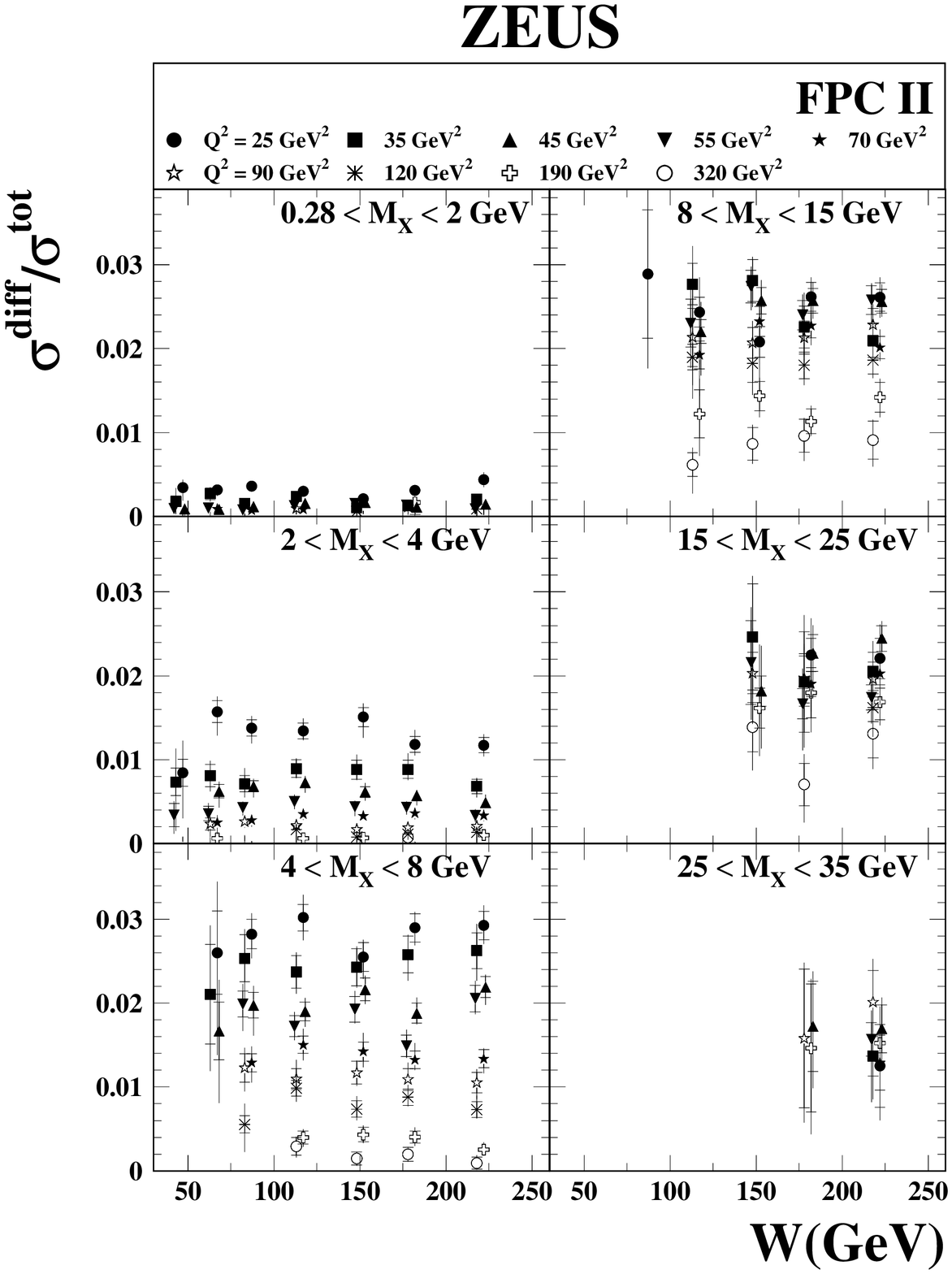,width=14.7cm,clip=}}
\caption{The ratio of the diffractive cross section, integrated over the $M_X$ intervals indicated, $\int^{M_b}_{M_a} dM_X d\sigma^{\rm diff}_{\gamma^* p \to XN, M_N < 2.3 {\rm GeV}}/dM_X$, to the total $\gamma ^{\ast}p$ cross section, $r^{\rm diff}_{\rm tot} = \sigma^{\rm diff}/\sigma^{\rm tot}_{\gamma^*p}$, as a function of $W$ for $Q^2$  between 25 and 320\GeV$^2$. The inner error bars show the statistical uncertainties and e full bars the statistical and systematic uncertainties added in quadrature; from ZEUS.}
\label{f:rdiftoth}
\vfill
\end{figure}

\begin{figure}[p]
\vfill
{\epsfig{file=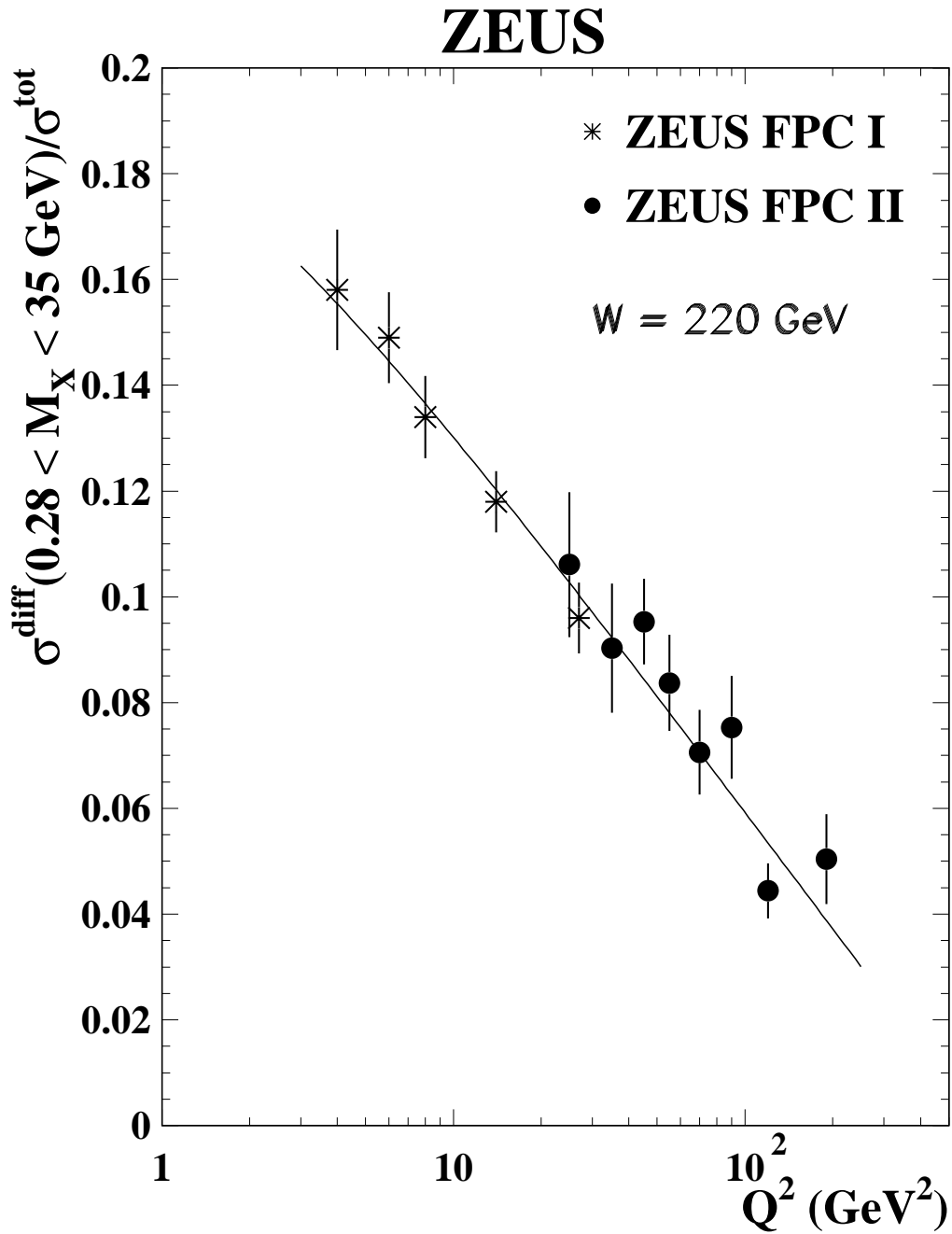,width=14.7cm,clip=}}
\caption{The ratio of the diffractive cross section, integrated over $0.28 < M_X < 35$\GeV, to the total $\gamma ^{\ast}p$ cross section, at $W = 220$\GeV as a function of $Q^2$. The line shows the result of fitting the data to the form $r  =  a -b \cdot \ln (1+Q^2)$; from ZEUS.}
\label{f:rdiftotlh220}
\vfill
\end{figure}

The ratio of the diffractive contribution to the total $\gamma^*p$ cross section is displayed in Figs.~\ref{f:rdiftotl},~\ref{f:rdiftoth} as a function of $W$, for different values of $Q^2$ and $M_X$. For fixed values of $Q^2$ the relative contribution of diffraction to the total $\gamma^*p$ cross section is approximately independent of $W$. It is substantial when $M^2_X > Q^2$. The ratio $r = \sigma^{\rm diff}(0.28 < M_X < 35 {\rm \; GeV}, M_N < 2.3 {\rm \; GeV})/\sigma^{\rm tot}$ is $15.8^{+1.1}_{-1.0}\%$ at $Q^2 = 4$ GeV$^2$, decreasing to $5.0 \pm 0.9 \%$ at $Q^2 =190$ GeV$^2$, see Fig.~\ref{f:rdiftotlh220}.
Since the optical theorem relates the forward amplitude for elastic scattering to the total cross section - which behaves, say, proportional to $W^{\delta}$ - one would naively expect that the diffractive cross section is proportional to $W^{2\delta}$. However, the measured ratio of the diffractive contribution to the total $\gamma^*p$ cross section is approximately independent of $W$, as shown in Figs.~\ref{f:rdiftotl},~\ref{f:rdiftoth}. This apparent contradiction is understood as the result of a strong growth of the gluon density (``color glass condensate'') such that gluon-gluon interactions become important and damp the rise of the diffractive cross section~\cite{epj:c43:3}. 

\begin{figure}[p]
\centerline
{\epsfig{file=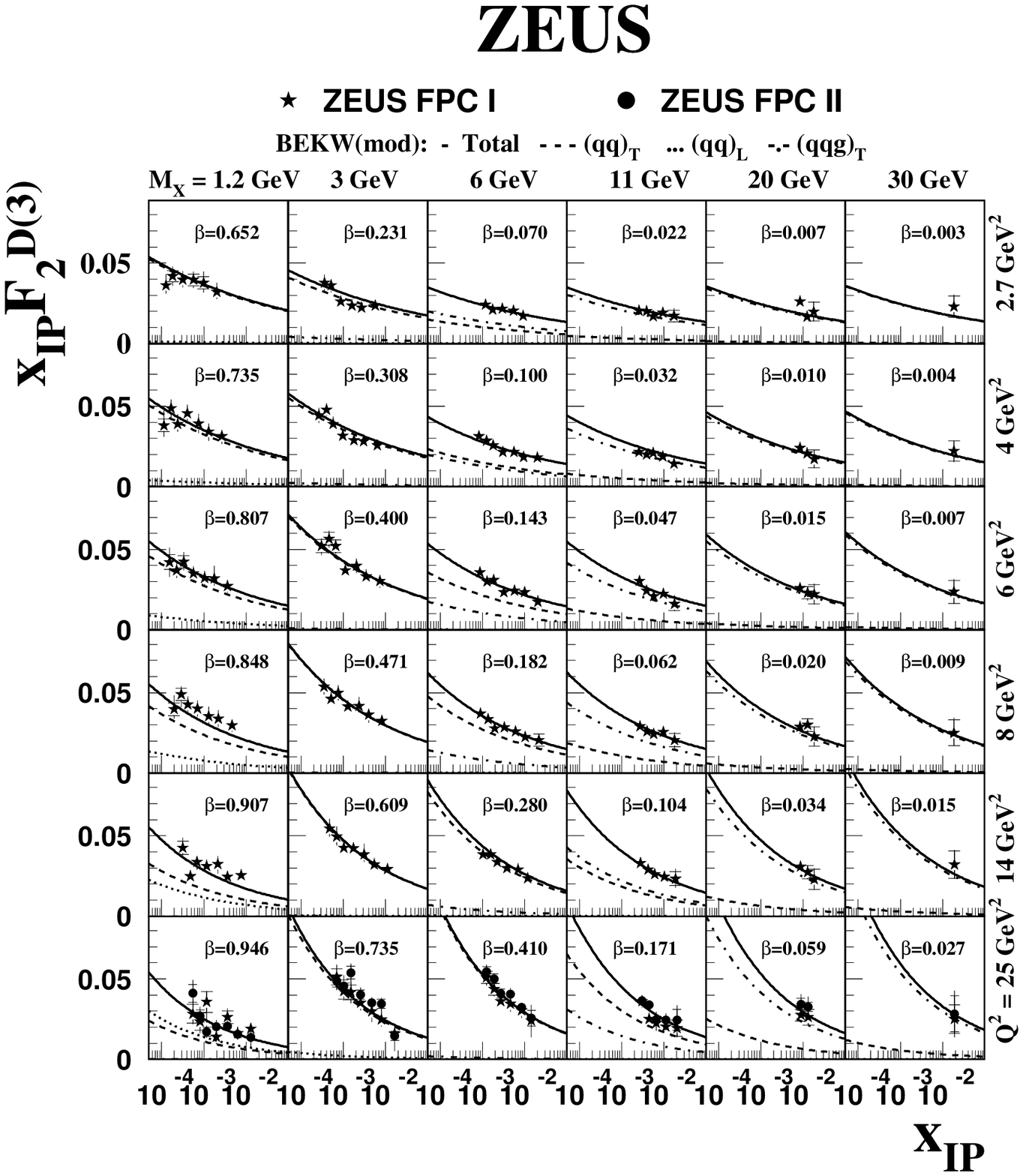,width=15cm}}
\caption{The diffractive structure function of the proton multiplied by $\xpom$, $\xpom F^{D(3)}_2$, for $\gamma^*p \to XN$, $M_N < 2.3$\GeV, as a function of $\xpom$ for different regions of $\beta$ and $Q^2 \le 25$\GeV$^2$. The curves show the results of the BEKW(mod) fit for the contributions from $(q \overline{q})$ for transverse (dashed) and longitudinal photons (dotted) and for the $(q \overline{q}g)$ contribution for transverse photons (dashed-dotted) together with the sum of all contributions (solid); from ZEUS.}
\label{f:f2d3vsxplh1}
\end{figure}

\begin{figure}[p]
\vspace*{-3.8cm}
\centerline
{\epsfig{file=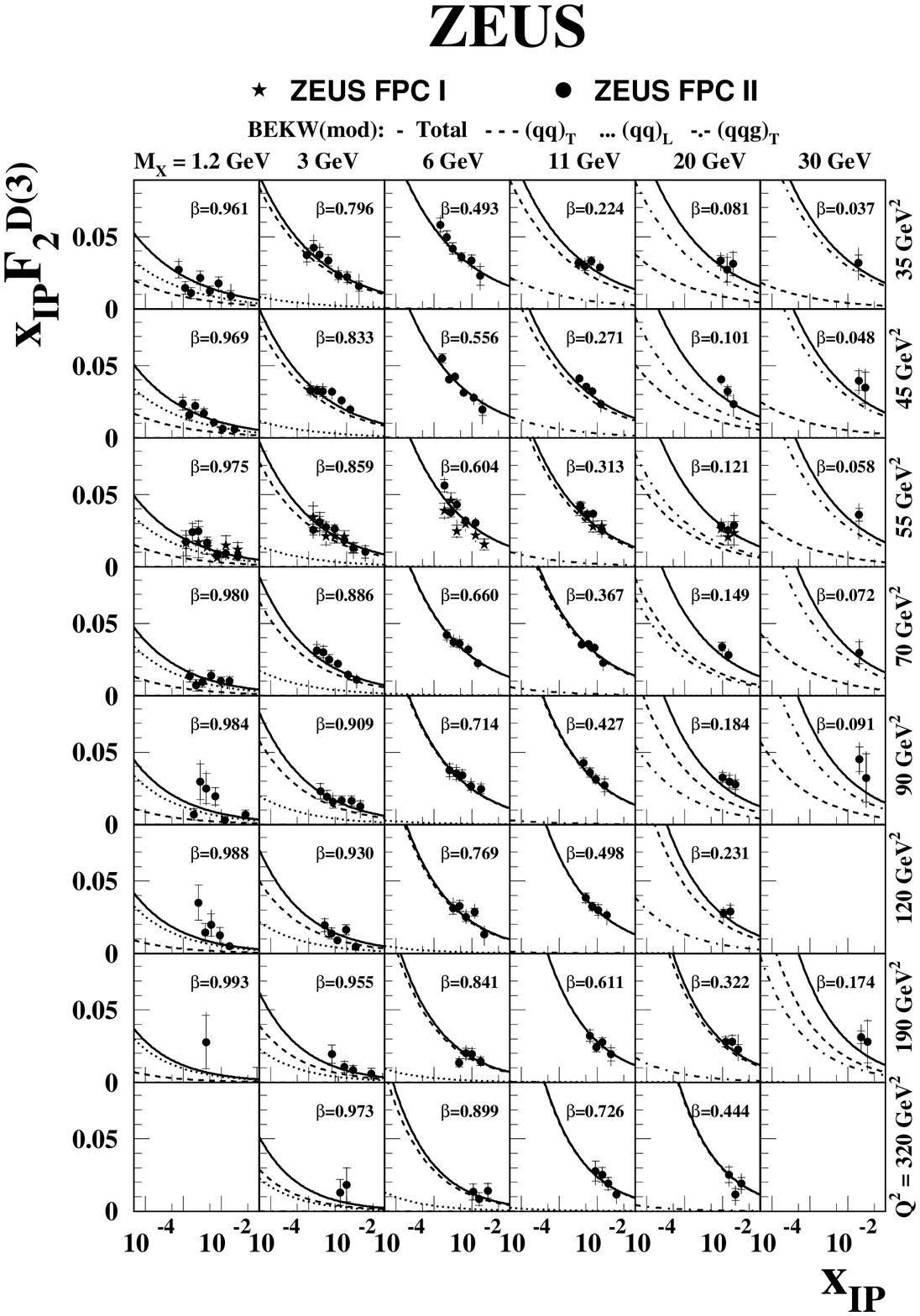,width=15cm}}
\caption{The diffractive structure function of the proton multiplied by $\xpom$, $\xpom F^{D(3)}_2$, for $\gamma^*p \to XN$, $M_N < 2.3$\GeV, as a function of $\xpom$ for different regions of $\beta$ and $Q^2 \ge 35$\GeV$^2$. The curves show the results of the BEKW(mod) fit for the contributions from $(q \overline{q})$ for transverse (dashed) and longitudinal photons (dotted) and for the $(q \overline{q}g)$ contribution for transverse photons (dashed-dotted) together with the sum of all contributions (solid); from ZEUS.} 
\label{f:f2d3vsxplh2}
\end{figure}

The diffractive structure function $\xpom F^{D(3)}_2$ for $\gamma^*p \to XN$, $M_N < 2.3{\rm \; GeV}$, is shown in Figs.~\ref{f:f2d3vsxplh1},~\ref{f:f2d3vsxplh2} as a function of $\xpom$ for fixed $Q^2, M_X$, or equivqalently for fixed $Q^2, \beta$: $\xpom F^{D(3)}_2$ rises approximately proportional to $\ln 1/\xpom$ as $\xpom \to 0$, reflecting the rise of the diffractive cross section $d\sigma^{\rm diff}/dM_X$ with increasing $W$. The rise is observed for most $Q^2$ values between 2.7 and 320 GeV$^2$.

\begin{figure}[p]
\centerline
{\epsfig{file=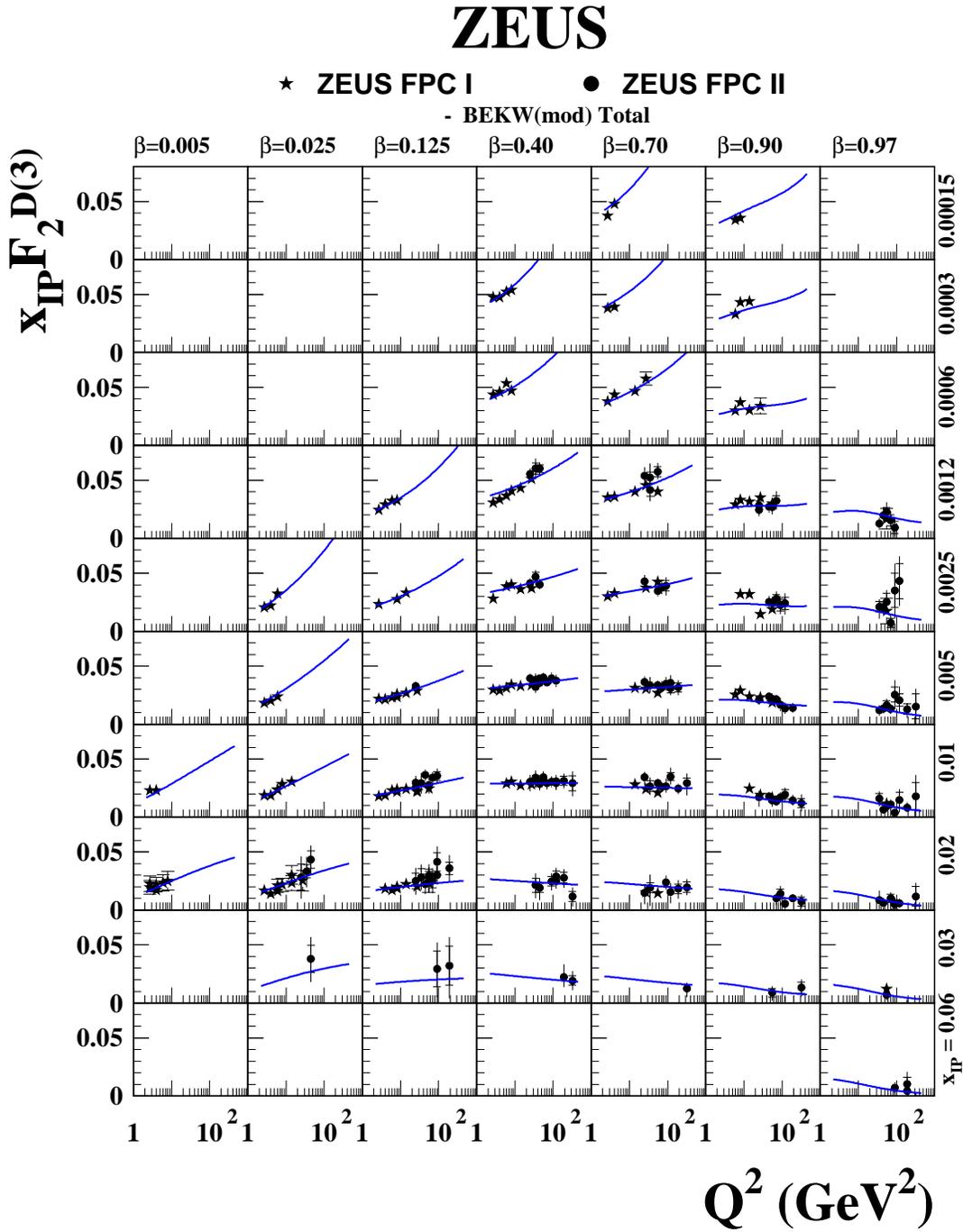,width=16.cm}}
\vspace*{-0.5cm}
\caption{The diffractive structure function of the proton multiplied by $\xpom$, $\xpom F^{D(3)}_2$, as a function of $Q^2$ for different regions of $\beta$ and $\xpom$. The curves show the result of the BEKW(mod) fit to the data; from ZEUS.}
\label{f:f2d3vsq2lh}
\end{figure}

The $Q^2$ dependence of $\xpom F^{D(3)}_2$ for fixed $\beta$ and $\xpom$ is shown in Fig.~\ref{f:f2d3vsq2lh}. The data are dominated by positive scaling violations for $\xpom \beta  =  x < 10^{-3}$, by negative scaling violations for $x \ge 5 \cdot 10^{-3}$, and by constancy in between. The data contradict the assumption of Regge factorisation, namely that $\xpom F^{D(3)}_2(\beta,\xpom,Q^2)$ factorises into a term that depends only on $\xpom$, and a second term that depends only on  $\beta$ and $Q^2$. 

\subsection{Leading proton production and the contribution from diffraction}
Deep inelastic diffractive scattering has also been studied in measurements where the scattered proton was detetced in a forward spectrometer. The first such measurement has been reported by ZEUS~\cite{epj:c1:81}. Figure~\ref{f:disdsdtz} shows $d\sigma/d|t|$ as a function of the four-momentum transfer squared $t$, for the range $5<Q^2<20$ GeV$^2$, $50<W<270$ GeV and $0.15 < \beta < 0.5$. A fit of the form $d\sigma/d|t| = A\cdot e^{-b|t|}$ yielded for the slope $b = 7.2 \pm 1.1({\rm{stat}})^{\rm{+0.7}}_{\rm{-0.9}}({\rm syst})$ GeV$^{-2}$. Further results on $b$ from~\cite{epj:c38:43,epj:c48:749} are also presented in Figure~\ref{f:disdsdtz}. For the $Q^2$-values selected, the transverse extension of the virtual photon is negligible compared to the radius of the proton.

The final measurement~\cite{np:b816:1}, based on an integrated luminosity of 32.6 pb$^{-1}$, provided results for the kinematic region $Q^2 > 2$ GeV$^2$, $W = 40-240$ GeV, $M_X > 2$ GeV. The diffractive peak covered the region $x_L >0.98$, where $x_L$ is the fraction of the momentum of the incoming proton carried by the scattered proton.

\begin{figure}
\begin{center}
\centerline
{\epsfig{file=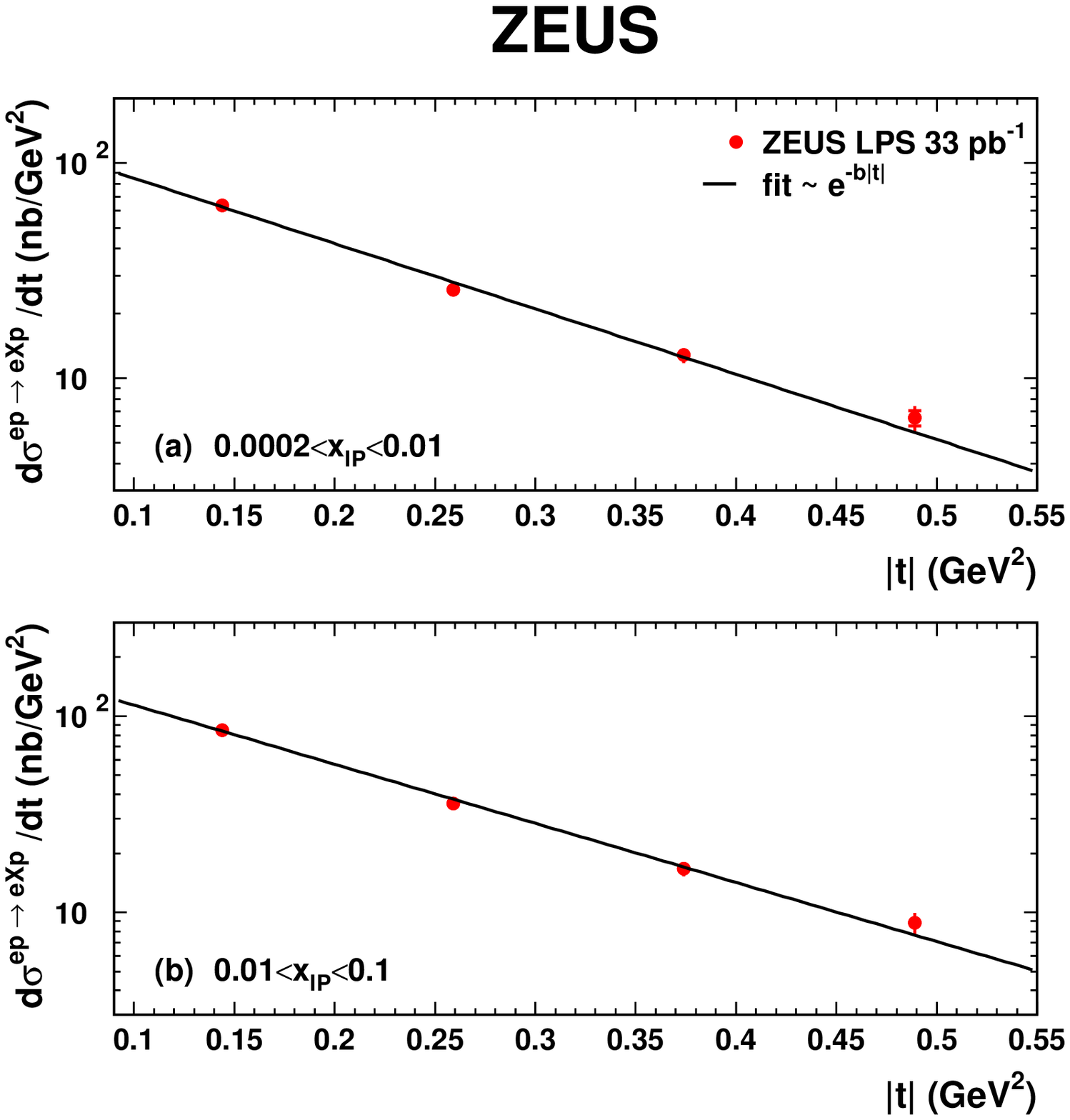,width=15cm}}
\caption{The differential cross section $d\sigma^{ep \to eXp}/d|t|$ for (a) $0.0002,\xpom<0.01$ and (b) $0.01 \xpom <0.1$; from ZEUS.}
\label{f:dsigdtepexp}
\end{center}
\end{figure}

The differential cross section $d\sigma^{ep \to eXp}/d|t|$ is shown in Fig.~\ref{f:dsigdtepexp} as a function of $|t|$ for $\xpom = 0.0002 - 0.01$ and $\xpom = 0.01 - 0.1$, respectively. The cross section is falling exponentially with $|t|$, $d\sigma^{ep \to eXp}/d|t| \propto e^{-b|t|}$, where $b = 7.0 \pm 0.3$ GeV$^{-2}$ for $\xpom = 0.0002 - 0.01$ and $b = 6.9 \pm 0.3$ GeV$^{-2}$ for $\xpom = 0.01 - 0.1$. Within errors the value of $b$ is independent of $\xpom$, $Q^2$ and $M_X$.  

The cross section for $ep \to eXp$ was also studied in terms of the azimuthal angle $\Phi$ between the positron scattering angle and the proton scattering plane. Within errors no significant dependence on $\Phi$ was observed for $\xpom = 0.0002 - 0.01$ and $\xpom = 0.01 - 0.1$, respectively. This is in contrast to $\rho^0$ production by deep-inelastic scattering, $ep \to e \rho^0 p$~\cite{pmc:a1:6}, where a substantial dependence on $\Phi$ had been observed, viz. $\frac{d\sigma}{d\Phi} \propto 1 +A_{LT} \cos \Phi + A_{TT} cos 2\Phi$, $A_{LT} = \sqrt{2\epsilon(1+\epsilon)}(r^5_{00} + 2r^5_{11}) = -0.256 \pm 0.030^{+0.032}_{-0.022}$. 

\begin{figure}
\begin{center}
\centerline
{\hspace*{1.0cm}\epsfig{file=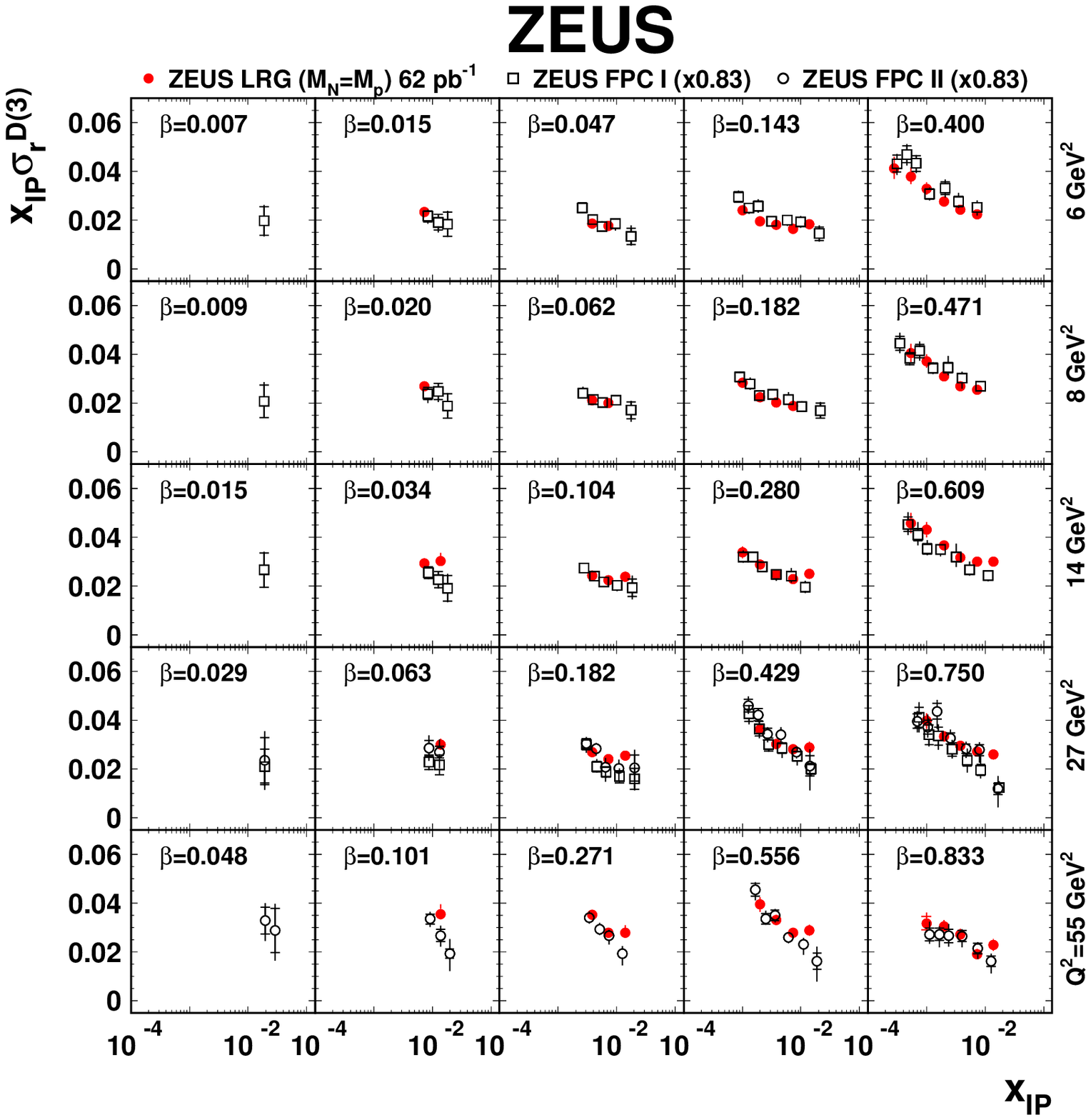,width=16cm}}
\caption{The diffractive structure function of the proton multiplied by $\xpom$, $\xpom F^{D(3)}_2$, as function of $\xpom$ for $Q^2 = 6 - 55$ {\rm GeV$^2$} (top) and $Q^2 = 55 - 190$ {\rm GeV$^2$} (bottom), for the $\beta$ values indicated; from ZEUS.}
\label{f:xpomsigd3loq}
\end{center}
\end{figure}

\begin{figure}
\begin{center}
\centerline
{\hspace*{1.0cm}\epsfig{file=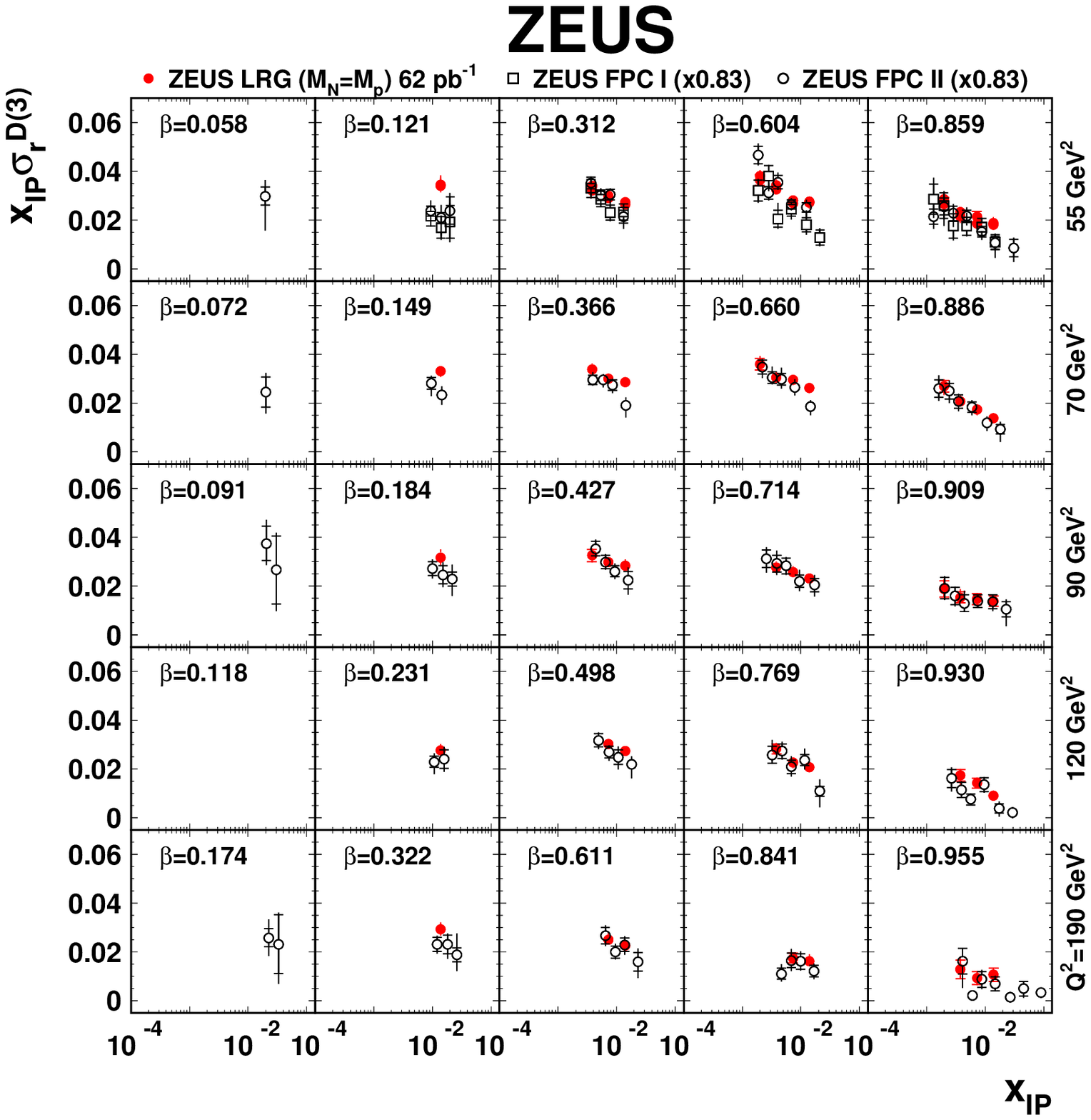,width=16cm}}
\caption{The diffractive structure function of the proton multiplied by $\xpom$, $\xpom F^{D(3)}_2$, as function of $\xpom$ for $Q^2 = 6 - 55$ {\rm GeV$^2$} (top) and $Q^2 = 55 - 190$ {\rm GeV$^2$} (bottom), for the $\beta$ values indicated; from ZEUS.}
\label{f:xpomsigd3hiq}
\end{center}
\end{figure}

The structure function $\xpom F^{D(3)}_2$, here denoted by $\xpom \sigma_r^{D(3)}$, is shown in Figs.~\ref{f:xpomsigd3loq},~\ref{f:xpomsigd3hiq} 
and compared with the FPCI,II measurements discussed before. The $\xpom F^{D(3)}_2$ data from the FPCI,II measurements have been scaled down by a factor of 0.83 to account for the extra contribution from $p \to N$ dissociation with $M_N < 2.3$ GeV. Good agreement between the FPC and LPS measurements is observed.

\subsubsection{Comparison with the BEKW parametrisation}

Further insight into the behavior of the diffractive structure function can be gained with the help of the BEKW parametrisation~\cite{epj:c7:443} which is guided by QCD. It considers the case where the incoming virtual photon fluctuates into a $q\overline{q}$ or $q\overline{q}g$ dipole, which in turn interacts with the target proton via the exchange of a two-gluon system, the Pomeron. The contributions from the transitions: transverse photon to $q\overline{q}$ and $q\overline{q}g$, respectively, and longitudinal photon to $q\overline{q}$, are taken into account. The $\beta$ spectrum and the scaling behaviour in $Q^2$ are derived in the non-perturbative limit from the wave functions of the incoming transverse ($T$) and longitudinal ($L$) photons on the light cone. The $\xpom$ dependence of the cross section is not predicted by BEKW and has to be determined by experiment. The BEKW parametrisation reads as follows: 

\begin{eqnarray}
\xpom F^{D(3)}_2(\beta,\xpom,Q^2) & = & c_T \cdot F^T_{q\overline{q}} + c_L \cdot F^L_{q\overline{q}} + c_g \cdot F^T_{q\overline{q}g},
\label{eq:bekw}
\end{eqnarray}
where 
\begin{eqnarray}
F^T_{q\overline{q}} & = & \left (\frac{x_0}{\xpom} \right )^{n_T(Q^2)}\cdot \beta(1 - \beta), \\
\label{eq:bekwqqT}
F^L_{q\overline{q}} & = & \left (\frac{x_0}{\xpom} \right )^{n_L(Q^2)} \cdot \frac{Q^2_0}{Q^2+Q^2_0} \cdot \left [\ln \left(\frac{7}{4} + \frac{Q^2}{4 \beta Q^2_0} \right) \right ]^2 \cdot \beta^3 (1 - 2\beta)^2, \\
\label{eq:bekwqqL}
F^T_{q\overline{q}g} & = & \left (\frac{x_0}{\xpom} \right )^{n_g(Q^2)} \cdot \ln \left(1+\frac{Q^2}{Q^2_0}\right)\cdot (1-\beta)^{\gamma}.
\label{eq:bekwqqg}
\end{eqnarray} 
In the ZEUS analysis, the powers $n_{L,T,g}$ were assumed to be the same for the contributions from $T,L$ and $g$, and of the form $n(Q^2) = n_1\cdot \ln(1+\frac{Q^2}{Q^2_0})$ with $Q^2_0 = 0.4$ GeV$^2$. 

The contribution from longitudinal photons coupling to $q \qbar$  is limited to $\beta$ values close to unity. The $q \overline{q}$ contribution from transverse photons is expected to have a broad maximum around $\beta = 0.5$, while the $q \overline{q} g$ contribution becomes important at small $\beta$, provided the power $\gamma$ is large. 

The original BEKW parametrisation also includes a higher-twist term for $q \overline{q}$ produced by transverse photons. Since the data were insensitive to this term, it was neglected. A fit of this modified BEKW form (BEKW(mod)) to the ZEUS data~\cite{np:b713:3,np:b800:1} gave an excellent description of the data with the following result for the five parameters left free: $c_T = 0.118 \pm 0.002$, $c_L = 0.087 \pm 0.005$, $c_g = 0.0090 \pm 0.0003$, $n_1 = 0.062 \pm 0.002$ and $\gamma = 8.22 \pm 0.46$.

The curves in Figs.~\ref{f:f2d3vsxplh1},~\ref{f:f2d3vsxplh2} show the individual contributions from the $(q \qbar)_T$, $(q \qbar)_L$ and $(q \qbar g)_T$ terms, and their sum. The  contribution from $(q \qbar)_L$ is important for $\beta \ge 0.96$;  the $(q \qbar)_T$ contribution is dominant for $0.14 < \beta < 0.9$, while gluon emission, $(q \qbar g)_T$, is important at small $\beta < 0.05 - 0.1$. The curves in Fig.~\ref{f:f2d3vsq2lh} show the sum of the contributions.

\begin{figure}
\begin{center}
\includegraphics[angle=90,totalheight=9.5cm]{{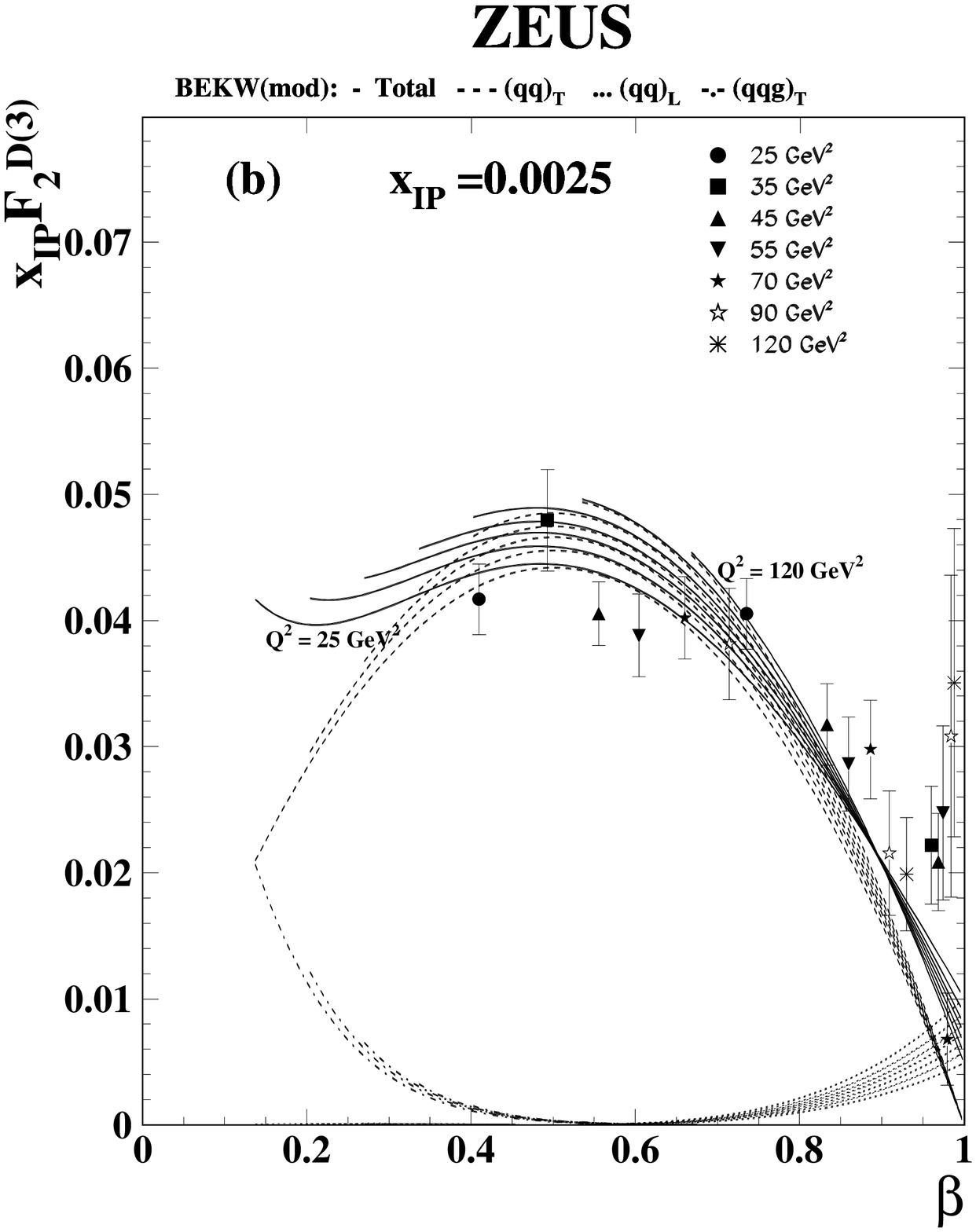}}
\includegraphics[angle=90,totalheight=9.5cm]{{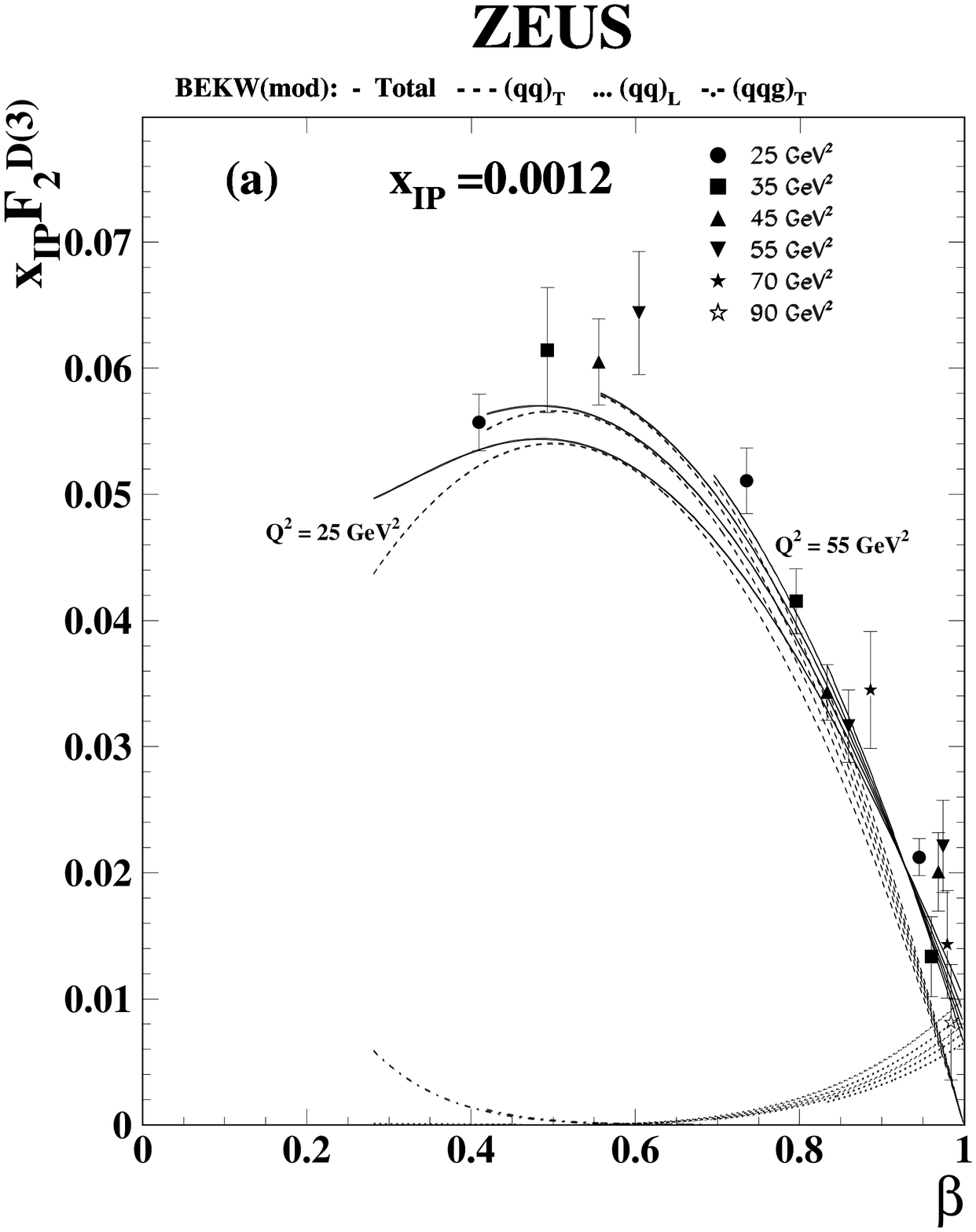}}
\caption{The diffractive structure function of the proton multiplied by
   $\xpom$, $\xpom F^{D(3)}_2$, as a function of $\beta$ for the $Q^2$ values indicated, at fixed (a) $\xpom = 0.0012$ and (b) $\xpom = 0.0025$. The curves show the results of the BEKW(mod) fit for the contributions from $(q \overline{q})$ for transverse (dashed) and longitudinal photons (dotted) and for the $(q \overline{q}g)$ contribution for transverse photons (dashed-dotted) together with the sum of all contributions (solid), for the $\beta$-region studied for diffractive scattering; from ZEUS.}
\label{f:f2d3vsbetah12}
\end{center}
\end{figure}

\begin{figure}
\begin{center}
\includegraphics[angle=90,totalheight=9.5cm]{{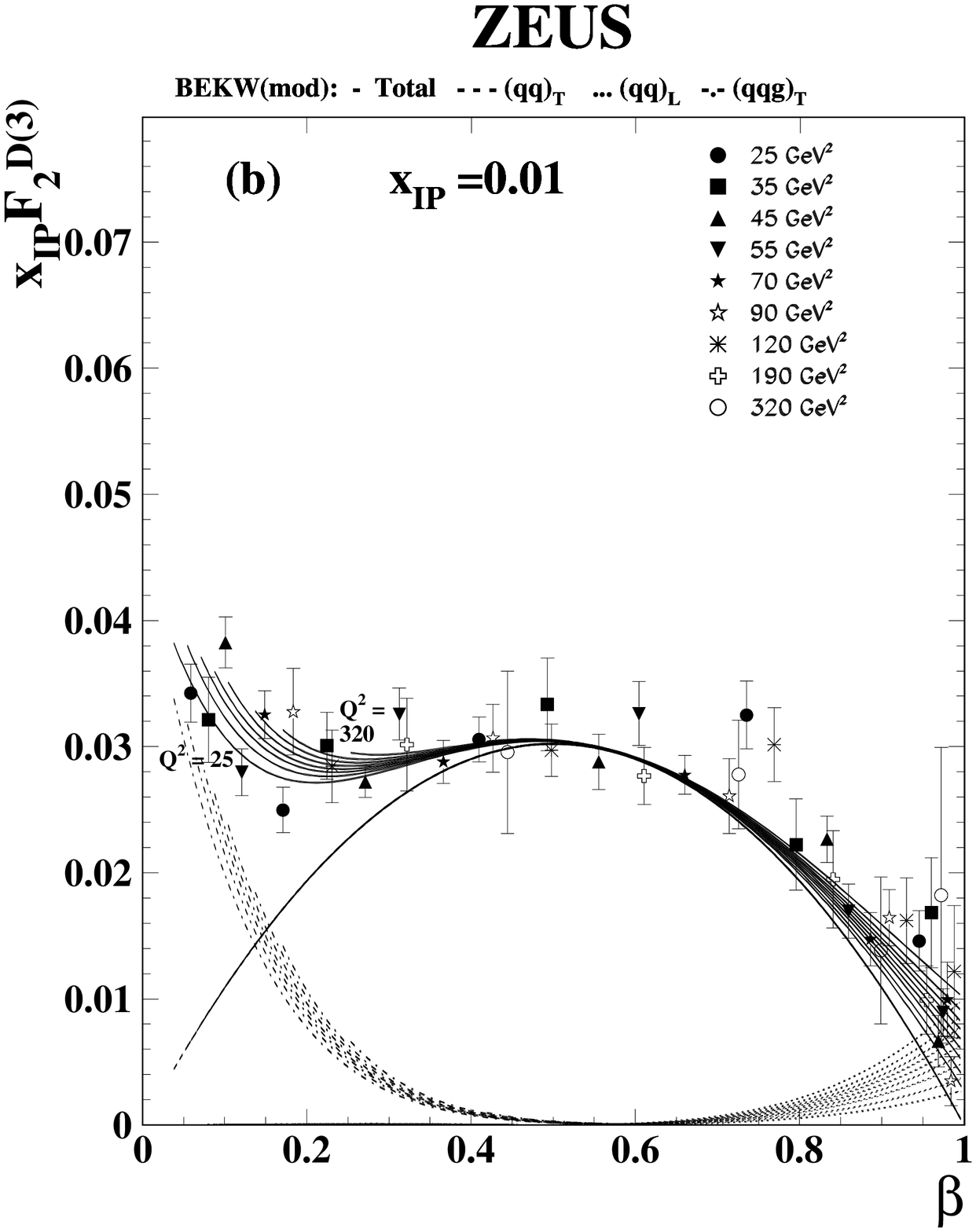}}
\includegraphics[angle=90,totalheight=9.5cm]{{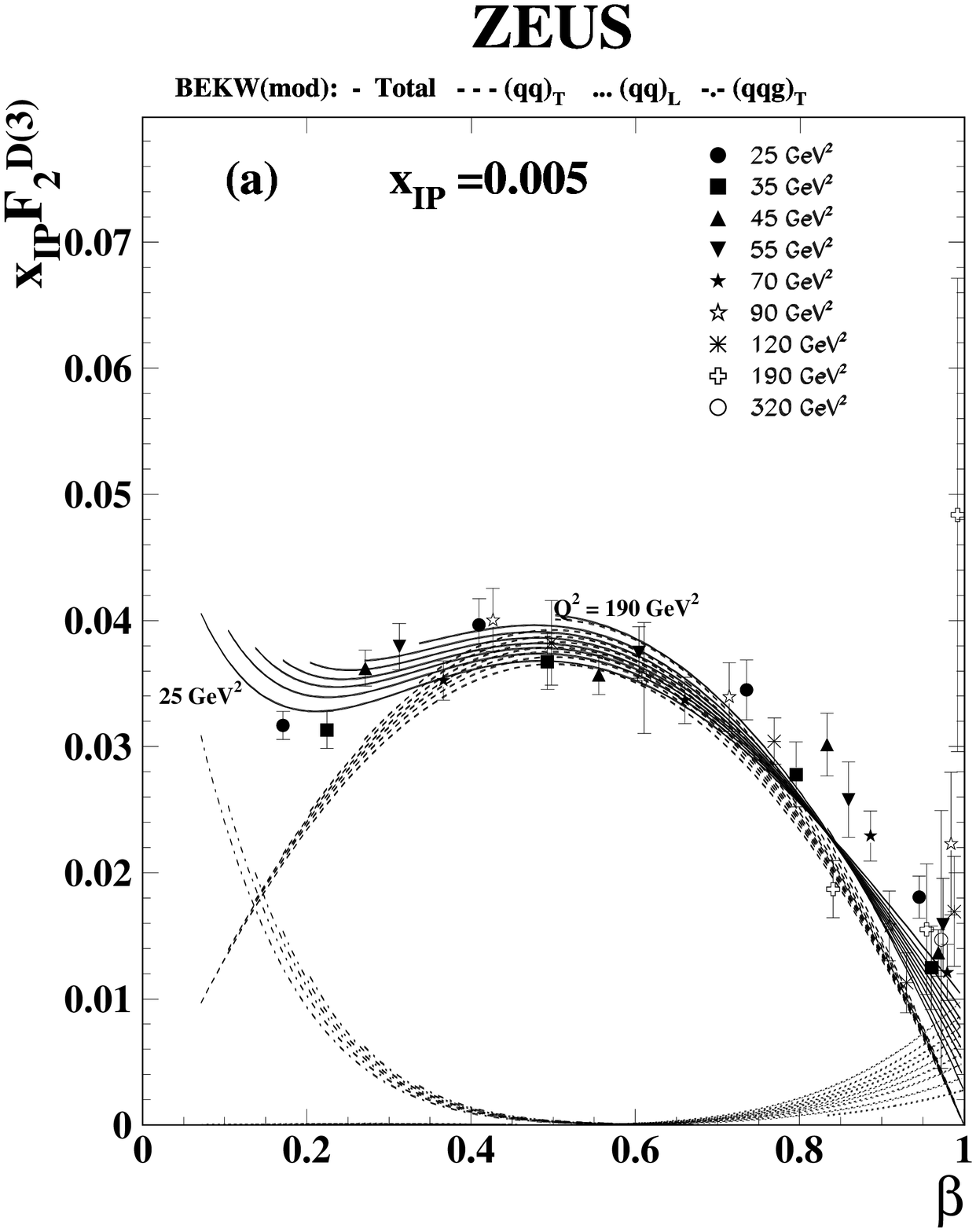}}
\caption{The diffractive structure function of the proton multiplied by
   $\xpom$, $\xpom F^{D(3)}_2$, as a function of $\beta$ for the $Q^2$ values indicated, at fixed (a) $\xpom = 0.005$ and (b) $\xpom = 0.01$. The curves show the results of the BEKW(mod) fit for the contributions from $(q \overline{q})$ for transverse (dashed) and longitudinal photons (dotted) and for the $(q \overline{q}g)$ contribution for transverse photons (dashed-dotted) together with the sum of all contributions (solid), for the $\beta$-region studied for diffractive scattering; from ZEUS}
\label{f:f2d3vsbetah34}
\end{center}
\end{figure}

\begin{figure}
\begin{center}
\includegraphics[angle=0,totalheight=12cm]{{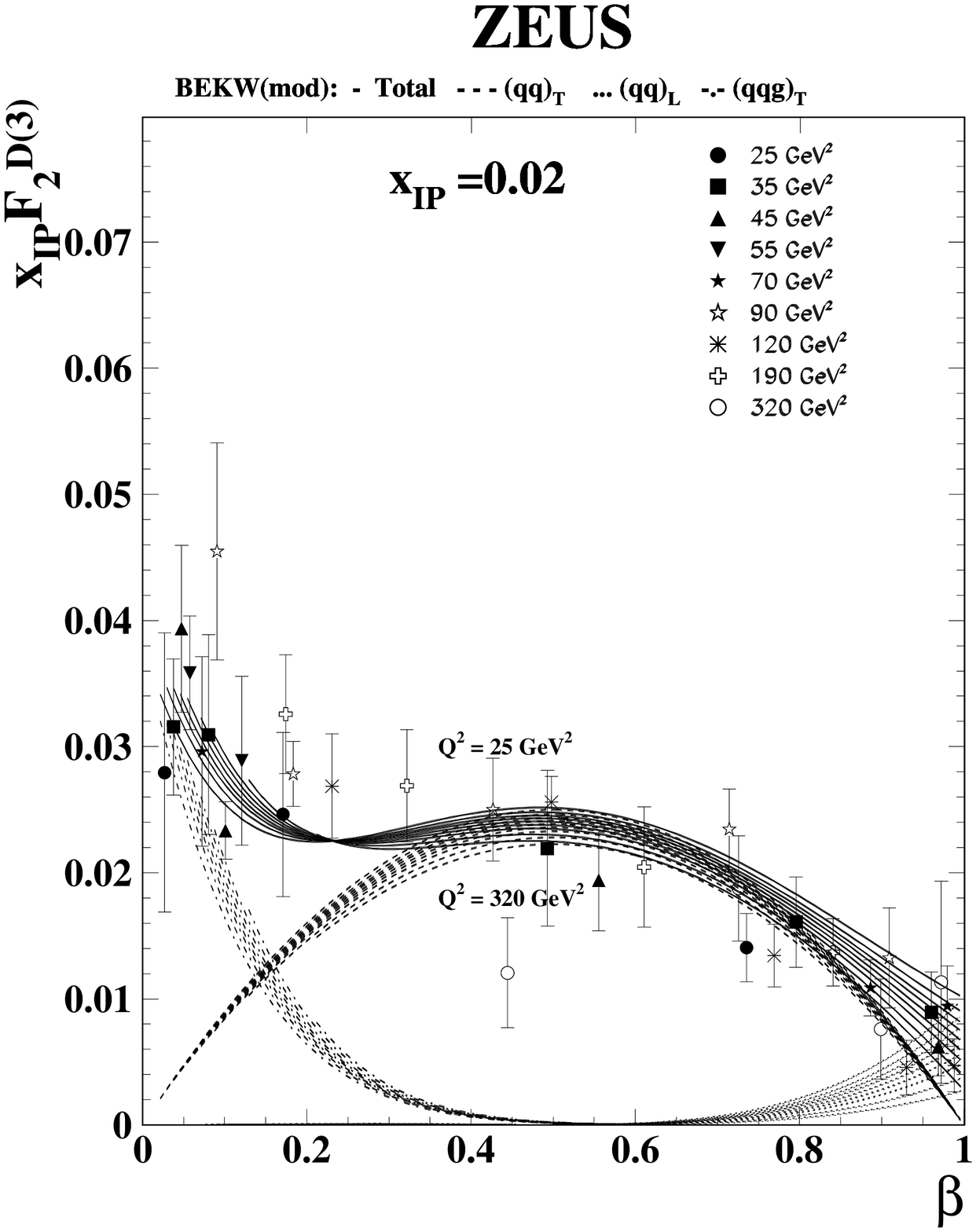}}
\caption{The diffractive structure function of the proton multiplied by
   $\xpom$, $\xpom F^{D(3)}_2$, as a function of $\beta$ for the $Q^2$ values indicated at fixed $\xpom = 0.02$. The curves show the results of the BEKW(mod) fit for the contributions from $(q \overline{q})$ for transverse (dashed) and longitudinal photons (dotted) and for the $(q \overline{q}g)$ contribution for transverse photons (dashed-dotted) together with the sum of all contributions (solid), for the $\beta$-region studied for diffractive scattering; from ZEUS.}
\label{f:f2d3vsbetah5}
\end{center}
\end{figure}

The $\beta$ dependence of $\xpom F^{D(3)}_2$  is shown in Figs.~\ref{f:f2d3vsbetah12},~\ref{f:f2d3vsbetah34},~\ref{f:f2d3vsbetah5} for fixed values of $\xpom$ and $Q^2$. In general, the $\xpom F^{D(3)}_2$ values measured at different $Q^2$  values cluster around a narrow band of the form $\beta \cdot (1 - \beta)$. The curves show the contributions from $(q \qbar)_T$, $(q \qbar g)_T$ and $(q \qbar)_L$, as determined by the BEKW(mod) fit. The dominant contribution is of the form $\beta(1-\beta)$ as predicted for $(q \qbar)_T$; the rise towards small $\beta \le 0.2$ is the result of gluon emission, $(q \qbar g)_T$; contributions from longitudinal photons, $(q \qbar)_L$, are important near $\beta = 1$ only. 

\section{Summary and conclusions}
HERA has opened the door to the study of diffraction in real and virtual photon proton scattering within a large range in total c.m. energy $W$ and photon virtuality $-Q^2$.  

A substantial fraction of diffractive scattering leads to the production of vectormesons, $\gamma^* p \to Vp$, where $V = \rho^0, \omega, \phi, J/\Psi, \Upsilon$. For fixed $Q^2$, the cross sections rise with c.m. energy proportional to $W^{\delta}$. In photoproduction, $\delta \approx 0.2$ for the light vectormesons, and $\delta = 0.8$ ($\approx 1.6$) for $J/\Psi$ ($\Upsilon)$. In virtual-photon proton scattering, the integrated cross sections at fixed $W = 75$ GeV lie on a universal curve, decreasing exponentially with $(Q^2 +M_V^2)$ provided $(Q^2 +M_V^2) > 3$ GeV$^2$. The differential cross sections $d\sigma/d|t|$ decrease with $|t|$ proportional to $e^{-b|t|}$, where $b \approx 10$ GeV$^{-2}$ for $(Q^2 + M^2_V) < 1$ GeV$^2$, and dropping rapidly to a constant value of $b \approx 5$ GeV$^{-2}$ when $(Q^2+M_V^2) \ge 5$ GeV$^2$.

The $Q^2$ dependence of $\sigma(\gamma^* p \to \rho^0 p)$ is reproduced well by a QCD based model where the photon fluctuates into a $q \qbar$ pair which interacts with the gluon cloud of the proton to produce the $\rho^0$.

The decay angular distributions of photoproduced $\rho^0$, $\omega$ and $\phi$ show that the vectormesons are transversely aligned and the helicity is conserved in the s-channel (SCHC). In virtual photon proton scattering, the decay angular distributions show that under the assumption of SCHC, $\rho^0, \phi$ and $J/\Psi$ are produced predominantly by longitudinal photons.

The production by real and virtual photons of events with a large rapidity gap and two or more jets gave evidence for hard diffractive scattering. The experimental results are reproduced by QCD calculations in NLO.

Inclusive diffraction, $\gamma^* p \to X N$, where $N$ is a proton or a low-mass nucleonic system, behaves similar to the production of vectormesons. For fixed $Q^2$ and sufficient coverage in $W$, the cross section differential in the mass $M_X$ of the system $X$, shows a rapid rise with $W$, $d\sigma^{diff}/dM_X \propto W^{\delta}$, where $\delta \approx 0.6 - 0.7$ for $Q^2 > 20$ GeV$^2$. It yields for the average of the Pomeron trajectory a value of $\bar{ \alpha_{\pom}} = 1.23 \pm 0.02(stat) \pm 0.04 (syst)$. The dependence on $t$ is of the form $d\sigma/d|t| \propto e^{-b|t|}$ with $b = 7.2 \pm 1.4$ GeV$^{-2}$.
 
The ratio of the diffractive cross section, integrated over $M_X < 35$ GeV, to the total cross section is approximately constant with $W$, being $15.8^{+1.1}_{-1.0}\%$ at $Q^2 = 4$ GeV$^2$ and decreasing to $5.0 \pm 0.9 \%$ at $Q^2 = 320$ GeV$^2$. 

The diffractive structure function of the proton, $\xpom F_2^{D(3)}(\beta,Q^2,\xpom)$, where $\beta = Q^2/(Q^2+M_X^2)$ and $\xpom = (Q^2+M_X^2)/(Q^2 + W^2)$, has been measured for $Q^2$ values between 2.7 and 1600 GeV$^2$. For fixed values of $\beta$ and $Q^2$, $\xpom F_2^{D(3)}$ is rising as $\xpom \to 0$. 

A QCD inspired model (BEKW), in which the incoming photon fluctuates into a $q \qbar$ or $q \qbar g$ dipole that interacts with the target proton by the exchange of a two-gluon system (the "Pomeron"), gives an excellent description of the data for inclusive diffraction, after adjusting the free parameters. The contribution from longitudinal photons via $q \qbar$ exchange is important for $\beta \ge 0.96$ while transverse photons dominate the regions $0.14 < \beta < 0.9$ by $q \qbar$ exchange, and $\beta < 0.1$ by $q \qbar g$ exchange.

\vspace*{1cm}
{\bf \Large Acknowledgements}\\
\\
The author is most grateful to Erich Lohrmann and Paul S\"oding for a critical review of the manuscript.

\include{padifrop-app}
\providecommand{\etal}{et al.\xspace}
\providecommand{\coll}{Coll.\xspace}
\catcode`\@=11\def\@bibitem#1{
\ifmc@bstsupport
  \mc@iftail{#1}%
    {;\newline\ignorespaces}%
    {\ifmc@first\else.\fi\orig@bibitem{#1}}
  \mc@firstfalse
\else
  \mc@iftail{#1}%
    {\ignorespaces}%
    {\orig@bibitem{#1}}
\fi}%
\catcode`\@=12

\include{padifrop-tab}

\clearpage

\clearpage

\newpage

\newpage

\clearpage

\clearpage

\clearpage 

\clearpage

 \clearpage

\clearpage

\clearpage

\clearpage

\clearpage 

\clearpage

\clearpage

\clearpage

\clearpage

\clearpage

\begin{figure}[p]
\begin{center}
\vfill
\vspace*{-4cm}
\hspace*{0.1cm}\epsfig{file=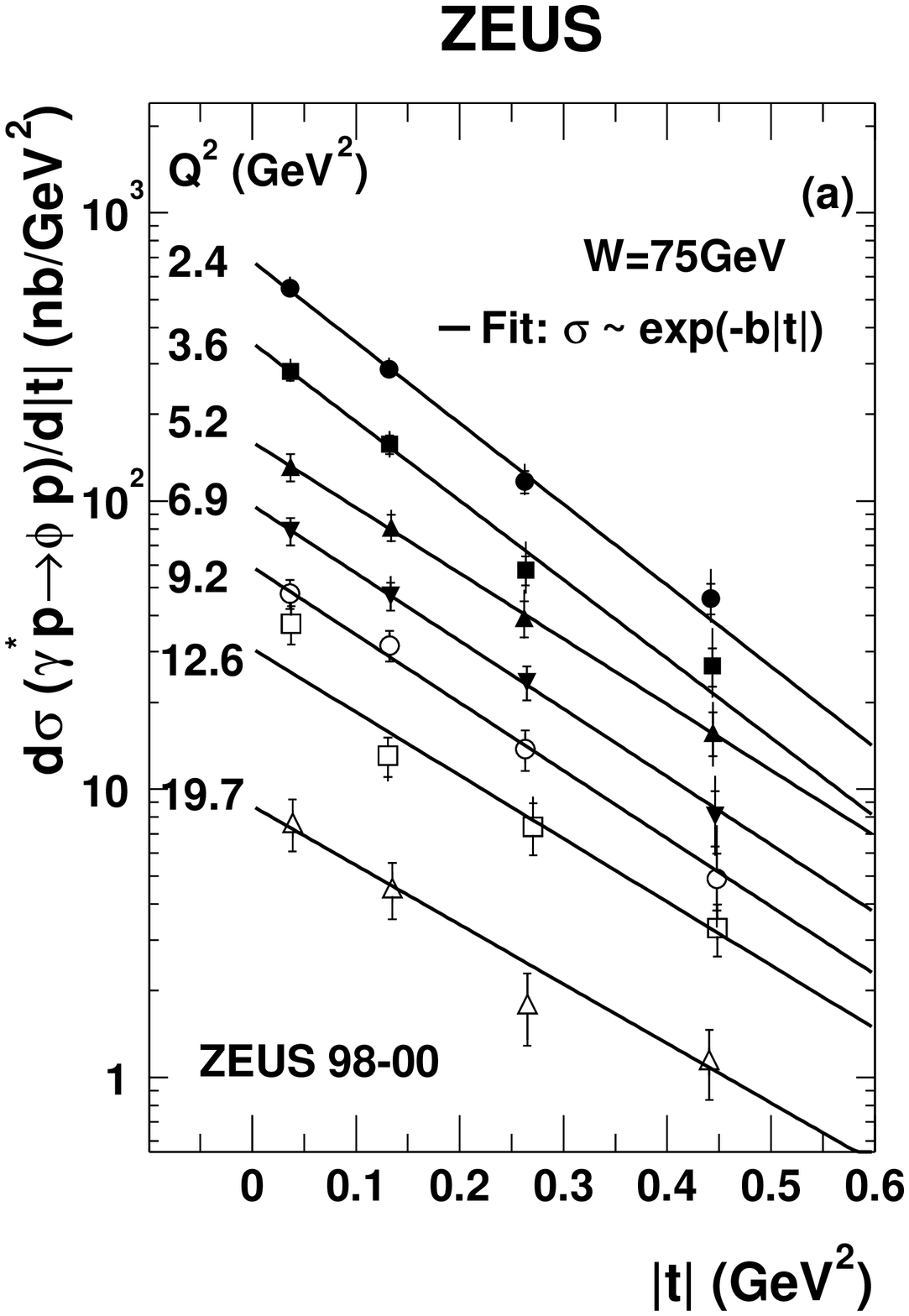,width=8cm}
\end{center} 
\end{figure}

\clearpage

\clearpage

\clearpage

\clearpage

\clearpage

\clearpage

\clearpage

\clearpage

\clearpage

\clearpage

\clearpage

\clearpage

\clearpage
 
\clearpage

\clearpage

\clearpage

\clearpage

\clearpage

\clearpage

\clearpage

\clearpage

\clearpage

\clearpage

\vfill
\clearpage

\vfill
\clearpage

\clearpage

\vfill

\clearpage

\vfill
\clearpage

\vfill
\clearpage

\vfill
\clearpage

\vfill
\clearpage

\vfill
\clearpage

\vfill
\clearpage

\vfill
\clearpage

\vfill

\vfill
\clearpage

\vfill
\clearpage

\vfill
\clearpage

\vfill
\clearpage

\vfill
\clearpage

\vfill
\clearpage

\vfill
\clearpage

\vfill
\clearpage

\vfill
\clearpage

\clearpage

\clearpage

\clearpage

\clearpage 

\clearpage

%
%
\end{document}